\documentclass[prd,twocolumn,superscriptaddress,amsfonts,amssymb,amsmath,showpacs,nofootinbib]{revtex4-2}
\usepackage{graphicx} 
\usepackage{setspace}
\usepackage{amsmath,amssymb}
\usepackage{graphicx}
\usepackage{dcolumn}
\usepackage{bm,color}
\usepackage[colorlinks=true, linkcolor=blue , citecolor=blue, urlcolor=blue]{hyperref}
\usepackage{accents}
\usepackage{amssymb,float}
\usepackage{amsmath}
\usepackage{booktabs}
\usepackage{multirow}
\usepackage{makecell}
\usepackage{tikz}
\usepackage{comment}
\usepackage[rightcaption]{sidecap}
\sidecaptionvpos{figure}{t}
\usepackage{enumerate}
\usepackage{booktabs}
\renewcommand{\arraystretch}{2}
\newcolumntype{P}[1]{>{\centering\arraybackslash}p{#1}}
\newcolumntype{M}[1]{>{\centering\arraybackslash}m{#1}}

\newcommand{\dd}{\mathrm{d}}

\begin{document}

\title{Hints of sign-changing scalar field energy density and a transient acceleration phase at $z\sim 2$ from model-agnostic reconstructions}

\author{\"{O}zg\"{u}r Akarsu}
\email{akarsuo@itu.edu.tr}
\affiliation{Department of Physics, Istanbul Technical University, Maslak 34469 Istanbul, T\"{u}rkiye}

\author{Maria Caruana}
\email{caruanamaria@itu.edu.tr}
\affiliation{Department of Physics, Istanbul Technical University, Maslak 34469 Istanbul, T\"{u}rkiye}

\author{Konstantinos F. Dialektopoulos}
\email{kdialekt@gmail.com}
\affiliation{Institute of Space Sciences and Astronomy, University of Malta, Malta, MSD 2080}

\author{Luis Escamilla}
\email{torresl@itu.edu.tr}
\affiliation{Department of Physics, Istanbul Technical University, Maslak 34469 Istanbul, T\"{u}rkiye}

\author{Emre O. Kahya}
\email{eokahya@itu.edu.tr}
\affiliation{Department of Physics, Istanbul Technical University, Maslak 34469 Istanbul, T\"{u}rkiye}

\author{Jackson Levi Said}
\email{jackson.said@um.edu.mt}
\affiliation{Institute of Space Sciences and Astronomy, University of Malta, Malta, MSD 2080}
\affiliation{Department of Physics, University of Malta, Malta}


\begin{abstract}
We present a data-driven reconstruction of the late-time expansion history and its implications for effective dark-energy dynamics. Modeling the reduced Hubble rate $E(z)\equiv H(z)/H_0$ with a node-based Gaussian-process-kernel interpolant, we constrain the reconstruction using cosmic chronometers, Pantheon+ Type~Ia supernovae, BAO measurements from SDSS and DESI, transversal BAO data, and external $H_0$ priors (SH0ES and H0DN). Assuming general relativity at the background level, we map the reconstructed kinematics onto an effective dark-energy fluid and an effective scalar-field description, yielding the total potential and kinetic contributions that reproduce the inferred $H(z)$. To interpret the reconstruction, we consider both a minimal single-field model (canonical or phantom) and a two-field (quintom) system consisting of one canonical and one phantom scalar field (or families). Within the GR-based effective-fluid mapping, the inferred dark-energy density changes sign for all dataset combinations explored, transitioning from $\rho_{\rm DE}<0$ at higher redshift to $\rho_{\rm DE}>0$ toward the present, and defining a transition redshift $z_\dagger$ by $\rho_{\rm DE}(z_\dagger)=0$. A single canonical scalar cannot realize such a smooth evolution during expansion, whereas a phantom field (as an effective description) or a two-field quintom framework can accommodate the required behavior; in particular, the two-field system permits smooth phantom-divide crossings at finite $\rho_{\rm DE}>0$ and distinguishes them from the separate notion of a density zero crossing. The reconstructed kinematics also admit intermediate-redshift structure in some combinations, including hints of an additional accelerated-expansion interval with $q(z)<0$ around $z\sim 1.7$--$2.3$. Bayesian evidence comparisons nevertheless favor the minimal flat $\Lambda$CDM baseline once model complexity is accounted for. Finally, the present-day equation of state remains close to a cosmological constant: combinations including supernovae give $w_0\simeq -1$, while combinations without supernovae but with an external $H_0$ prior show only a mild preference for $w_0<-1$ at the $\sim1.5$--$1.7\sigma$ level.
\end{abstract}

\maketitle

\section{Introduction}
\label{sec:intro}

Multiple independent observations establish that the expansion of the Universe is accelerating at late times.
Within general relativity (GR), the minimal concordance description is the spatially flat $\Lambda$ cold dark matter ($\Lambda$CDM) model~\cite{Peebles:2002gy,Copeland:2006wr}, which fits primary CMB temperature and polarization anisotropies and a broad set of late-time distance and growth probes---including baryon acoustic oscillations (BAO), weak lensing, and Type~Ia supernova (SN~Ia) Hubble diagrams---at high precision
\cite{Planck:2018vyg,Planck:2018nkj,AtacamaCosmologyTelescope:2025blo,SPT-3G:2025bzu,eBOSS:2020yzd,DESI:2025zgx,Scolnic:2021amr,Brout:2022vxf,Rubin:2023jdq,DES:2025sig}.
Yet the physical origin of the component driving acceleration remains unknown, and the cosmological-constant problem~\cite{Weinberg:1988cp,Sahni:2002kh} motivates continued tests of whether late-time data require a strictly constant $\Lambda$ or allow dynamics in an effective dark sector.
At the same time, the precision era has turned $\Lambda$CDM into a stringent target for internal-consistency tests across independent probes.

A major driver of current activity is the appearance of parameter discrepancies between datasets, which may arise from residual systematics, underestimated covariances, modeling assumptions, or new physics beyond $\Lambda$CDM~\cite{Perivolaropoulos:2021jda,Abdalla:2022yfr,DiValentino:2022fjm,Akarsu:2024qiq,CosmoVerseNetwork:2025alb}.
The best-known example is the \emph{Hubble tension}~\cite{Verde:2019ivm,DiValentino:2020zio,DiValentino:2021izs}:
distance-ladder measurements favor $H_0\simeq 73~{\rm km\,s^{-1}\,Mpc^{-1}}$~\cite{Riess:2021jrx,Breuval:2024lsv}, while early-Universe inferences within $\Lambda$CDM from CMB data prefer $H_0\simeq 67~{\rm km\,s^{-1}\,Mpc^{-1}}$~\cite{Planck:2018vyg,SPT-3G:2025bzu}.
As a representative comparison, the SPT-3G $\Lambda$CDM inference differs from the H0DN ``Local Distance Network'' value $H_0=73.50\pm0.81~{\rm km\,s^{-1}\,Mpc^{-1}}$ at $\simeq 7.1\sigma$~\cite{SPT-3G:2025bzu,H0DN:2025lyy}.
A related (and currently more survey- and modeling-dependent) issue concerns late-time clustering: Planck--$\Lambda$CDM predicts $S_8$ values higher than those preferred by several weak-lensing and large-scale-structure analyses~\cite{Planck:2018vyg,KiDS:2020suj,DES:2021vln,Wright:2025xka,DiValentino:2020vvd}.
Importantly, the quoted significance of any ``tension'' is not immutable: it can depend on how the background expansion is modeled or reconstructed and on how early-time calibrations---most notably the sound horizon---are propagated into late-time distances.
This motivates diagnostic approaches that reconstruct the expansion history as directly as possible at low and intermediate redshift and that separate \emph{kinematics} (constraints on $H(z)$ and its derivatives) from the subsequent \emph{dynamical interpretation} in terms of an effective dark sector.

Proposed solutions are commonly grouped into two broad categories by \emph{when} they act on the expansion history: \emph{early-time modifications}, which alter the expansion or energy content before recombination (e.g.\ early dark energy, EDE~\cite{Poulin:2018cxd,Karwal:2016vyq,Hill:2020osr,Kamionkowski:2022pkx,Ivanov:2020ril,Sakstein:2019fmf,Niedermann:2019olb,Niedermann:2020dwg,Poulin:2023lkg,Smith:2025grk,Poulin:2025nfb,SPT-3G:2025vyw}),
and \emph{late-time modifications}, which deform the post-recombination expansion while preserving high-redshift successes of the standard cosmology (e.g.\ interacting dark energy, IDE~\cite{Caprini:2016qxs,Nunes:2016dlj,Kumar:2017dnp,DiValentino:2017iww,Yang:2017ccc,Costa:2018aoy,vonMarttens:2018iav,Yang:2018euj,Yang:2018uae,Pan:2019gop,Kumar:2019wfs,DiValentino:2019jae,DiValentino:2019ffd,DiValentino:2020kpf,Gomez-Valent:2020mqn,Lucca:2020zjb,Pan:2020zza,Gao:2021xnk,Kumar:2021eev,Yang:2021hxg,Nunes:2022bhn,Bernui:2023byc,Escamilla:2023shf,Giare:2024smz,Li:2024qso,Sabogal:2025mkp,Silva:2025hxw,Yang:2025uyv,vanderWesthuizen:2025rip}).
A complementary classification groups departures by \emph{which sector} is modified, spanning non-standard dark-matter properties~\cite{Feng:2010gw,Dodelson:1993je,Joyce:2014kja,Abazajian:2012ys}, additional late-time dark-energy dynamics~\cite{Copeland:2006wr,Benisty:2021gde,Benisty:2020otr,Bamba:2012cp}, and modifications of gravity on cosmological scales~\cite{Clifton:2011jh,CANTATA:2021ktz,Bahamonde:2021gfp,AlvesBatista:2021gzc,Addazi:2021xuf,Capozziello:2011et}.
In parallel, direct detection of particle dark matter remains elusive~\cite{Baudis:2016qwx,Bertone:2004pz}, keeping open the possibility that some of the emerging phenomenology reflects new physics in the dark sector and/or gravity.

In this context, recent results from DESI~\cite{DESI:2024mwx,DESI:2025zgx} have sharpened the observational picture, further motivating systematic explorations of late-time departures from $\Lambda$CDM.
In particular, a growing body of work shows that extensions such as \textit{dynamical dark energy} (DDE) can substantially improve the joint consistency of BAO and supernova data relative to $\Lambda$CDM~\cite{Giare:2024gpk,Gialamas:2024lyw,RoyChoudhury:2024wri,Dinda:2024kjf,Giare:2024oil,RoyChoudhury:2025dhe,RoyChoudhury:2025iis,Scherer:2025esj,Pang:2025lvh,Roy:2024kni,Ormondroyd:2025iaf,Li:2025cxn,Cortes:2024lgw,Najafi:2024qzm,Wang:2024dka,Giare:2025pzu,Kessler:2025kju,Teixeira:2025czm,Specogna:2025guo,Sabogal:2025jbo,Cheng:2025lod,Herold:2025hkb,Cheng:2025hug,Ozulker:2025ehg,Lee:2025pzo,Silva:2025twg,Fazzari:2025lzd}.
It has also long been recognized that simple phantom-like effective descriptions can mitigate the $H_0$ tension, while models featuring phantom-divide-line (PDL) crossing---first suggested phenomenologically in a form now often referred to as DMS20~\cite{DiValentino:2020naf}---can yield even larger shifts in inferred $H_0$.
Recent analyses of the DMS20 model~\cite{DiValentino:2020naf,Adil:2023exv,Specogna:2025guo}, interpreted as an embodiment of omnipotent DE~\cite{Adil:2023exv}, have emphasized that its ability to attain negative effective densities for $z\gtrsim 2$ (resembling a negative cosmological constant at higher redshifts), together with a PDL crossing around $z\sim 0.1$, plays a central role in its phenomenology.
Along related lines, model-independent reconstructions of IDE kernels do not rule out negative effective DE densities at $z\gtrsim 2$~\cite{Escamilla:2023shf}.

A particularly economical realization of this broader phenomenology is the $\Lambda_{\rm s}$CDM framework (also known as the \emph{sign-switching cosmological constant})~\cite{Akarsu:2019hmw,Akarsu:2021fol,Akarsu:2022typ,Akarsu:2023mfb}, which posits a rapid (smooth or abrupt) transition of the effective vacuum energy from anti-de Sitter (AdS) to de Sitter (dS) around a characteristic redshift $z_\dagger\sim 2$, as originally conjectured phenomenologically in Ref.~\cite{Akarsu:2019hmw} and shown to be promising for jointly mitigating the $H_0$ and $S_8$ tensions (among others) in subsequent analyses~\cite{Akarsu:2021fol,Akarsu:2022typ,Akarsu:2023mfb,Escamilla:2025imi}. The abrupt limit of $\Lambda_{\rm s}$CDM, the simplest phenomenological realization of this framework, has been studied extensively under the assumption of GR; see, e.g., Refs.~\cite{Akarsu:2021fol,Akarsu:2022typ,Akarsu:2023mfb,Yadav:2024duq,Akarsu:2024eoo,Escamilla:2025imi,Paraskevas:2024ytz,Akarsu:2025ijk,Akarsu:2025gwi,Akarsu:2025nns}.
More broadly, a large literature explores related scenarios invoking negative cosmological constants or effective DE sectors admitting negative energy densities at intermediate/high redshift, as well as model-agnostic reconstructions pointing in that direction; see Refs.~\cite{Sahni:2002dx,Vazquez:2012ag,BOSS:2014hwf,Sahni:2014ooa,BOSS:2014hhw,DiValentino:2017rcr,Mortsell:2018mfj,Poulin:2018zxs,Capozziello:2018jya,Wang:2018fng,Banihashemi:2018oxo,Dutta:2018vmq,Banihashemi:2018has,Li:2019yem,Akarsu:2019ygx,Visinelli:2019qqu,Ye:2020btb,Perez:2020cwa,Akarsu:2020yqa,Ruchika:2020avj,DiValentino:2020naf,Calderon:2020hoc,Ye:2020oix,DeFelice:2020cpt,Paliathanasis:2020sfe,Bonilla:2020wbn,Acquaviva:2021jov,Bag:2021cqm,Bernardo:2021cxi,Escamilla:2021uoj,Sen:2021wld,Ozulker:2022slu,DiGennaro:2022ykp,Akarsu:2022lhx,Moshafi:2022mva,Bernardo:2022pyz,vandeVenn:2022gvl,Ong:2022wrs,Tiwari:2023jle,Malekjani:2023ple,Vazquez:2023kyx,Escamilla:2023shf,Adil:2023exv,Alexandre:2023nmh,Adil:2023ara,Paraskevas:2023itu,Gomez-Valent:2023uof,Wen:2023wes,Wen:2024orc,Medel-Esquivel:2023nov,DeFelice:2023bwq,Anchordoqui:2023woo,Menci:2024rbq,Anchordoqui:2024gfa,Akarsu:2024qsi,Gomez-Valent:2024tdb,DESI:2024aqx,Bousis:2024rnb,Wang:2024hwd,Colgain:2024ksa,Tyagi:2024cqp,Toda:2024ncp,Sabogal:2024qxs,Dwivedi:2024okk,Escamilla:2024ahl,Anchordoqui:2024dqc,Akarsu:2024nas,Gomez-Valent:2024ejh,Manoharan:2024thb,Souza:2024qwd,Pai:2024ydi,Paraskevas:2024ytz,Akarsu:2025ijk,Keeley:2025stf,Mukherjee:2025myk,Giare:2025pzu,Akarsu:2025gwi,Soriano:2025gxd,Sabogal:2025mkp,Mukherjee:2025ytj,Efstratiou:2025xou,Escamilla:2025imi,Silva:2025hxw,Specogna:2025guo,Gomez-Valent:2025mfl,Scherer:2025esj,Wang:2025dtk,Bouhmadi-Lopez:2025ggl,Tamayo:2025xci,Gonzalez-Fuentes:2025lei,Bouhmadi-Lopez:2025spo,Hogas:2025ahb,Yadav:2025vpx,Lehnert:2025izp,Tan:2025xas,Pedrotti:2025ccw,Forconi:2025gwo,Nyergesy:2025lyi,Ghafari:2025eql,Akarsu:2025nns}.

Scalar fields provide an economical effective language for late-time departures from $\Lambda$CDM, since they can realize a wide range of background histories while remaining under theoretical control.
They are central to early-Universe inflation~\cite{1982PhLB..108..389L,1982PhRvL..49.1110G} and have been explored extensively as pre-recombination solutions to the $H_0$ tension (e.g.\ early dark energy)~\cite{Poulin:2023lkg}.
At late times, canonical scalar fields define the standard quintessence paradigm, while phantom-like scalars (wrong-sign kinetic term) should be regarded as effective descriptions and treated with care when discussing fundamental stability~\cite{Deffayet:2009mn,Deffayet:2009wt}.
More broadly, Horndeski gravity provides the most general single-scalar framework with second-order field equations~\cite{Kobayashi:2019hrl}, and multifield generalizations add freedom that can become relevant if the data prefer transitions between qualitatively different dynamical regimes~\cite{Ohashi:2015fma,Iacconi:2023slv}.
In this work, scalar fields are used in precisely this spirit: as a controlled \emph{background-level} classifier of the effective behaviors suggested by reconstructed $H(z)$.

The observational situation at intermediate redshift has sharpened rapidly.
Modern BAO measurements from SDSS and DESI extend precision distance information to $z\gtrsim 2$ and constrain combinations of distances and expansion rates (in units of the sound horizon) at the percent level, providing powerful leverage on departures from a strictly constant-$\Lambda$ late-time history~\cite{eBOSS:2020yzd,DESI:2025zgx,DESI:2025fii,DESI:2025qqy}.
Combined with high-redshift SNe~Ia (Pantheon+)~\cite{Scolnic:2021amr,Brout:2022vxf} and direct expansion-rate measurements from cosmic chronometers~\cite{Zhang:2012mp,Jimenez:2003iv,Simon:2004tf,Moresco:2012by,Moresco:2016mzx,Ratsimbazafy:2017vga,Moresco:2015cya}, this enables incisive tests of the expansion history out to $z\simeq 2.3$.

A complementary route to model comparison is to reconstruct the late-time expansion history non-parametrically, letting the data determine $H(z)$ with minimal assumptions.
Gaussian-process (GP) methods and related techniques have a long history as tools to infer $H(z)$~\cite{Dialektopoulos:2023jam} and, crucially, its derivatives, enabling kinematic diagnostics such as the deceleration parameter $q(z)$ without committing to a particular $w(z)$ ansatz~\cite{Seikel:2012uu}.
Because distances constrain integrals of $H^{-1}(z)$, derivative-based diagnostics are often where subtle but physically meaningful departures first become visible; conversely, overly restrictive dark-energy parameterizations can wash out localized structure or inadvertently generate spurious features.
We therefore employ a node-based GP-kernel interpolant, which retains the smoothness advantages of GP kernels while allowing the data to constrain a small set of interpretable amplitudes that can be propagated straightforwardly into derivative-based inferences.

A particularly relevant phenomenology that has emerged in both parametric fits and non-parametric reconstructions is that, within GR-based effective-fluid mappings, the inferred dark-energy density can become small, vanish, or even take negative values at intermediate redshift while still approaching a $\Lambda$-like state at low $z$~\cite{DiValentino:2020naf,Adil:2023exv,Escamilla:2023shf,Sabogal:2024qxs,Escamilla:2024ahl,CosmoVerseNetwork:2025alb}.
In such circumstances the commonly plotted ratio $w_{\rm DE}(z)=p_{\rm DE}(z)/\rho_{\rm DE}(z)$ can exhibit large excursions or apparent singularities even when $H(z)$ is perfectly regular, simply because the ratio becomes ill-defined at $\rho_{\rm DE}=0$.
Moreover, flexible reconstructions can accommodate localized departures in $H(z)$ with compensating deviations elsewhere, potentially ``hiding'' the compensation in redshift ranges with weaker anchoring~\cite{Akarsu:2022lhx,Escamilla:2024ahl}.
These points motivate working directly with reconstructed kinematics and with well-defined fluid variables $(\rho_{\rm DE},p_{\rm DE})$, and distinguishing a phantom-divide crossing at $w_{\rm DE}=-1$ (at finite $\rho_{\rm DE}\neq 0$) from a density zero crossing $\rho_{\rm DE}=0$.

Guided by these considerations, we adopt a two-step strategy.
We first reconstruct the reduced Hubble rate $E(z)\equiv H(z)/H_0$ using a node-based GP-kernel interpolant and constrain it with combinations of cosmic chronometers~\cite{Zhang:2012mp,Jimenez:2003iv,Simon:2004tf,Moresco:2012by,Moresco:2016mzx,Ratsimbazafy:2017vga,Moresco:2015cya}, Pantheon+ SNe~Ia~\cite{Scolnic:2021amr,Brout:2022vxf}, BAO measurements from SDSS and DESI~\cite{eBOSS:2020yzd,DESI:2025zgx,DESI:2025fii,DESI:2025qqy}, transverse (angular) BAO compilations~\cite{Sanchez:2010zg,Carvalho:2015ica,Alcaniz:2016ryy,Carvalho:2017tuu,deCarvalho:2017xye,Menote:2021jaq}, and external Gaussian priors on $H_0$ from SH0ES and the Local Distance Network~\cite{Breuval:2024lsv,H0DN:2025lyy}.
From the reconstructed $H(z)$ we obtain $H'(z)$ and the deceleration parameter $q(z)=(1+z)H'(z)/H(z)-1$.
Assuming GR at the background level, we then map the reconstructed kinematics onto an effective dark-energy fluid and, subsequently, onto an effective scalar-field description, yielding the total effective kinetic and potential contributions that reproduce the inferred expansion history.

It is important to separate what is reconstructed from what is interpreted.
The most direct data product is $H(z)$ (or $E(z)$) and derivative-based kinematic diagnostics.
By contrast, $\rho_{\rm DE}(z)$, $p_{\rm DE}(z)$, and $w_{\rm DE}(z)$ additionally depend on the GR background relations, on the assumed matter-sector specification (e.g.\ $\Omega_{m0}$), and on the BAO ruler calibration through the sound horizon.
Accordingly, when we refer to a sign change of $\rho_{\rm DE}$ we mean a sign change \emph{within this effective GR-based mapping}, and we define the derived transition redshift $z_\dagger$ by $\rho_{\rm DE}(z_\dagger)=0$.
In this sense the reconstruction points toward an \emph{omnipotent} effective dark sector~\cite{Adil:2023exv}, whose inferred behavior can change character with redshift---including transitions between quintessence-like and phantom-like regimes and even a change in the sign of the effective dark-energy density.

The scalar-field step provides a minimal theoretical language for diagnosing which effective behaviors can be realized in simple models.
We therefore analyze (i) a single-field description in which the dark-energy sector is represented by either a canonical (quintessence) scalar or a phantom scalar (as an effective background-level description), and (ii) a two-field (quintom) system composed of one canonical and one phantom scalar.
The two-field system is the minimal scalar framework that can accommodate smooth phantom-divide crossings at finite $\rho_{\rm DE}>0$ (through a sign change of the net kinetic contribution) and that cleanly distinguishes such crossings from the separate notion of a density zero crossing $\rho_{\rm DE}=0$.
These effective descriptions connect naturally to sign-switching vacuum-energy scenarios, such as $\Lambda_{\rm s}$CDM-like constructions in which an effective dark-energy component transitions from negative to positive values at a characteristic epoch~\cite{Akarsu:2019hmw,Akarsu:2021fol,Akarsu:2022typ,Akarsu:2023mfb,Akarsu:2024eoo}.
With this framework, our goal is to use reconstructed kinematics as a diagnostic of which effective dark-sector behaviors are suggested by current data and which are artifacts of the mapping assumptions.

The paper is organized as follows.
In Sec.~\ref{sec:theory} we present the single- and two-field scalar frameworks and derive the relations that map reconstructed kinematics onto effective fluid variables and scalar-field diagnostics.
In Sec.~\ref{sec:reconstruction} we describe the reconstruction methodology and datasets and present the reconstructed kinematics (Fig.~\ref{fig:H_and_q}), the implied effective DE-fluid evolution (Fig.~\ref{fig:rhode_pde_wde}), and the corresponding scalar-field diagnostics (Fig.~\ref{fig:Keff_and_V}), together with parameter constraints and model-comparison statistics (Table~\ref{tab:results}).
We discuss interpretation, robustness, and connections to sign-switching DE scenarios in Sec.~\ref{sec:disc}, and we summarize our conclusions in Sec.~\ref{sec:conc}.

\section{Dark Energy Sector Driven by Scalar Fields}\label{sec:theory} 

This work aims to reconstruct dark energy within the framework of General Relativity by mapping the inferred dark-energy sector onto a minimally coupled scalar-field description and characterizing its dynamics through effective kinetic and potential contributions. To achieve this goal, we consider two complementary modeling approaches that differ in their level of complexity and dynamical freedom. In Sec.~\ref{subsec:sin_sca_theory}, the dark energy sector is modeled by a single scalar field, taken to be either canonical (quintessence) or phantom. This minimal setup provides a phenomenological description of dynamical dark energy but, by construction (with a fixed-sign kinetic term), does not allow the coexistence or a smooth transition between the two regimes. In Sec.~\ref{subsec:twofield-DE}, we extend this framework by introducing a two-scalar-field system consisting of one quintessence and one phantom field. This richer description retains the scalar-field interpretation while enabling smooth transitions between dynamical regimes and accommodating a broader range of behaviors.

\subsection{A Single Field Dark Energy Sector: Phantom vs.\ Quintessence}
\label{subsec:sin_sca_theory}

\noindent Consider the action in which the scalar field $\phi$ represents either a canonical (quintessence) field or a phantom field,
\begin{align} \label{eq:action_single}
    \mathcal{S}
    = \int d^4x \,\sqrt{-g}\,\bigg[
        \frac{1}{2\kappa^2} R
        - \frac{1}{2} \epsilon (\nabla \phi)^2 - U(\phi)  \bigg]
    + \mathcal{S}_{\rm m}\,,
\end{align}
where $\kappa^2 = 8\pi G$, $U(\phi)$ is the potential, and $\mathcal{S}_{\rm m}$ is the matter action for a perfect fluid.
Here and throughout we adopt the metric signature $(-,+,+,+)$ and use the shorthand
$(\nabla\phi)^2 \equiv g^{\mu\nu}\nabla_\mu\phi\nabla_\nu\phi$.
The constant $\epsilon$ selects a quintessence field for $\epsilon=+1$ and a phantom field for $\epsilon=-1$, i.e.\ it fixes the sign of the kinetic term.\footnote{A phantom field corresponds to a wrong-sign kinetic term and therefore to a ghost at the level of a fundamental field theory. In the present work it is employed as an effective description at the level of the homogeneous background, which is sufficient for our reconstruction-based discussion.}
Varying the action with respect to $g_{\mu\nu}$ and $\phi$ yields the metric field equations and the Klein--Gordon equation,
\begin{equation}
\begin{aligned} \label{eq:met_var_single}
    G_{\mu\nu}
    &=\kappa^2\Big[
        \epsilon\,\nabla_{\mu}\phi \nabla_{\nu}\phi
        \\ 
    &\qquad 
    - g_{\mu\nu}\Big(\tfrac12 \epsilon (\nabla \phi)^2 + U(\phi)\Big)
    + T^{\rm (m)}_{\mu\nu}
    \Big] \,,
\end{aligned}
\end{equation}
\begin{equation}
    \label{eq:sca_var_single}
     \epsilon\,\Box\phi - U_{,\phi} = 0\,,
\end{equation}
where $\Box \equiv g^{\mu\nu}\nabla_\mu\nabla_\nu$ and
$T^{\rm (m)}_{\mu\nu} \equiv -\frac{2}{\sqrt{-g}}\frac{\delta \mathcal{S}_{\rm m}}{\delta g^{\mu\nu}}$.

In a spatially flat Friedmann--Lema\^{i}tre--Robertson--Walker (FLRW) cosmology,
\begin{align}
    \dd s^2 = - \dd t^2 + a(t)^2 \delta_{ij}\,\dd x^i \dd x^j \,,
\end{align}
where the lapse has been fixed to unity (cosmic time gauge), the background equations obtained from
Eqs.~(\ref{eq:met_var_single}--\ref{eq:sca_var_single}) read
\begin{align}
     \label{eq:1stFriedmann_single}
     3 H^2 &= \kappa^2 \left(U + \frac{\epsilon}{2} \dot{\phi}^2 + \rho_{\rm m} \right) \,, \\
    \label{eq:2ndFriedmann_single}
    3 H^2 + 2 \dot{H} &= \kappa^2 \left(U - \frac{\epsilon}{2} \dot{\phi}^2 - p_{\rm m}\right) \,, \\
    \label{eq:scalarfield_single}
     \ddot{\phi} + 3 H \dot{\phi}  &= - \frac{1}{\epsilon} U_{,\phi}\,,
\end{align}
where $\rho_{\rm m}$ and $p_{\rm m}$ are the matter energy density and pressure, overdots denote derivatives with respect to $t$,
and commas denote derivatives with respect to the argument.

Let us examine the Klein--Gordon equation~\eqref{eq:scalarfield_single} in more detail.
The term $\ddot{\phi}$ denotes the field acceleration, while the damping term $3H\dot{\phi}$ provides Hubble friction due to cosmic expansion.
The potential contribution appears as the driving term $-\;U_{,\phi}/\epsilon$, which sets the direction of the \emph{acceleration} sourced by the potential.
Since $\dot{\phi}$ can initially have either sign, this term does not by itself fix the instantaneous direction of motion; rather, it biases the evolution of $\dot{\phi}$, while in an expanding Universe ($H>0$) the friction term progressively damps large initial velocities.
In the friction-dominated (overdamped/attractor) regime, $|\ddot{\phi}|\ll 3H|\dot{\phi}|$, Eq.~\eqref{eq:scalarfield_single} implies $3H\dot{\phi}\simeq -U_{,\phi}/\epsilon$, and hence $\dot{\phi}\simeq -U_{,\phi}/(3H\,\epsilon)$.
Thus, once transients have been damped, for quintessence ($\epsilon=+1$) the attractor drift is toward decreasing $U(\phi)$, so the field is naturally driven toward a minimum of the potential, whereas for a phantom field ($\epsilon=-1$) the drift is toward increasing $U(\phi)$, so the homogeneous dynamics tends toward a maximum (equivalently, the system behaves as a canonical field evolving in the inverted potential $-U$).
This extremum selection can be made explicit by linearizing about a critical point $U_{,\phi}(\phi_\star)=0$, which yields $\delta\ddot{\phi}+3H\delta\dot{\phi}=-(U_{,\phi\phi}/\epsilon)\,\delta\phi$:
for $\epsilon=+1$ stability requires $U_{,\phi\phi}(\phi_\star)>0$ (a minimum), while for $\epsilon=-1$ stability requires $U_{,\phi\phi}(\phi_\star)<0$ (a maximum), i.e.\ minima are unstable and maxima can act as attractors (at the level of homogeneous dynamics with Hubble friction).
More generally, the field may also approach an asymptotically flat plateau where $U_{,\phi}\to 0$, in which case the field gradually freezes and its dynamics approaches an approximately cosmological-constant-like behavior at late times.

The effective stress--energy of the scalar sector can be written in perfect-fluid form, yielding the dark-energy density and pressure
\begin{align}
     \label{eq:rhoDE_single}
     \rho_{\text{DE}} &= U + \frac{\epsilon}{2}\dot{\phi}^2 \,, \\
    \label{eq:pDE_single}
    p_{\text{DE}} &= -U + \frac{\epsilon}{2}\dot{\phi}^2 \,,
\end{align}
so that the corresponding equation-of-state parameter is
\begin{align}
\label{eq:EoS_eff_single}
    w_{\text{DE}} \equiv \frac{p_{\text{DE}}}{\rho_{\text{DE}}}
    = \frac{-U + \frac{\epsilon}{2}\dot{\phi}^2}{U + \frac{\epsilon}{2}\dot{\phi}^2}\,.
\end{align}
A useful identity is $\rho_{\text{DE}}+p_{\text{DE}}=\epsilon\dot{\phi}^2$, and hence
$w_{\text{DE}}+1=(\rho_{\text{DE}}+p_{\text{DE}})/\rho_{\text{DE}}=\epsilon\dot{\phi}^2/\rho_{\text{DE}}$.
Therefore, for a fixed kinetic sign $\epsilon$, the sign of $\rho_{\text{DE}}$ determines on which side of the dark NEC boundary $w_{\text{DE}}=-1$ the scalar sector lies.
In this minimal single-field setup, an apparent ``crossing'' of $w_{\text{DE}}=-1$ can only occur through a sign change of $\rho_{\text{DE}}$, with $w_{\text{DE}}$ becoming ill-defined at $\rho_{\text{DE}}=0$.
The corresponding branches are summarized in Table~\ref{tab:branches_single_scalar}.

\begin{table}[t]
    \centering
    \setlength{\tabcolsep}{10pt}
    \renewcommand{\arraystretch}{1.15}
    \begin{tabular}{cccc}
        \toprule
        $\epsilon$ & $\rho_{\text{DE}}$ & $w_{\text{DE}}$ & Classification \\
        \midrule
        $+1$ & $>0$ & $>-1$ & $p$-quintessence \\
        $+1$ & $<0$ & $<-1$ & $n$-quintessence \\
        \addlinespace
        $-1$ & $>0$ & $<-1$ & $p$-phantom \\
        $-1$ & $<0$ & $>-1$ & $n$-phantom \\
        \bottomrule
    \end{tabular}
    \caption{Classification of branches according to the kinetic sign $\epsilon$ and the sign of the effective energy density $\rho_{\text{DE}}$. The prefixes $p$ and $n$ denote $\rho_{\text{DE}}>0$ and $\rho_{\text{DE}}<0$, respectively.}
    \label{tab:branches_single_scalar}
\end{table} 

A dynamical behavior of $w_{\text{DE}}$, e.g.\ $w_{\text{DE}}>-1$ at present and $w_{\text{DE}}<-1$ at earlier times, can thus be accommodated in the minimal single-field setup~\eqref{eq:action_single} (real scalar and fixed-sign kinetic term) only if the effective energy density changes sign: for a canonical scalar ($\epsilon=+1$), $w_{\text{DE}}<-1$ necessarily corresponds to $\rho_{\text{DE}}<0$, whereas $w_{\text{DE}}>-1$ corresponds to $\rho_{\text{DE}}>0$; the converse holds for $\epsilon=-1$. This is crucial because the initial choice of $\epsilon$ in the action~\eqref{eq:action_single} fixes the kinetic sign and forbids a \emph{smooth} transition between canonical and phantom dynamics in the usual sense (i.e.\ a continuous crossing of $w_{\text{DE}}=-1$ at finite $\rho_{\text{DE}}>0$). For this reason, within the single-field framework we will only consider transitions between $n$-quintessence (phantom) and $p$-quintessence (phantom), where $n$ corresponds to $\rho_{\text{DE}}<0$ and $p$ corresponds to $\rho_{\text{DE}}>0$.\footnote{The $p$-- and $n$-- convention is based on the sign of the effective energy density $\rho_{\text{DE}}$, while the quintessence and phantom fields are differentiated by the coefficient of the kinetic term $\epsilon$, as done in Ref.~\cite{Akarsu:2025dmj}. This convention may differ in works without the scalar field, where quintessence is defined by $w_{\text{DE}}>-1$ and phantom by $w_{\text{DE}}<-1$, such as in Ref.~\cite{Adil:2023exv}.} The viability of these transitions is explored later in this section, including the scenario where $\rho_{\text{DE}}$ passes through a null value.

We will henceforth refer to the line $w_{\rm DE} = -1$ as the \emph{dark NEC boundary}, defined by the condition $\rho_{\rm DE} + p_{\rm DE} = 0$, rather than the phantom divide line (PDL). As shown in Table~\ref{tab:branches_single_scalar}, the classification into phantom or quintessence is determined by the sign of the kinetic term. However, since the sign of $\rho_{\rm DE}$ can vary, the quantity $w_{\rm DE} + 1$ may be either positive or negative regardless of this classification. Consequently, the region $w_{\rm DE} > -1$ does not necessarily correspond to quintessence, nor does $w_{\rm DE} < -1$ always indicate phantom behavior. Moreover, it is theoretically possible for a source to evolve across the regions $w_{\rm DE} > -1$ and $w_{\rm DE} < -1$ while remaining entirely phantom. This demonstrates that the PDL is an appropriate definition only when $\rho_{\rm DE} > 0$. In contrast, the dark NEC boundary provides a more general and consistent characterization that encompasses all realized branches.

Finally, for a minimally coupled scalar of the form \eqref{eq:action_single}, the scalar sound speed is luminal.
Defining the standard kinetic variable $X\equiv -\frac12 g^{\mu\nu}\partial_\mu\phi\,\partial_\nu\phi$ (so $X=\dot\phi^2/2$ for a homogeneous field),
the Lagrangian takes the form $p(\phi,X)=\epsilon X-U(\phi)$, implying
\begin{equation}
\label{eq:cs2_single}
    c_{\text{s}}^2 \equiv \frac{p_{,X}}{\rho_{,X}} = \frac{\epsilon}{\epsilon} = 1\,.
\end{equation}
The potential $U$ does not enter $c_{\text{s}}^2$ for a minimally coupled canonical/phantom scalar; it controls the background and the effective mass of perturbations.

The conservation of the total energy--momentum tensor follows from the Bianchi identity and the Einstein equations,
\begin{align}
    0 = \nabla^{\mu} T_{\mu\nu}
    = \nabla^{\mu} T_{\mu\nu}^{(\text{DE})} + \nabla^{\mu} T_{\mu\nu}^{\rm (m)}\,,
\end{align}
where $T_{\mu\nu}^{(\text{DE})}$ is the energy--momentum tensor of the scalar sector.
In the minimally coupled action~\eqref{eq:action_single} there is no direct coupling between the scalar field and the matter fields, and each sector is separately diffeomorphism invariant; hence the two sectors are separately conserved,
$\nabla^{\mu} T_{\mu\nu}^{(\text{DE})}=0$ and $\nabla^{\mu} T_{\mu\nu}^{\rm (m)}=0$.
At the FLRW background level this yields the continuity equations
\begin{align}
    \label{eq:TDE_continuity}
    \dot{\rho}_{\text{DE}}
    &= -3 H \left(\rho_{\text{DE}} + p_{\text{DE}}\right)
     = -3 H \rho_{\text{DE}}\left(1+w_{\text{DE}}\right) \,,
    \\
    \label{eq:Tmatter_continuity}
    \dot{\rho}_{\rm m}
    &= -3 H \left(\rho_{\rm m} + p_{\rm m}\right)\,.
\end{align}
Moreover, substituting Eqs.~\eqref{eq:rhoDE_single} and~\eqref{eq:pDE_single} into Eq.~\eqref{eq:TDE_continuity} reproduces the scalar-field equation~\eqref{eq:scalarfield_single}.
Using $\rho_{\text{DE}}+p_{\text{DE}}=\epsilon\dot{\phi}^2$ (with $\dot{\phi}^2\ge 0$ for a real field), Eq.~\eqref{eq:TDE_continuity} can be written in the particularly transparent form
\begin{align}
    \label{eq:rhoDE_monotonic}
    \dot{\rho}_{\text{DE}} = -3H\,\epsilon\,\dot{\phi}^2\,.
\end{align}
For an expanding Universe, $H>0$, this implies that $\rho_{\text{DE}}(t)$ is monotonic in the sense of being non-increasing for a canonical field ($\epsilon=+1$) and non-decreasing for a phantom field ($\epsilon=-1$), with equality only when $\dot{\phi}=0$.
We will restrict ourselves to $\rho_{\text{DE}}(z=0)>0$, in line with observations, and examine whether a transition from $\rho_{\text{DE}}<0$ at earlier times to $\rho_{\text{DE}}>0$ toward the present epoch is dynamically allowed within the single-field setup.

\paragraph*{Canonical field ($\epsilon=+1$).}
If the scalar is canonical, Eq.~\eqref{eq:rhoDE_monotonic} gives $\dot{\rho}_{\text{DE}}=-3H\dot{\phi}^2\le 0$ for $H>0$,
so $\rho_{\text{DE}}(t)$ is non-increasing throughout any expanding FLRW phase (with equality only when $\dot{\phi}=0$).
In particular, if $\rho_{\text{DE}}<0$ at some epoch during expansion, it becomes more negative toward the future and therefore cannot approach, cross, and become positive.
Equivalently, writing the continuity equation as $\dot{\rho}_{\text{DE}}=-3H\rho_{\text{DE}}(1+w_{\text{DE}})$ and using $w_{\text{DE}}+1=\dot{\phi}^2/\rho_{\text{DE}}$ shows that $\rho_{\text{DE}}(1+w_{\text{DE}})=\dot{\phi}^2>0$ even when $\rho_{\text{DE}}<0$,
so $\dot{\rho}_{\text{DE}}$ cannot flip sign while the background remains regular and expanding.
Thus, within the minimal canonical single-field framework, a smooth sign change $\rho_{\text{DE}}<0 \rightarrow \rho_{\text{DE}}>0$ during an expanding FLRW epoch is dynamically forbidden.
Formally, connecting a negative branch in the past to a positive branch at later times would require a breakdown of the regular expanding FLRW evolution---e.g.\ a turnaround to $H=0$ (followed by contraction) and/or a blow-up in $|\rho_{\text{DE}}|$ (and hence in $H^2\propto \rho_{\text{tot}}$), so that the FLRW description ceases to be valid---rather than a smooth passage through $\rho_{\text{DE}}=0$.

\paragraph*{Phantom field ($\epsilon=-1$).}
Once again, we start off with $\rho_{\text{DE}}<0$, which for a phantom scalar corresponds to $w_{\text{DE}}>-1$ (see Table~\ref{tab:branches_single_scalar}).
In an expanding Universe ($H>0$), Eq.~\eqref{eq:rhoDE_monotonic} gives $\dot{\rho}_{\text{DE}}=+3H\dot{\phi}^2>0$ (unless $\dot\phi=0$), so a negative energy density becomes less negative as it approaches zero and can in principle cross to positive values.
Specializing to $\epsilon=-1$, one has $\rho_{\text{DE}}=U-\dot{\phi}^2/2$ and $p_{\text{DE}}=-U-\dot{\phi}^2/2$.
Near the crossing it is convenient to parameterize the two sides by $\rho_{\rm DE}=s\,\delta$ with $\delta>0$ and $s=\pm1$, which is equivalent to writing $U=\dot\phi^2/2+s\,\delta$.
It then follows that $p_{\rm DE}=-\dot\phi^2-s\,\delta$ and hence, for $\delta>0$,
$w_{\rm DE}=p_{\rm DE}/\rho_{\rm DE}=-1-\dot\phi^2/(s\,\delta)$:
as $\delta\to0^+$, $w_{\rm DE}\to-\infty$ on the $\rho_{\rm DE}>0$ side ($s=+1$) and $w_{\rm DE}\to+\infty$ on the $\rho_{\rm DE}<0$ side ($s=-1$),
while at $\delta=0$ the ratio $w_{\rm DE}$ is ill-defined.
Thus, the divergence at $\rho_{\text{DE}}=0$ is a kinematic artifact of the ratio $w_{\text{DE}}=p_{\text{DE}}/\rho_{\text{DE}}$ rather than a physical singularity of the background:
$\rho_{\text{DE}}$ passes smoothly through zero while $p_{\text{DE}}$ remains finite (and correspondingly $H$ and curvature scalars need not diverge).
In particular, the crossing corresponds to the perfectly regular condition $U=\dot{\phi}^2/2$ at the crossing.

Away from the transition, whenever the scalar behaves in a potential-dominated regime with $\dot{\phi}^2 \ll 2|U|$ (equivalently $|\rho_{\text{DE}}|\gg |\rho_{\text{DE}}+p_{\text{DE}}|=\dot{\phi}^2$),
one has $w_{\text{DE}}\simeq -1$ on either side of the crossing.
It should be emphasized that the evolution through $\rho_{\text{DE}}=0$ is therefore characterized by $w_{\text{DE}}$ diverging to $-\infty$ as $\rho_{\text{DE}}\to 0^{+}$ (approaching the crossing from the $\rho_{\text{DE}}>0$ side),
and reappearing from $+\infty$ as $\rho_{\text{DE}}\to 0^{-}$ (on the $\rho_{\text{DE}}<0$ side), before relaxing back toward finite values away from the crossing.

Next, Eqs.~\eqref{eq:1stFriedmann_single} and~\eqref{eq:2ndFriedmann_single} can be combined to isolate the potential $U$ and the kinetic contribution,
\begin{align}
    \label{eq:reconstruction_potential_single}
     \kappa^2 U  &= 3 H^2 + \dot{H} - \frac{\kappa^2}{2} \left(\rho_{\rm m} - p_{\rm m}\right)  \,, \\
    \label{eq:reconstruction_kinetic_single}
     \kappa^2\Delta\mathcal{X} &\equiv \frac{1}{2}\epsilon \kappa^2 \dot{\phi}^2
     =  - \dot{H}  -  \frac{\kappa^2}{2}\left(\rho_{\rm m} +  p_{\rm m}\right) \,.
\end{align}
For a real single field one has $\dot{\phi}^2\ge 0$, so the sign of $\Delta\mathcal{X}$ is fixed by the choice of $\epsilon$: $\Delta\mathcal{X}\ge 0$ for $\epsilon=+1$ (quintessence) and $\Delta\mathcal{X}\le 0$ for $\epsilon=-1$ (phantom), with equality only when $\dot\phi=0$.
Thus, within the minimal single-field framework the effective kinetic contribution cannot change sign, and a smooth transition between quintessence-like and phantom-like dynamics is not possible.
Alternatively, in Sec.~\ref{subsec:twofield-DE} we consider a dark-energy sector composed of two scalar fields, for which the net kinetic contribution can change sign, enabling smooth transitions between quintessence-like and phantom-like regimes.

\subsection{Two-field Dark Energy Sector: Minimally Coupled Phantom and Quintessential Scalars}
\label{subsec:twofield-DE}

We model the dark-energy sector with two minimally coupled scalars: a canonical field $Q$ and a phantom field $P$.
The following results can be generalized to $q$ canonical (quintessence) fields $Q_i$ ($i=1,\dots,q$) and $p$ phantom fields $P_j$ ($j=1,\dots,p$).
At the level of the homogeneous background (which is our focus), multiple canonical/phantom fields can be packaged into effective single canonical/phantom degrees of freedom along the background trajectory (while additional entropy modes may appear at the perturbation level).
In full generality the scalar potential is an arbitrary function of both fields, $U\equiv U(Q,P)$; here we restrict to the \textit{separable} case
$U(Q,P)=U_1(Q)+U_2(P)$, for which the scalar equations decouple in the potential (with the two fields still coupled through gravity via $H$). Accordingly, action~\eqref{eq:action_single} is rewritten as
\begin{equation}
\begin{aligned}
\label{eq:action_double}
    \mathcal{S}
    = \int d^4x \,\sqrt{-g}\,\bigg[&
        \frac{1}{2\kappa^2} R
        - \frac{1}{2}(\nabla Q)^2 - U_{1}(Q)
        \\
        &+ \frac{1}{2}(\nabla P)^2 - U_{2}(P)
    \bigg]
    + \mathcal{S}_{\rm m}\,,
\end{aligned}
\end{equation}
such that the Einstein equations and the Klein--Gordon equations read
\begin{equation}
\begin{aligned}
    G_{\mu\nu} &= \kappa^2\Big[
        \nabla_{\mu}Q \nabla_{\nu}Q
        - g_{\mu\nu}\Big(\tfrac12(\nabla Q)^2 + U_1\Big)
        \\
        & \, - \nabla_{\mu}P \nabla_{\nu}P
        + g_{\mu\nu}\Big(\tfrac12(\nabla P)^2 - U_2\Big)
        + T^{\rm (m)}_{\mu\nu}
    \Big]\,,
\end{aligned}
\end{equation}
\vskip -0.8cm
\begin{equation}
    \Box Q = U_{1,Q}\,, \qquad
    \Box P = - U_{2,P}\,.
\end{equation}

For the separable potential the single-field Klein--Gordon equation~\eqref{eq:sca_var_single} is replaced by two decoupled equations, one for each scalar.
Without separability, the field equations remain two coupled equations through the mixed derivatives of $U(Q,P)$.
Note also that a single complex \emph{canonical} scalar is equivalent to two real canonical scalars; this is not the case here because one of the degrees of freedom carries a wrong-sign kinetic term.

Once again, specializing to a spatially flat FLRW background yields
\begin{equation}
\begin{aligned}
     \label{eq:1stFriedmann_double}
     3 H^2 &= \kappa^2 \Big[U_{1}(Q) + U_{2}(P) 
     \\ & \qquad
     + \tfrac{1}{2} \big(\dot{Q}^2 - \dot{P}^2\big) + \rho _{\rm m}\Big] \,,
\end{aligned}
\end{equation}
\vskip -0.8cm
\begin{equation}
\begin{aligned}
    \label{eq:2ndFriedmann_double}
    3 H^2 + 2 \dot{H} &= \kappa^2 \Big[U_{1}(Q) + U_{2}(P) 
    \\ & \qquad 
    - \tfrac{1}{2} \big(\dot{Q}^2 - \dot{P}^2\big) - p_{\rm m}\Big] \,, 
\end{aligned}
\end{equation}
\vskip -0.8cm
\begin{align}
    \label{eq:scalarfield_double_Q}
    \ddot Q + 3H\dot Q  &= - U_{1,Q}\,, \\
    \label{eq:scalarfield_double_P}
     \ddot P + 3H\dot P &= U_{2,P} \,.
\end{align}
It is then convenient to introduce the total potential and the net (effective) kinetic contribution,
\begin{equation}
\begin{aligned}
    \label{eq:reconstruction_potential_double}
     \kappa^2 U  &\equiv \kappa^2 \left(U_{1}(Q) + U_{2}(P)\right)
     \\ &= 3 H^2 + \dot{H} - \frac{\kappa^2}{2}\left(\rho_{\rm m} - p_{\rm m}\right)  \,,
\end{aligned}
\end{equation}
\vskip -0.8cm
\begin{equation}
\begin{aligned}
    \label{eq:reconstruction_kinetic_double}
     \kappa^2 \Delta\mathcal{X} &\equiv \frac{\kappa^2}{2}\left( \dot{Q}^2 - \dot{P}^2\right)
     \\ &=  - \dot{H}  -  \frac{\kappa^2}{2} \left(\rho _{\rm m}+  p_{\rm m}\right) \,.
\end{aligned}
\end{equation}
Thus, $U$ receives contributions from both component potentials, while the sign of $\Delta\mathcal{X}$ tracks which kinetic term dominates at the background level:
$\Delta\mathcal{X}>0$ corresponds to $\dot Q^2>\dot P^2$ (canonical-kinetic dominance), whereas $\Delta\mathcal{X}<0$ corresponds to $\dot P^2>\dot Q^2$ (phantom-kinetic dominance).

The dark energy sector can be expressed in terms of $Q$ and $P$ as
\begin{align}
    \label{eq:rhoDE_double}
    \rho_{\rm DE} &= \frac{1}{2}\left(\dot Q^2 - \dot P^2\right) + U_1(Q) + U_2(P) \,, \\
    \label{eq:pDE_double}
    p_{\rm DE}  &= \frac{1}{2}\left(\dot Q^2 - \dot P^2\right) - U_1(Q) - U_2(P) \,,
\end{align}
with $w_{\rm DE}=p_{\rm DE}/\rho_{\rm DE}$, so that $\rho_{\rm DE}+p_{\rm DE}=\dot Q^2-\dot P^2=2\Delta\mathcal{X}$.
Since the matter sector remains minimally coupled, the continuity equations
Eqs.~(\ref{eq:TDE_continuity}--\ref{eq:Tmatter_continuity}) still hold; in particular,
\begin{align}
    \dot \rho_{\rm DE}
    = -3H\big(\rho_{\rm DE} + p_{\rm DE}\big)
    = -6H\,\Delta\mathcal{X} \,.
\end{align}
For a separable potential $U(Q,P)=U_1(Q)+U_2(P)$, its time derivative is
\begin{align}
    \label{eq:potential_timederivative_double}
    \dot{U} = U_{1,Q} \dot{Q} + U_{2,P} \dot{P}\,.
\end{align}
Although $\dot\rho_{\rm DE}$ is controlled by the sign of $\Delta\mathcal{X}$, the evolution of $U$ is not fixed by this sign alone.
Using Eq.~\eqref{eq:potential_timederivative_double} together with the Klein--Gordon equations~(\ref{eq:scalarfield_double_Q}--\ref{eq:scalarfield_double_P}) yields the exact identity
\begin{equation}
\begin{aligned}
    \label{eq:Udot-identity}
    \dot{U}
      &= \frac{1}{2}\frac{d}{dt}\big(\dot P^{2}-\dot Q^{2}\big)
        + 3H\big(\dot P^{2}-\dot Q^{2}\big)
      \\ &= - \frac{\text{d}\,}{\text{d}t}\Delta\mathcal{X} - 6H\,\Delta\mathcal{X}
      = -\,a^{-6}\frac{d}{dt}\big(a^{6}\Delta\mathcal{X}\big)\,,
\end{aligned}
\end{equation}
which makes explicit how the rate of change of the total potential is tied to both $\Delta\mathcal{X}$ and its time variation.
In particular, even in an expanding Universe one may have $\dot\rho_{\rm DE}>0$ (i.e.\ $\Delta\mathcal{X}<0$) while $\dot U<0$, depending on the relative size of $\dot{\Delta\mathcal{X}}$.

\textit{Plateau total potential:}
First, consider a near-plateau regime in which $\dot{U}\simeq 0$.
Then Eq.~\eqref{eq:Udot-identity} implies $\frac{d}{dt}(a^{6}\Delta\mathcal{X})\simeq 0$, hence $\Delta\mathcal{X}\propto a^{-6}\to 0$ as the Universe expands.
From Eqs.~\eqref{eq:rhoDE_double} and~\eqref{eq:pDE_double}, this drives $\rho_{\rm DE}\simeq U$ and $p_{\rm DE}\simeq -U$, so the equation of state approaches $w_{\rm DE}\to -1$ irrespective of whether the evolution passes through quintessence-like ($\Delta\mathcal{X}>0$) or phantom-like ($\Delta\mathcal{X}<0$) stages.

\textit{At the dark NEC boundary:}
At $w_{\rm DE}=-1$ one has $\rho_{\rm DE}+p_{\rm DE}=0$, i.e.\ $\Delta\mathcal{X}=0$ (equivalently $\dot Q^2=\dot P^2$).
At this point Eq.~\eqref{eq:Udot-identity} reduces to $\dot{U}=-\,\frac{d}{dt}\Delta\mathcal{X}$.
Therefore, if $\dot U<0$ at the crossing then $\dot{\Delta\mathcal{X}}>0$ and the system is driven toward the $\Delta\mathcal{X}>0$ (quintessence-like) side, whereas if $\dot U>0$ then $\dot{\Delta\mathcal{X}}<0$ and the evolution is directed toward the $\Delta\mathcal{X}<0$ (phantom-like) side.

\textit{Kinetic-term sign bias:}
From Eq.~\eqref{eq:Udot-identity}, $\dot U=-\dot{\Delta\mathcal{X}}-6H\Delta\mathcal{X}$, so the sign of $\dot U$ is not fixed by the sign of $\Delta\mathcal{X}$ alone.
However, in regimes where $\Delta\mathcal{X}$ varies slowly on a Hubble time, $|\dot{\Delta\mathcal{X}}|\ll 6H|\Delta\mathcal{X}|$, one has $\dot U\simeq -6H\Delta\mathcal{X}$.
In that quasi-adiabatic limit, phantom-kinetic dominance ($\Delta\mathcal{X}<0$) typically corresponds to $\dot\rho_{\rm DE}>0$ and $\dot U>0$, while quintessence-kinetic dominance ($\Delta\mathcal{X}>0$) corresponds to $\dot\rho_{\rm DE}<0$ and $\dot U<0$.

We now analyze the neighborhood of a zero crossing of the effective dark-energy density, $\rho_{\rm DE}=0$.
In the two-field system one has $\rho_{\rm DE}=\Delta\mathcal{X}+U$ and $p_{\rm DE}=\Delta\mathcal{X}-U$, where
$U\equiv U_1(Q)+U_2(P)$ and $\Delta\mathcal{X}\equiv \frac12(\dot Q^2-\dot P^2)$.
Thus $\rho_{\rm DE}=0$ corresponds to the condition $U=-\Delta\mathcal{X}$.

To parameterize small departures from the crossing, write
\begin{align}
\label{eq:near_cross_param}
\rho_{\rm DE}= s\,\delta,\qquad
U=-\Delta\mathcal{X}+s\,\delta,
\end{align}
where $\delta\ge 0$ measures the magnitude of the departure from the crossing and (for $\delta>0$) $s=\pm 1$ labels the sign of $\rho_{\rm DE}$.
In the single-field phantom case, $\Delta\mathcal{X}$ has fixed sign and this reduces to the behavior discussed Sec.~\ref{subsec:sin_sca_theory};
in the present two-field setup the situation is richer because $\Delta\mathcal{X}$ can change sign.

We therefore also write
\begin{align}
\label{eq:DXi_def}
\Delta\mathcal{X}=\sigma\,\xi,\qquad \xi\ge 0,\qquad \sigma=\pm 1,
\end{align}
where $\sigma=\mathrm{sgn}(\Delta\mathcal{X})$ distinguishes canonical-kinetic dominance ($\sigma=+1$) from phantom-kinetic dominance ($\sigma=-1$).
Using $p_{\rm DE}=\Delta\mathcal{X}-U$ together with Eq.~\eqref{eq:near_cross_param} gives
$p_{\rm DE}=2\sigma\xi-s\,\delta$ and hence
\begin{align}
\label{eq:w_near_cross}
w_{\rm DE} \equiv \frac{p_{\rm DE}}{\rho_{\rm DE}}
= -1 + 2\,\frac{\sigma}{s}\,\frac{\xi}{\delta}\,,
\quad (\delta>0),
\end{align}
while at $\delta=0$ the ratio $w_{\rm DE}=p_{\rm DE}/\rho_{\rm DE}$ is ill-defined.
Equivalently, one may note the exact identity $w_{\rm DE}+1=(\rho_{\rm DE}+p_{\rm DE})/\rho_{\rm DE}=2\Delta\mathcal{X}/\rho_{\rm DE}$, so the side of the NEC boundary of the dark energy sector is set by the relative sign of $\Delta\mathcal{X}$ and $\rho_{\rm DE}$:
$\mathrm{sgn}(w_{\rm DE}+1)=\mathrm{sgn}(\Delta\mathcal{X}/\rho_{\rm DE})$.
In particular, a dark NEC boundary crossing with finite $\rho_{\rm DE}\neq 0$ corresponds simply to $\Delta\mathcal{X}=0$ (i.e.\ $\dot Q^2=\dot P^2$), which can occur smoothly in the two-field system even when $\rho_{\rm DE}>0$.

Equation~\eqref{eq:w_near_cross} immediately yields the four branches:
(i) $\rho_{\rm DE}>0$ and $\Delta\mathcal{X}>0$ ($s=\sigma=+1$) gives $w_{\rm DE}>-1$ (``$p$-quintessence-like'');
(ii) $\rho_{\rm DE}<0$ and $\Delta\mathcal{X}>0$ ($s=-1,\sigma=+1$) gives $w_{\rm DE}<-1$ (``$n$-quintessence-like'');
(iii) $\rho_{\rm DE}>0$ and $\Delta\mathcal{X}<0$ ($s=+1,\sigma=-1$) gives $w_{\rm DE}<-1$ (``$p$-phantom-like'');
(iv) $\rho_{\rm DE}<0$ and $\Delta\mathcal{X}<0$ ($s=\sigma=-1$) gives $w_{\rm DE}>-1$ (``$n$-phantom-like'').

Finally, the limiting behavior as $\delta\to 0^+$ is controlled by the ratio $\xi/\delta$:
if $\xi/\delta\to\infty$ (e.g.\ $\xi$ tends to a nonzero constant while $\delta\to 0^+$), then $w_{\rm DE}$ diverges;
if $\xi/\delta\to c$ (e.g.\ $\xi=c\,\delta$), then $w_{\rm DE}\to -1+2(\sigma/s)c$ is finite;
and if $\xi/\delta\to 0$ (e.g.\ $\xi\ll \delta$), then $w_{\rm DE}\to -1$ on either side.
Thus, a $\rho_{\rm DE}=0$ crossing generically produces a kinematic divergence of $w_{\rm DE}$ (to $\pm\infty$ on one side and $\mp\infty$ on the other), whereas the NEC boundary crossing for dark energy at $w_{\rm DE}=-1$ can occur smoothly at finite $\rho_{\rm DE}$ through $\Delta\mathcal{X}=0$.
This illustrates explicitly why the two-field system admits a broader set of behaviors near $\rho_{\rm DE}=0$ than the single-field case.

\section{Reconstruction methodology, datasets and results}\label{sec:reconstruction}

We first infer the late-time expansion history in a largely model-independent way by reconstructing $E(z)\equiv H(z)/H_0$ from the data, and we present the key kinematical diagnostics---$H(z)$, the reduced expansion rate $H(z)/(1+z)$, and the deceleration parameter $q(z)$---in Fig.~\ref{fig:H_and_q}.
Assuming GR at the background level, we then map the reconstructed kinematics onto an effective dark-energy fluid, yielding $\rho_{\rm DE}(z)$, $p_{\rm DE}(z)$ and $w_{\rm DE}(z)$ (Fig.~\ref{fig:rhode_pde_wde}).
Finally, adopting a scalar-field interpretation of this effective fluid, we translate the same background relations into an effective kinetic contribution and a total potential, $\Delta\mathcal{X}(z)$ and $U(z)$, whose evolution is shown in Fig.~\ref{fig:Keff_and_V}.
Importantly, the algebraic relations used to reconstruct $\Delta\mathcal{X}(z)$ and $U(z)$ from $H(z)$ have the same form in the single-field and two-field cases [cf. Eqs.~\eqref{eq:reconstruction_kinetic_single}, \eqref{eq:reconstruction_kinetic_double} and \eqref{eq:reconstruction_potential_single}, \eqref{eq:reconstruction_potential_double}], so the reconstruction can be carried out at the level of background kinematics independently of the underlying scalar-field interpretation.
Below we first describe the reconstruction methodology and datasets (Sec.~\ref{subsec:methodology}), then summarize the main reconstruction results (Sec.~\ref{subsec:results}). The broader interpretation and physical implications are discussed in Sec.~\ref{sec:disc}.

\subsection{Methodology}
\label{subsec:methodology}

Our reconstruction efforts target the reduced expansion history $E(z)\equiv H(z)/H_0$, so that $H(z)=H_0\,E(z)$.
In this way we obtain a data-driven description of the late-time expansion history, which has the potential to reveal dynamics preferred by the data.

The reconstruction method employed here makes use of Gaussian Processes (GP) \cite{Velazquez:2024aya}, but not in the usual regression sense.
A GP is the generalization of a Gaussian distribution: at every position $x$ one has a random variable $f(x)$ characterized by a mean function $\mu(x)$ and a covariance
$\sigma^2K(x,x')$, where $\sigma^2$ is the variance and $K(x,x')$ is the kernel encoding correlations between $f(x)$ and $f(x')$.
For an arbitrary set of positions $x_1,\dots,x_n$, the corresponding function values follow a multivariate Gaussian distribution,
\begin{equation}
    \Bar{f}=[f(x_1),\dots,f(x_n)] \sim \mathcal{N}\!\big(\Bar{\mu},\,\sigma^2K(\Bar{x},\Bar{x}')\big)\,,
\end{equation}
where $\Bar{\mu}=[\mu(x_1),\dots,\mu(x_n)]$, and
\begin{equation}
K(\Bar{x},\Bar{x}')=
\begin{pmatrix}
K(x_1,x_1) & K(x_1,x_2) & \cdots & K(x_1,x_n) \\
K(x_2,x_1) & K(x_2,x_2) & \cdots & K(x_2,x_n) \\
\vdots     & \vdots     & \ddots & \vdots     \\
K(x_n,x_1) & K(x_n,x_2) & \cdots & K(x_n,x_n)
\end{pmatrix}.
\end{equation}
We adopt the squared-exponential (Gaussian/RBF) kernel,
\begin{equation}
    K(x,x')=\exp\!\big[-\theta(x-x')^2\big]\,,
\end{equation}
where the hyperparameter $\theta$ controls the correlation strength (smoothness).
In our node-based implementation we keep $\theta$ fixed, treating it as a smoothness/correlation-strength scale that controls how strongly neighboring nodes are coupled; this avoids introducing additional hyperparameter degeneracies and retains an infinitely differentiable interpolant, which is particularly convenient when reconstructing derivative-based quantities such as $H'(z)$ and $q(z)$.

In this context, the GP kernel is used as an interpolation tool between a set of fixed redshift nodes, which act as the free parameters in a Bayesian parameter-estimation procedure.
The $i$th node is located at redshift $z_i$; these $z_i$ values remain fixed once chosen.
The free parameters of the reconstruction are the amplitudes $E(z_i)=E_i$ and the value of the Hubble constant today, $H(z=0)\equiv H_0$.
A key advantage of this approach, in contrast to applying GP regression directly to the observables, is that derived quantities involving derivatives of the expansion history
(e.g.\ $E'(z)$ and hence $H'(z)$) can be obtained straightforwardly by differentiating the interpolant for each posterior sample, so that uncertainties propagate naturally through the sampling.
Standard GP regression libraries such as GAPP \cite{Seikel:2012uu} provide analytic expressions for derivative covariances of a GP predictive kernel; here the corresponding derivative uncertainty is instead captured directly by the posterior distribution of the node amplitudes.

This approach of using a GP as an interpolant has been previously used in the literature, typically within the context of the $\Lambda$CDM paradigm.
For example, Ref.~\cite{Gerardi:2019obr} used it to reconstruct the equation of state of a dark energy component, while Ref.~\cite{Escamilla:2023shf} applied it to a hypothetical interaction between dark energy and dark matter.

To investigate the late-time dynamics encoded in our data-driven reconstruction, we rely on several low-redshift probes.
Since our approach is designed to extract information directly from the data, it is essential to include measurements that trace the expansion history in the range $z\lesssim 2.5$.
For this purpose, we make use of the following datasets:
\begin{itemize}
\item \textbf{Cosmic chronometers (CC):} direct, model-independent measurements of $H(z)$ \cite{Zhang:2012mp,Jimenez:2003iv,Simon:2004tf,Moresco:2012by,Moresco:2016mzx,Ratsimbazafy:2017vga,Moresco:2015cya}.
This dataset consists of 31 points spanning $0.07<z<1.96$. We refer to this dataset as \textbf{CC}.

\item \textbf{Type Ia supernovae (SN):} the Pantheon+ compilation \cite{Scolnic:2021amr,Brout:2022vxf}, providing luminosity-distance information.
It contains 1701 lightcurves corresponding to 1550 distinct SNe~Ia and spans $0.01<z<2.26$.
When using this dataset we denote it by \textbf{SN}.

\item \textbf{BAO measurements (SDSS/DESI):} BAO measurements from SDSS \cite{eBOSS:2020yzd} and DESI \cite{DESI:2025zgx,DESI:2025fii,DESI:2025qqy}, which constrain the expansion rate and distance scales.
These datasets constrain distances and expansion-rate combinations normalized by the sound horizon scale $r_d$; following the DESI analysis, $r_d$ is calibrated using Big Bang Nucleosynthesis \cite{DESI:2025zgx}.
SDSS and DESI contain 14 and 13 measurements, respectively, and both cover $0.295<z<2.34$.
When in use, the legend \textbf{SDSS} or \textbf{DESI} will be added.

\item \textbf{Transversal BAO (BAOtr):} angular-distance BAO measurements \cite{Sanchez:2010zg} from the Observat\'orio Nacional (ON) \cite{Carvalho:2015ica,Alcaniz:2016ryy,Carvalho:2017tuu,deCarvalho:2017xye}
and from Menote \& Marra (MnM) \cite{Menote:2021jaq}.
They are referred to as \textbf{ON-BAOtr} and \textbf{MnM-BAOtr} and consist of 15 and 14 data points, respectively, covering $0.11<z<2.2$ and $0.35<z<0.63$, respectively.

\item \textbf{Gaussian prior on the Hubble constant:} an external Gaussian prior on $H_0$ (in km\,s$^{-1}$\,Mpc$^{-1}$), which anchors the absolute distance scale and acts effectively as a data point at $z=0$.
When used, \textbf{SH0ES} or \textbf{H0DN} is written alongside the rest of the datasets, depending on whether the prior comes from SH0ES \cite{Breuval:2024lsv} or from the Local Distance Network (H0DN) \cite{H0DN:2025lyy}.
\end{itemize}

On the technical side of the reconstruction, we use the sampler code \texttt{SimpleMC} \cite{simplemc} and implement nested sampling \cite{skilling} via the \texttt{python} library \texttt{dynesty} \cite{speagle2020dynesty} for parameter estimation and evidence evaluation.
The convergence criterion for each run is $0.01$ and the number of live points is $500$, which yields stable posteriors and reliable Bayesian evidence estimates.
We settle for five nodes, meaning that the total number of free parameters is six: $\{H_0,E_1,\dots,E_5\}$.
The nodes are evenly distributed in the range $0.6\leq z \leq 3.0$, namely $z_{1,\dots,5}=(0.6,\,1.2,\,1.8,\,2.4,\,3.0)$, while the $z=0$ normalization is governed solely by $H_0$.
An important caveat is that no data lie in the interval $2.4<z<3.0$, so the node at $z=3$ is expected to be weakly constrained and to act effectively as an extrapolation anchor; accordingly, any reconstructed behavior in this regime should be interpreted with caution, particularly for derivative-based quantities.

The agnostic priors used are $H_0\in[40,90]$ km\,s$^{-1}$\,Mpc$^{-1}$ and $E_i\in[0.5,4.0]$.

Given that it is of great interest to compare the reconstruction against the standard model, we also perform parameter inference for a minimal flat $\Lambda$CDM baseline using the same dataset combinations.
Since we are only interested in background dynamics, the radiation component is omitted and spatial flatness is assumed.
Only two free parameters are inferred: $H_0\in[40,90]$ km\,s$^{-1}$\,Mpc$^{-1}$ and $\Omega_{m0}\in[0.1,0.9]$.

\subsection{Results of the data-driven reconstruction}
\label{subsec:results}

The main parameter constraints for the reconstruction and for the minimal flat $\Lambda$CDM baseline are reported in Table~\ref{tab:results}.
We consider 10 base dataset combinations (listed in the first column of Table~\ref{tab:results}), and for each we analyze three cases: no $H_0$ prior (``none''), a SH0ES prior, and an H0DN prior.

\begin{table*}[t!]
\centering
\caption{
Parameter constraints for all dataset combinations.
For each base dataset combination (first column), we report results without an external $H_0$ prior (row ``none''), with the SH0ES prior (row ``+SH0ES''), and with the H0DN prior (row ``+H0DN'').
For the reconstruction we report constraints on $(H_0,q_0,w_0,z_\dagger)$, where $q_0\equiv q(z=0)$ and $w_0\equiv w_{\rm DE}(z=0)$ are derived from the reconstructed $H(z)$ and its derivative, and $z_\dagger$ is defined by the zero crossing $\rho_{\rm DE}(z_\dagger)=0$ in the scalar-field mapping.
For the minimal flat $\Lambda$CDM baseline we quote only $H_0$ and the corresponding derived value of $q_0$ for the best-fit $\Lambda$CDM solution to the same dataset combination.
Model-comparison statistics are $\Delta\chi^2_{\min}\equiv \chi^2_{\min,\,\rm rec}-\chi^2_{\min,\,\Lambda{\rm CDM}}$ and $\Delta\ln B_{1,2}\equiv \ln(Z_{\Lambda{\rm CDM}}/Z_{\rm rec})$,
where $Z$ denotes the Bayesian evidence computed within the same inference setup.
Under this sign convention, $\Delta\chi^2_{\min}<0$ indicates a better best-fit (lower $\chi^2$) for the reconstruction, while $\Delta\ln B_{1,2}<0$ favors the reconstruction and $\Delta\ln B_{1,2}>0$ favors $\Lambda$CDM.
}
\label{tab:results}

\scriptsize
\setlength{\tabcolsep}{2.8pt}
\renewcommand{\arraystretch}{1.08}

\begin{tabular*}{\textwidth}{@{\extracolsep{\fill}} l c cc cc c c c c @{}}
\toprule
& & \multicolumn{2}{c}{$H_0\,[{\rm km\,s^{-1}\,Mpc^{-1}}]$}
& \multicolumn{2}{c}{$q_0$}
& $w_0$ & $z_\dagger$ & $\Delta\chi^2_{\min}$ & $\Delta\ln B_{1,2}$ \\
\cmidrule(lr){3-4}\cmidrule(lr){5-6}
\makecell[l]{Dataset} & Prior
& Rec. & $\Lambda$CDM
& Rec. & $\Lambda$CDM
&  &  &  &  \\
\midrule

\multirow{3}{*}{CC}
& none
& $55.58 \pm 8.78$ & $67.61 \pm 4.31$
& $0.4^{+1.2}_{-1.5}$ & $-0.501^{+0.076}_{-0.11}$
& $-0.37^{+0.80}_{-1.0}$ & $2.01^{+0.30}_{-0.44}$ & $-2.24$ & $5.01$ \\
& +SH0ES
& $73.13 \pm 1.95$ & $72.32 \pm 1.82$
& $-0.85^{+0.36}_{-0.41}$ & $-0.581^{+0.049}_{-0.062}$
& $-1.32\pm 0.20$ & $1.72^{+0.22}_{-0.24}$ & $-3.11$ & $7.52$ \\
& +H0DN
& $73.21 \pm 1.85$ & $72.88 \pm 1.69$
& $-0.85^{+0.36}_{-0.41}$ & $-0.584^{+0.044}_{-0.057}$
& $-1.28\pm 0.17$ & $1.74^{+0.21}_{-0.23}$ & $-3.14$ & $7.71$ \\
\addlinespace[2pt]
\midrule

\multirow{3}{*}{CC+SN}
& none
& $69.76 \pm 1.77$ & $67.62 \pm 2.76$
& $-0.64\pm 0.26$ & $-0.505^{+0.024}_{-0.028}$
& $-1.09\pm 0.25$ & $2.04\pm 0.36$ & $-0.23$ & $9.74$ \\
& +SH0ES
& $69.93 \pm 1.43$ & $71.20 \pm 1.54$
& $-0.62\pm 0.22$ & $-0.517\pm 0.024$
& $-1.07\pm 0.21$ & $2.01\pm 0.38$ & $-0.12$ & $10.29$ \\
& +H0DN
& $70.04 \pm 1.19$ & $71.18 \pm 1.24$
& $-0.61\pm 0.21$ & $-0.512\pm 0.023$
& $-1.07\pm 0.19$ & $2.02\pm 0.35$ & $-0.42$ & $9.63$ \\
\addlinespace[2pt]
\midrule

\multirow{3}{*}{CC+DESI}
& none
& $65.96 \pm 3.80$ & $68.50 \pm 0.53$
& $-0.35^{+0.27}_{-0.31}$ & $-0.552^{+0.012}_{-0.013}$
& $-0.81^{+0.26}_{-0.30}$ & $2.51^{+0.17}_{-0.10}$ & $-3.47$ & $10.76$ \\
& +SH0ES
& $72.12 \pm 2.13$ & $68.74 \pm 0.52$
& $-0.65\pm 0.21$ & $-0.553^{+0.011}_{-0.013}$
& $-1.09\pm 0.20$ & $2.22^{+0.30}_{-0.059}$ & $-6.54$ & $11.07$ \\
& +H0DN
& $72.41 \pm 1.81$ & $68.92 \pm 0.42$
& $-0.61\pm 0.20$ & $-0.551^{+0.011}_{-0.011}$
& $-1.02\pm 0.19$ & $2.16^{+0.27}_{-0.05}$ & $-6.88$ & $10.59$ \\
\addlinespace[2pt]
\midrule

\multirow{3}{*}{CC+SN+DESI}
& none
& $68.05 \pm 1.63$ & $68.54 \pm 0.52$
& $-0.56^{+0.16}_{-0.22}$ & $-0.543\pm 0.011$
& $-1.01^{+0.15}_{-0.21}$ & $2.56^{+0.034}_{-0.079}$ & $-6.61$ & $12.23$ \\
& +SH0ES
& $69.96 \pm 1.40$ & $68.78 \pm 0.52$
& $-0.57^{+0.12}_{-0.20}$ & $-0.542\pm 0.011$
& $-1.02^{+0.11}_{-0.19}$ & $2.48^{+0.048}_{-0.088}$ & $-7.28$ & $12.51$ \\
& +H0DN
& $70.12 \pm 1.31$ & $68.92 \pm 0.42$
& $-0.57^{+0.11}_{-0.15}$ & $-0.548\pm 0.011$
& $-1.01^{+0.08}_{-0.20}$ & $2.45^{+0.042}_{-0.069}$ & $-7.24$ & $12.12$ \\
\addlinespace[2pt]
\midrule

\multirow{3}{*}{CC+SDSS}
& none
& $65.92 \pm 4.79$ & $68.27 \pm 0.84$
& $-0.55^{+0.33}_{-0.43}$ & $-0.536\pm 0.023$
& $-1.00^{+0.31}_{-0.41}$ & $2.32^{+0.27}_{-0.098}$ & $-4.71$ & $8.04$ \\
& +SH0ES
& $72.44 \pm 2.25$ & $68.83 \pm 0.73$
& $-0.91^{+0.24}_{-0.29}$ & $-0.537\pm 0.022$
& $-1.34^{+0.23}_{-0.28}$ & $1.97\pm 0.35$ & $-8.88$ & $8.12$ \\
& +H0DN
& $72.71 \pm 1.91$ & $69.12 \pm 0.66$
& $-0.88^{+0.22}_{-0.25}$ & $-0.542 \pm 0.021$
& $-1.31^{+0.22}_{-0.24}$ & $1.93\pm 0.42$ & $-8.92$ & $7.99$ \\
\addlinespace[2pt]
\midrule

\multirow{3}{*}{CC+SN+SDSS}
& none
& $67.45 \pm 1.73$ & $68.29 \pm 0.77$
& $-0.57^{+0.23}_{-0.26}$ & $-0.522\pm 0.018$
& $-1.01^{+0.21}_{-0.26}$ & $2.38^{+0.09}_{-0.08}$ & $-5.03$ & $11.79$ \\
& +SH0ES
& $69.73 \pm 1.38$ & $68.77 \pm 0.70$
& $-0.57^{+0.17}_{-0.19}$ & $-0.524\pm 0.017$
& $-1.01^{+0.15}_{-0.21}$ & $2.24^{+0.12}_{-0.11}$ & $-5.43$ & $12.03$ \\
& +H0DN
& $70.51 \pm 1.21$ & $69.52\pm 0.66$
& $-0.56\pm 0.16$ & $-0.532 \pm 0.017$
& $-1.02^{+0.15}_{-0.19}$ & $2.25 \pm 0.11$ & $-5.22$ & $11.52$ \\
\addlinespace[2pt]
\midrule

\multirow{3}{*}{CC+ON-BAOtr}
& none
& $62.71 \pm 5.75$ & $72.96 \pm 1.78$
& $0.57^{+0.37}_{-0.59}$ & $-0.607^{+0.025}_{-0.029}$
& $-1.02^{+0.35}_{-0.50}$ & $1.92\pm 0.34$ & $-3.47$ & $5.94$ \\
& +SH0ES
& $72.28 \pm 2.06$ & $73.25 \pm 1.41$
& $-1.03\pm 0.30$ & $-0.606\pm 0.026$
& $-1.46\pm 0.28$ & $1.74^{+0.16}_{-0.37}$ & $-2.25$ & $8.95$ \\
& +H0DN
& $72.75 \pm 1.56$ & $73.34 \pm 1.02$
& $-0.99\pm 0.31$ & $-0.613 \pm 0.025$
& $-1.37\pm 0.26$ & $1.73^{+0.21}_{-0.36}$ & $-2.61$ & $8.53$ \\
\addlinespace[2pt]
\midrule

\multirow{3}{*}{CC+SN+ON-BAOtr}
& none
& $67.75 \pm 1.89$ & $74.55 \pm 1.56$
& $-0.63\pm 0.24$ & $-0.550\pm 0.020$
& $-1.07^{+0.28}_{-0.33}$ & $2.06\pm 0.37$ & $-9.13$ & $7.46$ \\
& +SH0ES
& $69.91 \pm 1.45$ & $74.33 \pm 1.32$
& $-0.68\pm 0.21$ & $-0.551\pm 0.019$
& $-1.03^{+0.21}_{-0.25}$ & $2.02\pm 0.25$ & $-5.36$ & $9.43$ \\
& +H0DN
& $70.22 \pm 1.29$ & $74.22 \pm 1.24$
& $-0.70\pm 0.21$ & $-0.549 \pm 0.020$
& $-1.02^{+0.22}_{-0.23}$ & $2.01\pm 0.26$ & $-5.11$ & $10.01$ \\
\addlinespace[2pt]
\midrule

\multirow{3}{*}{CC+MnM-BAOtr}
& none
& $56.60 \pm 5.48$ & $69.10 \pm 1.42$
& $-0.13\pm 0.54$ & $-0.531\pm 0.030$
& $-0.60\pm 0.51$ & $1.99^{+0.26}_{-0.18}$ & $-2.29$ & $5.85$ \\
& +SH0ES
& $72.44 \pm 1.92$ & $70.38 \pm 1.22$
& $-1.06\pm 0.31$ & $-0.511\pm 0.029$
& $-1.48\pm 0.29$ & $1.71^{+0.14}_{-0.36}$ & $-3.42$ & $7.93$ \\
& +H0DN
& $72.65 \pm 1.77$ & $70.85 \pm 1.02$
& $-1.01\pm 0.30$ & $-0.531 \pm 0.029$
& $-1.39\pm 0.28$ & $1.72^{+0.16}_{-0.34}$ & $-3.23$ & $8.19$ \\
\addlinespace[2pt]
\midrule

\multirow{3}{*}{CC+SN+MnM-BAOtr}
& none
& $67.70 \pm 1.80$ & $69.62 \pm 1.24$
& $-0.63\pm 0.23$ & $-0.515\pm 0.021$
& $-1.08\pm 0.22$ & $2.06^{+0.29}_{-0.45}$ & $-1.11$ & $10.91$ \\
& +SH0ES
& $69.99 \pm 1.48$ & $70.48 \pm 1.08$
& $-0.62\pm 0.23$ & $-0.507\pm 0.019$
& $-1.07\pm 0.21$ & $1.99\pm 0.38$ & $-0.09$ & $12.03$ \\
& +H0DN
& $70.12 \pm 1.21$ & $70.55 \pm 0.97$
& $-0.65\pm 0.22$ & $-0.511\pm 0.019$
& $-1.06\pm 0.22$ & $1.95\pm 0.41$ & $-0.91$ & $11.51$ \\
\bottomrule
\end{tabular*}
\end{table*}

\begin{figure*}
    \centering
    \includegraphics[width=0.32\linewidth]{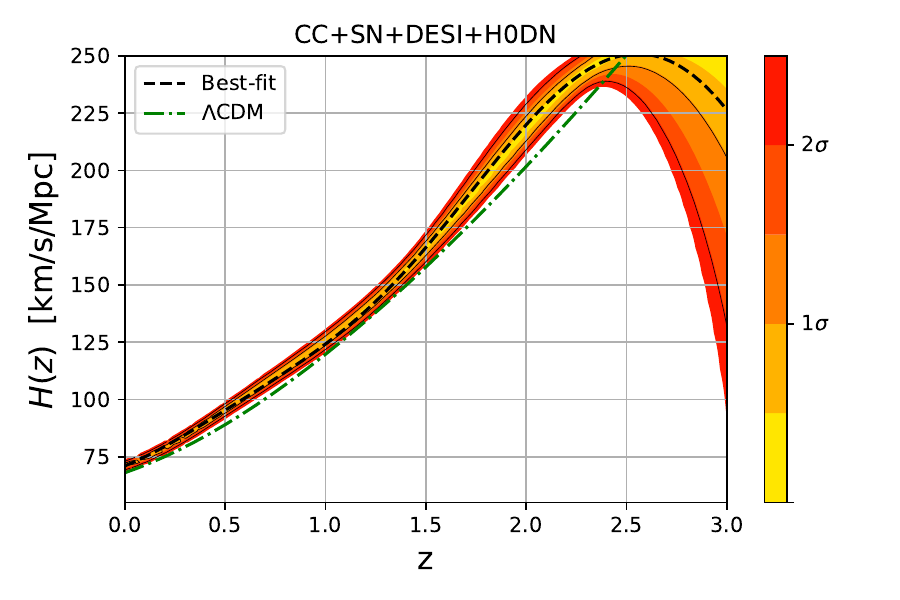}
    \includegraphics[width=0.32\linewidth]{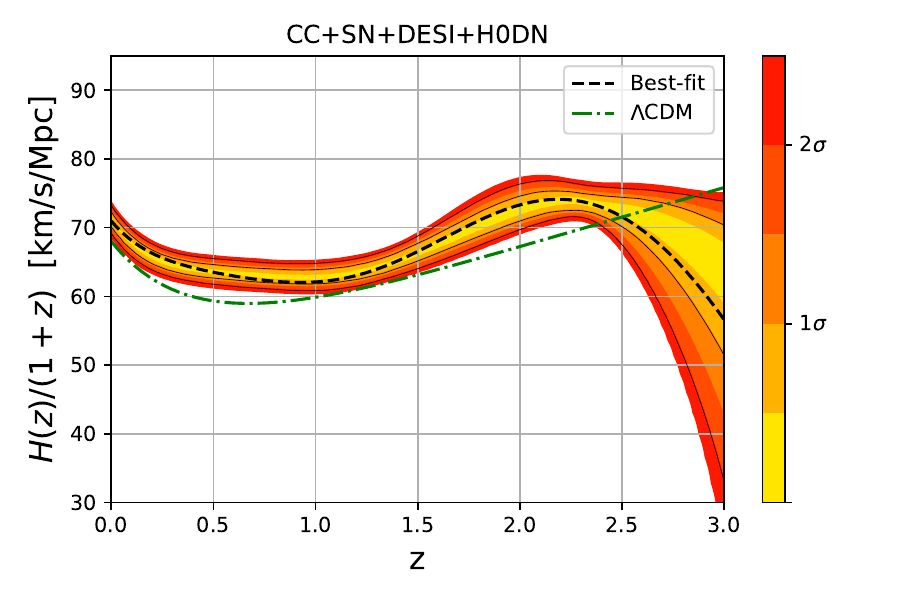}
    \includegraphics[width=0.32\linewidth]{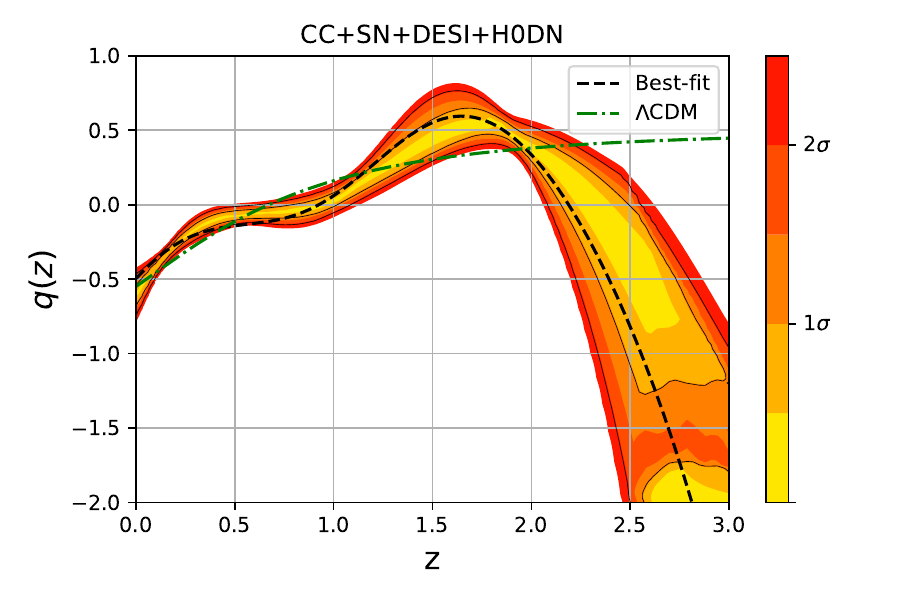}
    
    \includegraphics[width=0.32\linewidth]{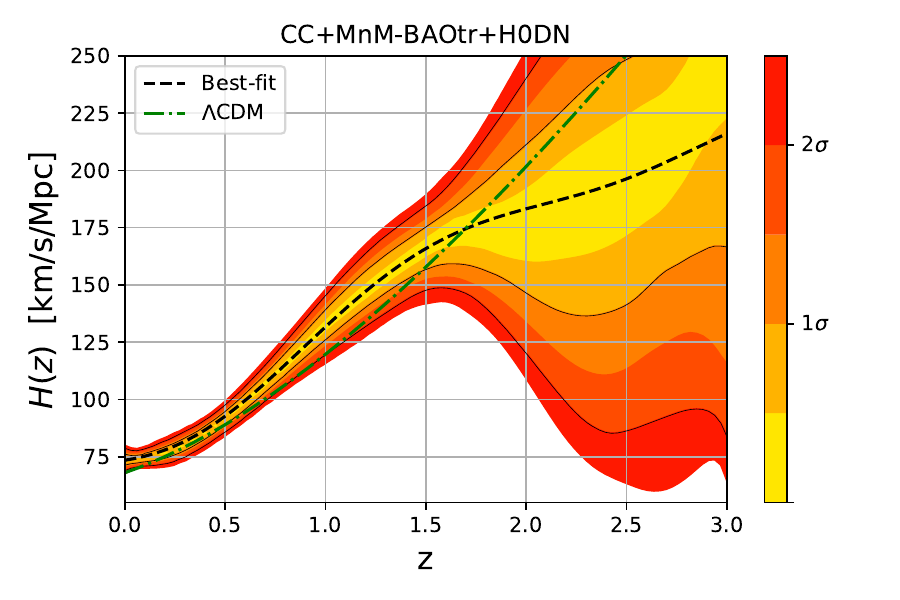}
    \includegraphics[width=0.32\linewidth]{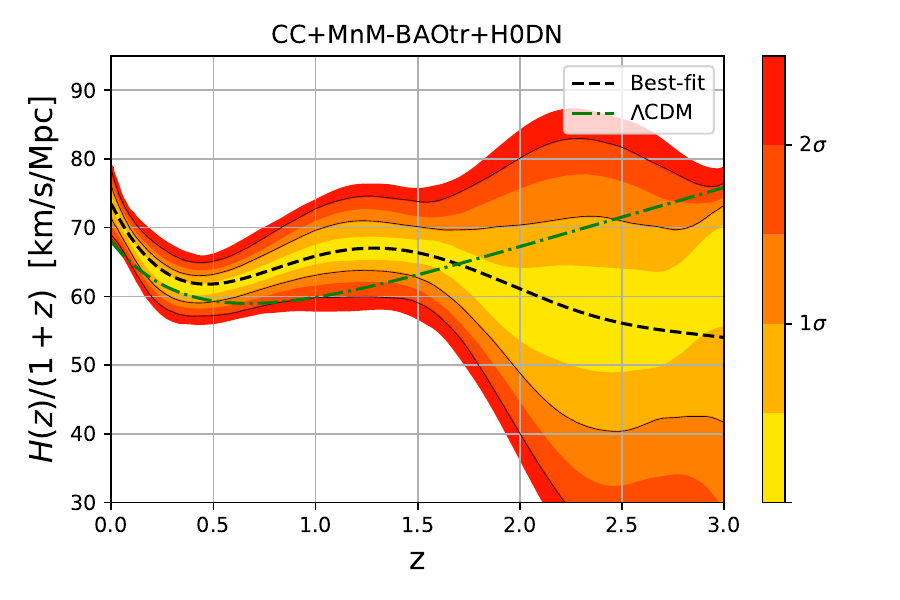}
    \includegraphics[width=0.32\linewidth]{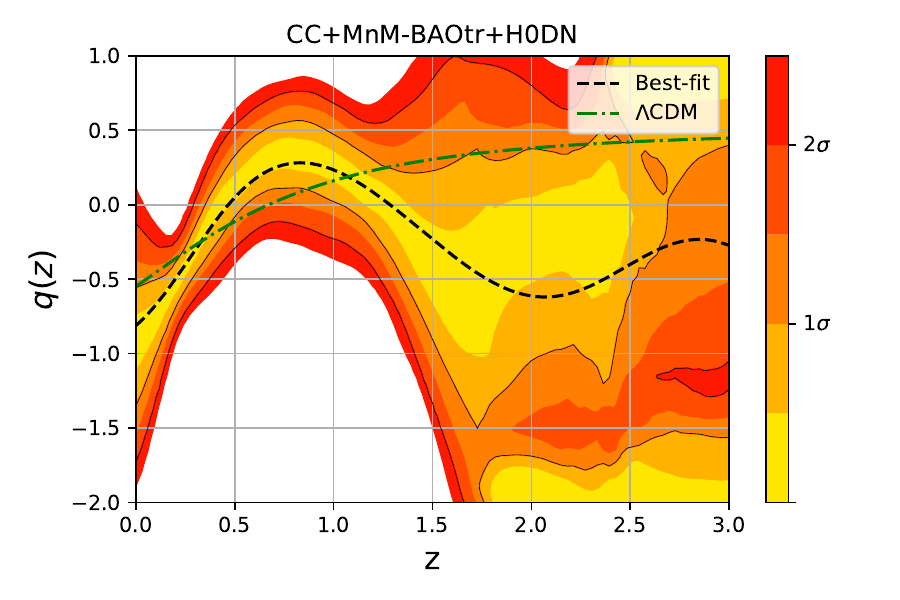}
    
    \includegraphics[width=0.32\linewidth]{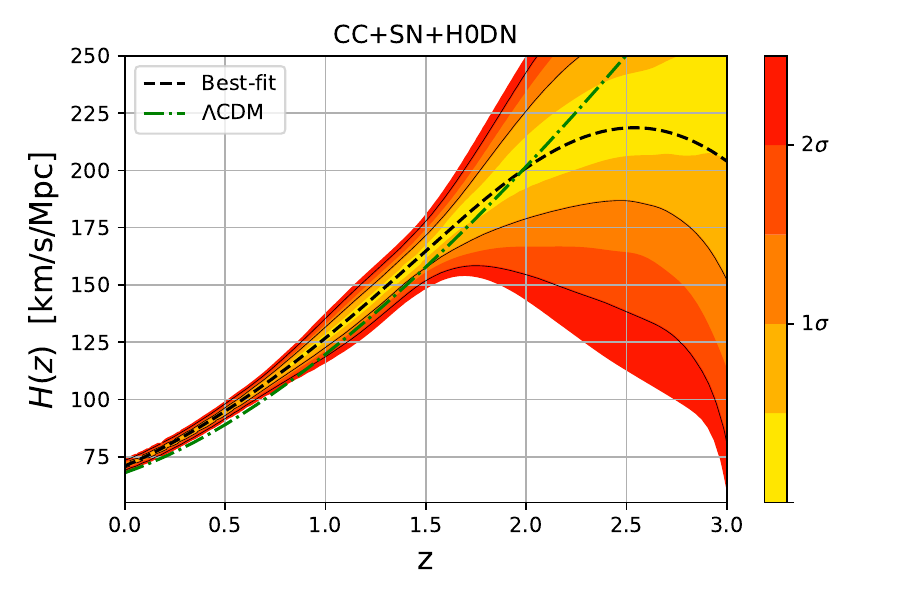}
    \includegraphics[width=0.32\linewidth]{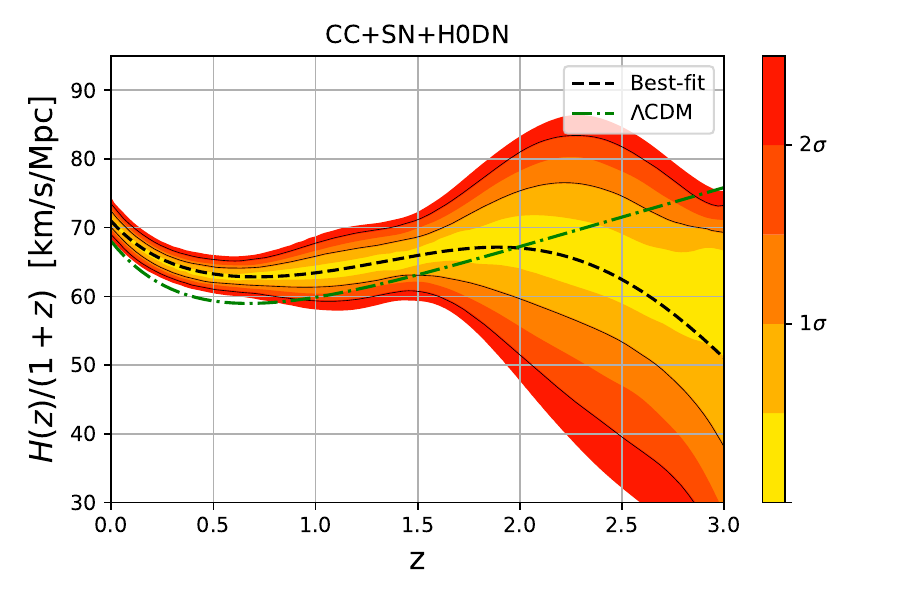}
    \includegraphics[width=0.32\linewidth]{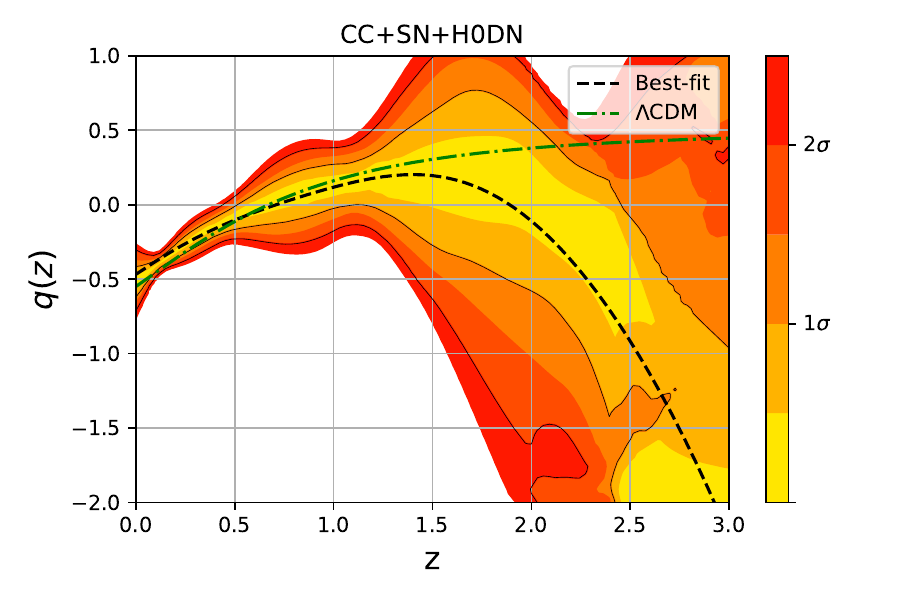}
    \caption{
    Posterior predictive bands for the reconstructed kinematic quantities $H(z)$, $H(z)/(1+z)$, and $q(z)$ for three illustrative dataset combinations (top to bottom rows): CC+SN+DESI+H0DN, CC+MnM-BAOtr+H0DN, and CC+SN+H0DN.
    The color shading encodes the $\sigma$-equivalent credible level around the best-fit reconstruction, as indicated by the color bar in each panel (up to $\sim2.5\sigma$); for a Gaussian posterior, the $1\sigma$ and $2\sigma$ levels correspond approximately to 68\% and 95\% credible regions.
    The black dotted curve shows the best-fit reconstruction, while the green dotted curve shows the best-fit flat $\Lambda$CDM baseline for the same dataset combination.
    Since the highest-redshift node is at $z=3$ and there are no data in $2.4<z<3.0$, behavior in this interval should be interpreted cautiously, especially for derivative-based quantities.
    }
    \label{fig:H_and_q}
\end{figure*}

\begin{figure*}
    \centering
    \includegraphics[width=0.32\linewidth]{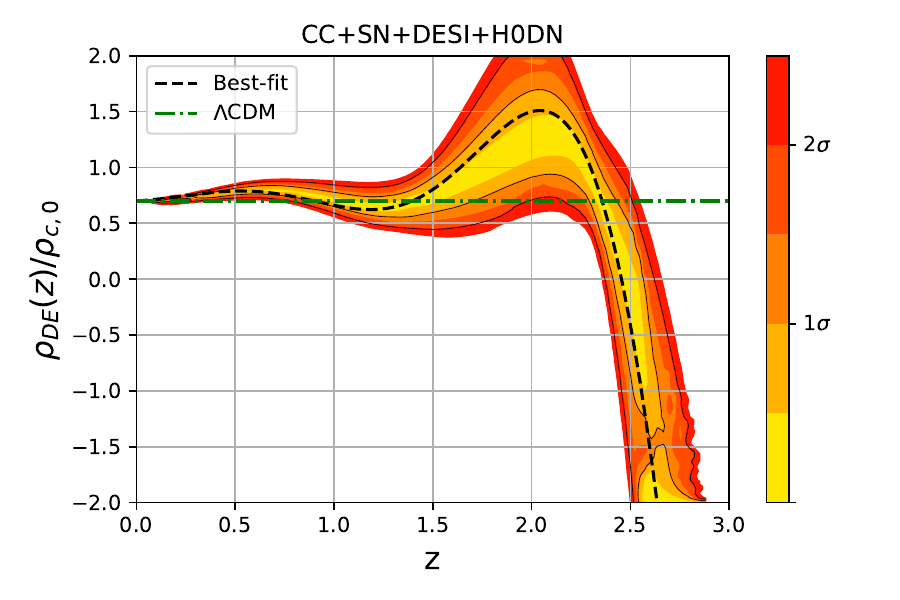}
    \includegraphics[width=0.32\linewidth]{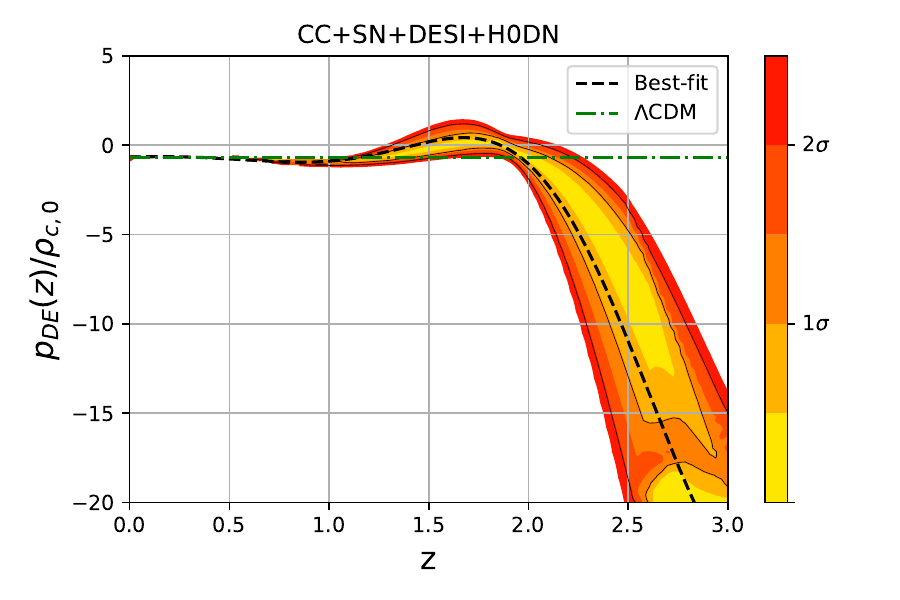}
    \includegraphics[width=0.32\linewidth]{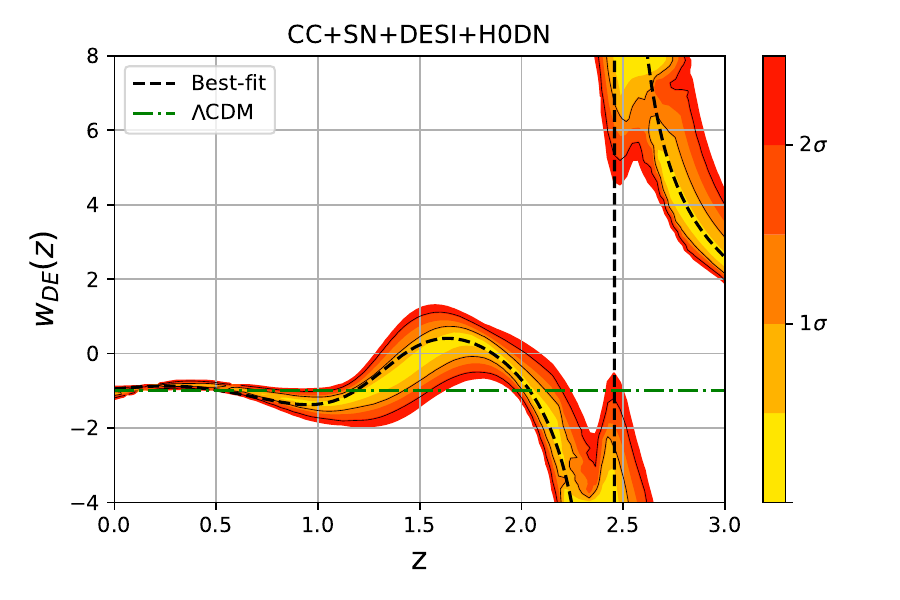}
    
    \includegraphics[width=0.32\linewidth]{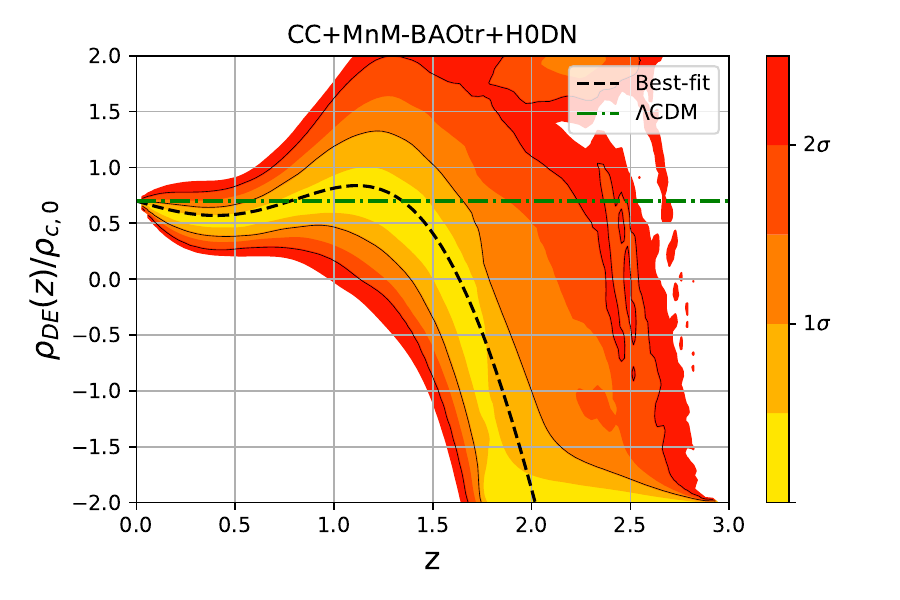}
    \includegraphics[width=0.32\linewidth]{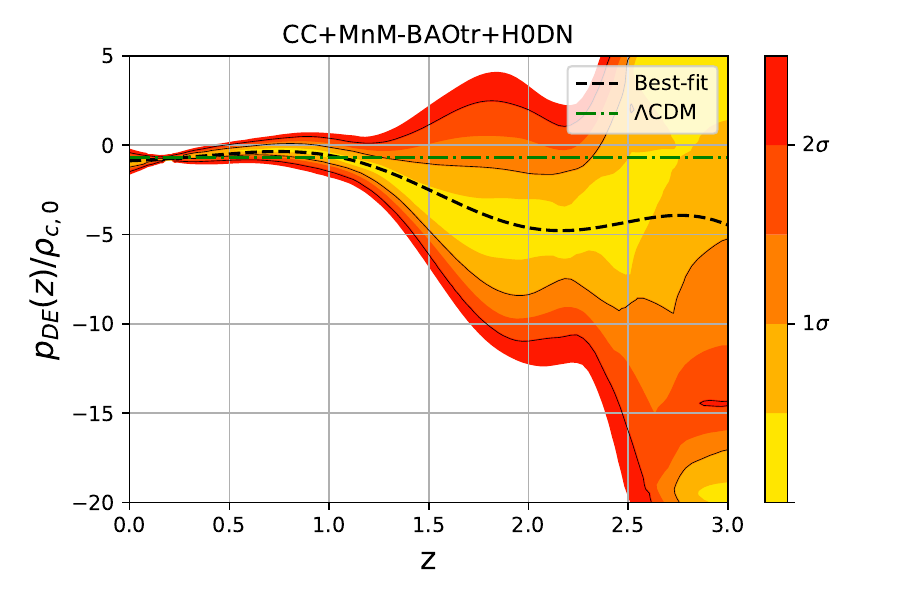}
    \includegraphics[width=0.32\linewidth]{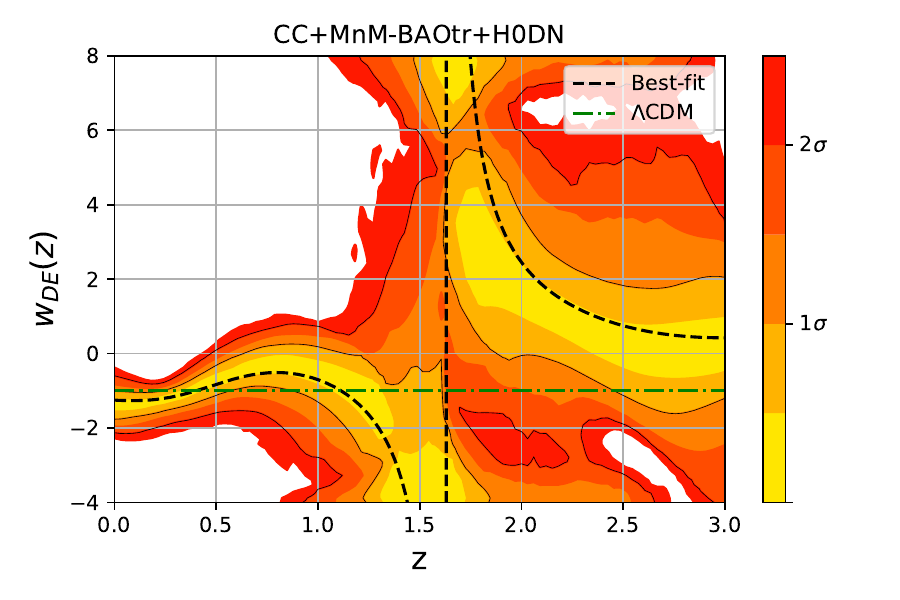}
    
    \includegraphics[width=0.32\linewidth]{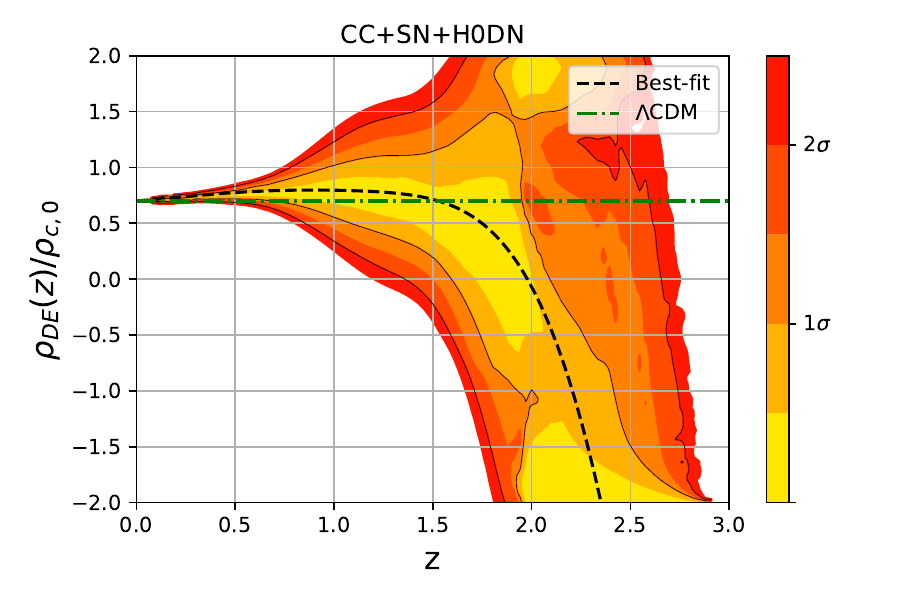}
    \includegraphics[width=0.32\linewidth]{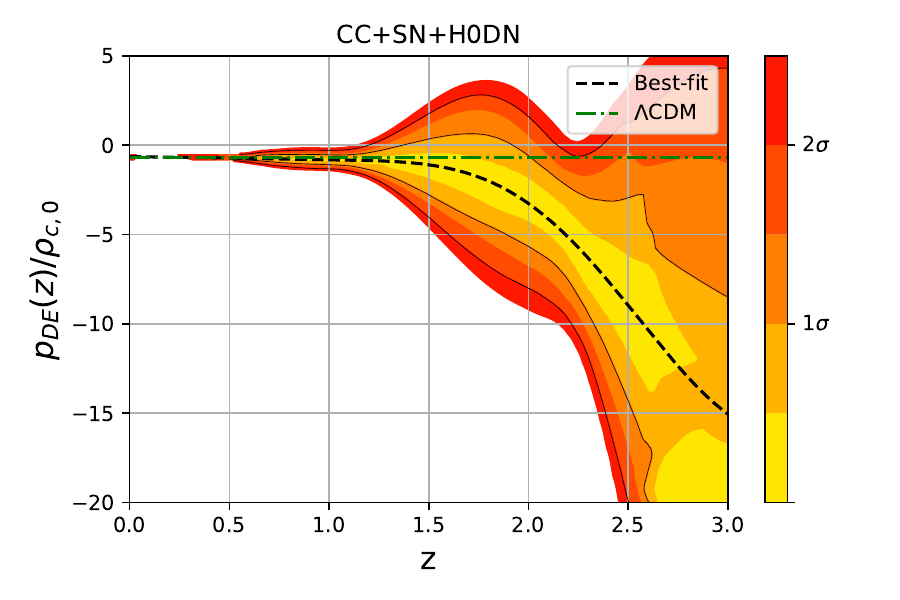}
    \includegraphics[width=0.32\linewidth]{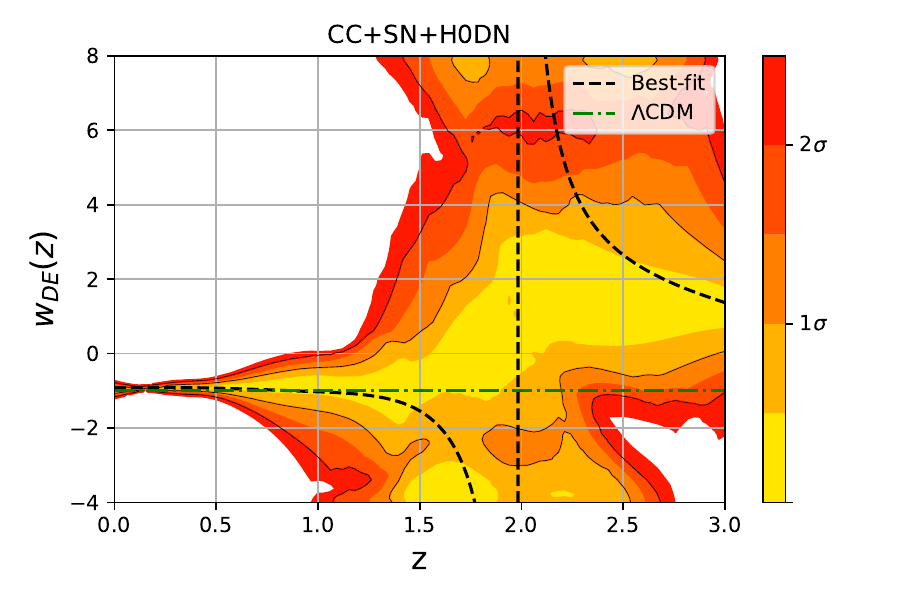}
    \caption{
    Posterior predictive bands for the effective dark-energy density $\rho_{\rm DE}(z)$, pressure $p_{\rm DE}(z)$ (both shown normalized to the present-day critical density as in the axis labels), and the equation of state $w_{\rm DE}(z)=p_{\rm DE}/\rho_{\rm DE}$ for the same three dataset combinations as in Fig.~\ref{fig:H_and_q} (top to bottom rows): CC+SN+DESI+H0DN, CC+MnM-BAOtr+H0DN, and CC+SN+H0DN.
    The color shading encodes the $\sigma$-equivalent credible level around the best-fit reconstruction, as indicated by the color bar in each panel (up to $\sim2.5\sigma$).
    The black dotted curve shows the best-fit reconstruction, while the green dotted curve shows the best-fit flat $\Lambda$CDM baseline for the same dataset combination.
    The divergence of $w_{\rm DE}$ occurs when $\rho_{\rm DE}$ crosses zero and reflects the kinematic ratio $p_{\rm DE}/\rho_{\rm DE}$ rather than a singularity in $H(z)$.
    Since there are no data in $2.4<z<3.0$, behavior in this interval should be interpreted cautiously.
    }
    \label{fig:rhode_pde_wde}
\end{figure*}

A clear trend emerges for the Hubble constant.
When neither SN nor an external $H_0$ prior is included, $H_0$ is only weakly constrained, reflecting the limited information anchoring the expansion rate near $z\simeq 0$ in those combinations.
Including SN substantially sharpens the reconstruction over $0.01\lesssim z\lesssim 2.26$, and when combined with CC and/or BAO it typically yields $H_0$ in the $\sim 68$--70~km\,s$^{-1}$\,Mpc$^{-1}$ range (and very close to $\sim 70$ once an external $H_0$ prior is applied).
As expected, imposing an external $H_0$ prior drives the inferred value of $H_0$ toward the prior mean, most clearly in combinations without SN.
A particularly instructive example is provided by ON-BAOtr: in the $\Lambda$CDM baseline, CC+SN+ON-BAOtr prefers a comparatively high $H_0$ (Table~\ref{tab:results}),
whereas the reconstruction can accommodate the same combination with a substantially lower $H_0$, contributing to a large improvement in the best-fit $\chi^2$ (i.e.\ a more negative $\Delta\chi^2_{\min}$; Table~\ref{tab:results}) for that case.

Reconstructing $H(z)$ also allows us to obtain its redshift derivative $H'(z)$ and hence the deceleration parameter
$q(z)=(1+z)\,H'(z)/H(z)-1$.
Representative functional posteriors for $H(z)$, $H(z)/(1+z)$, and $q(z)$ are shown in Fig.~\ref{fig:H_and_q} for three illustrative combinations, while the full atlas is provided in Appendix~\ref{app:reconstructions}.
The present-day value $q_0\equiv q(z=0)$ is reported in Table~\ref{tab:results}.
In most combinations, the reconstructed $q_0$ is consistent with the corresponding $\Lambda$CDM baseline values quoted in the same table within $\sim1\sigma$.
A subset of combinations without SN but with an $H_0$ prior (most notably those involving BAOtr) yield mean values closer to $q_0\simeq -1$, indicating a stronger late-time acceleration in those reconstructions, while still remaining compatible with the $\Lambda$CDM expectations at the $\sim2\sigma$ level given their uncertainties.

We remind the reader that accelerated expansion corresponds to $q(z)<0$, with $q=-1$ for exact de Sitter.
Moreover, $q=-1-\dot H/H^2$, so $q(z)<-1$ indicates ``super-acceleration'' ($\dot H>0$), which in GR corresponds to an effective violation of the null energy condition by the \emph{total} cosmic fluid.
In our reconstructions, values $q<-1$ appear primarily in regimes where the reconstruction is weakly anchored: at very low redshift in combinations lacking direct low-$z$ information for the derivative $H'(z)$, and at high redshift ($z\gtrsim 2.4$) where no direct data are present and the boundary node at $z=3$ influences the interpolation/extrapolation.
Accordingly, while such behavior is mathematically allowed in the reconstruction, it should not be over-interpreted physically; we discuss its theoretical interpretation and consistency requirements in Sec.~\ref{sec:disc}.

Using the scalar-field mapping, we infer an effective present-day dark-energy equation of state $w_0\equiv w_{\rm DE}(z=0)$.
To quantify consistency with a cosmological constant, we define $N_\sigma \equiv |w_0+1|/\sigma_{w_0}$, where for asymmetric posteriors we take $\sigma_{w_0}$ as the mean of the upper and lower $1\sigma$ errors.
All dataset combinations that include SN are fully consistent with $w_0=-1$, typically at the $\lesssim0.5\sigma$ level.
A mild preference for $w_0<-1$ appears only when an external $H_0$ prior is imposed \emph{without} SN, reaching at most the $\sim1.5$--$1.7\sigma$ level.
Thus, deviations of $w_0$ from $-1$ today are not the primary signature of the reconstructed dynamics.

\begin{figure*}[t!]
    \centering
    \makebox[9cm][c]{
        \includegraphics[trim=0mm 34mm 0mm 20mm, clip, width=4.4cm, height=4.cm]{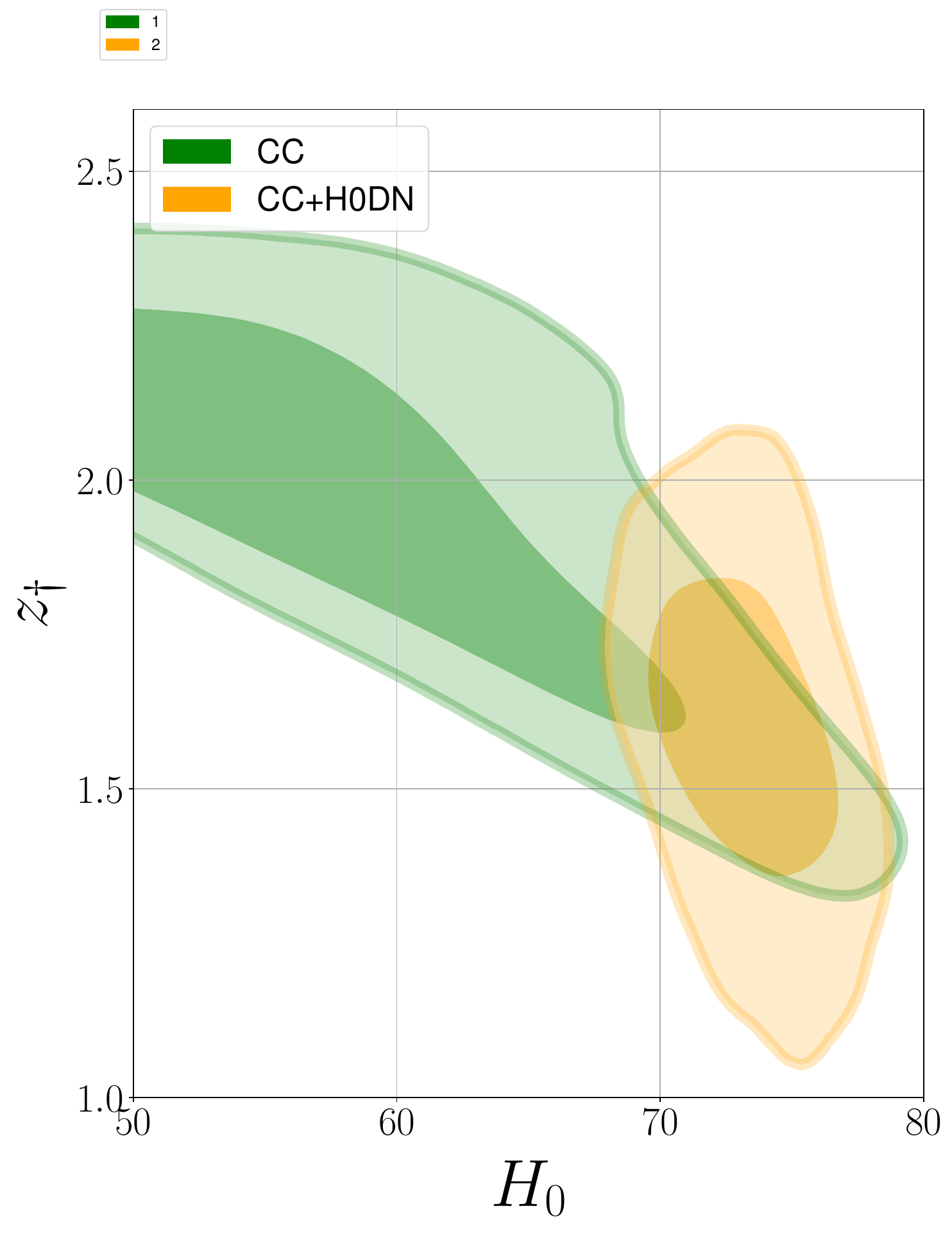}
        \includegraphics[trim=35mm 35mm 0mm 20mm, clip, width=3.5cm, height=4.cm]{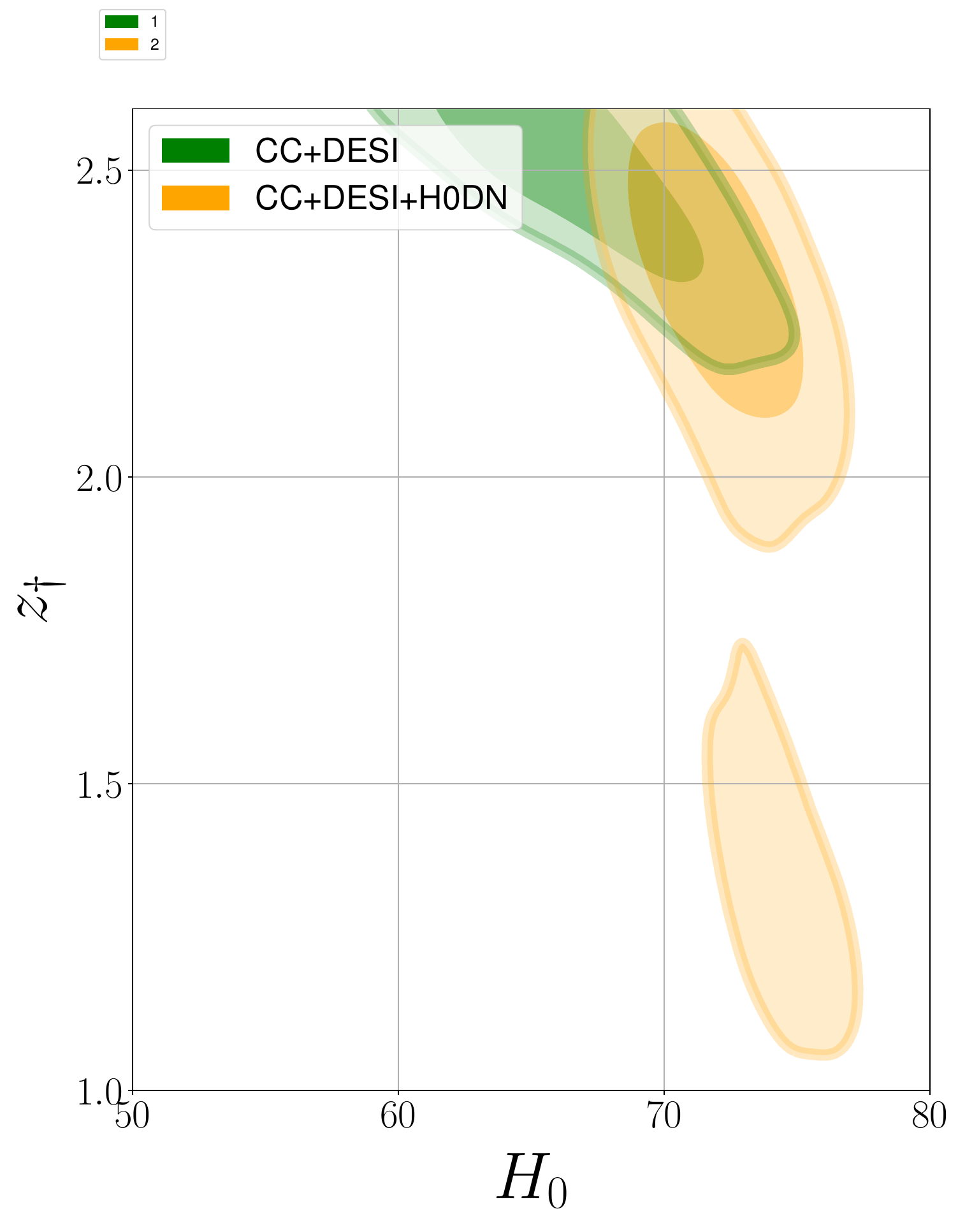}
        \includegraphics[trim=35mm 35mm 0mm 20mm, clip, width=3.5cm, height=4.cm]{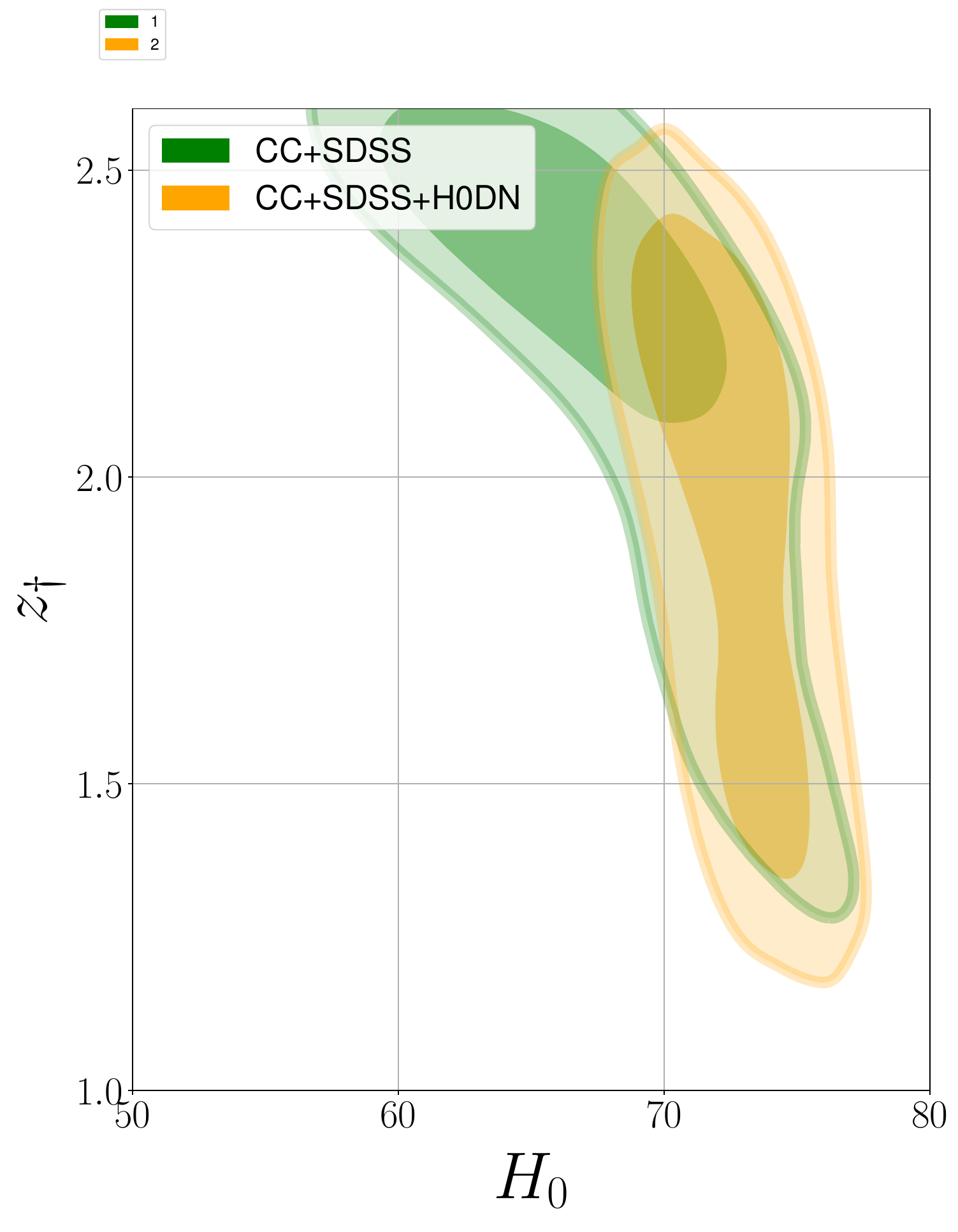}
        \includegraphics[trim=35mm 35mm 0mm 20mm, clip, width=3.5cm, height=4.cm]{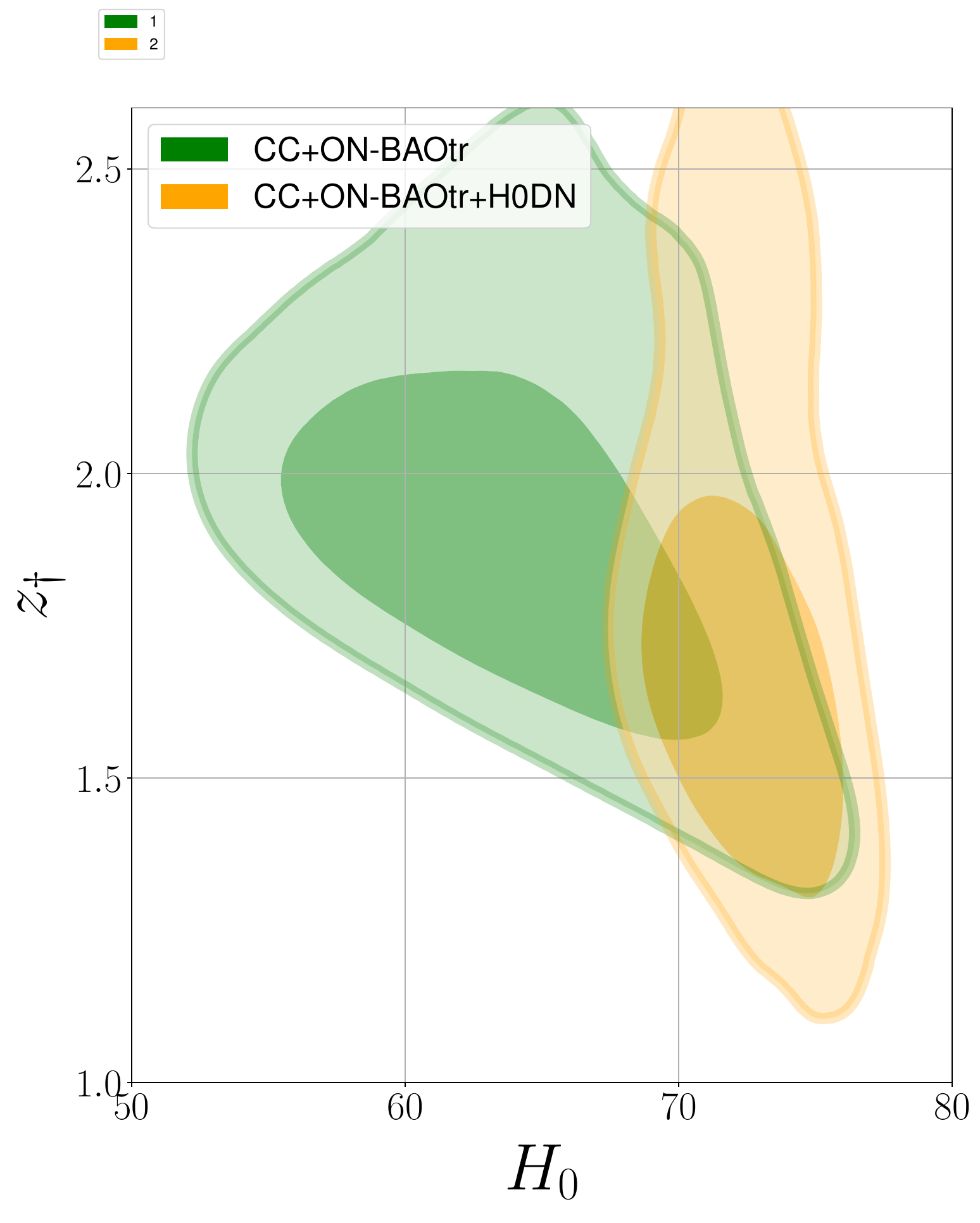}
        \includegraphics[trim=35mm 35mm 0mm 20mm, clip, width=3.5cm, height=4.cm]{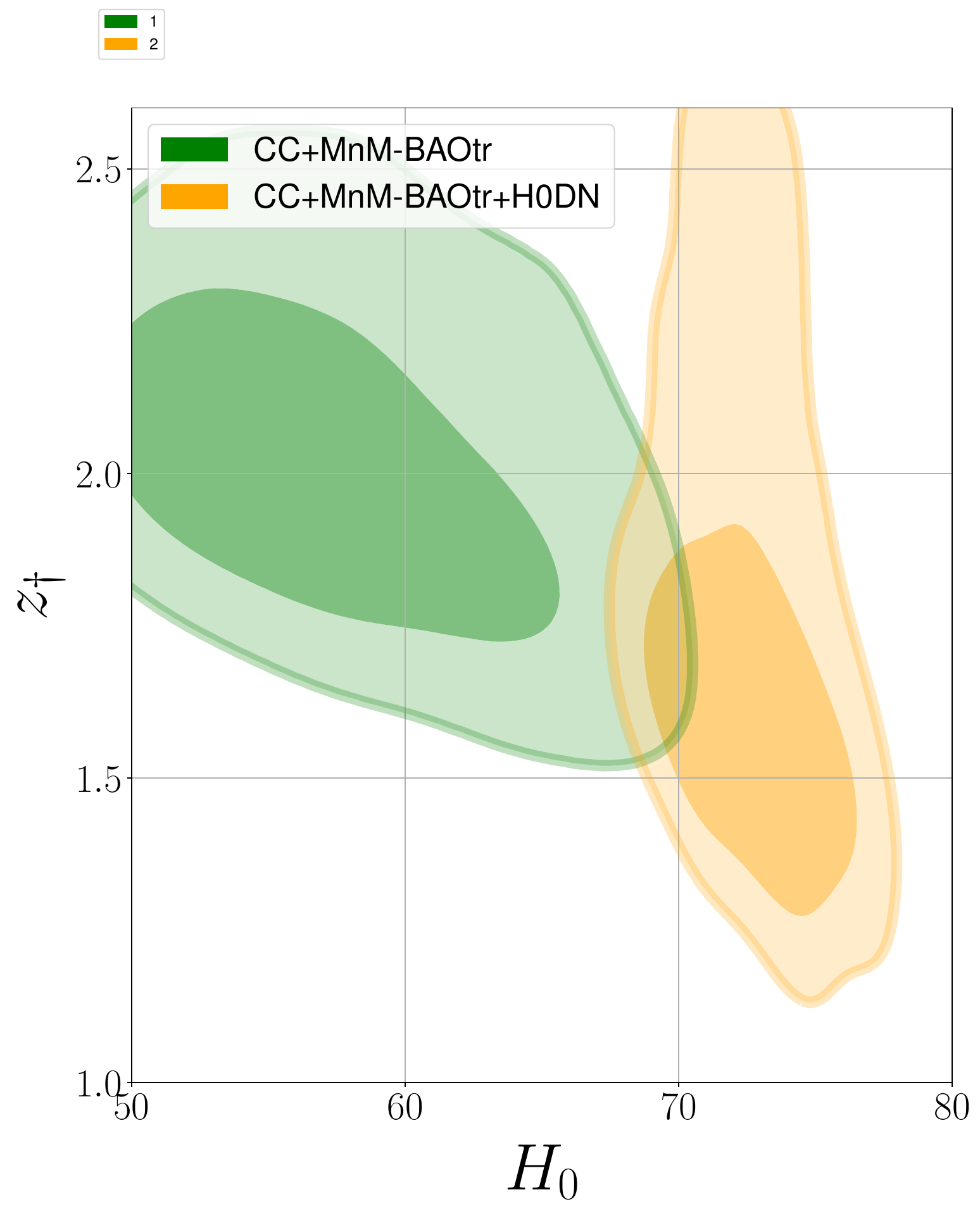}
    }
    \makebox[9cm][c]{
        \includegraphics[trim=0mm 0mm 0mm 20mm, clip, width=4.3cm, height=4.5cm]{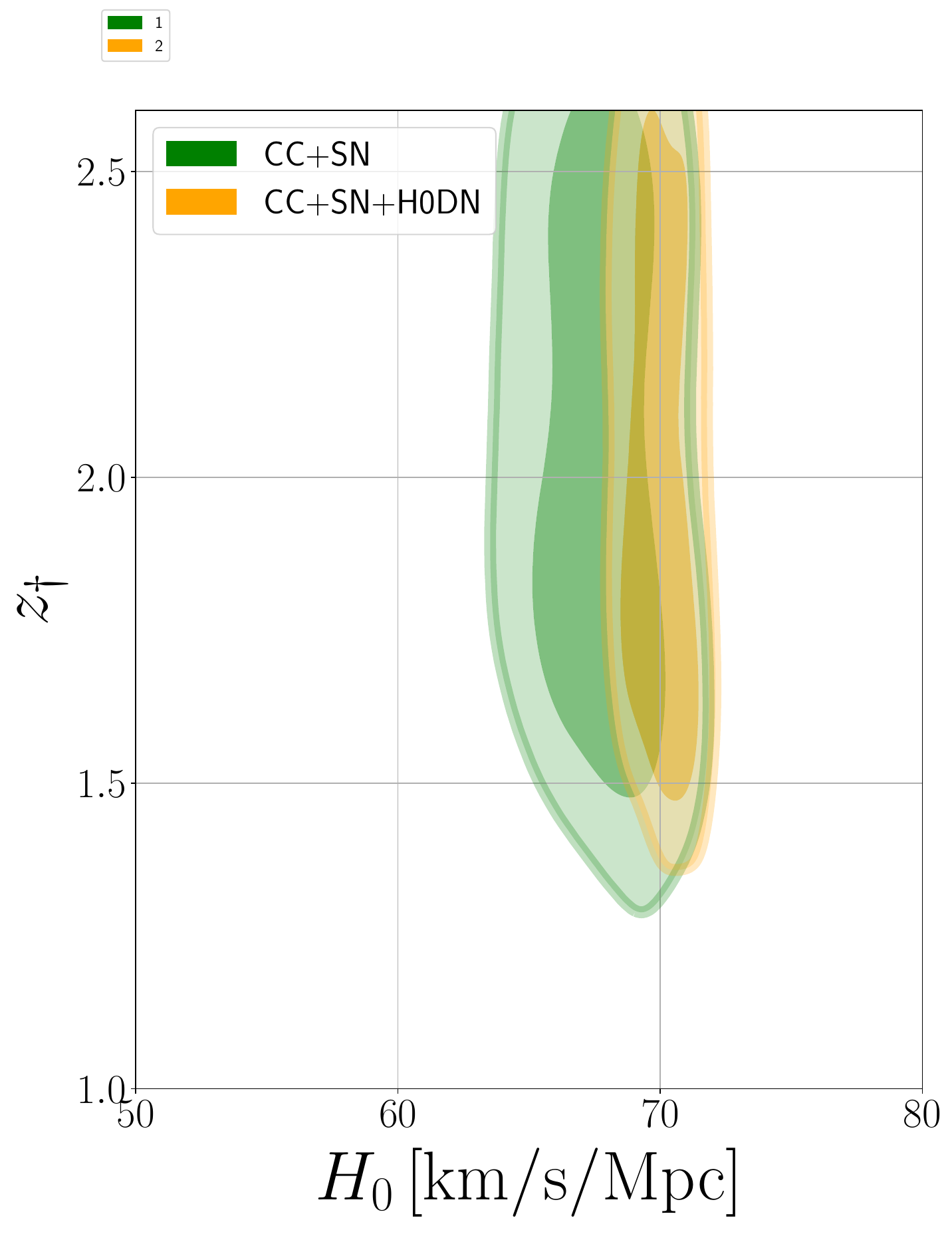}
        \includegraphics[trim=34mm 0mm 0mm 20mm, clip, width=3.5cm, height=4.5cm]{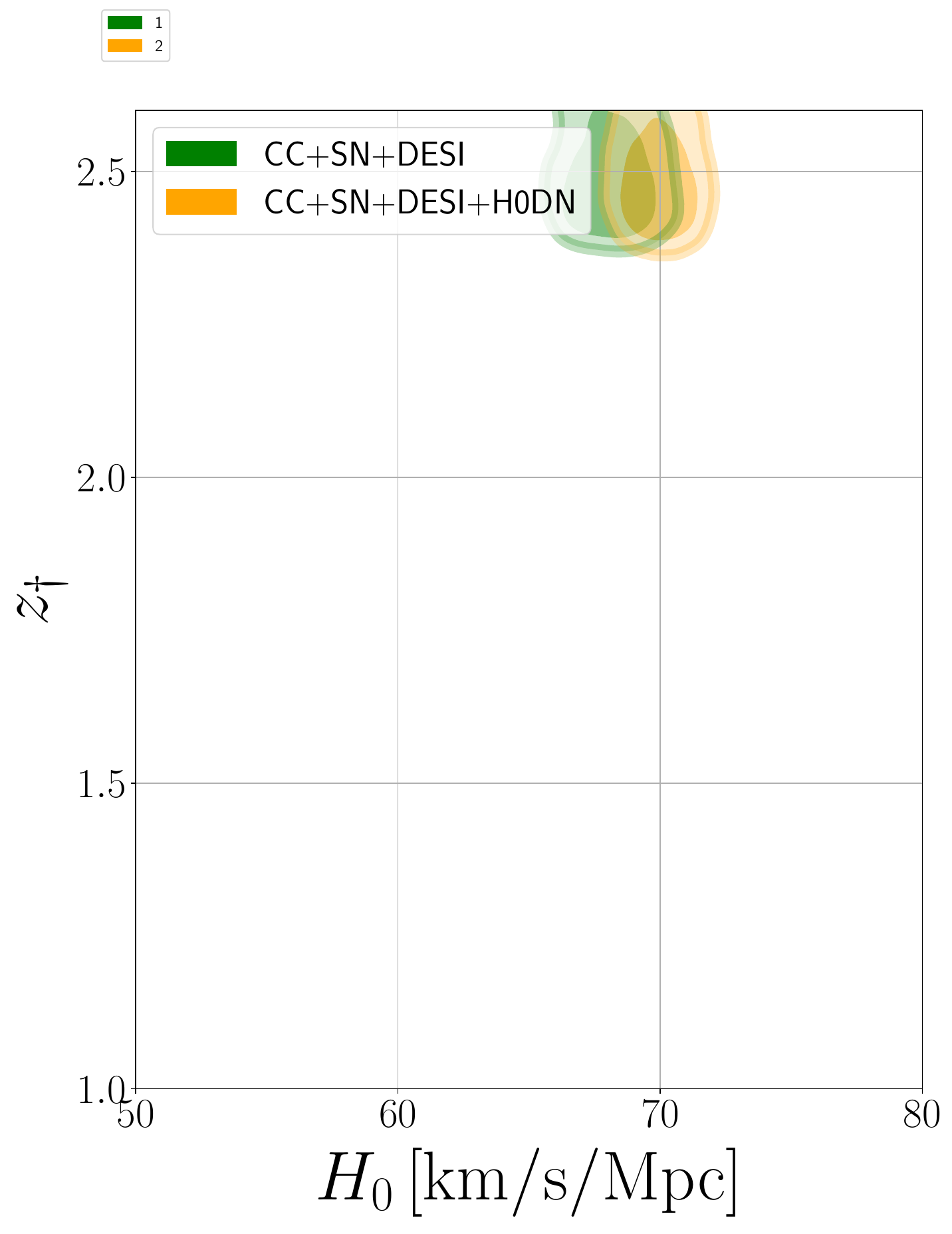}
        \includegraphics[trim=34mm 0mm 0mm 20mm, clip, width=3.5cm, height=4.5cm]{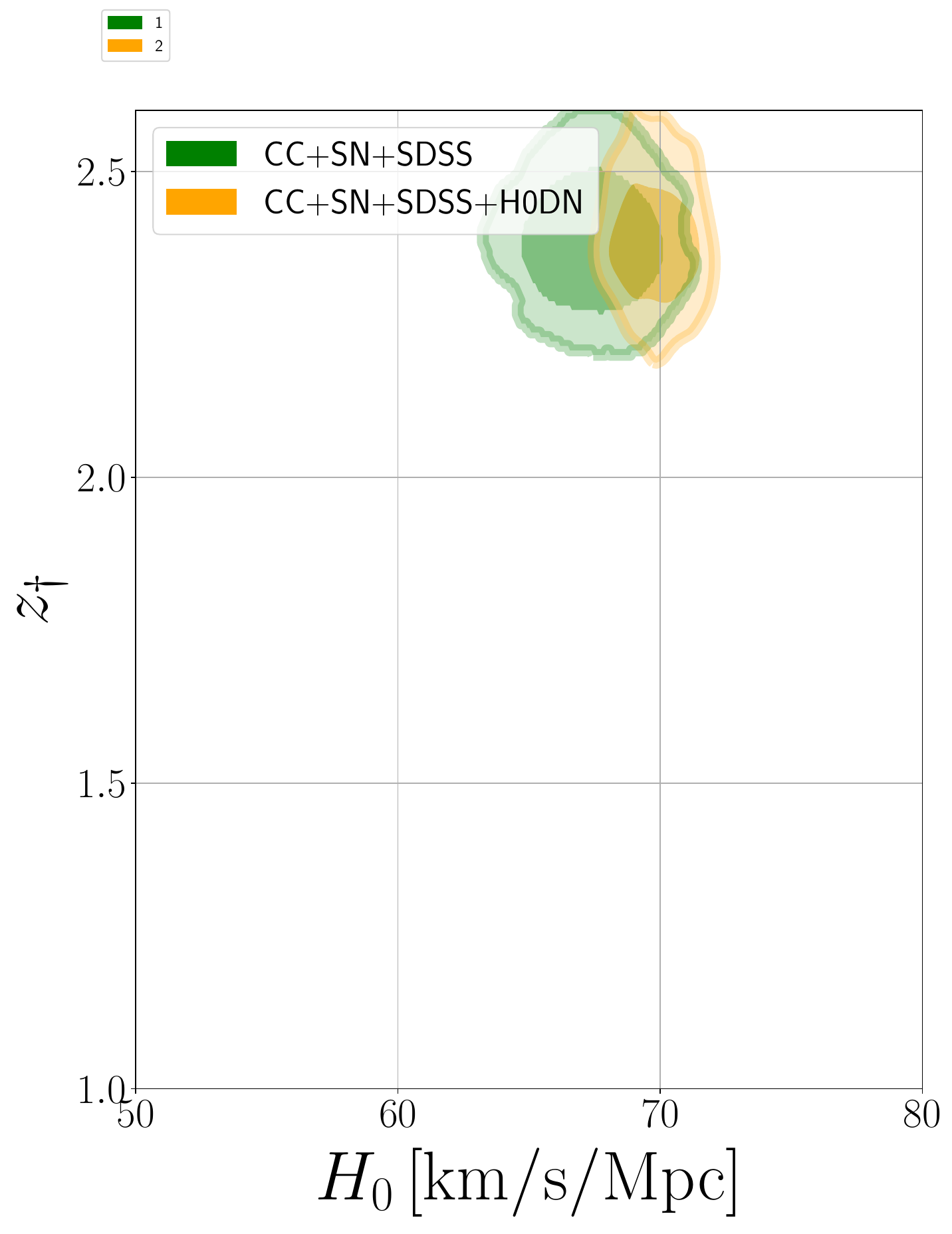}
        \includegraphics[trim=34mm 0mm 0mm 20mm, clip, width=3.5cm, height=4.5cm]{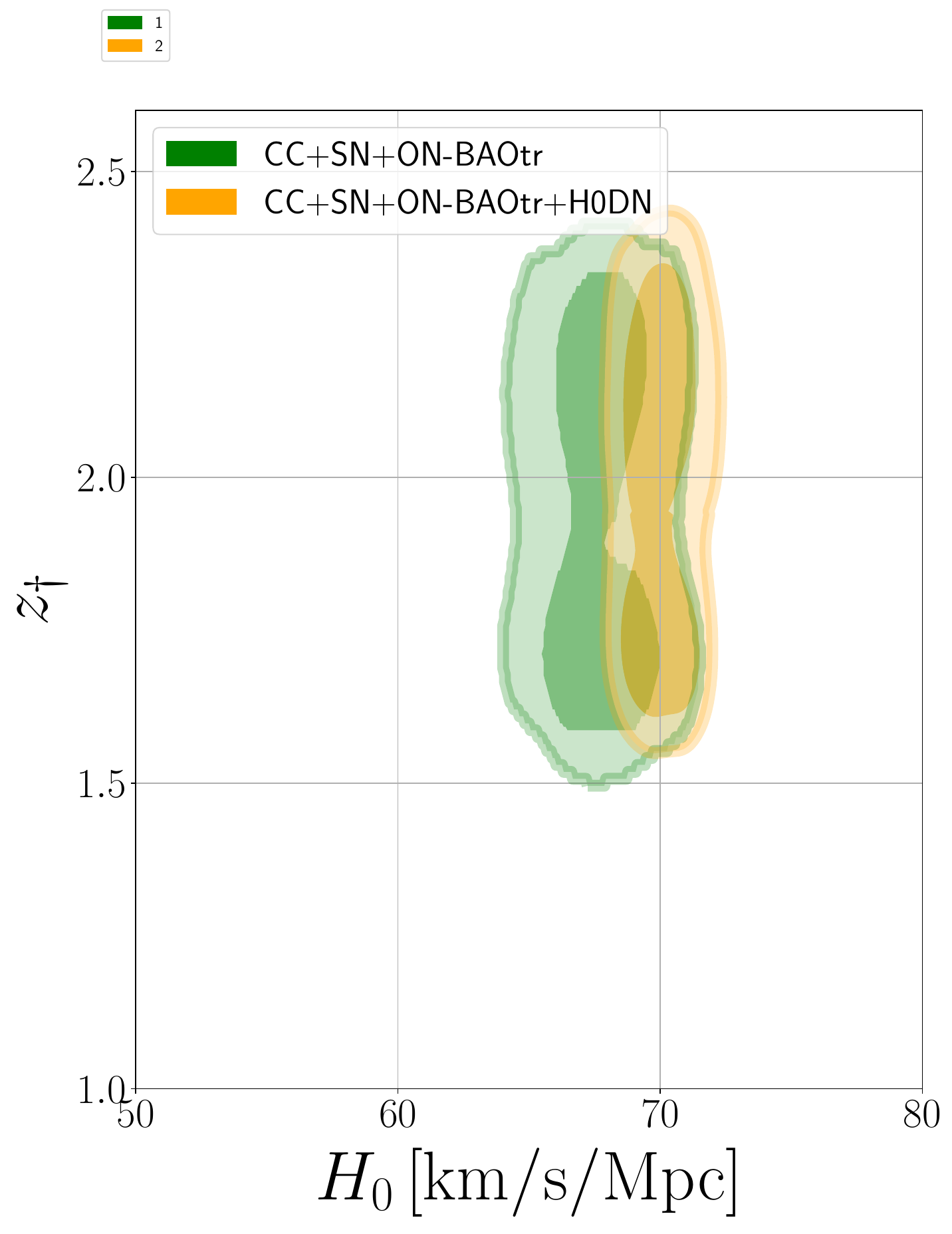}
        \includegraphics[trim=34mm 0mm 0mm 20mm, clip, width=3.5cm, height=4.5cm]{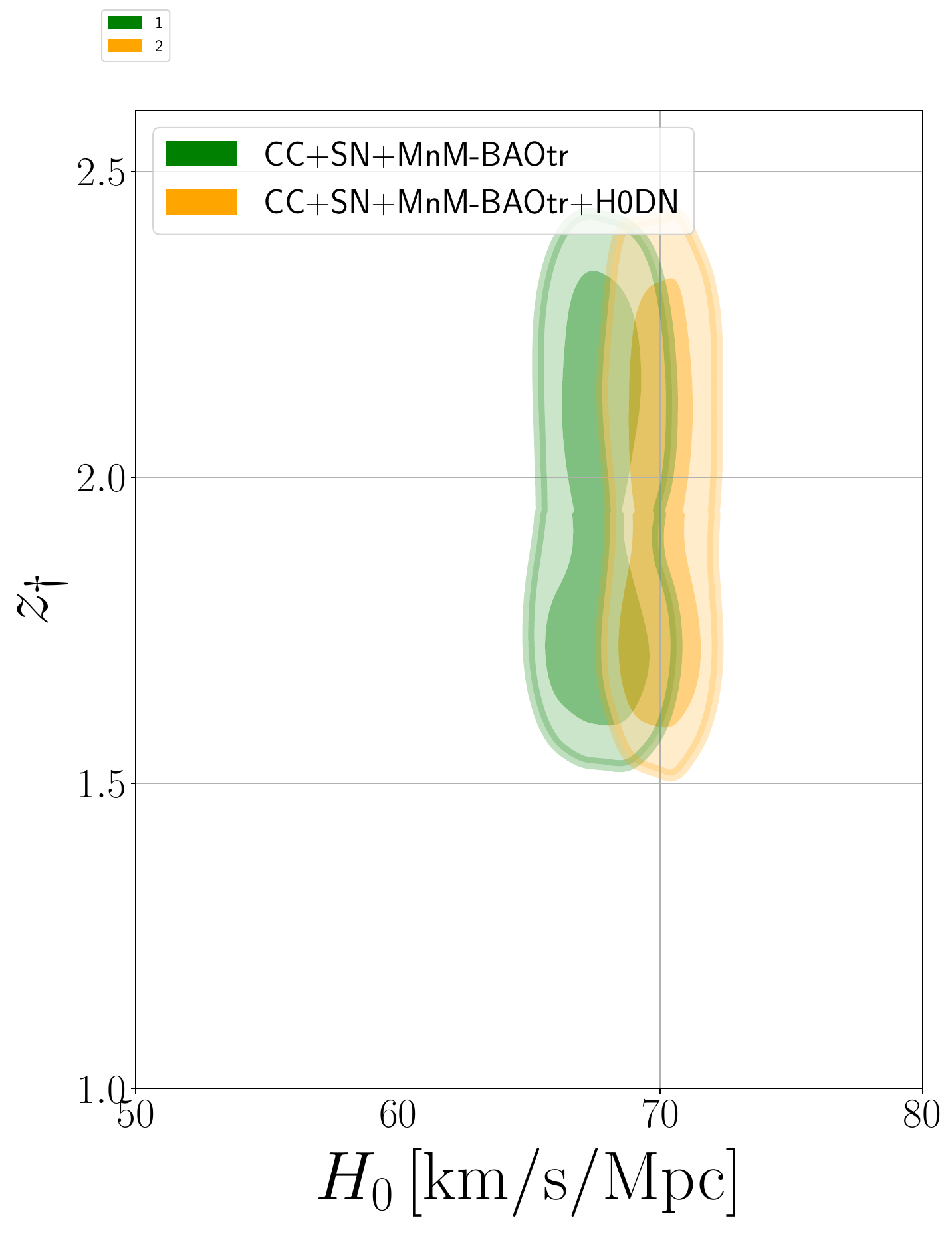}
    }
\caption{
Two-dimensional marginalized posterior distributions for the derived transition redshift $z_\dagger$ versus $H_0$ (shown here for H0DN; SH0ES cases are nearly indistinguishable in our pipeline).
Contours denote the 68\% and 95\% credible regions.
The top row shows combinations without SN (as labeled in the panels), and the bottom row shows the corresponding combinations including SN.
Combinations without SN exhibit a pronounced anticorrelation between $z_\dagger$ and $H_0$ in several cases, while adding SN significantly reduces this degeneracy and compresses the $H_0$ posterior toward $\sim70~{\rm km\,s^{-1}\,Mpc^{-1}}$.
}

    \label{fig:z_vs_h}
\end{figure*}

Interpreting the reconstructed expansion history within the scalar-field framework yields an effective dark-energy density $\rho_{\rm DE}(z)$ and pressure $p_{\rm DE}(z)$, and hence $w_{\rm DE}(z)=p_{\rm DE}/\rho_{\rm DE}$ (Fig.~\ref{fig:rhode_pde_wde}).
Across all dataset combinations we find a sign change of $\rho_{\rm DE}$, with $\rho_{\rm DE}<0$ at higher redshift and $\rho_{\rm DE}>0$ toward the present.
We define the transition redshift $z_\dagger$ by $\rho_{\rm DE}(z_\dagger)=0$ and report it in Table~\ref{tab:results}.
The inferred $z_\dagger$ typically lies at $z_\dagger\gtrsim2$, while a subset of cases without SN but with an $H_0$ prior (notably CC+$H_0$, CC+ON-BAOtr+$H_0$, and CC+MnM-BAOtr+$H_0$) favor a lower transition around $z_\dagger\sim1.7$.
These lower-$z_\dagger$ cases also coincide with higher inferred $H_0$ ($\sim73$~km\,s$^{-1}$\,Mpc$^{-1}$) and a stronger present-day phantom preference, consistent with the anticorrelation between $z_\dagger$ and $H_0$ visible in Fig.~\ref{fig:z_vs_h}.
Adding SN generally reduces this degeneracy and pulls the reconstruction toward $H_0\simeq70$~km\,s$^{-1}$\,Mpc$^{-1}$.

A further qualitative feature visible in several reconstructions is an additional interval with $q(z)<0$ at intermediate redshift, around $z\sim 1.7$--$2.3$, in addition to the familiar late-time acceleration at $z\lesssim 0.5$--$0.7$.
In the examples shown in Fig.~\ref{fig:H_and_q} this behavior is most pronounced for the BAOtr-driven case and for CC+SN+H0DN, while for other combinations the credible intervals remain compatible with $q(z)\ge 0$ at comparable redshifts.
In some combinations this intermediate acceleration occurs in the vicinity of the inferred transition epoch $z_\dagger$ and coincides with rapid evolution of the effective DE-fluid variables, including large excursions of $w_{\rm DE}$ as $\rho_{\rm DE}$ approaches zero (a kinematic effect of the ratio $p_{\rm DE}/\rho_{\rm DE}$).
We therefore regard it as an intriguing hint whose robustness should be tested with improved high-redshift distance measurements and dedicated stability checks of the reconstruction assumptions.

We emphasize that when $z_\dagger$ lies near or beyond the upper edge of the data-supported redshift range (e.g.\ $z_\dagger\gtrsim2.4$ in several DESI-containing combinations), the transition is not directly localized by data points and becomes more sensitive to the reconstruction smoothness assumptions and the extrapolation between the last data-supported node and the boundary node at $z=3$ (cf. Sec.~\ref{subsec:methodology} and Appendix~\ref{app:reconstructions}).
Moreover, the inferred values of $z_\dagger$ (and slightly $w_0$) exhibit a mild dependence on the assumed $\Omega_{m0}$ used to translate the reconstructed kinematics into an effective $\rho_{\rm DE}(z)$ (see Appendix~A).

Figure~\ref{fig:Keff_and_V} shows representative posteriors for the effective kinetic contribution and total potential inferred from the reconstructed background.
Dataset combinations that favor a sign change in the effective kinetic contribution (as quantified by $\Delta\mathcal{X}$) are naturally interpreted within the two-field (quintom) framework developed in Sec.~\ref{sec:theory}, whereas a single fixed-sign kinetic scalar cannot realize such behavior.
These reconstructions therefore provide a direct background-level motivation for the two-field interpretation when the inferred $\Delta\mathcal{X}$ evolution requires it.

Finally, the last two columns of Table~\ref{tab:results} report $\Delta\chi^2_{\min}\equiv \chi^2_{\min,\,\rm rec}-\chi^2_{\min,\,\Lambda{\rm CDM}}$ and
$\Delta\ln B_{1,2}\equiv \ln(Z_{\Lambda{\rm CDM}}/Z_{\rm rec})$.
As expected, the reconstruction typically achieves a lower best-fit $\chi^2$ (i.e.\ $\Delta\chi^2_{\min}<0$) due to its higher flexibility (six parameters versus two for $\Lambda$CDM).
However, for all dataset combinations we find $\Delta\ln B_{1,2}>0$, indicating an overall preference for $\Lambda$CDM under our sign convention: the improvement in best fit is not sufficient to overcome the Occam penalty associated with the additional degrees of freedom.

\begin{figure*}
    \centering
    \includegraphics[width=0.34\linewidth]{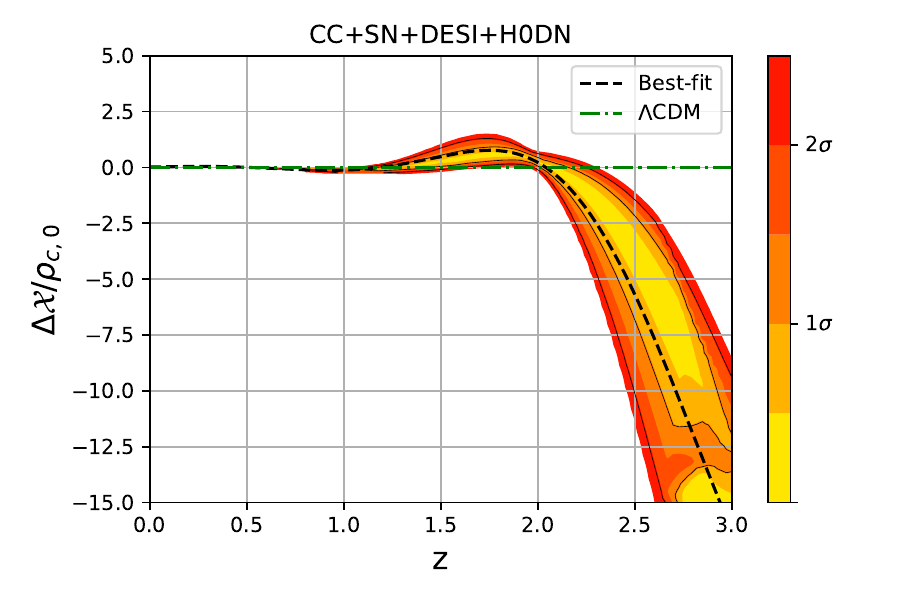}
    \includegraphics[width=0.34\linewidth]{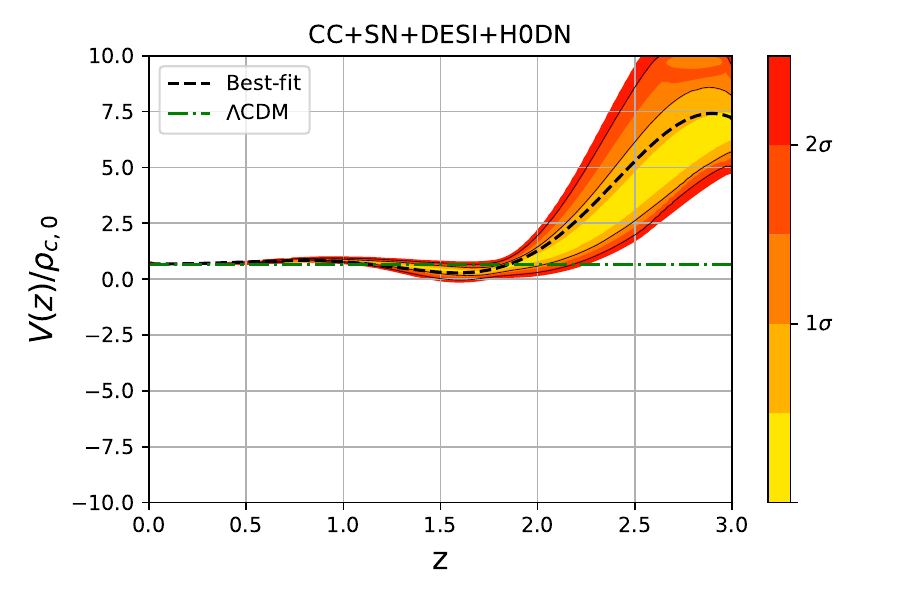}
    
    \includegraphics[width=0.34\linewidth]{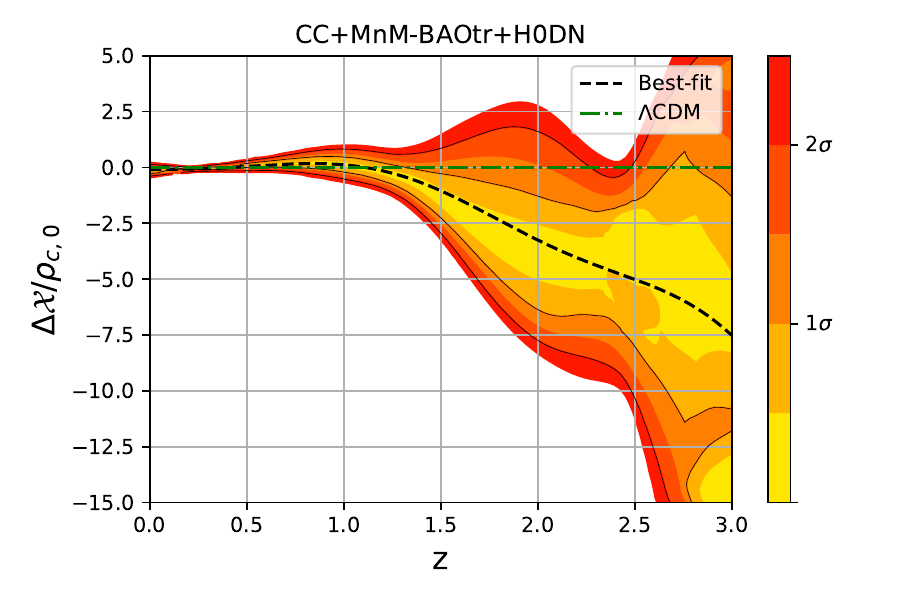}
    \includegraphics[width=0.34\linewidth]{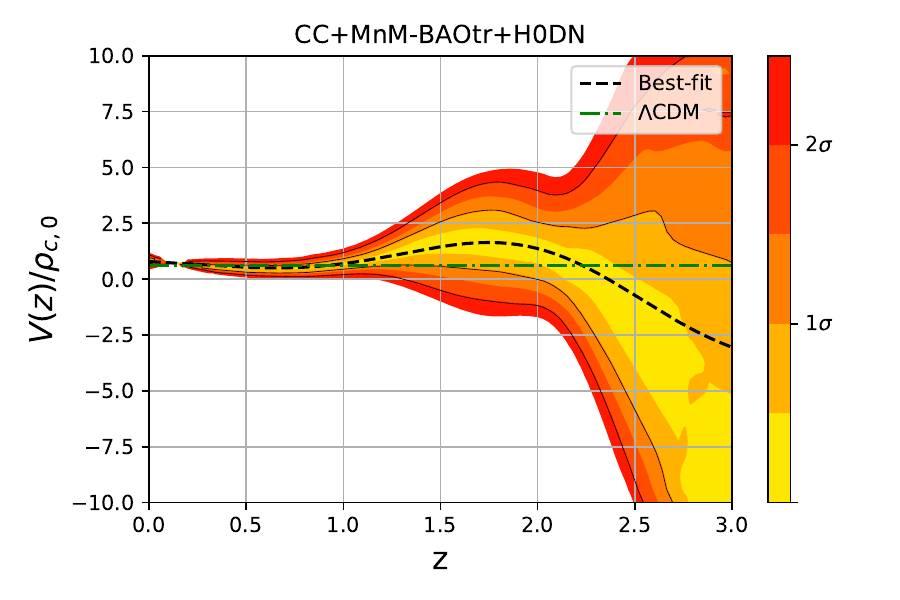}
    
    \includegraphics[width=0.34\linewidth]{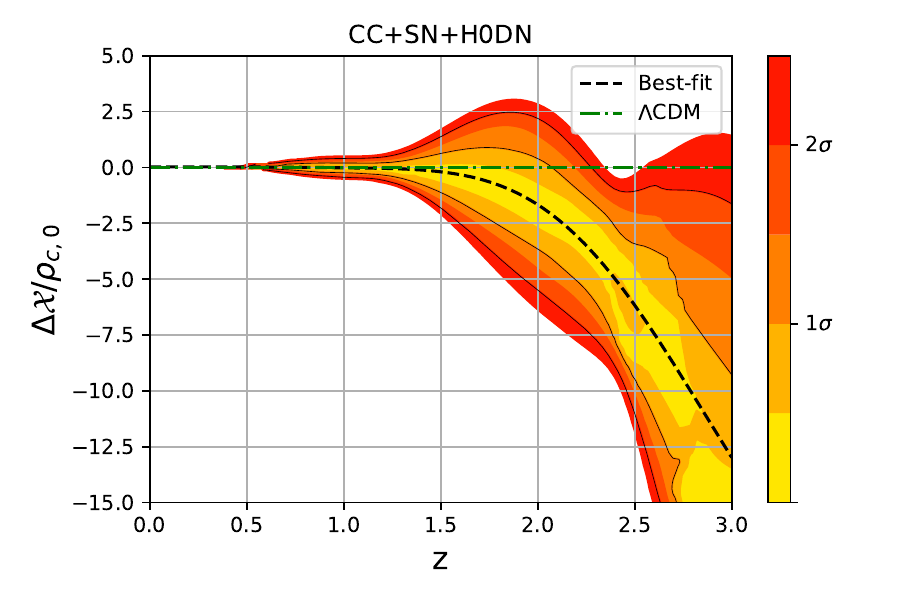}
    \includegraphics[width=0.34\linewidth]{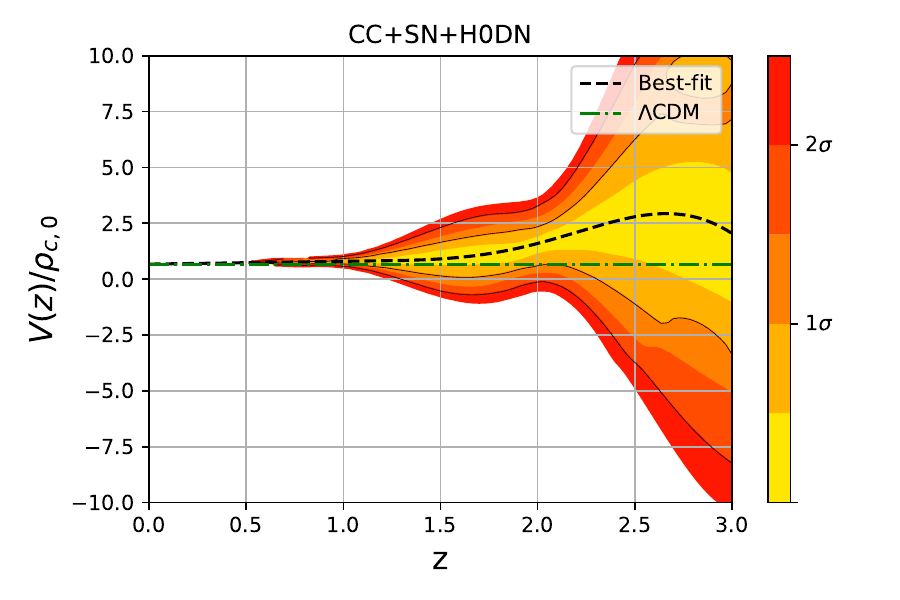}
    \caption{
    Posterior predictive bands for the effective kinetic contribution $\Delta\mathcal{X}(z)$ and the total effective potential (denoted $V(z)$ in the plot), inferred from the reconstructed background for the same three dataset combinations as in Fig.~\ref{fig:H_and_q} (top to bottom rows): CC+SN+DESI+H0DN, CC+MnM-BAOtr+H0DN, and CC+SN+H0DN.
    The color shading encodes the $\sigma$-equivalent credible level around the best-fit reconstruction, as indicated by the color bar in each panel (up to $\sim2.5\sigma$).
    The black dotted curve shows the best-fit reconstruction, while the green dotted curve shows the best-fit flat $\Lambda$CDM baseline for the same dataset combination.
    A change in the sign of $\Delta\mathcal{X}$ is naturally interpreted within the two-field (quintom) framework, in which the net kinetic contribution can change sign.
    Since there are no data in $2.4<z<3.0$, behavior in this interval should be interpreted cautiously.
    }
    \label{fig:Keff_and_V}
\end{figure*}

\section{Discussion}
\label{sec:disc}

The reconstruction presented in Sec.~\ref{sec:reconstruction} is best viewed as a diagnostic pipeline: the data constrain the expansion history $H(z)$ (or equivalently $E(z)$), and assuming GR at the background level we then map the reconstructed kinematics onto an effective dark-energy fluid and, subsequently, onto an effective scalar-field description.
This separation is essential for interpretation.
The kinematical reconstruction is the most direct data product, whereas quantities such as $\rho_{\rm DE}(z)$, $w_{\rm DE}(z)$, and the inferred transition redshift $z_\dagger$ additionally depend on the GR mapping and on the adopted matter-sector specification (most notably the assumed $\Omega_{m0}$ used in the reconstruction-to-fluid translation, and the BAO calibration through $r_d$).

A robust outcome of the analysis is that the reconstructed $H(z)$ admits nontrivial intermediate-redshift structure while remaining fully compatible with standard late-time acceleration at $z\lesssim 0.5$--$0.7$.
In several dataset combinations, the best-fit $q(z)$ suggests an additional interval with $q(z)<0$ around $z\sim 1.7$--$2.3$ (Sec.~\ref{subsec:results}).
At the qualitative level this is intriguing because it points to the possibility that the expansion history may contain a transient phase of accelerated expansion beyond the canonical late-time epoch.
At the same time, the statistical significance of this feature is dataset dependent and decreases as it approaches the upper edge of the data-supported range; it should therefore be regarded as a hint rather than a detection with the present data.
A particularly useful way to interpret this behavior is through the DE-fluid mapping: the intermediate-redshift feature tends to occur near the inferred transition epoch $z_\dagger$ and is accompanied by rapid evolution in the effective fluid variables, including large excursions of $w_{\rm DE}$ whenever $\rho_{\rm DE}$ becomes small.
Consistent with this picture, Fig.~\ref{fig:z_vs_h} shows that several combinations without SN exhibit a pronounced anticorrelation between $z_\dagger$ and $H_0$, while adding SN significantly reduces this degeneracy and compresses the $H_0$ posterior toward $\sim70~{\rm km\,s^{-1}\,Mpc^{-1}}$.
In particular, the BAOtr+$H_0$-prior combinations (e.g.\ CC+ON-BAOtr+$H_0$ and CC+MnM-BAOtr+$H_0$) tend to prefer $z_\dagger\sim 1.7$ and mean values closer to $q_0\simeq -1$ (Table~\ref{tab:results}), accompanied by the strongest present-day phantom preference among the cases considered; adding SN weakens these degeneracies.

The reconstruction-to-fluid mapping generically yields a sign change in the effective dark-energy density, with $\rho_{\rm DE}<0$ at higher redshift and $\rho_{\rm DE}>0$ toward the present, leading to the derived transition parameter $z_\dagger$ defined by $\rho_{\rm DE}(z_\dagger)=0$.
This is a striking phenomenological outcome; however, its interpretation requires care.
First, the location of $z_\dagger$ is not equally well localized by all dataset combinations: when $z_\dagger$ lies near or beyond the highest-redshift distance information, the inferred crossing becomes sensitive to the smoothness assumptions and boundary conditions of the reconstruction (notably the extrapolation between the last data-supported node and the boundary node at $z=3$).
Second, $z_\dagger$ depends systematically on the assumed $\Omega_{m0}$ used to translate $H(z)$ into $\rho_{\rm DE}(z)$ (Appendix~\ref{omegam_zdag_corr}): shifting $\Omega_{m0}$ changes the normalization $1-\Omega_{m0}$ and therefore moves the redshift at which $\rho_{\rm DE}$ crosses zero.
This dependence is illustrated explicitly in Appendix~\ref{omegam_zdag_corr} (Fig.~\ref{fig:omega_correlation}): for the representative case CC+SN+DESI+H0DN, lowering the assumed $\Omega_{m0}$ from $0.30$ to $0.27$ shifts the inferred transition to higher redshift, $z_\dagger: 2.48^{+0.04}_{-0.07}\rightarrow 2.59^{+0.06}_{-0.10}$, and correspondingly shifts the inferred $w_0$ slightly toward less negative values, $w_0: -1.03\pm0.06 \rightarrow -1.005\pm0.059$.
For these reasons, the most conservative statement is that our pipeline robustly identifies an effective sign-changing $\rho_{\rm DE}$ within the assumed mapping, while the precise localization of $z_\dagger$ should be interpreted as a derived, model- and assumption-dependent quantity, especially when it lies close to the data boundary.
Because BAO constrain distances in units of $r_d$, the assumed calibration/anchor for $r_d$ can propagate into the inferred intermediate-redshift structure and hence into the localization of $z_\dagger$.

It is also useful to place the reconstructed phenomenology in the context of sign-switching vacuum-energy scenarios, most notably the $\Lambda_{\rm s}$CDM framework and closely related extensions, in which the effective cosmological term changes sign at a transition epoch (often idealized as an AdS-to-dS switch) characterized by a transition redshift.
In that class of models the transition scale is not a cosmetic parameter: it governs where the expansion history departs from a strictly $\Lambda$-like evolution and therefore controls how efficiently one can reconcile early- and late-time distance information, with a ``sweet spot'' around $z_\dagger\sim{\cal O}(1.5$--$2)$ often found in joint analyses as the regime where multiple late-time discrepancies can be simultaneously mitigated (see, e.g., Refs.~\cite{Akarsu:2019hmw,Akarsu:2021fol,Akarsu:2022typ,Akarsu:2023mfb,Akarsu:2024eoo}).
While the present work does not perform a direct parameter inference within $\Lambda_{\rm s}$CDM, it is notable that the model-agnostic reconstruction repeatedly maps onto an effective sign change of $\rho_{\rm DE}(z)$ and yields transition scales that cluster either near $z_\dagger\sim 1.7$ (in the high-$H_0$ mode realized most clearly in BAOtr+$H_0$ combinations) or at somewhat higher redshift once SN are included.
This is qualitatively consistent with the degeneracy structure expected in sign-switching scenarios: moving the transition to lower redshift increases the leverage of late-time distances to raise the inferred $H_0$, and Fig.~\ref{fig:z_vs_h} shows this anticorrelation directly.
Moreover, in $\Lambda_{\rm s}$CDM the vacuum-like relation $p_{\Lambda_{\rm s}}=-\rho_{\Lambda_{\rm s}}$ implies $w\simeq-1$ away from the crossing, while $w$ becomes ill-defined at $\rho_{\Lambda_{\rm s}}=0$; correspondingly, the large excursions of $w_{\rm DE}$ near $z_\dagger$ in our reconstruction should be understood as a kinematic consequence of $\rho_{\rm DE}\to 0$, not as evidence for a violently dynamical equation of state.
Taken together, these points suggest that $\Lambda_{\rm s}$CDM-like sign-switching parametrizations provide a particularly sharp benchmark for interpreting our data-driven results and motivate a targeted follow-up in which a minimal $\Lambda_{\rm s}$CDM template is fit directly to the same dataset combinations under the same calibration assumptions. This should be read as contextual benchmarking, not as a detection of a specific model.

Within the scalar-field interpretation, it is crucial to distinguish two conceptually different ``crossings.''
A NEC boundary crossing $w_{\rm DE}=-1$ at finite $\rho_{\rm DE}\neq 0$ corresponds to $\rho_{\rm DE}+p_{\rm DE}=0$ and therefore to $\Delta\mathcal{X}=0$.
This can occur smoothly in the two-field (quintom) framework because the net kinetic contribution $\Delta\mathcal{X}\propto \dot Q^2-\dot P^2$ can change sign.
By contrast, a zero crossing $\rho_{\rm DE}=0$ renders the ratio $w_{\rm DE}=p_{\rm DE}/\rho_{\rm DE}$ ill-defined, producing the large excursions seen in the reconstructed $w_{\rm DE}(z)$.
As emphasized in Sec.~\ref{sec:theory}, this divergence is kinematic rather than a singularity of the background expansion: $\rho_{\rm DE}$ can pass smoothly through zero while $p_{\rm DE}$ remains finite.
This perspective helps interpret the intermediate-redshift behavior: large $|w_{\rm DE}|$ in the vicinity of $z_\dagger$ does not necessarily imply a violent physical event; rather, it signals that the effective fluid decomposition is passing through a point where $\rho_{\rm DE}$ is small.

The effective kinetic diagnostics in Fig.~\ref{fig:Keff_and_V} provide a direct background-level motivation for the two-field interpretation whenever $\Delta\mathcal{X}(z)$ requires a sign change.
A single real scalar with fixed-sign kinetic term cannot realize such an evolution without pathology; the sharpest obstruction is the monotonicity $\dot\rho_{\rm DE}=-3H\dot\phi^2\le0$ for a canonical single field, which forbids a smooth evolution from $\rho_{\rm DE}<0$ to $\rho_{\rm DE}>0$ during expansion.
By contrast, a quintom pair can accommodate the required sign changes of $\Delta\mathcal{X}$ while keeping the reconstructed $H(z)$ regular.
At the same time, one should regard the reconstructed ``potential'' as an effective quantity: in the two-field system it corresponds to the sum $U_1(Q)+U_2(P)$, and its decomposition into component potentials is not unique.
Accordingly, apparent features such as negative values of the total effective potential in some reconstructions should not be over-interpreted as excluding a consistent multi-field description.

A useful diagnostic emerging from the results is the appearance of $q(z)<-1$ (or mean values $q_0\lesssim -1$ in some combinations).
Since $q=-1-\dot H/H^2$, this corresponds to $\dot H>0$, which within GR implies $\rho_{\rm tot}+p_{\rm tot}<0$ for the total cosmic fluid, i.e.\ an effective violation of the null energy condition.
While such behavior can arise in effective descriptions, it is theoretically delicate and often points to instability or to an incomplete modeling of the underlying degrees of freedom.
In our reconstructions, $q<-1$ is most prominent precisely where the reconstruction is least constrained: at very low redshift in combinations lacking direct low-$z$ leverage for $H'(z)$, and at high redshift approaching the boundary node at $z=3$.
This strongly suggests that the present indications of super-acceleration are dominated by limited anchoring and boundary effects rather than representing a robust physical inference.
A decisive assessment requires higher-precision low-$z$ measurements of $H(z)$ and improved high-$z$ distance information, together with dedicated robustness tests of the reconstruction choices (node placement and smoothness).

From a statistical standpoint, the model-comparison results are also informative.
The reconstruction, by construction, is flexible and therefore achieves a better best-fit $\chi^2_{\min}$ than the two-parameter $\Lambda$CDM baseline across all combinations.
However, the Bayesian evidence uniformly favors $\Lambda$CDM once the additional degrees of freedom are penalized (Table~\ref{tab:results}).
This should be read as ``the data do not yet require the additional freedom,'' rather than as a statement that the reconstructed features are ruled out.
In this sense, the reconstruction identifies where deviations \emph{could} be accommodated and where future data would be most decisive: intermediate redshifts around $z\sim 1.7$--$2.3$, and those dataset combinations (notably involving BAOtr) that exhibit the strongest degeneracies between $H_0$ and $z_\dagger$.

Several concrete robustness checks are suggested by the present analysis.
First, shifting the boundary node from $z=3$ to the highest-redshift data point (or adding an additional node at $z\simeq 2.34$) would directly test whether the inferred behavior near $z\gtrsim 2.4$ is extrapolation driven.
Second, varying the kernel smoothness parameter $\theta$ within a reasonable range would quantify the sensitivity of derivative-based inferences, including the intermediate $q<0$ window and any excursions to $q<-1$.
Third, marginalizing over $\Omega_{m0}$ (rather than fixing it) would propagate matter-density uncertainty into the inferred $\rho_{\rm DE}(z)$ and $z_\dagger$, and would allow a cleaner assessment of which aspects of the sign change are truly data-driven.
Finally, a mock-data reconstruction based on a fiducial $\Lambda$CDM expansion history, evaluated at the same redshifts with comparable uncertainties, would provide an important null test for spurious zero crossings and intermediate-redshift acceleration features.

In summary, the reconstruction identifies a coherent phenomenological pattern: a late-time expansion history consistent with standard acceleration at low redshift, accompanied in some dataset combinations by hints of additional intermediate-redshift structure that maps onto a sign-changing effective $\rho_{\rm DE}$ and motivates a two-field (quintom) scalar interpretation at the background level.
The present data favor $\Lambda$CDM in Bayesian model comparison, but the reconstruction highlights where improved measurements (particularly at $z\sim 2$ and at the low-$z$ anchor) will most strongly test the persistence of these features.

\section{Conclusions}
\label{sec:conc}

We have presented a data-driven reconstruction of the late-time expansion history by modeling the reduced Hubble rate $E(z)\equiv H(z)/H_0$ with a node-based Gaussian-process-kernel interpolant and constraining it with combinations of cosmic chronometers, Type~Ia supernovae (Pantheon+), BAO measurements from SDSS and DESI, transversal BAO data, and external $H_0$ priors (SH0ES and H0DN).
This framework is designed to infer the kinematics directly from the data without imposing a specific dark-energy parametrization; assuming GR at the background level, we then map the reconstructed $H(z)$ onto an effective dark-energy fluid and, subsequently, onto an effective scalar-field description.

At the kinematical level, the reconstruction remains fully consistent with standard late-time acceleration at $z\lesssim 0.5$--$0.7$.
At the same time, several dataset combinations admit nontrivial intermediate-redshift structure, including hints of an additional interval with $q(z)<0$ around $z\sim 1.7$--$2.3$.
Given current uncertainties and the proximity of this feature to the upper edge of the data-supported range in some combinations, we regard it as suggestive rather than decisive; improved high-redshift distance measurements and a denser low-$z$ anchor for $H(z)$ and its derivatives will be essential to assess its robustness.

Within the GR-based effective-fluid mapping, a central phenomenological outcome is the repeated appearance of an effective sign change in the reconstructed dark-energy density, with $\rho_{\rm DE}<0$ at higher redshift and $\rho_{\rm DE}>0$ toward the present.
We characterized the transition by the derived redshift $z_\dagger$ defined through $\rho_{\rm DE}(z_\dagger)=0$.
While the emergence of a sign change is robust within the adopted mapping, we emphasized that the localization of $z_\dagger$ is a derived, dataset- and assumption-dependent quantity: it shifts systematically with the assumed $\Omega_{m0}$ and becomes increasingly sensitive to reconstruction/extrapolation choices when it lies near the boundary of the data-supported range.
Correspondingly, the large excursions of $w_{\rm DE}(z)=p_{\rm DE}/\rho_{\rm DE}$ near $z_\dagger$ should be interpreted as a kinematic consequence of $\rho_{\rm DE}\to 0$, not as a singularity in the reconstructed expansion history.

The scalar-field interpretation clarifies the minimal theoretical requirements implied by the reconstructed phenomenology.
A single \emph{canonical} scalar field cannot realize a smooth evolution from $\rho_{\rm DE}<0$ to $\rho_{\rm DE}>0$ during an expanding phase, since $\dot\rho_{\rm DE}=-3H\dot\phi^2\le0$ for $H>0$.
By contrast, a single \emph{phantom} scalar satisfies $\dot\rho_{\rm DE}=+3H\dot\phi^2\ge0$ and can therefore accommodate a zero crossing $\rho_{\rm DE}=0$ and a subsequent transition to $\rho_{\rm DE}>0$, with the understanding that $w_{\rm DE}=p_{\rm DE}/\rho_{\rm DE}$ is ill-defined at the crossing and that a phantom field should be regarded as an effective description.
More generally, when the reconstructed evolution requires the effective kinetic contribution to change sign (or equivalently permits a smooth phantom-divide crossing at finite $\rho_{\rm DE}>0$), a two-field (quintom) framework provides the minimal scalar-field interpretation: the net kinetic contribution $\Delta\mathcal{X}\propto \dot Q^2-\dot P^2$ can change sign while the background remains regular, and it naturally distinguishes a dark NEC boundary crossing ($\Delta\mathcal{X}=0$ at $\rho_{\rm DE}\neq0$) from the separate notion of a density zero crossing ($\rho_{\rm DE}=0$). The dark NEC boundary allows a consistent description of the crossing between $w_{\rm DE} > -1$ and $w_{\rm DE} < -1$, and reduces to a PDL crossing when restricted to $\rho_{\rm DE}(z) > 0$.

From a statistical perspective, the flexible reconstruction improves the best-fit $\chi^2_{\min}$ relative to the two-parameter flat $\Lambda$CDM baseline across all dataset combinations, as expected.
However, Bayesian evidence comparisons favor $\Lambda$CDM once the additional degrees of freedom are penalized, indicating that current data do not yet \emph{require} the extra freedom of the reconstruction.
The appropriate reading is therefore not that the reconstructed features are ruled out, but that the present data allow them while preferring the minimal model under Occam’s razor.

Several immediate robustness and follow-up analyses are well motivated by the present results: (i) varying the boundary-node placement and reconstruction smoothness to diagnose extrapolation-driven features, particularly near $z\gtrsim 2.4$ and in derivative-based quantities; (ii) marginalizing over $\Omega_{m0}$ (and, where relevant, BAO calibration assumptions) to propagate matter-sector uncertainty self-consistently into $(\rho_{\rm DE},w_{\rm DE},z_\dagger)$; and (iii) performing mock-$\Lambda$CDM null tests to quantify the rate of spurious zero crossings under the same pipeline.
With forthcoming improvements in high-redshift distance information and low-redshift expansion-rate measurements, the reconstruction strategy developed here provides a systematic route to testing whether the intermediate-redshift phenomenology highlighted by current data persists or disappears, and to sharpening the theoretical interpretation of any such departures from the minimal $\Lambda$CDM expansion history.

\begin{acknowledgments}
Project BridgingCosmology is financed by Xjenza Malta and the Scientific and Technological Research Council of T\"{U}B\.{I}TAK, through the Xjenza Malta--T\"{U}B\.{I}TAK 2024 Joint Call for R\&I projects.
\"{O}.A.\ acknowledges support from the Turkish Academy of Sciences through the Outstanding Young Scientist Award programme (T\"{U}BA-GEB\.{I}P).
This work was supported by T\"{U}B\.{I}TAK under Grant No.~124N627.
M.C.\ and L.E.\ acknowledge support from T\"{U}B\.{I}TAK through postdoctoral researcher fellowships associated with Grant No.~124N627.
The authors thank T\"{U}B\.{I}TAK for their support.
This article is based upon work from COST Action CA21136 \emph{Addressing observational tensions in cosmology with systematics and fundamental physics} (CosmoVerse), supported by COST (European Cooperation in Science and Technology).
This initiative is part of the PRIMA Programme supported by the European Union.
\end{acknowledgments}

\bibliography{references}

\appendix
\section{Effect of the assumed \texorpdfstring{$\Omega_{m0}$}{}}
\label{omegam_zdag_corr}

In standard cosmological parameter inference, the present-day matter density parameter $\Omega_{m0}$ (often reported in the form $\Omega_{m0}h^2$) is typically sampled together with the background parameters.
In this work, however, we reconstruct the expansion history directly through $E(z)$, which removes the need to sample $\Omega_{m0}$ at the reconstruction stage.
Nevertheless, to interpret the reconstructed kinematics within the GR-based effective-fluid and scalar-field mapping, a value of $\Omega_{m0}$ must be specified.
Since this choice sets the present-day normalization of the matter contribution, it can propagate into derived quantities inferred from the mapping.
Here we illustrate how varying the assumed $\Omega_{m0}$ affects our results.

\begin{figure}[t!]
    \centering
    \includegraphics[trim=0mm 0mm 0mm 20mm, clip, width=0.88\linewidth]{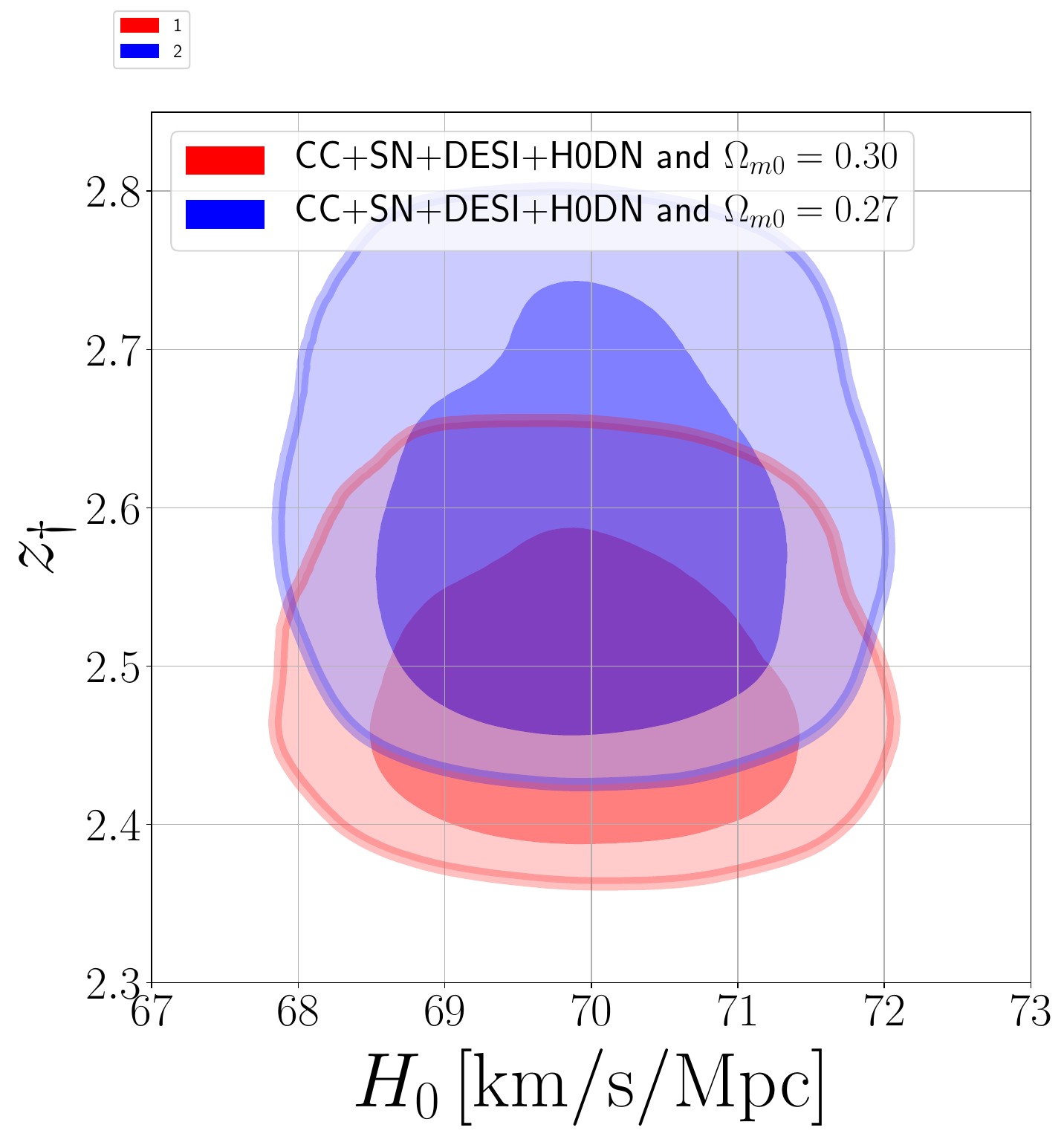}
    \caption{
    Two-dimensional marginalized posterior distribution for $z_\dagger$ versus $H_0$ illustrating the sensitivity of the inferred transition epoch to the assumed $\Omega_{m0}$ (shown here for the same dataset combination and $H_0$ prior as used in the underlying run).
    Lower $\Omega_{m0}$ shifts the inferred transition to higher $z_\dagger$, and vice versa.
    }
    \label{fig:omega_correlation}
\end{figure}

The most significant impact is observed in the derived transition redshift $z_\dagger$ and, to a lesser extent, in the inferred present-day equation-of-state parameter $w_0$.
For $z_\dagger$, the effect is straightforward: although the sign change of $\rho_{\rm DE}(z)$ persists, larger (smaller) values of $\Omega_{m0}$ shift the inferred transition to lower (higher) redshift.
This occurs because the overall shape of $\rho_{\rm DE}(z)$ is largely preserved, while its normalization is shifted by the change in the present-day matter contribution, which scales with $\Omega_{m0}$.
As a consequence, the redshift at which $\rho_{\rm DE}$ crosses zero moves systematically.
As an explicit example, Fig.~\ref{fig:omega_correlation} shows that changing the assumed value from $\Omega_{m0}=0.30$ to $\Omega_{m0}=0.27$ shifts the inferred transition from
$z_\dagger=2.48^{+0.04}_{-0.07}$ to $z_\dagger=2.59^{+0.06}_{-0.10}$ for the representative CC+SN+DESI+H0DN case.

For $w_0$, the dependence can be understood from the scalar-field mapping.
For a fixed reconstructed expansion history, increasing the present-day dark-energy density (which occurs for a smaller assumed $\Omega_{m0}$) requires a redistribution between the effective kinetic and potential contributions in the scalar-field interpretation.
This typically shifts the inferred $w_0$ toward less negative values.
For the same representative case, we find $w_0=-1.03\pm0.06$ for $\Omega_{m0}=0.30$ and $w_0=-1.005\pm0.059$ for $\Omega_{m0}=0.27$.\\

\section{Complete atlas of reconstructions}
\label{app:reconstructions}

For completeness, we provide the full atlas of posterior predictive reconstructions for all dataset combinations considered in this work.
For each combination we show the reconstructed kinematical quantities $H(z)$, $H(z)/(1+z)$, and $q(z)$, together with the corresponding effective dark-energy-fluid variables $\rho_{\rm DE}(z)$, $p_{\rm DE}(z)$, and $w_{\rm DE}(z)$ inferred from the GR-based mapping, and the effective scalar-field diagnostics $\Delta\mathcal{X}(z)$ and $U(z)$ (denoted $V(z)$ in the plots).
In all panels, the color shading encodes the $\sigma$-equivalent credible level around the best-fit reconstruction as indicated by the color bar, the black dotted curve shows the best-fit reconstruction, and the green dotted curve shows the best-fit flat $\Lambda$CDM baseline for the same dataset combination.

We remind the reader that our node-based reconstruction includes a boundary node at $z=3$, while the highest-redshift distance information in the datasets used here lies at $z\simeq 2.3$ (BAO) and $z\simeq 2.26$ (SN).
Accordingly, behavior in the interval $2.4<z<3.0$ is effectively extrapolation-dominated and should be interpreted with caution, especially for derivative-based quantities such as $q(z)$ and for derived fluid variables that depend on the mapping.

 \begin{figure*}[t!]
     \centering
       \makebox[10cm][c]{
      \includegraphics[trim = 0mm  0mm 0mm 0mm, clip, width=8.9cm, height=5.3cm]{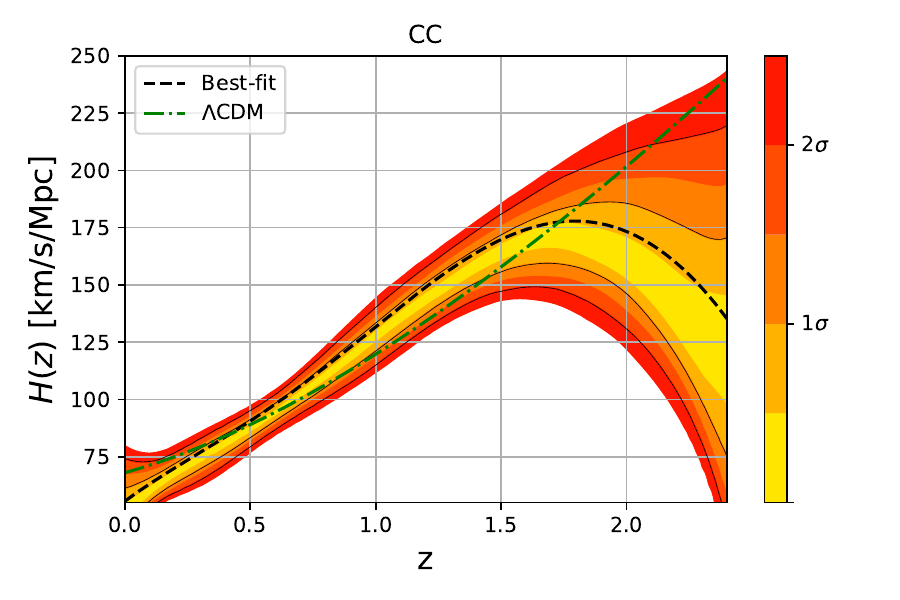}
      \includegraphics[trim = 0mm  0mm 0mm 0mm, clip, width=8.9cm, height=5.3cm]{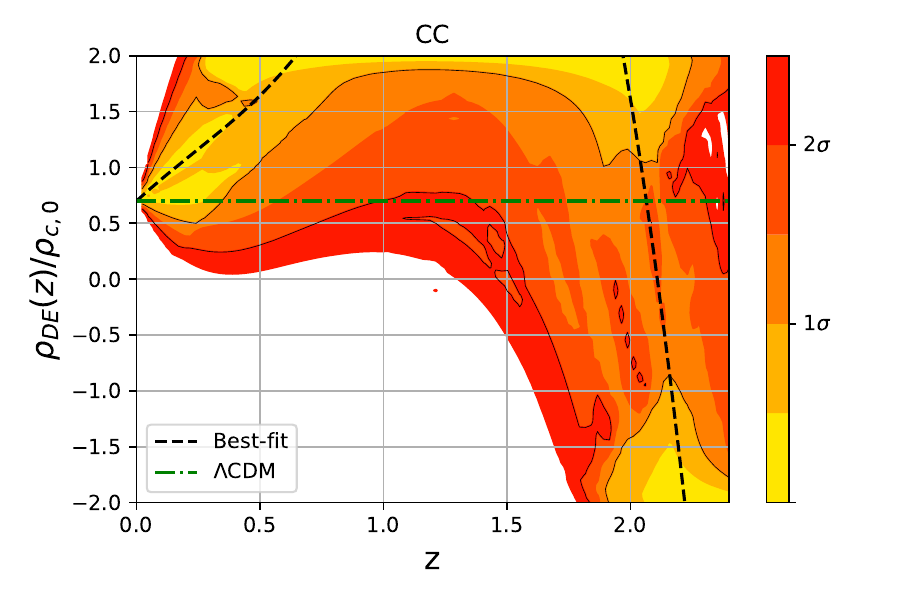}
      }
     \makebox[10cm][c]{
      \includegraphics[trim = 0mm  0mm 0mm 0mm, clip, width=8.9cm, height=5.3cm]{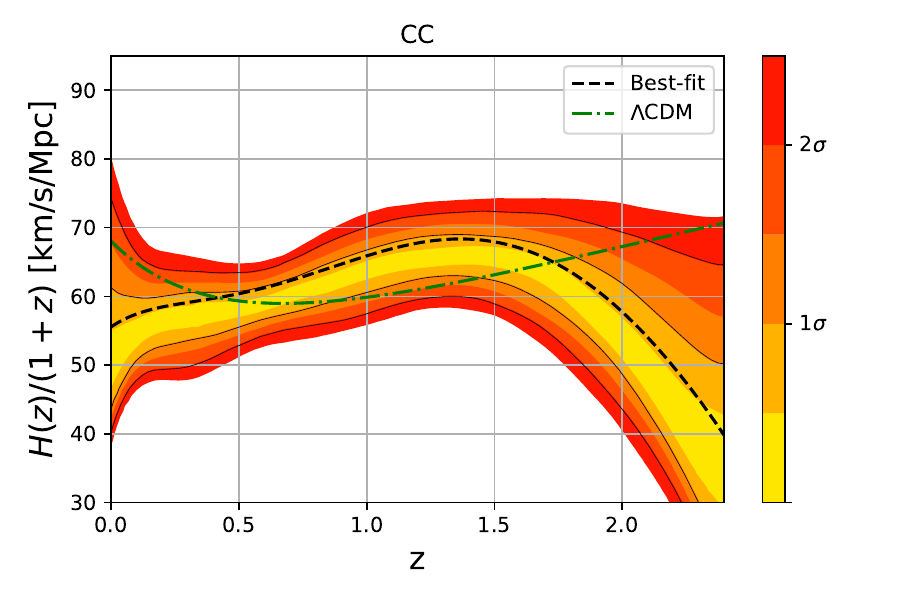}
      \includegraphics[trim = 0mm  0mm 0mm 0mm, clip, width=8.9cm, height=5.3cm]{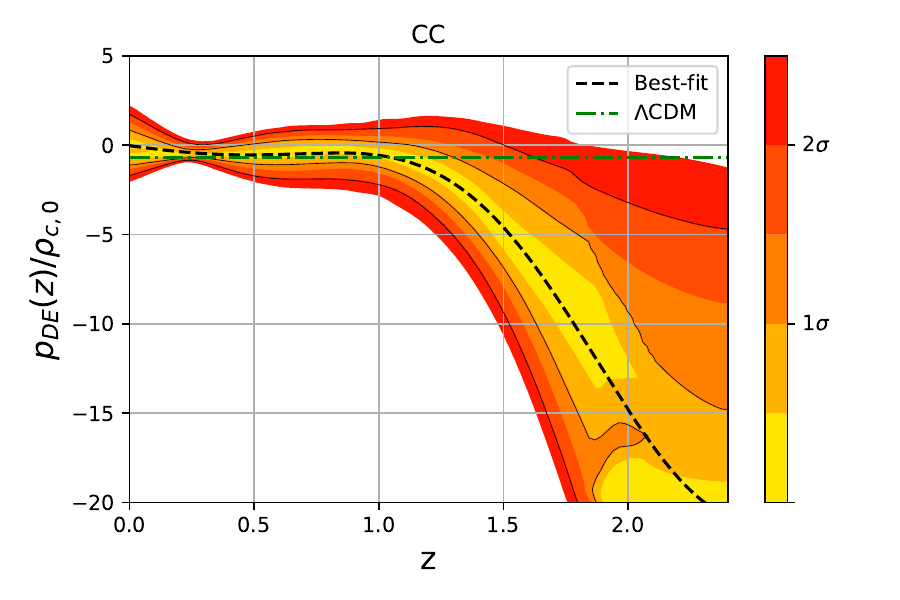}
      }
     \makebox[10cm][c]{
      \includegraphics[trim = 0mm  0mm 0mm 0mm, clip, width=8.9cm, height=5.3cm]{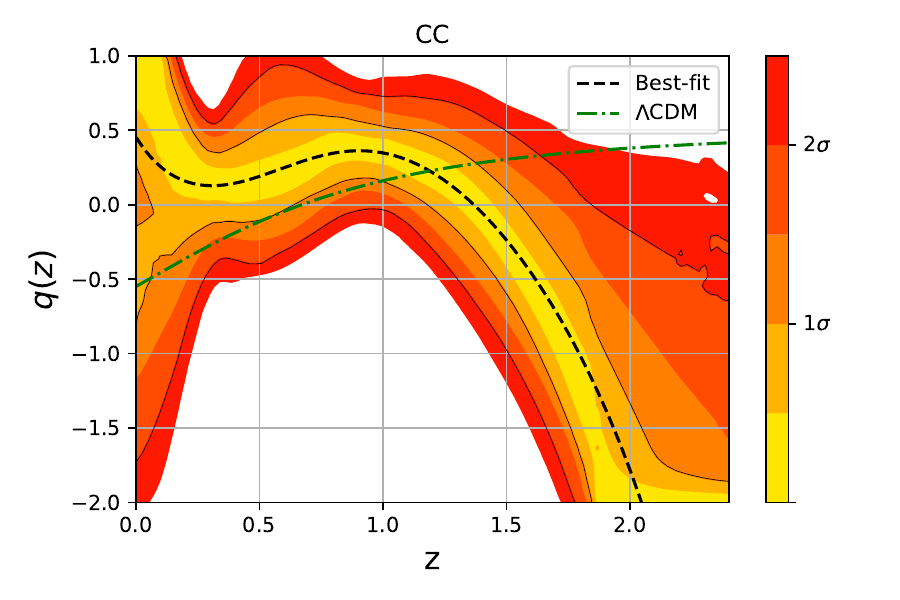}
      \includegraphics[trim = 0mm  0mm 0mm 0mm, clip, width=8.9cm, height=5.3cm]{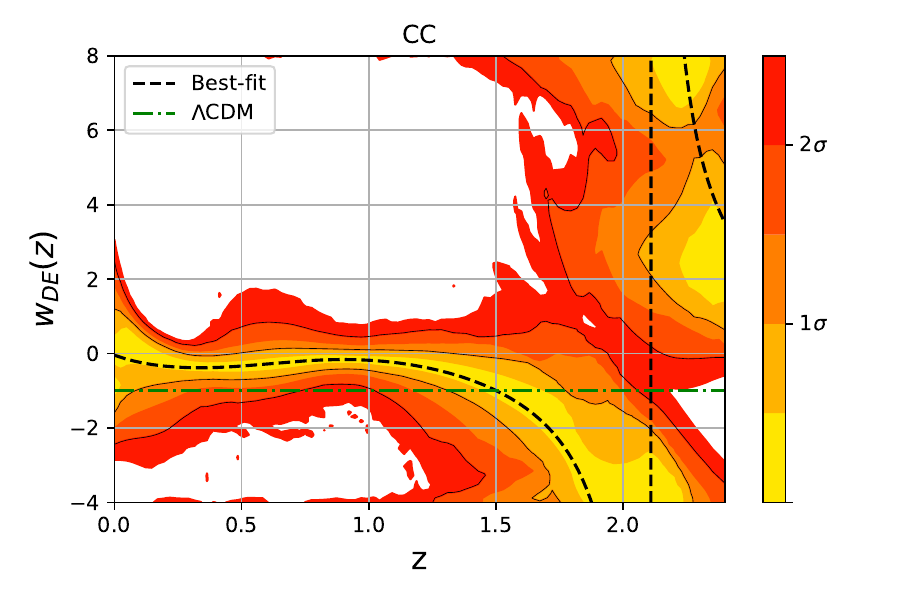}
      }
      \makebox[10cm][c]{
      \includegraphics[trim = 0mm  0mm 0mm 0mm, clip, width=8.9cm, height=5.3cm]{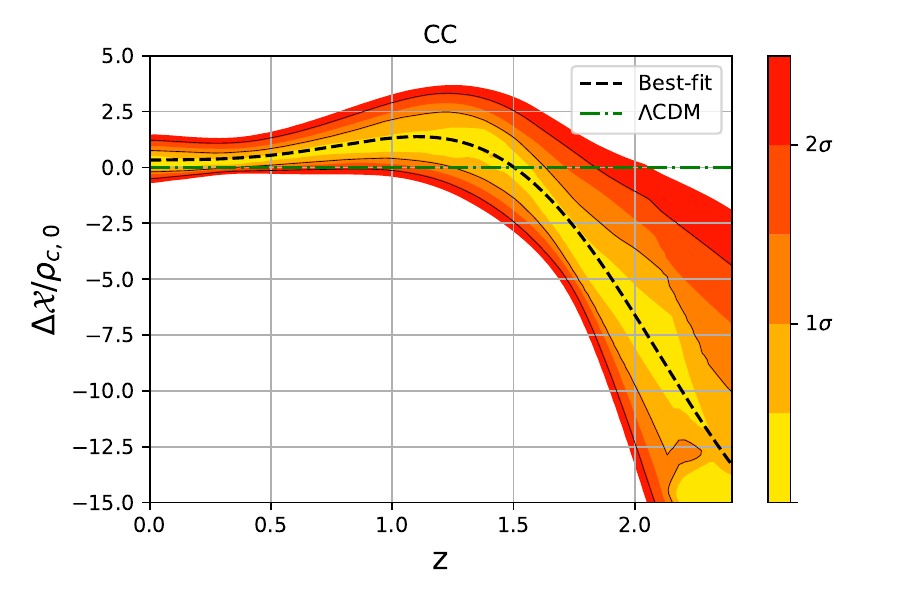}
      \includegraphics[trim = 0mm  0mm 0mm 0mm, clip, width=8.9cm, height=5.3cm]{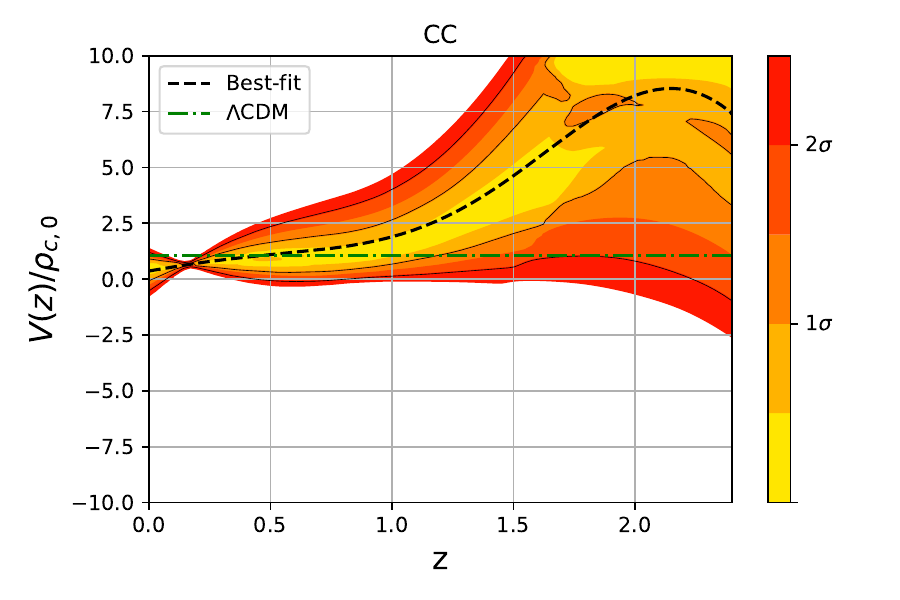}
    }
  \caption{Results of the reconstruction of $H(z)$ for the dataset combination of CC. From top-to-bottom and left-to-right we have: $H(z)$, $H(z)/(1+z)$, $q(z)$, $\Delta\mathcal{X} / \rho_{c,0}$, $\rho_{DE}/\rho_{c,0}$, $p_{DE}/\rho_{c,0}$, $w_{DE}$, and $V(z) / \rho_{c,0}$. An important thing to note and clarify is that the last node is located at $z=3.0$, which means that it cannot be constrained by data. As such, high-redshift results around this region should be taken as merely statistical noise. }\label{fig:GP_cc}
 \end{figure*}

 \begin{figure*}[t!]
     \centering
       \makebox[10cm][c]{
      \includegraphics[trim = 0mm  0mm 0mm 0mm, clip, width=8.9cm, height=5.3cm]{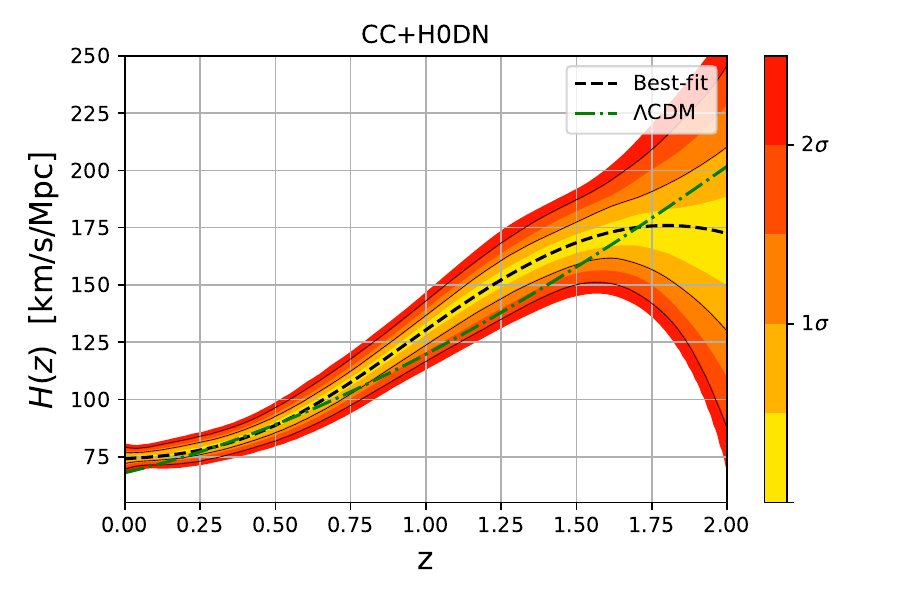}
      \includegraphics[trim = 0mm  0mm 0mm 0mm, clip, width=8.9cm, height=5.3cm]{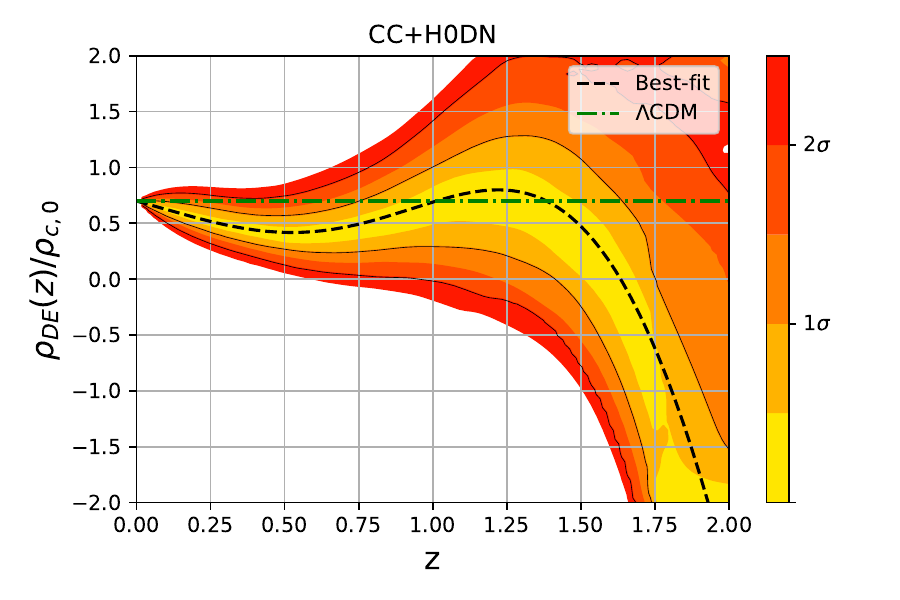}
      }
     \makebox[10cm][c]{
      \includegraphics[trim = 0mm  0mm 0mm 0mm, clip, width=8.9cm, height=5.3cm]{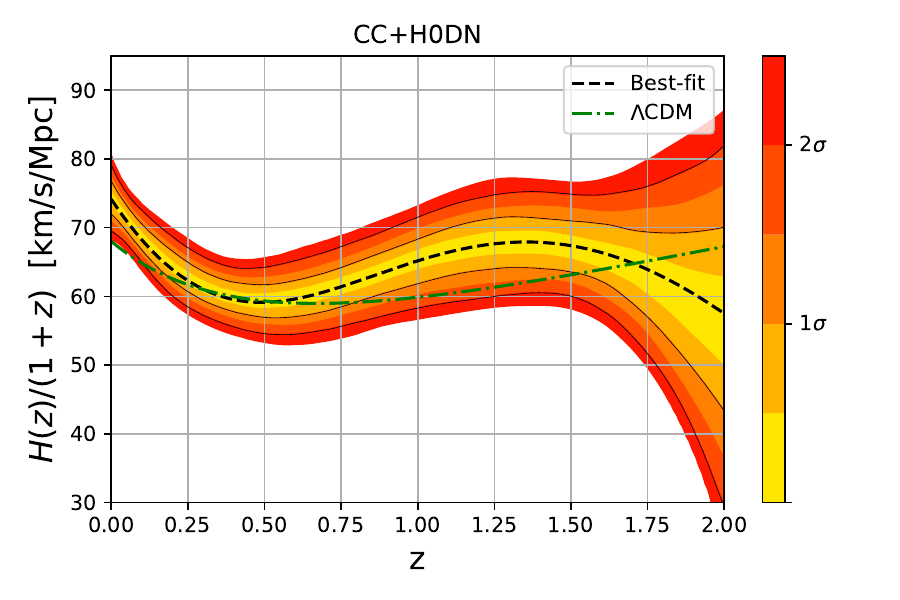}
      \includegraphics[trim = 0mm  0mm 0mm 0mm, clip, width=8.9cm, height=5.3cm]{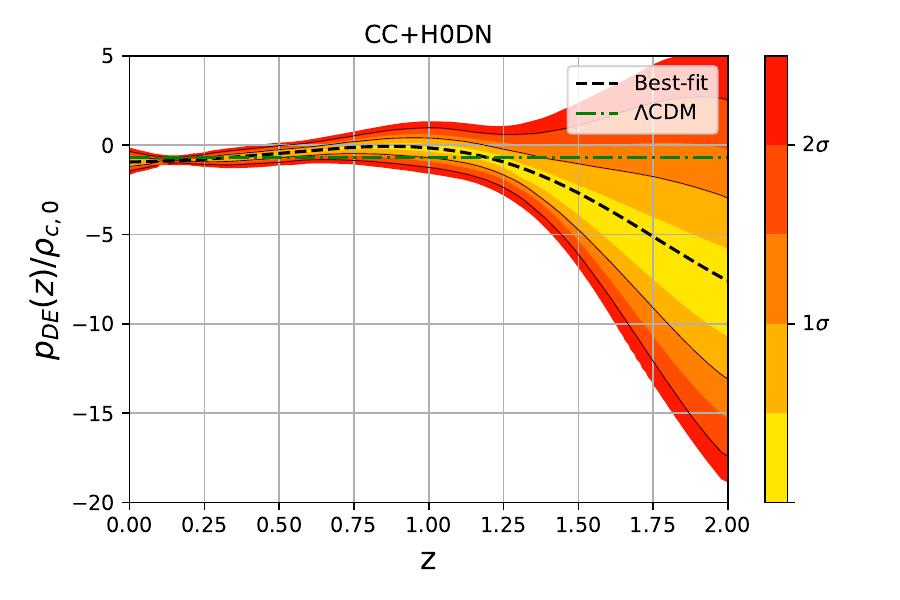}
      }
     \makebox[10cm][c]{
      \includegraphics[trim = 0mm  0mm 0mm 0mm, clip, width=8.9cm, height=5.3cm]{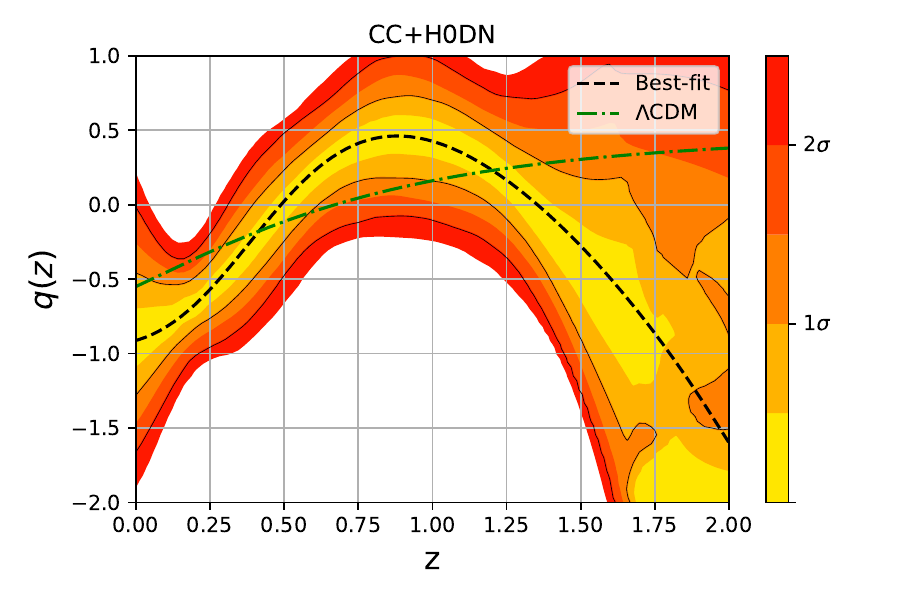}
      \includegraphics[trim = 0mm  0mm 0mm 0mm, clip, width=8.9cm, height=5.3cm]{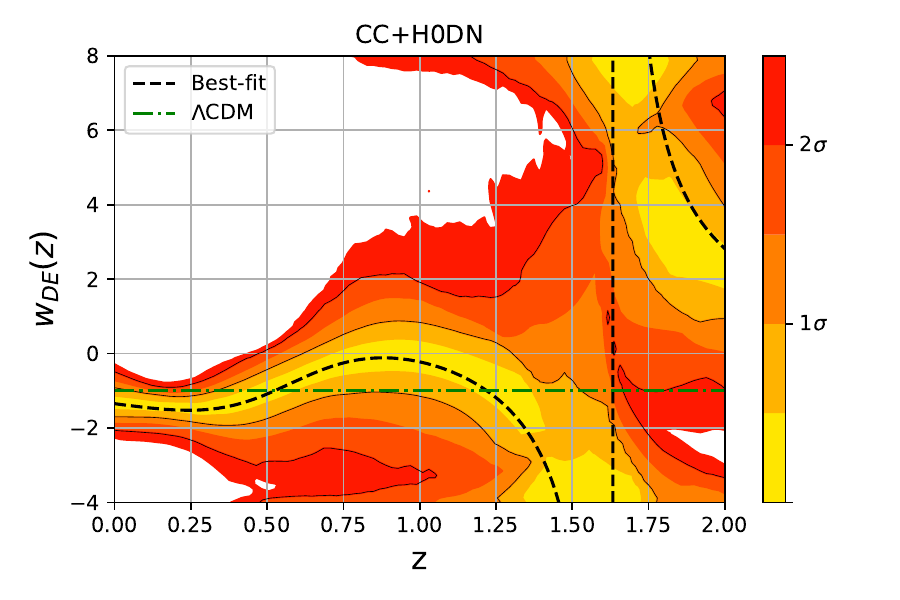}
      }
      \makebox[10cm][c]{
      \includegraphics[trim = 0mm  0mm 0mm 0mm, clip, width=8.9cm, height=5.3cm]{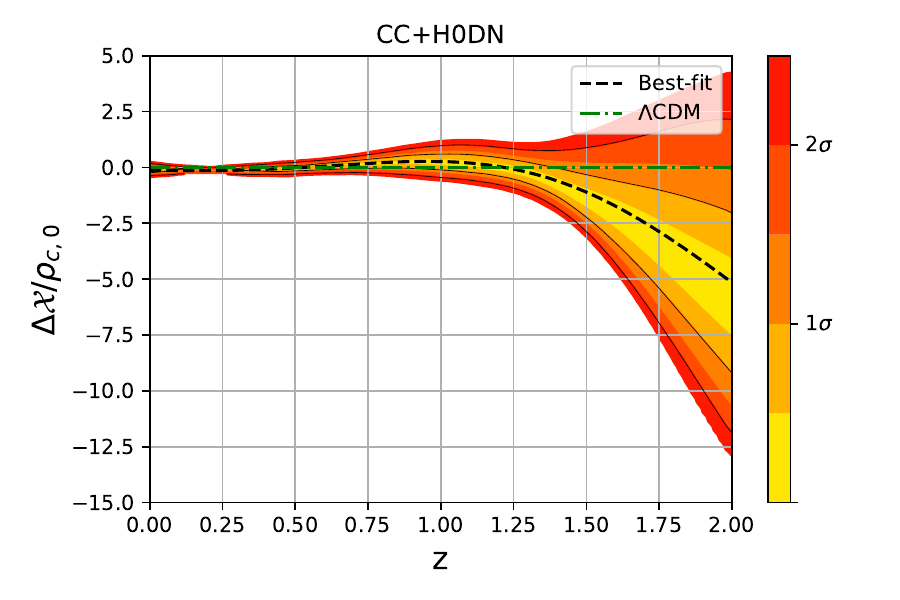}
      \includegraphics[trim = 0mm  0mm 0mm 0mm, clip, width=8.9cm, height=5.3cm]{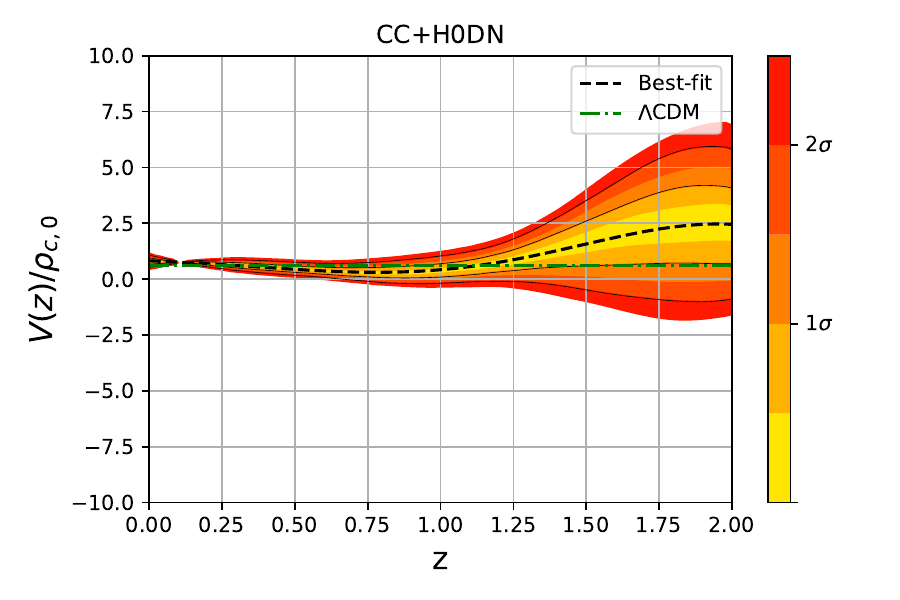}
      }
  \caption{Results of the reconstruction of $H(z)$ for the dataset combination of CC+H0DN. From top-to-bottom and left-to-right we have: $H(z)$, $H(z)/(1+z)$, $q(z)$, $\Delta\mathcal{X} / \rho_{c,0}$, $\rho_{DE}/\rho_{c,0}$, $p_{DE}/\rho_{c,0}$, $w_{DE}$, and $V(z) / \rho_{c,0}$. An important thing to note and clarify is that the last node is located at $z=3.0$, which means that it cannot be constrained by data. As such, high-redshift results around this region should be taken as merely statistical noise. }\label{fig:GP_cc_h0dn}
 \end{figure*}

 \begin{figure*}[t!]
     \centering
       \makebox[10cm][c]{
      \includegraphics[trim = 0mm  0mm 0mm 0mm, clip, width=8.9cm, height=5.3cm]{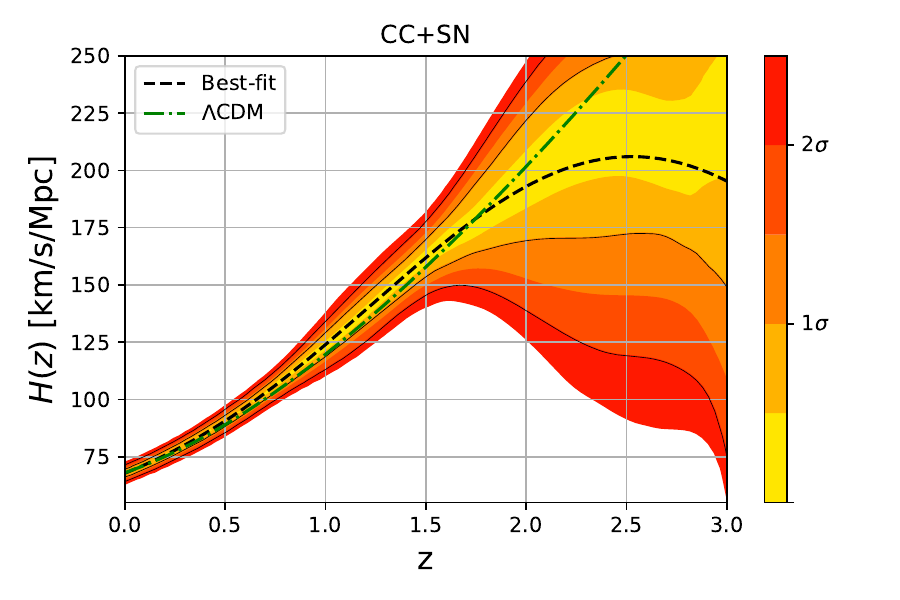}
      \includegraphics[trim = 0mm  0mm 0mm 0mm, clip, width=8.9cm, height=5.3cm]{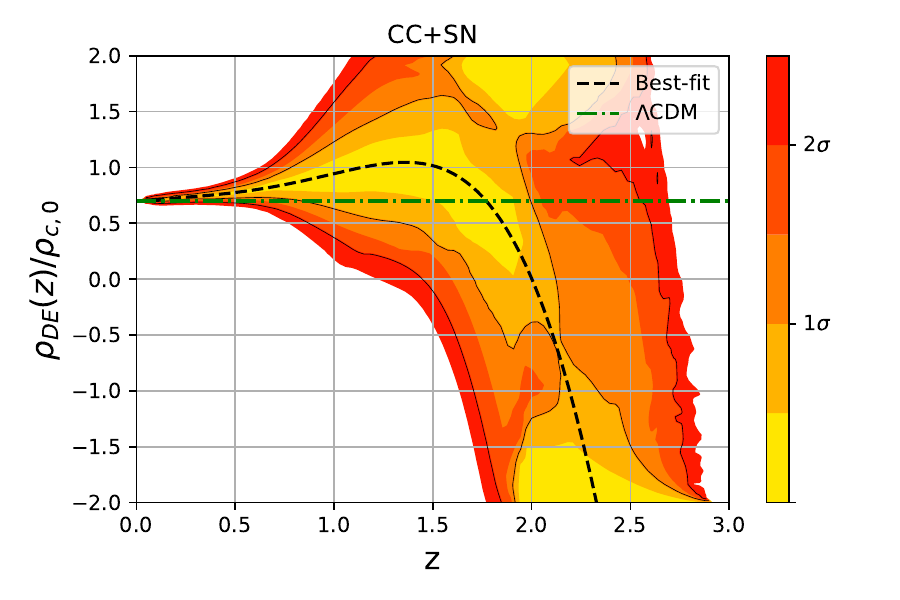}
      }
     \makebox[10cm][c]{
      \includegraphics[trim = 0mm  0mm 0mm 0mm, clip, width=8.9cm, height=5.3cm]{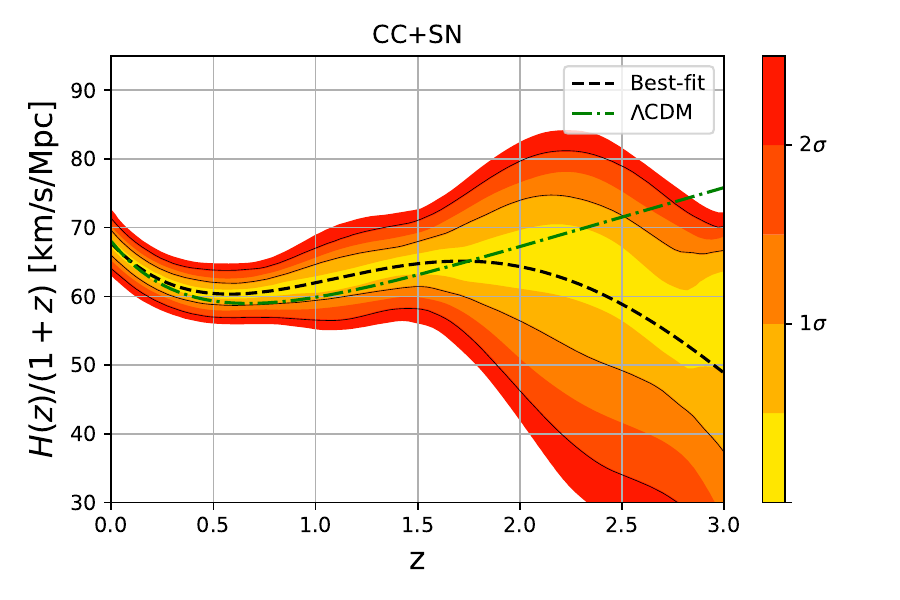}
      \includegraphics[trim = 0mm  0mm 0mm 0mm, clip, width=8.9cm, height=5.3cm]{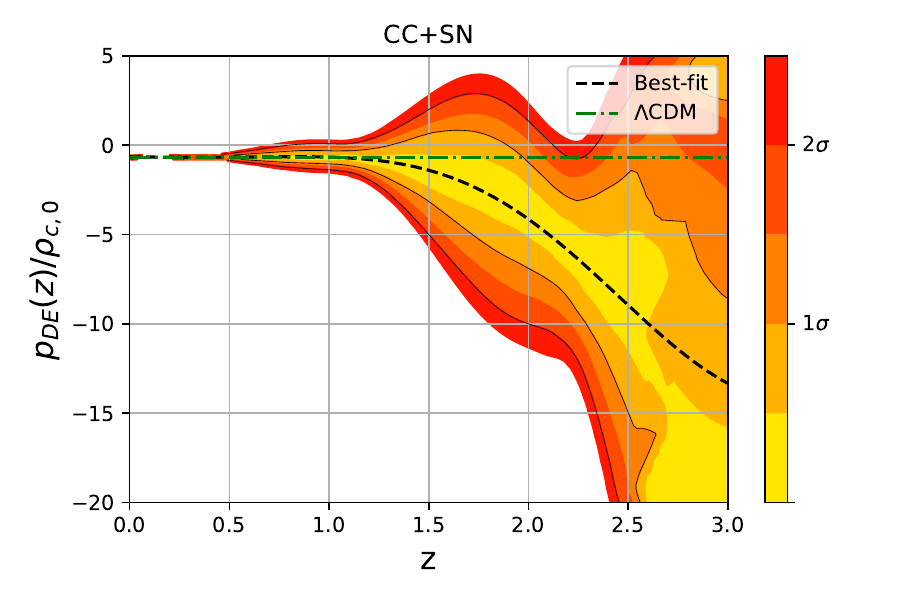}
      }
     \makebox[10cm][c]{
      \includegraphics[trim = 0mm  0mm 0mm 0mm, clip, width=8.9cm, height=5.3cm]{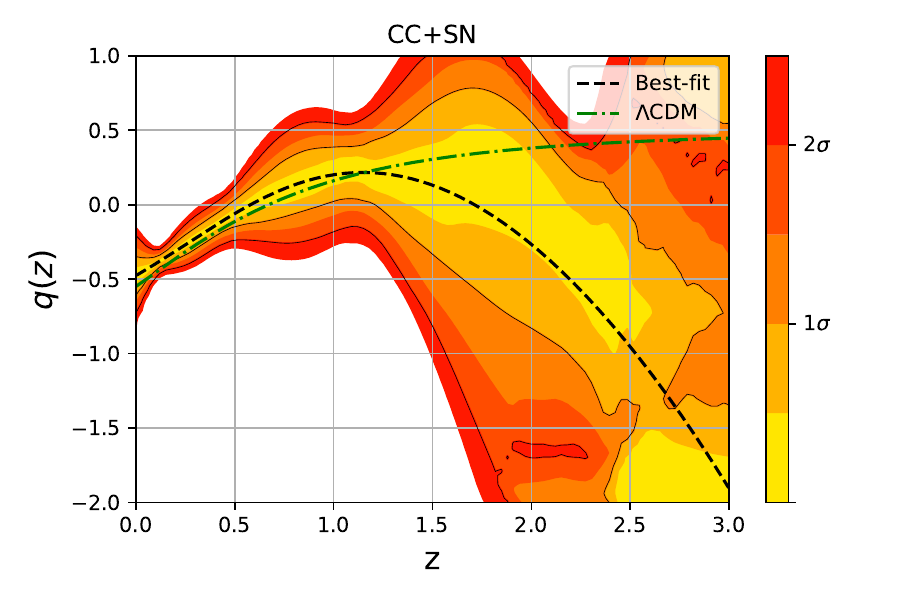}
      \includegraphics[trim = 0mm  0mm 0mm 0mm, clip, width=8.9cm, height=5.3cm]{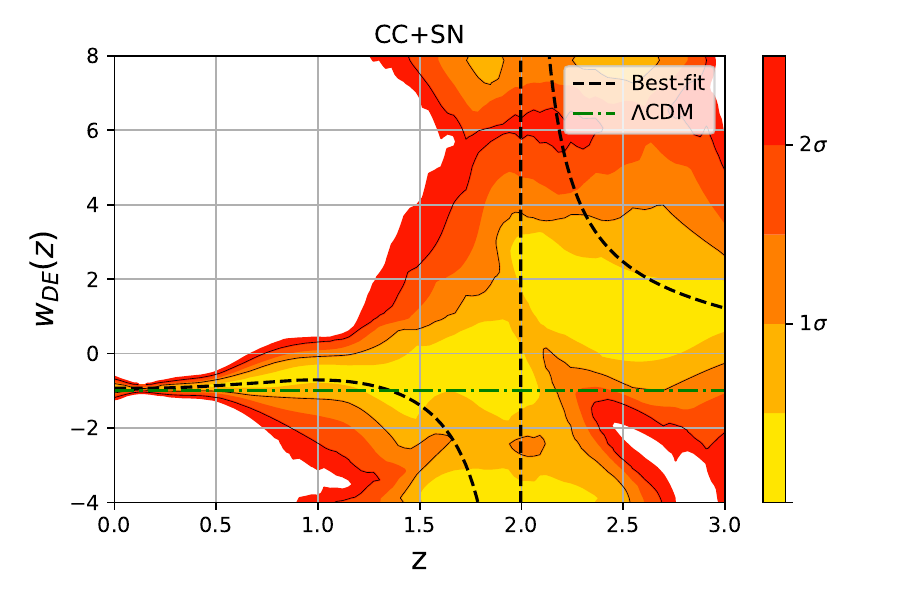}
      }
      \makebox[10cm][c]{
      \includegraphics[trim = 0mm  0mm 0mm 0mm, clip, width=8.9cm, height=5.3cm]{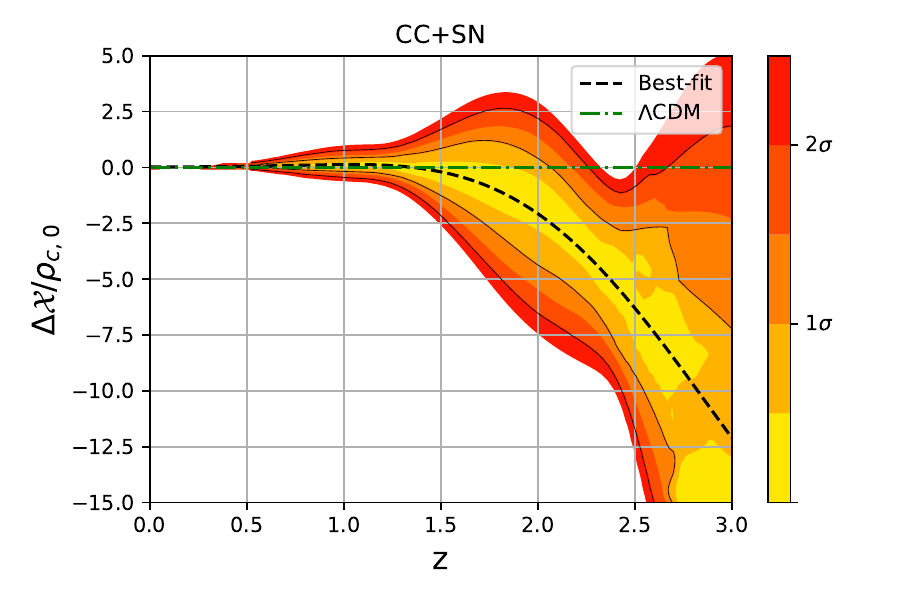}
      \includegraphics[trim = 0mm  0mm 0mm 0mm, clip, width=8.9cm, height=5.3cm]{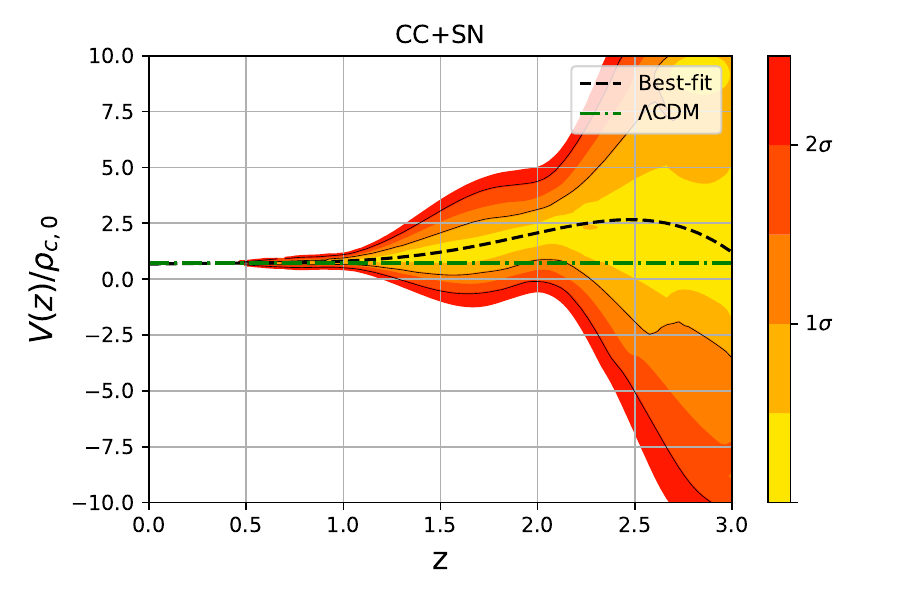}
      }
 \caption{Results of the reconstruction of $H(z)$ for the dataset combination of CC+SN. From top-to-bottom and left-to-right we have: $H(z)$, $H(z)/(1+z)$, $q(z)$, $\Delta\mathcal{X} / \rho_{c,0}$, $\rho_{DE}/\rho_{c,0}$, $p_{DE}/\rho_{c,0}$, $w_{DE}$, and $V(z) / \rho_{c,0}$. An important thing to note and clarify is that the last node is located at $z=3.0$, which means that it cannot be constrained by data. As such, high-redshift results around this region should be taken as merely statistical noise. }\label{fig:GP_cc_sn}
 \end{figure*}

 \begin{figure*}[t!]
     \centering
       \makebox[10cm][c]{
      \includegraphics[trim = 0mm  0mm 0mm 0mm, clip, width=8.9cm, height=5.3cm]{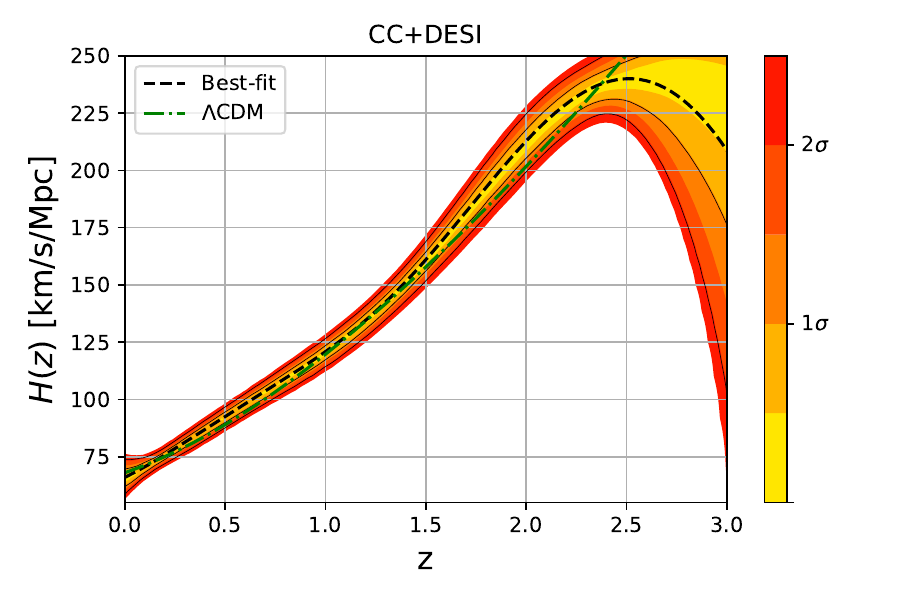}
      \includegraphics[trim = 0mm  0mm 0mm 0mm, clip, width=8.9cm, height=5.3cm]{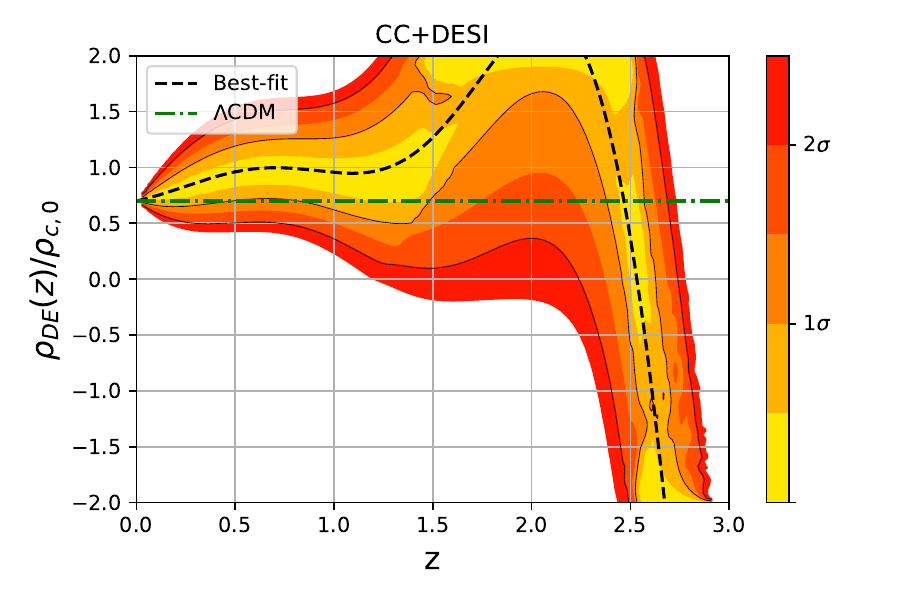}
      }
     \makebox[10cm][c]{
      \includegraphics[trim = 0mm  0mm 0mm 0mm, clip, width=8.9cm, height=5.3cm]{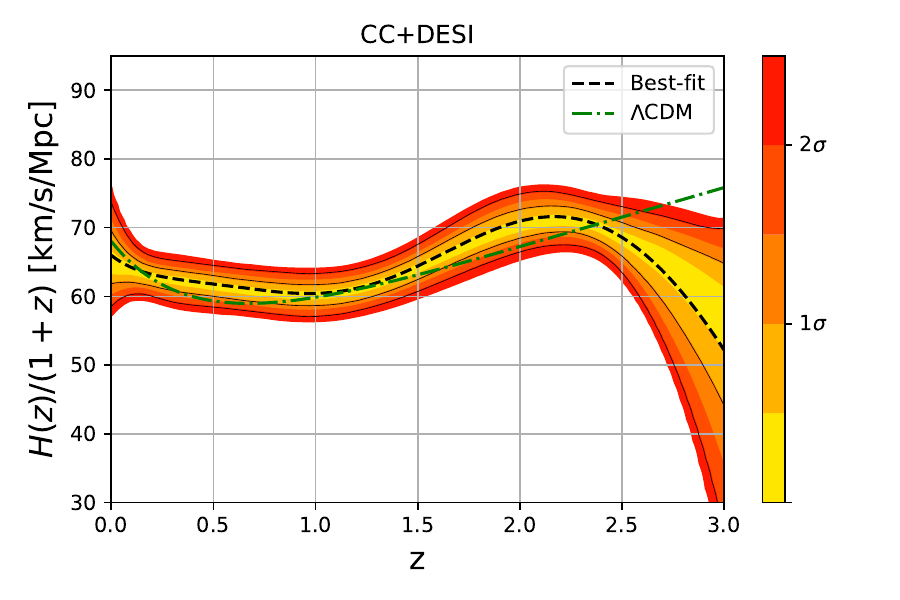}
      \includegraphics[trim = 0mm  0mm 0mm 0mm, clip, width=8.9cm, height=5.3cm]{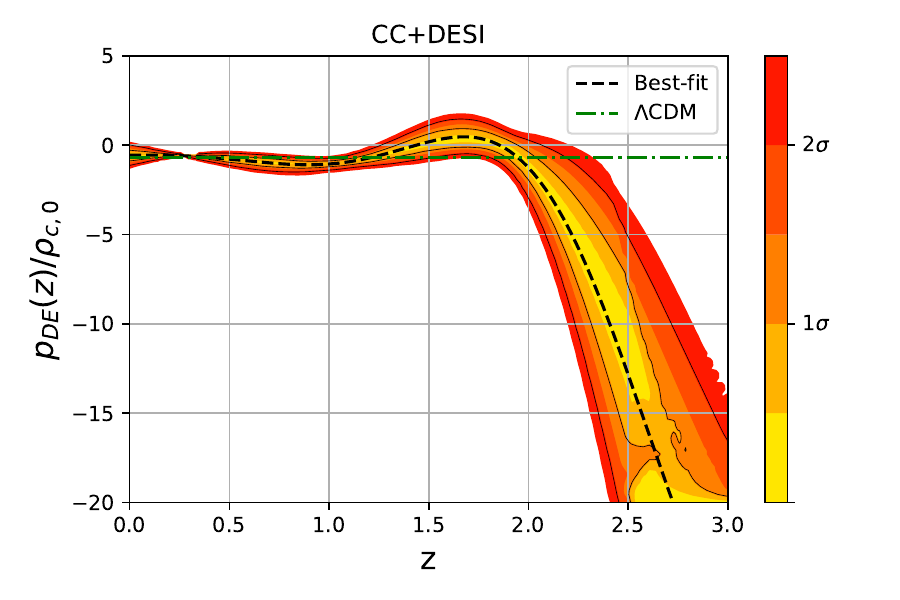}
      }
     \makebox[10cm][c]{
      \includegraphics[trim = 0mm  0mm 0mm 0mm, clip, width=8.9cm, height=5.3cm]{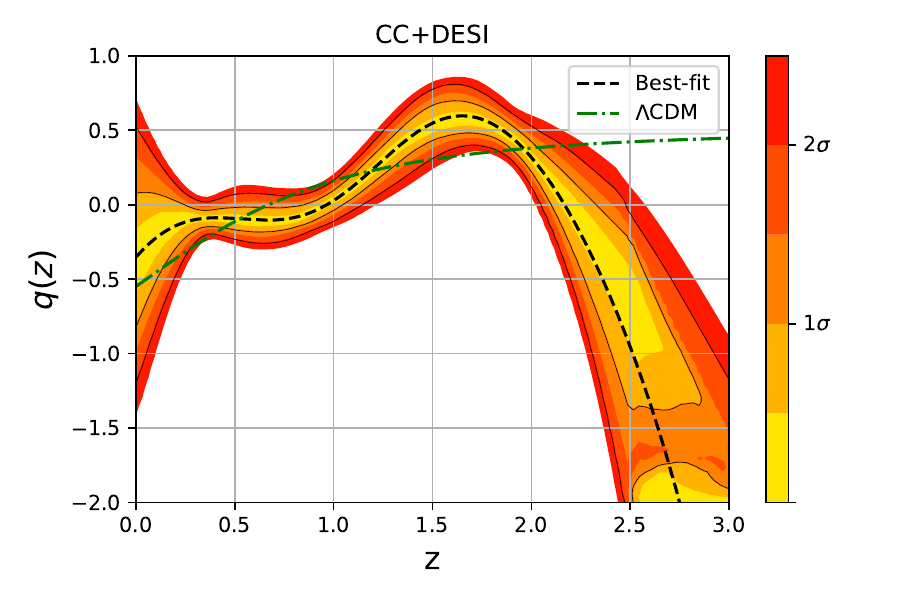}
      \includegraphics[trim = 0mm  0mm 0mm 0mm, clip, width=8.9cm, height=5.3cm]{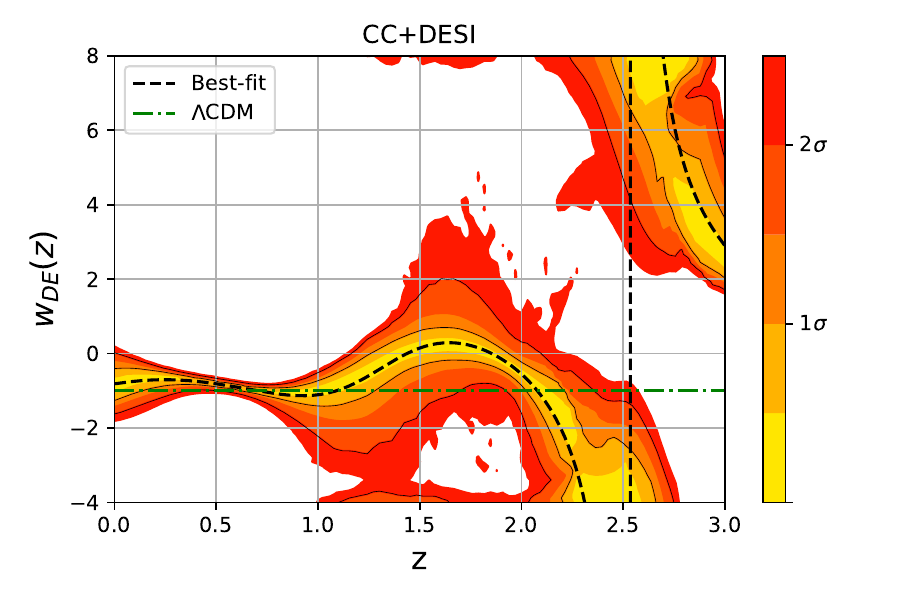}
      }
      \makebox[10cm][c]{
      \includegraphics[trim = 0mm  0mm 0mm 0mm, clip, width=8.9cm, height=5.3cm]{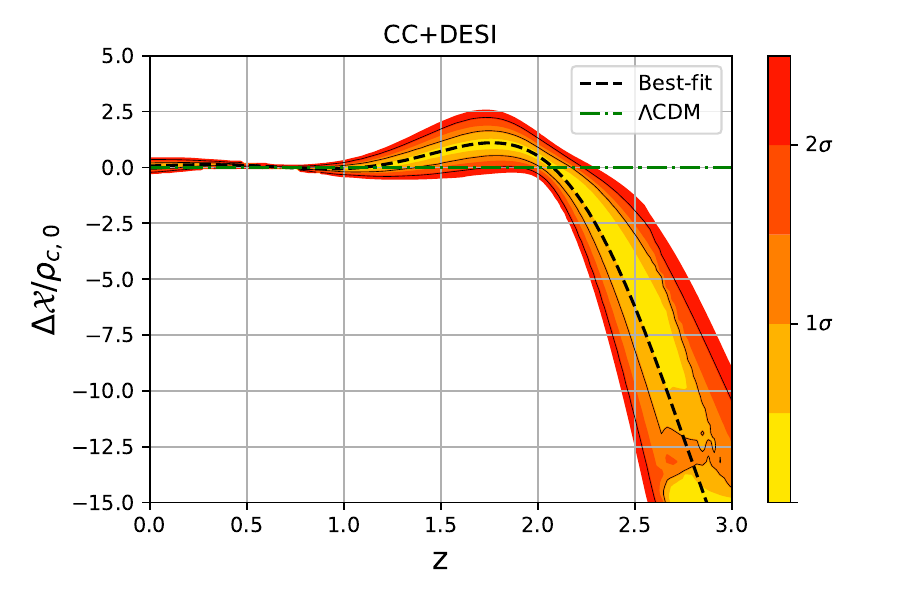}
      \includegraphics[trim = 0mm  0mm 0mm 0mm, clip, width=8.9cm, height=5.3cm]{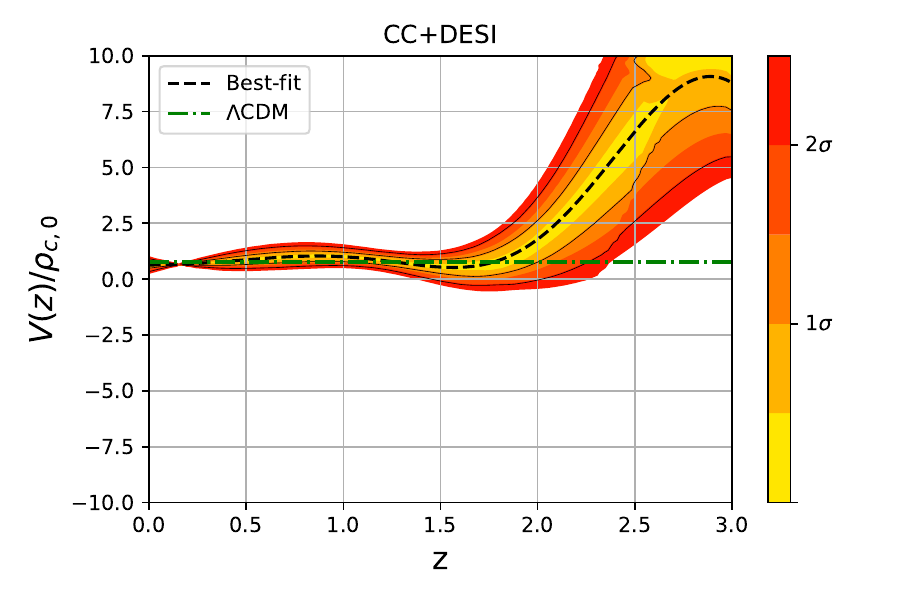}
      }
 \caption{Results of the reconstruction of $H(z)$ for the dataset combination of CC+DESI. From top-to-bottom and left-to-right we have: $H(z)$, $H(z)/(1+z)$, $q(z)$, $\Delta\mathcal{X} / \rho_{c,0}$, $\rho_{DE}/\rho_{c,0}$, $p_{DE}/\rho_{c,0}$, $w_{DE}$, and $V(z) / \rho_{c,0}$. An important thing to note and clarify is that the last node is located at $z=3.0$, which means that it cannot be constrained by data. As such, high-redshift results around this region should be taken as merely statistical noise. }\label{fig:GP_cc_desi}
 \end{figure*}

 \begin{figure*}[t!]
     \centering
       \makebox[10cm][c]{
      \includegraphics[trim = 0mm  0mm 0mm 0mm, clip, width=8.9cm, height=5.3cm]{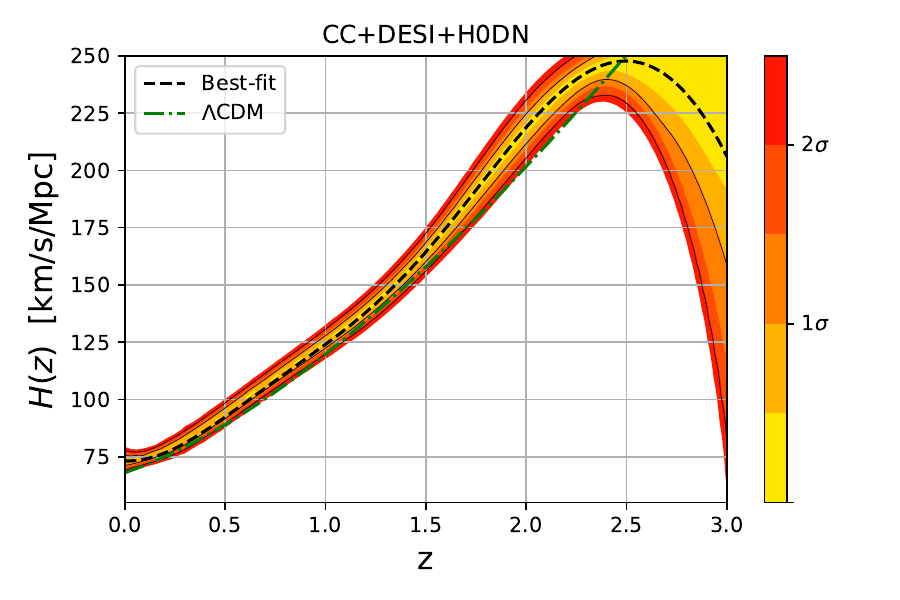}
      \includegraphics[trim = 0mm  0mm 0mm 0mm, clip, width=8.9cm, height=5.3cm]{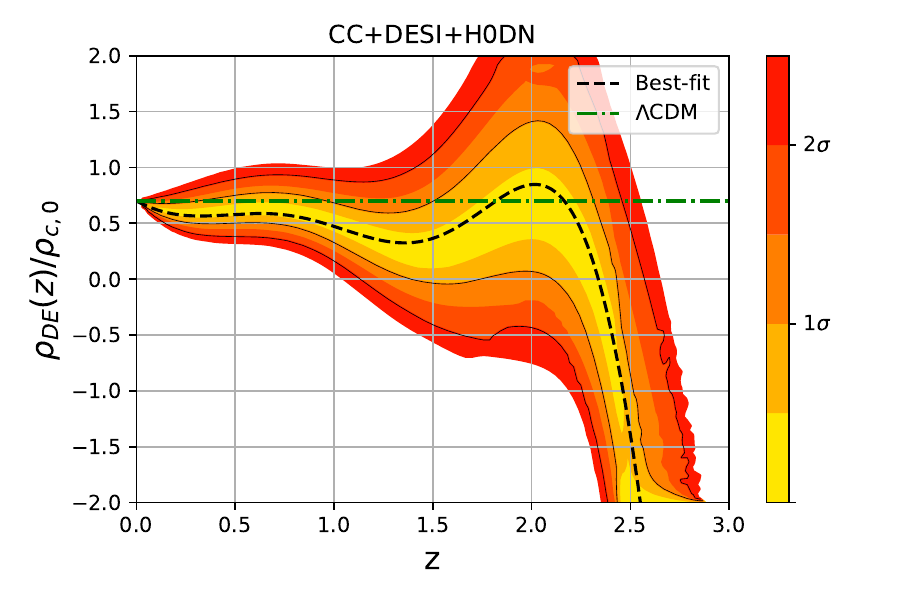}
      }
     \makebox[10cm][c]{
      \includegraphics[trim = 0mm  0mm 0mm 0mm, clip, width=8.9cm, height=5.3cm]{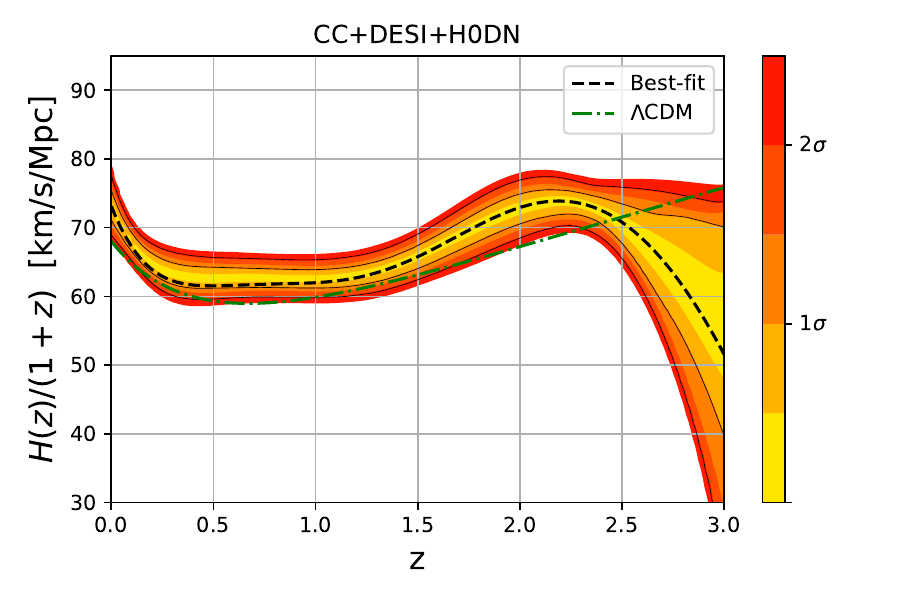}
      \includegraphics[trim = 0mm  0mm 0mm 0mm, clip, width=8.9cm, height=5.3cm]{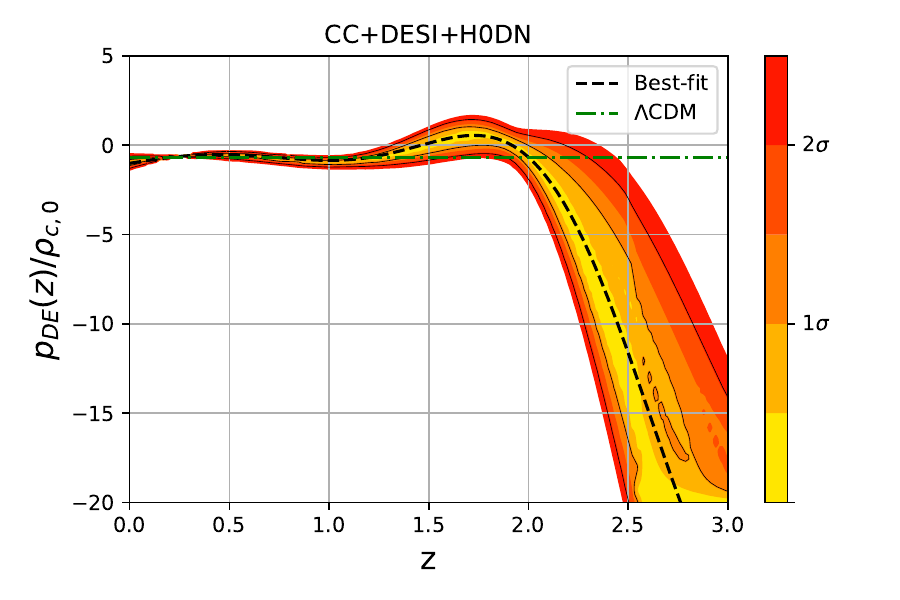}
      }
     \makebox[10cm][c]{
      \includegraphics[trim = 0mm  0mm 0mm 0mm, clip, width=8.9cm, height=5.3cm]{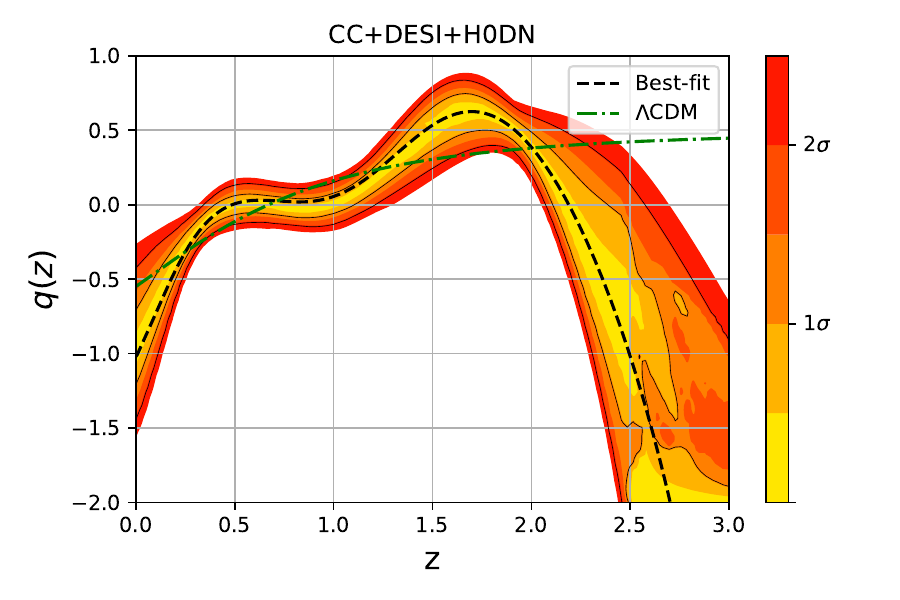}
      \includegraphics[trim = 0mm  0mm 0mm 0mm, clip, width=8.9cm, height=5.3cm]{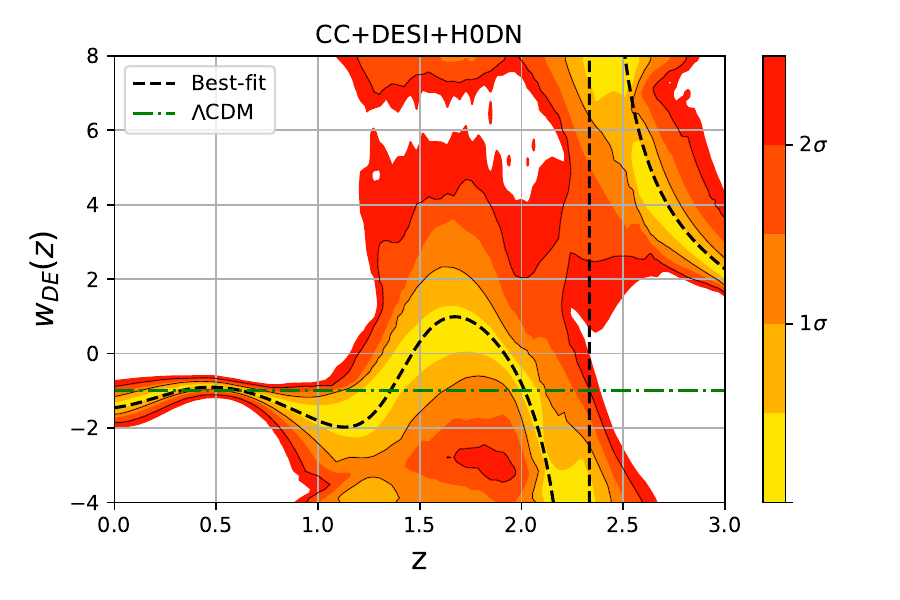}
      }
      \makebox[10cm][c]{
      \includegraphics[trim = 0mm  0mm 0mm 0mm, clip, width=8.9cm, height=5.3cm]{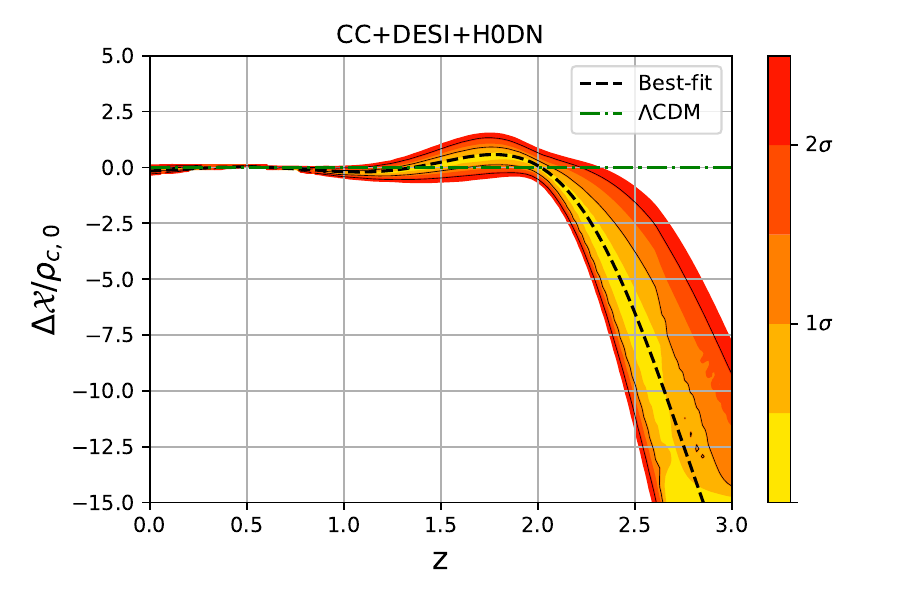}
      \includegraphics[trim = 0mm  0mm 0mm 0mm, clip, width=8.9cm, height=5.3cm]{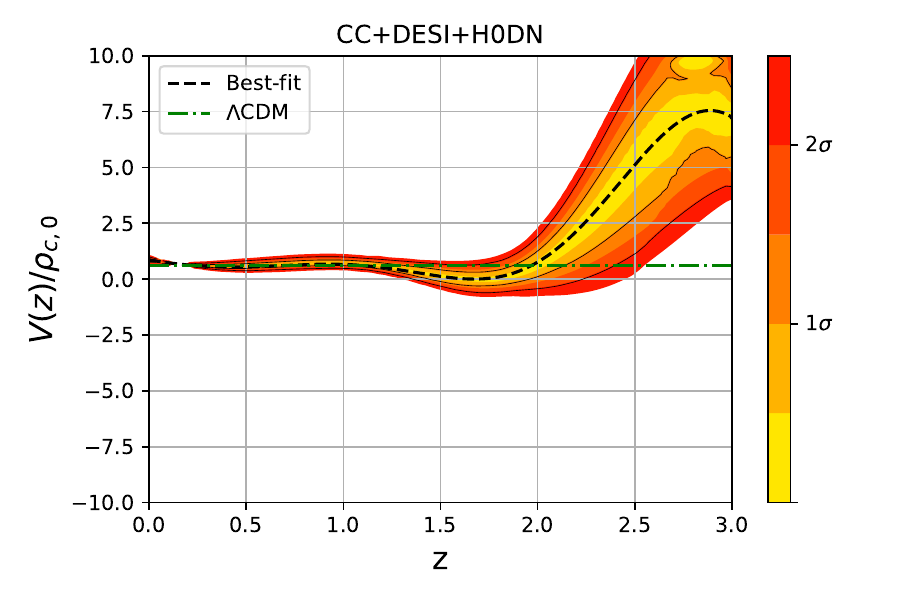}
      }
 \caption{Results of the reconstruction of $H(z)$ for the dataset combination of CC+DESI+H0DN. From top-to-bottom and left-to-right we have: $H(z)$, $H(z)/(1+z)$, $q(z)$, $\Delta\mathcal{X} / \rho_{c,0}$, $\rho_{DE}/\rho_{c,0}$, $p_{DE}/\rho_{c,0}$, $w_{DE}$, and $V(z) / \rho_{c,0}$. An important thing to note and clarify is that the last node is located at $z=3.0$, which means that it cannot be constrained by data. As such, high-redshift results around this region should be taken as merely statistical noise. }\label{fig:GP_cc_desi_h0dn}
 \end{figure*}

 \begin{figure*}[t!]
     \centering
       \makebox[10cm][c]{
      \includegraphics[trim = 0mm  0mm 0mm 0mm, clip, width=8.9cm, height=5.3cm]{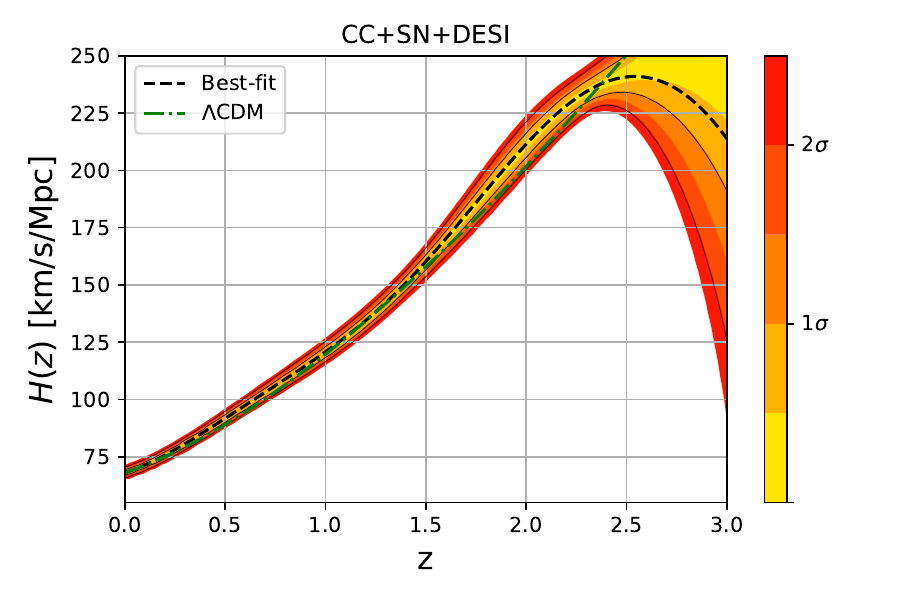}
      \includegraphics[trim = 0mm  0mm 0mm 0mm, clip, width=8.9cm, height=5.3cm]{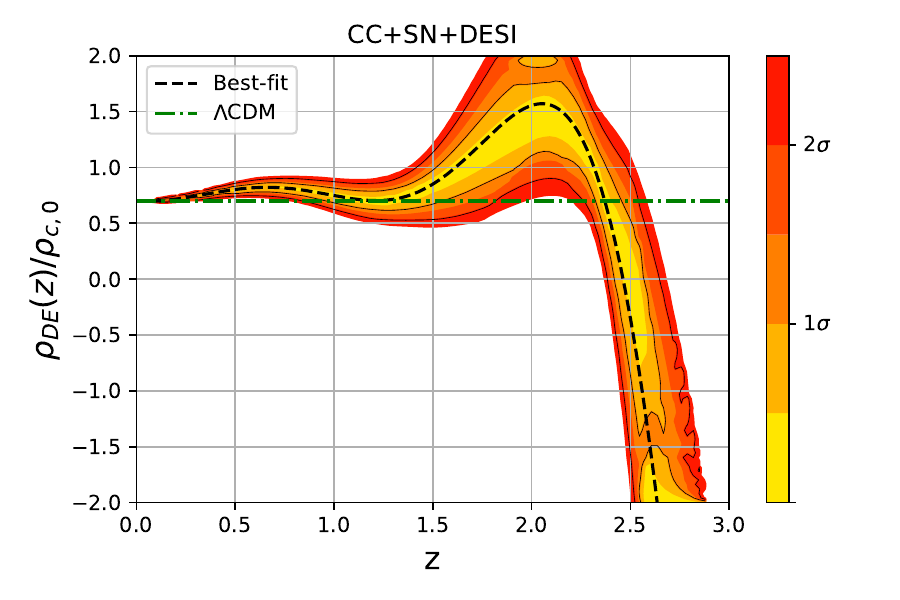}
      }
     \makebox[10cm][c]{
      \includegraphics[trim = 0mm  0mm 0mm 0mm, clip, width=8.9cm, height=5.3cm]{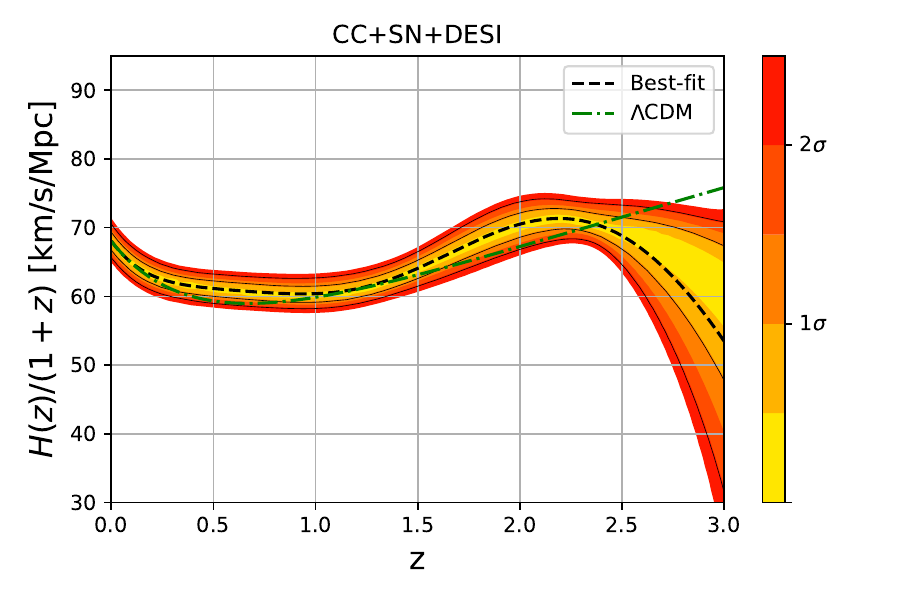}
      \includegraphics[trim = 0mm  0mm 0mm 0mm, clip, width=8.9cm, height=5.3cm]{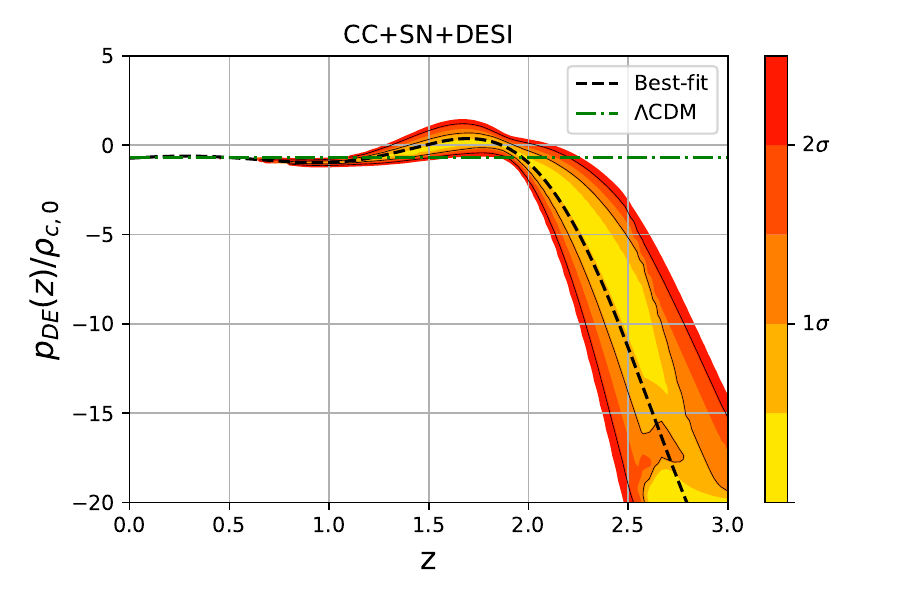}
      }
     \makebox[10cm][c]{
      \includegraphics[trim = 0mm  0mm 0mm 0mm, clip, width=8.9cm, height=5.3cm]{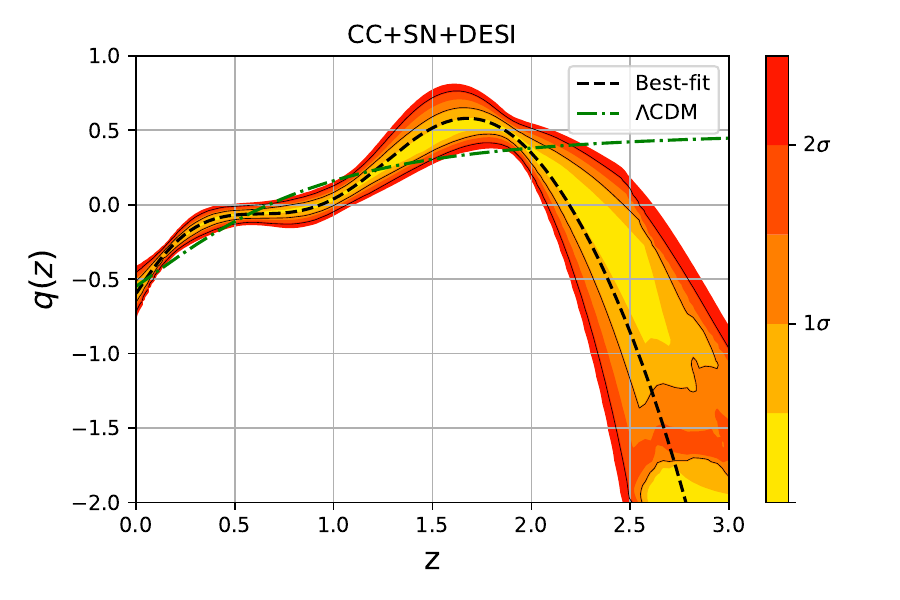}
      \includegraphics[trim = 0mm  0mm 0mm 0mm, clip, width=8.9cm, height=5.3cm]{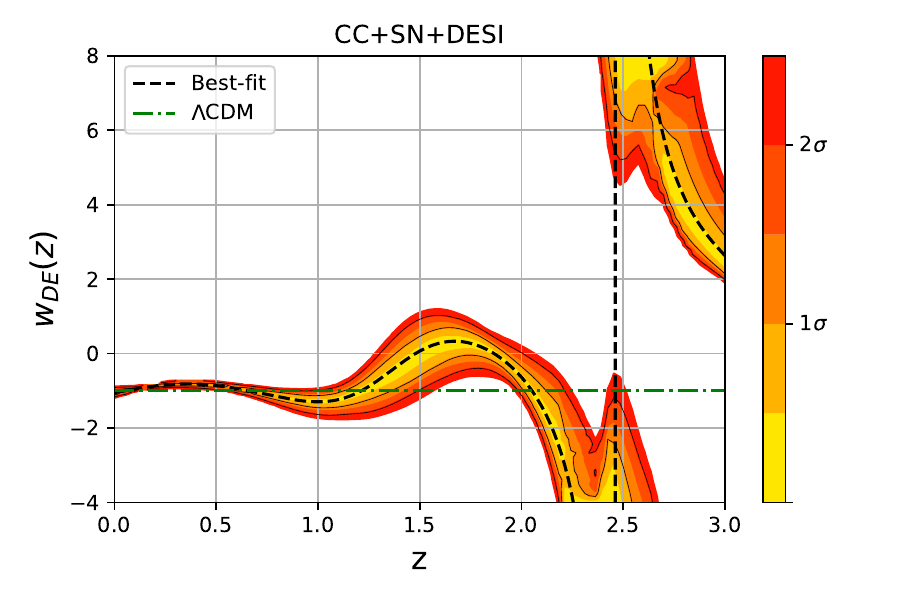}
      }
      \makebox[10cm][c]{
      \includegraphics[trim = 0mm  0mm 0mm 0mm, clip, width=8.9cm, height=5.3cm]{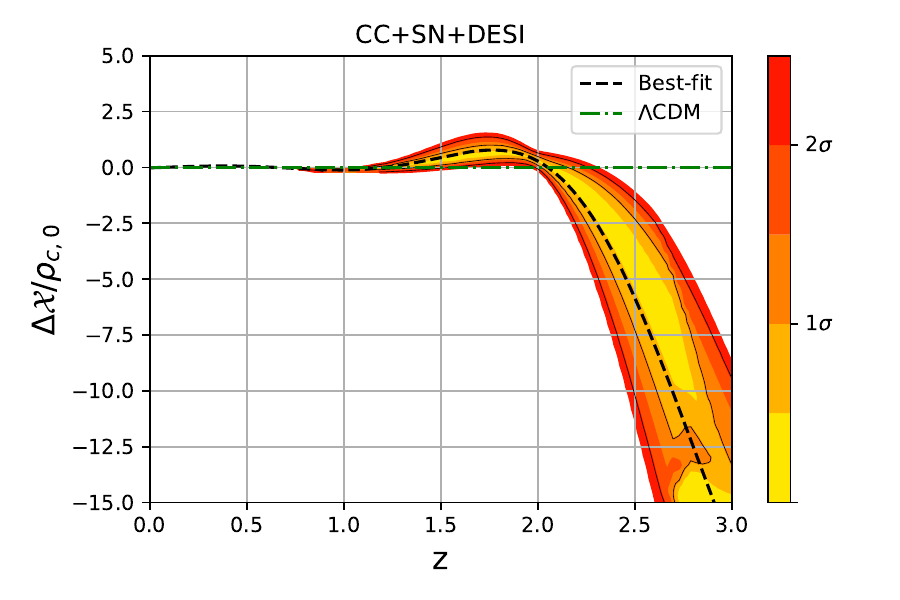}
      \includegraphics[trim = 0mm  0mm 0mm 0mm, clip, width=8.9cm, height=5.3cm]{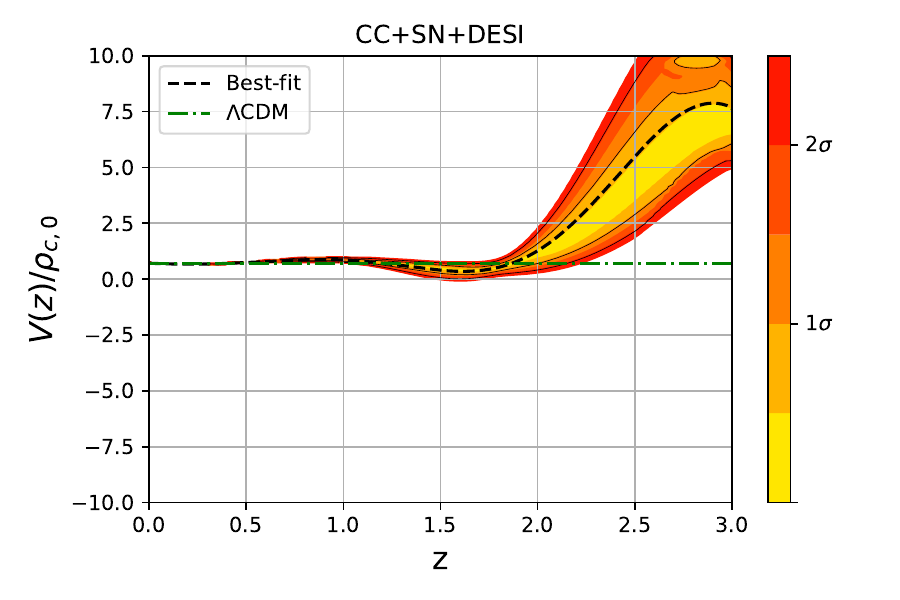}
      }
 \caption{Results of the reconstruction of $H(z)$ for the dataset combination of CC+SN+DESI. From top-to-bottom and left-to-right we have: $H(z)$, $H(z)/(1+z)$, $q(z)$, $\Delta\mathcal{X} / \rho_{c,0}$, $\rho_{DE}/\rho_{c,0}$, $p_{DE}/\rho_{c,0}$, $w_{DE}$, and $V(z) / \rho_{c,0}$. An important thing to note and clarify is that the last node is located at $z=3.0$, which means that it cannot be constrained by data. As such, high-redshift results around this region should be taken as merely statistical noise. }\label{fig:GP_cc_sn_desi}
 \end{figure*}

 \begin{figure*}[t!]
     \centering
       \makebox[10cm][c]{
      \includegraphics[trim = 0mm  0mm 0mm 0mm, clip, width=8.9cm, height=5.3cm]{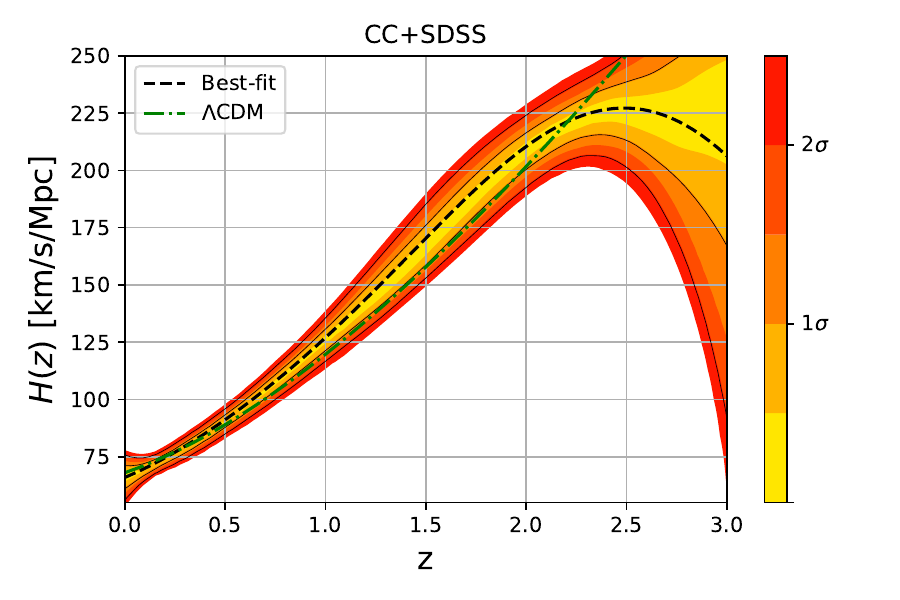}
      \includegraphics[trim = 0mm  0mm 0mm 0mm, clip, width=8.9cm, height=5.3cm]{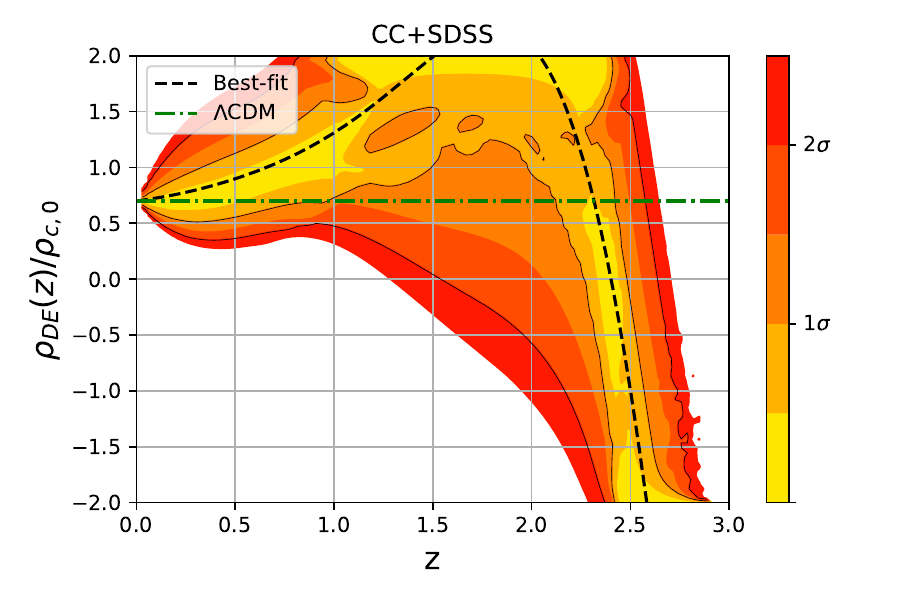}
      }
     \makebox[10cm][c]{
      \includegraphics[trim = 0mm  0mm 0mm 0mm, clip, width=8.9cm, height=5.3cm]{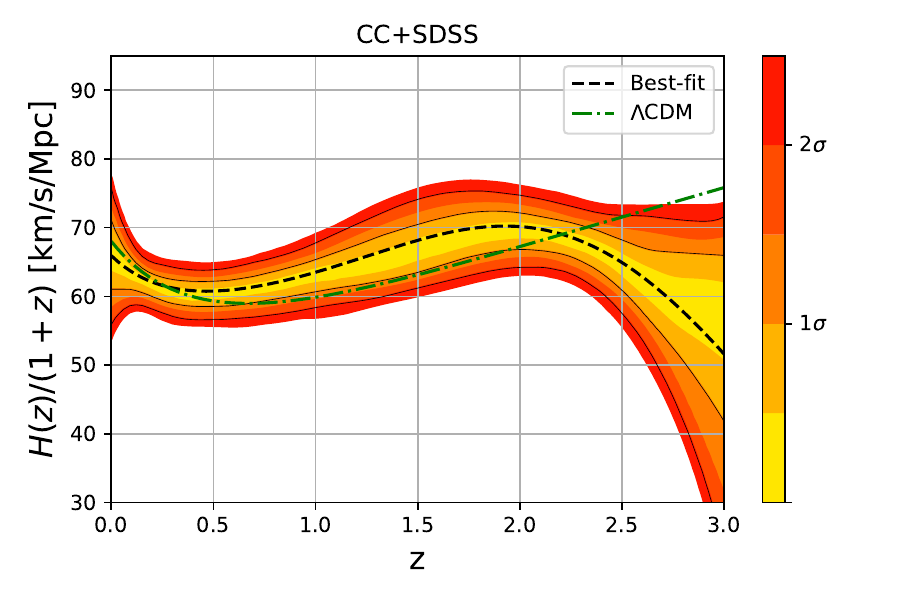}
      \includegraphics[trim = 0mm  0mm 0mm 0mm, clip, width=8.9cm, height=5.3cm]{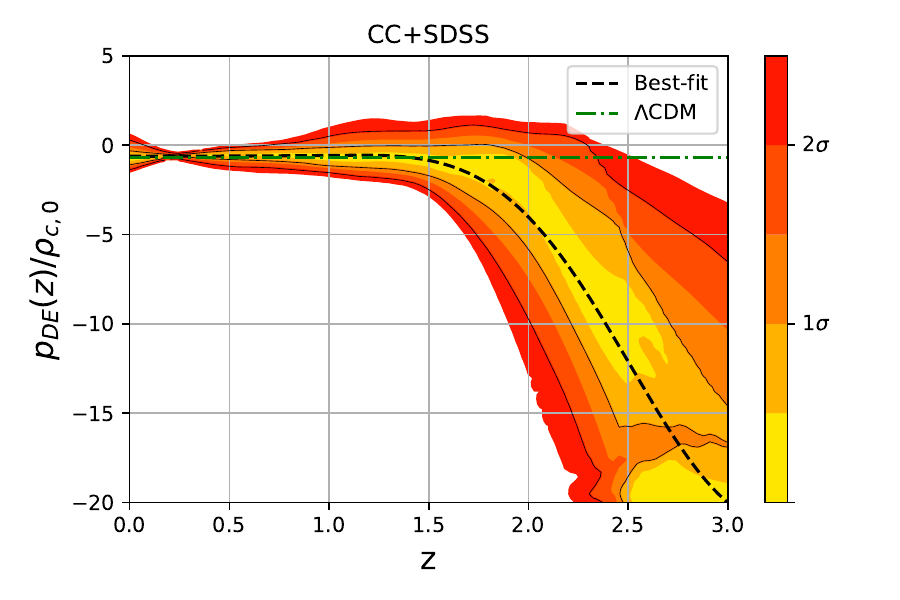}
      }
     \makebox[10cm][c]{
      \includegraphics[trim = 0mm  0mm 0mm 0mm, clip, width=8.9cm, height=5.3cm]{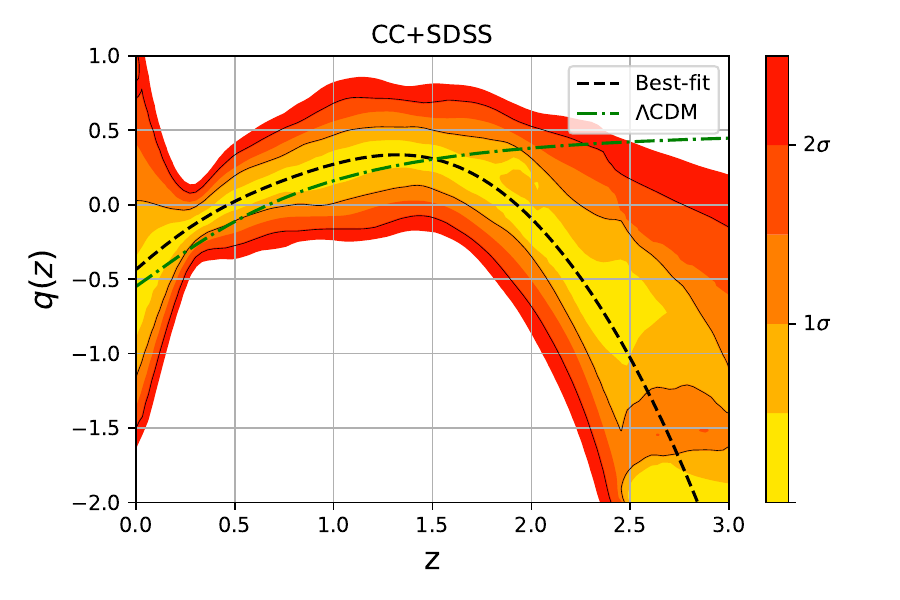}
      \includegraphics[trim = 0mm  0mm 0mm 0mm, clip, width=8.9cm, height=5.3cm]{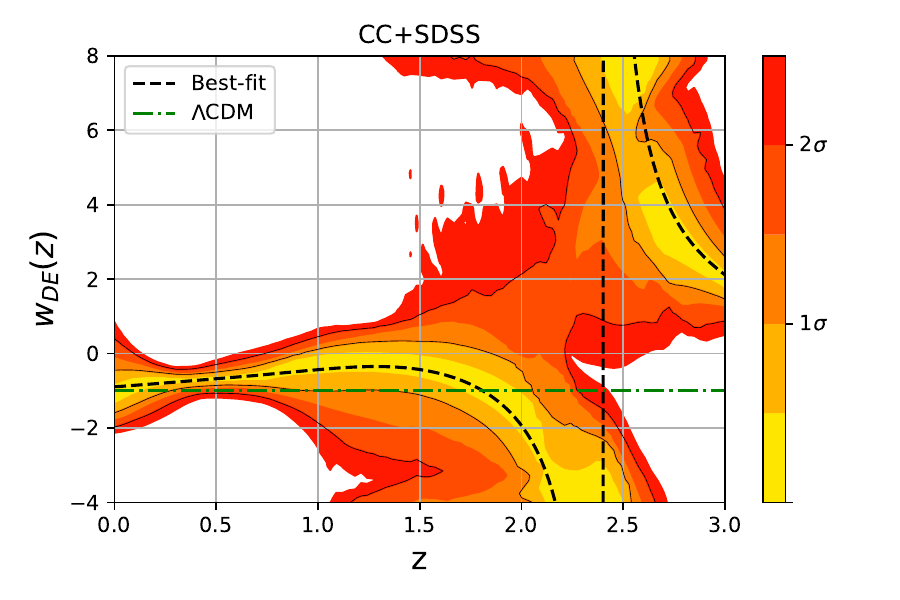}
      }
      \makebox[10cm][c]{
      \includegraphics[trim = 0mm  0mm 0mm 0mm, clip, width=8.9cm, height=5.3cm]{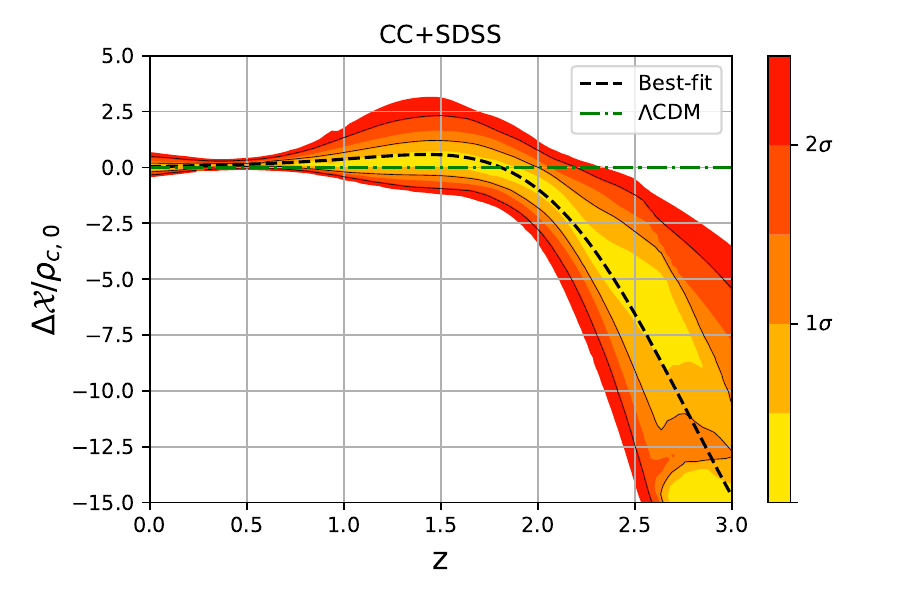}
      \includegraphics[trim = 0mm  0mm 0mm 0mm, clip, width=8.9cm, height=5.3cm]{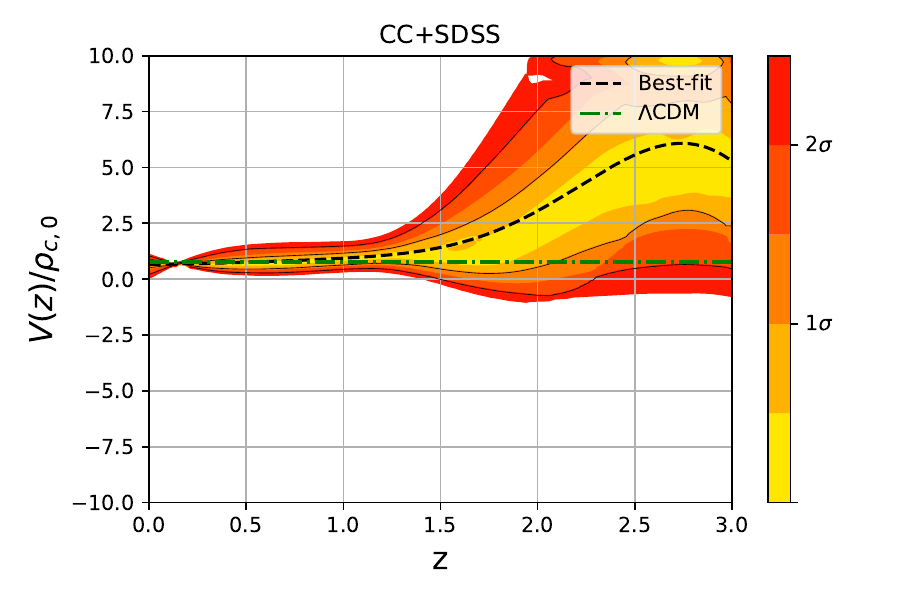}
      }
 \caption{Results of the reconstruction of $H(z)$ for the dataset combination of CC+SDSS. From top-to-bottom and left-to-right we have: $H(z)$, $H(z)/(1+z)$, $q(z)$, $\Delta\mathcal{X} / \rho_{c,0}$, $\rho_{DE}/\rho_{c,0}$, $p_{DE}/\rho_{c,0}$, $w_{DE}$, and $V(z) / \rho_{c,0}$. An important thing to note and clarify is that the last node is located at $z=3.0$, which means that it cannot be constrained by data. As such, high-redshift results around this region should be taken as merely statistical noise. }\label{fig:GP_cc_sdss}
 \end{figure*}

 \begin{figure*}[t!]
     \centering
       \makebox[10cm][c]{
      \includegraphics[trim = 0mm  0mm 0mm 0mm, clip, width=8.9cm, height=5.3cm]{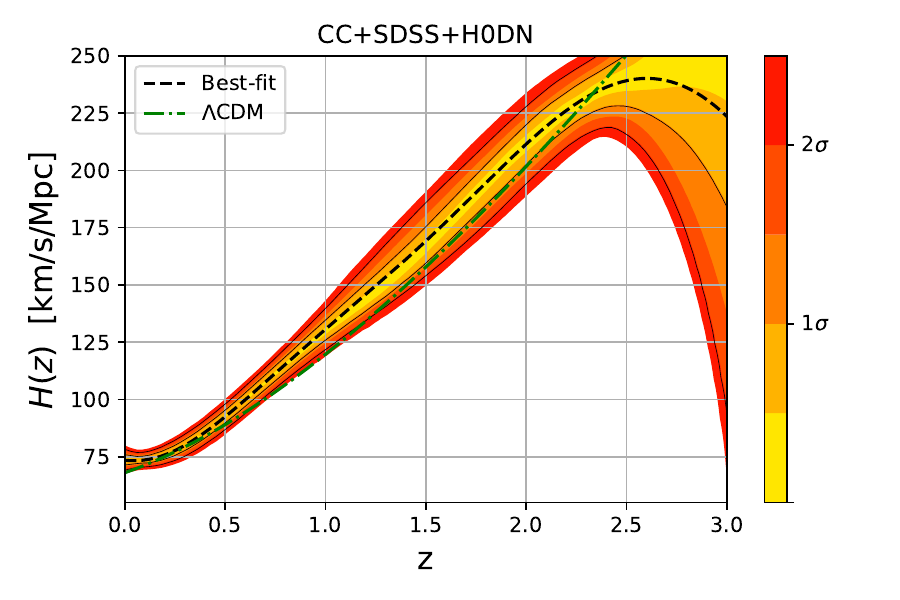}
      \includegraphics[trim = 0mm  0mm 0mm 0mm, clip, width=8.9cm, height=5.3cm]{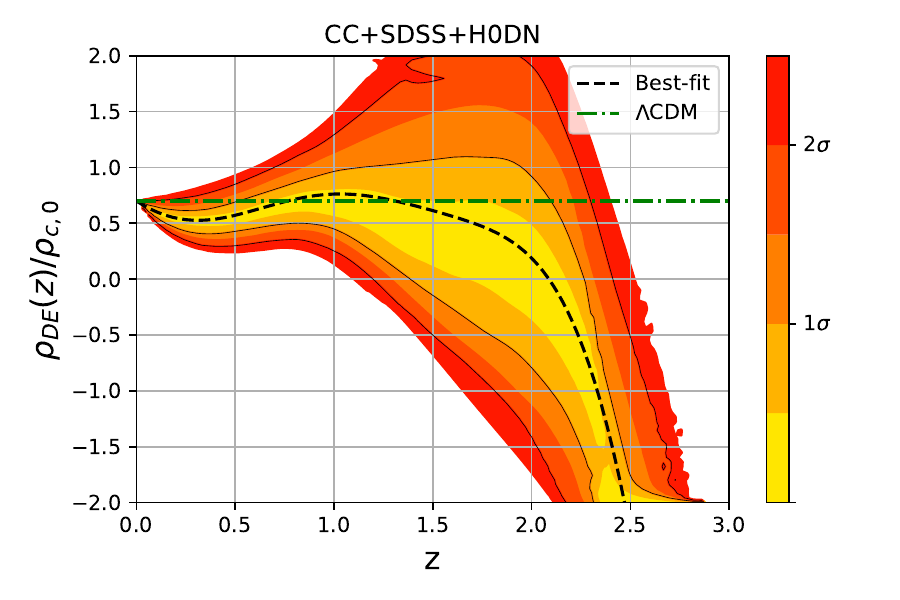}
      }
     \makebox[10cm][c]{
      \includegraphics[trim = 0mm  0mm 0mm 0mm, clip, width=8.9cm, height=5.3cm]{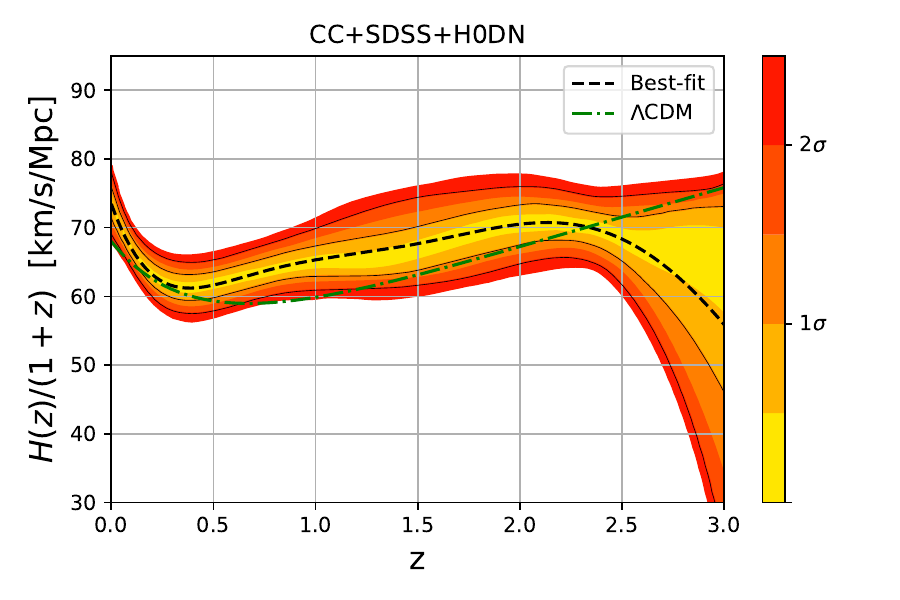}
      \includegraphics[trim = 0mm  0mm 0mm 0mm, clip, width=8.9cm, height=5.3cm]{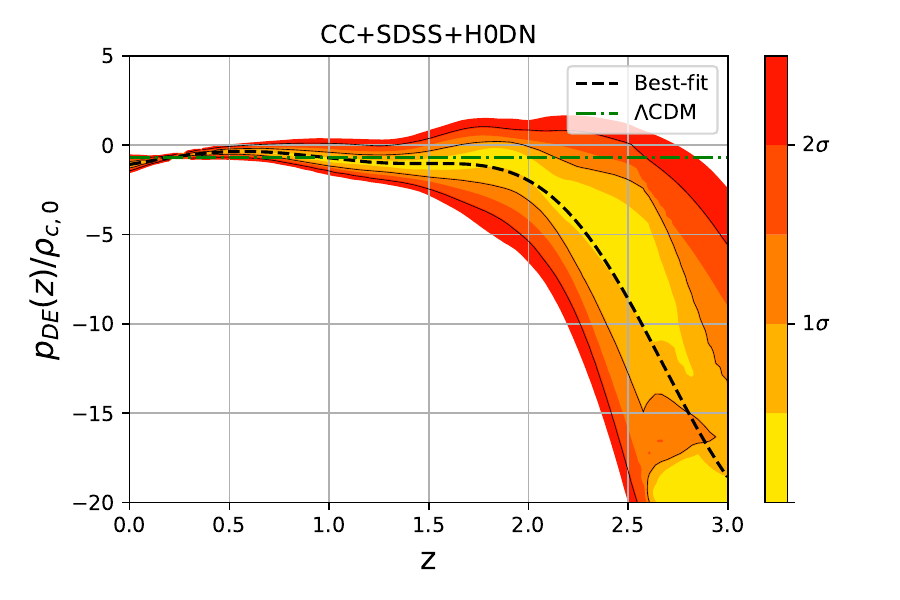}
      }
     \makebox[10cm][c]{
      \includegraphics[trim = 0mm  0mm 0mm 0mm, clip, width=8.9cm, height=5.3cm]{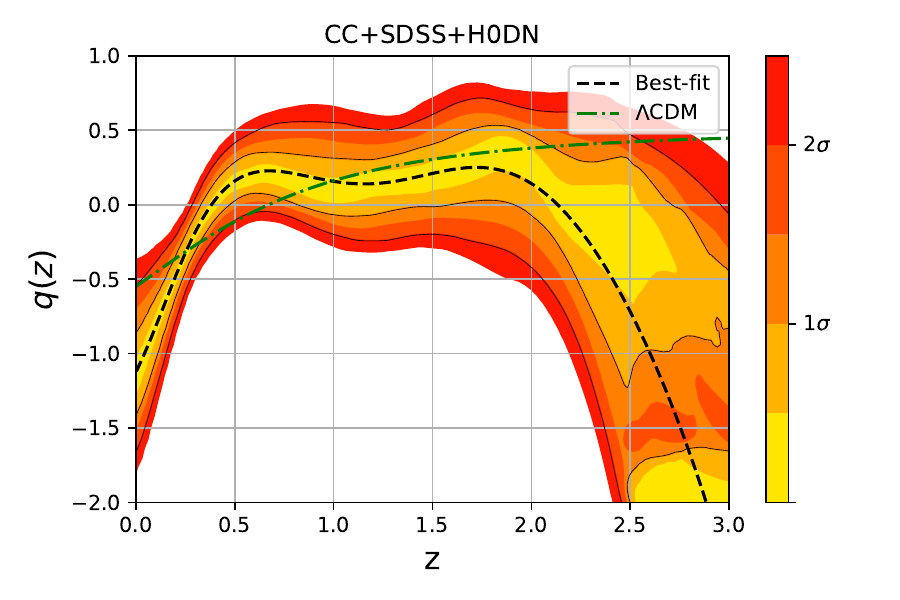}
      \includegraphics[trim = 0mm  0mm 0mm 0mm, clip, width=8.9cm, height=5.3cm]{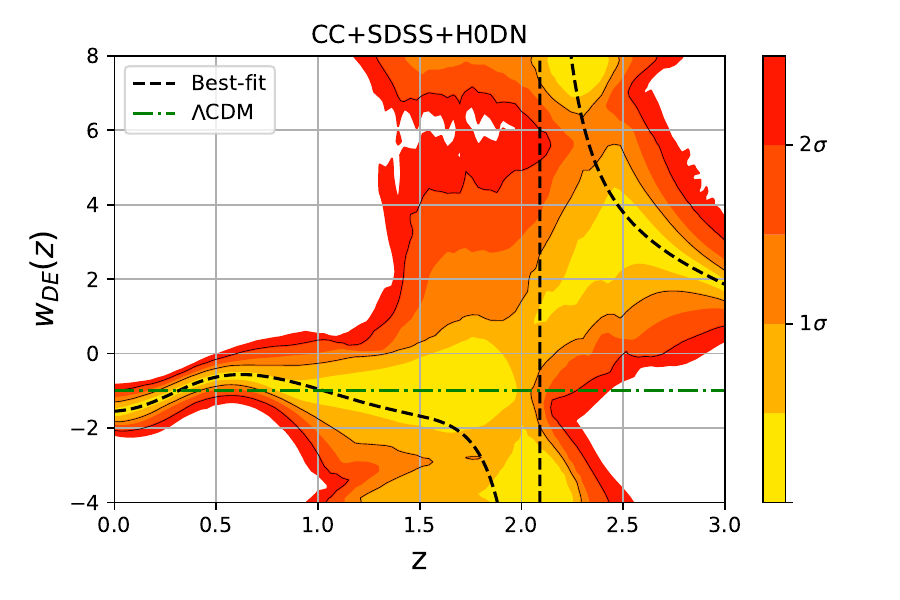}
      }
      \makebox[10cm][c]{
      \includegraphics[trim = 0mm  0mm 0mm 0mm, clip, width=8.9cm, height=5.3cm]{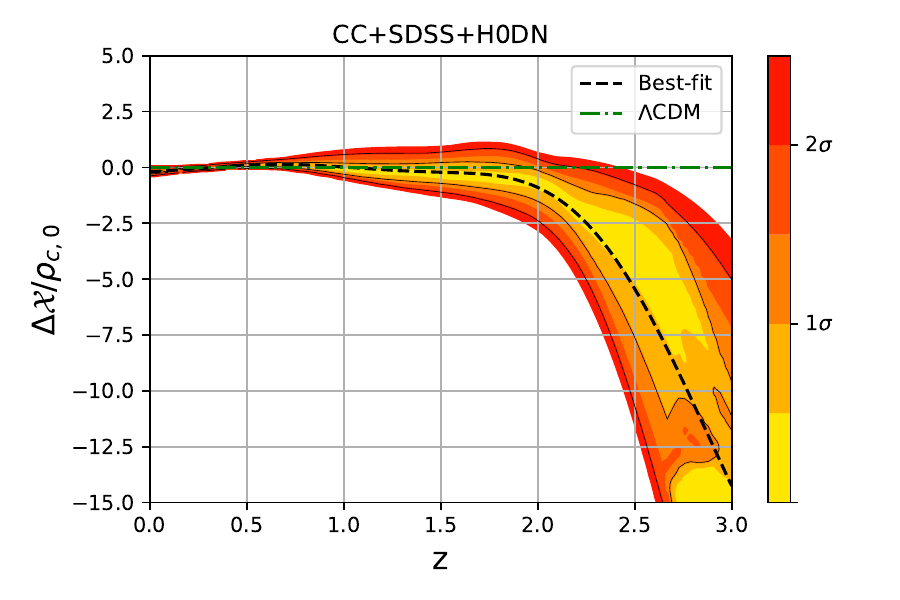}
      \includegraphics[trim = 0mm  0mm 0mm 0mm, clip, width=8.9cm, height=5.3cm]{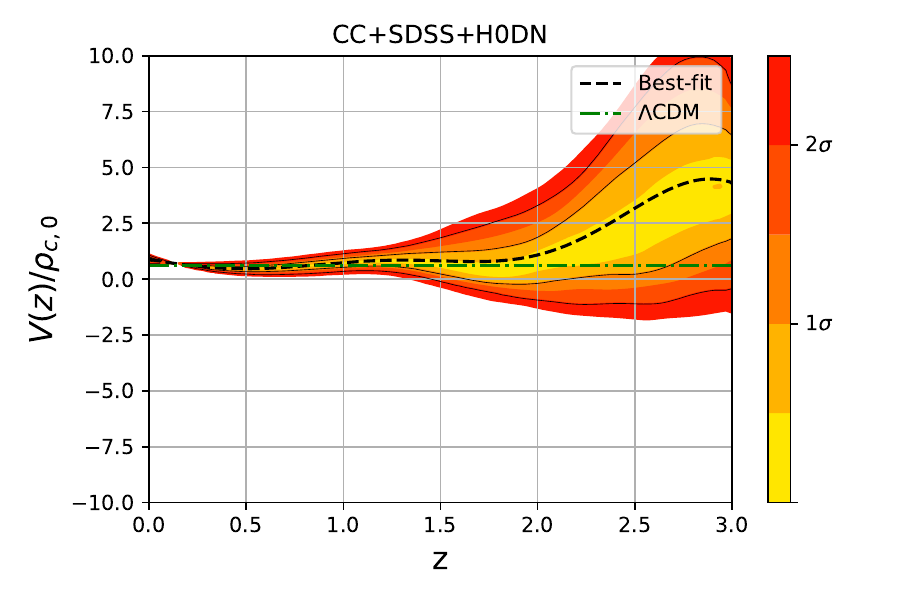}
      }
 \caption{Results of the reconstruction of $H(z)$ for the dataset combination of CC+SDSS+H0DN. From top-to-bottom and left-to-right we have: $H(z)$, $H(z)/(1+z)$, $q(z)$, $\Delta\mathcal{X} / \rho_{c,0}$, $\rho_{DE}/\rho_{c,0}$, $p_{DE}/\rho_{c,0}$, $w_{DE}$, and $V(z) / \rho_{c,0}$. An important thing to note and clarify is that the last node is located at $z=3.0$, which means that it cannot be constrained by data. As such, high-redshift results around this region should be taken as merely statistical noise. }\label{fig:GP_cc_sdss_h0dn}
 \end{figure*}

 \begin{figure*}[t!]
     \centering
       \makebox[10cm][c]{
      \includegraphics[trim = 0mm  0mm 0mm 0mm, clip, width=8.9cm, height=5.3cm]{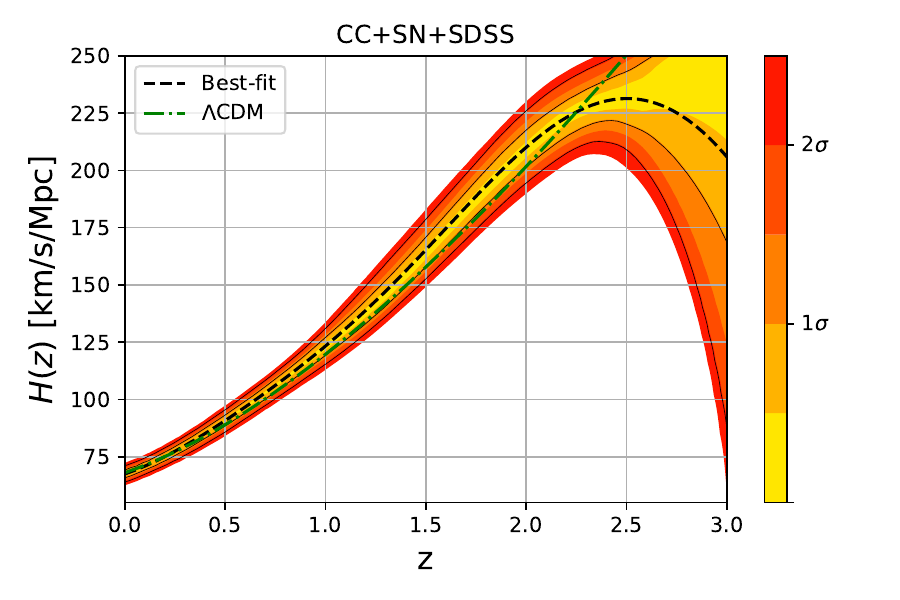}
      \includegraphics[trim = 0mm  0mm 0mm 0mm, clip, width=8.9cm, height=5.3cm]{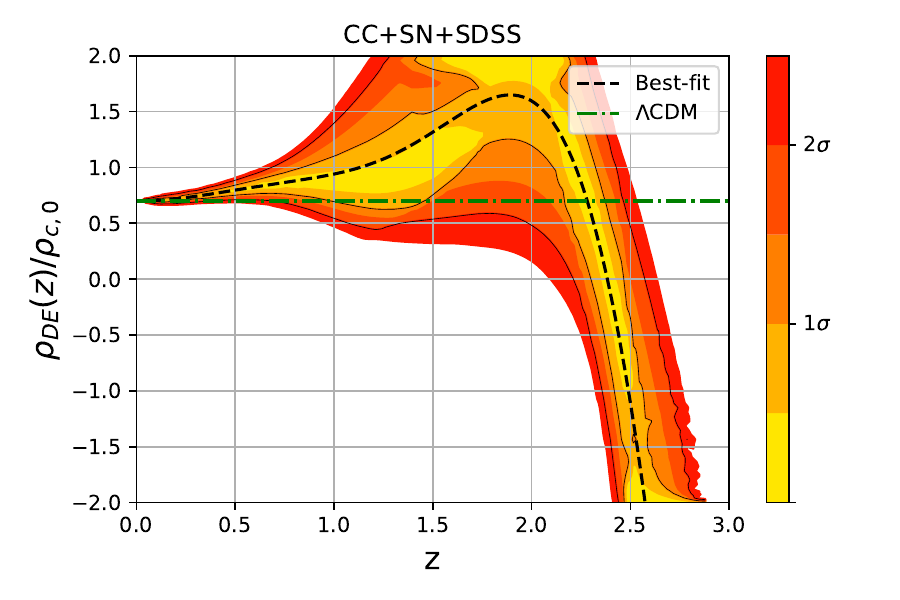}
      }
     \makebox[10cm][c]{
      \includegraphics[trim = 0mm  0mm 0mm 0mm, clip, width=8.9cm, height=5.3cm]{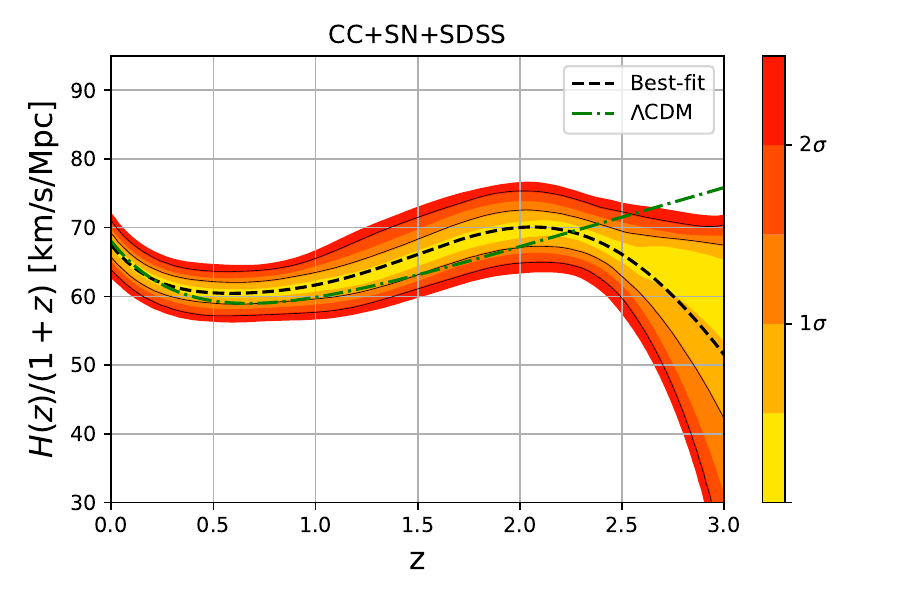}
      \includegraphics[trim = 0mm  0mm 0mm 0mm, clip, width=8.9cm, height=5.3cm]{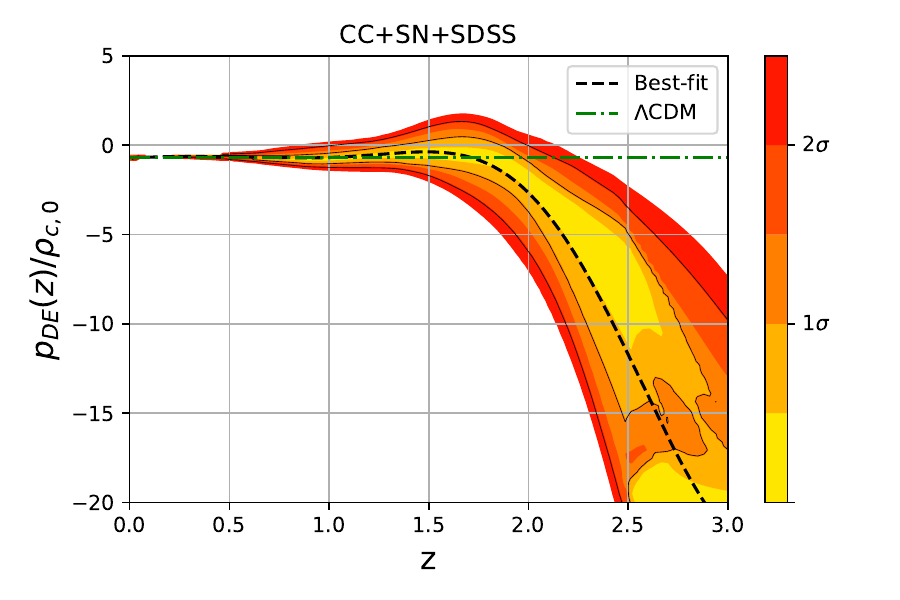}
      }
     \makebox[10cm][c]{
      \includegraphics[trim = 0mm  0mm 0mm 0mm, clip, width=8.9cm, height=5.3cm]{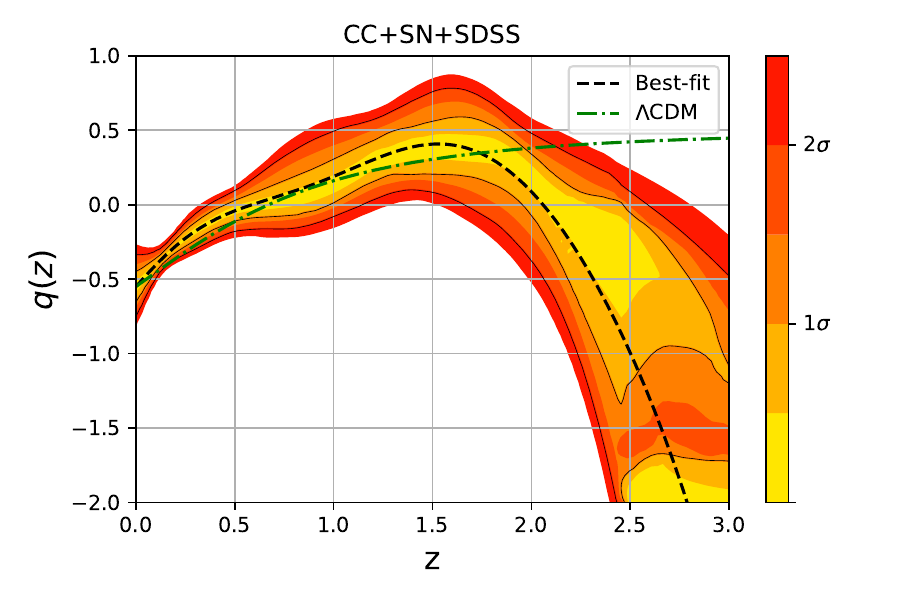}
      \includegraphics[trim = 0mm  0mm 0mm 0mm, clip, width=8.9cm, height=5.3cm]{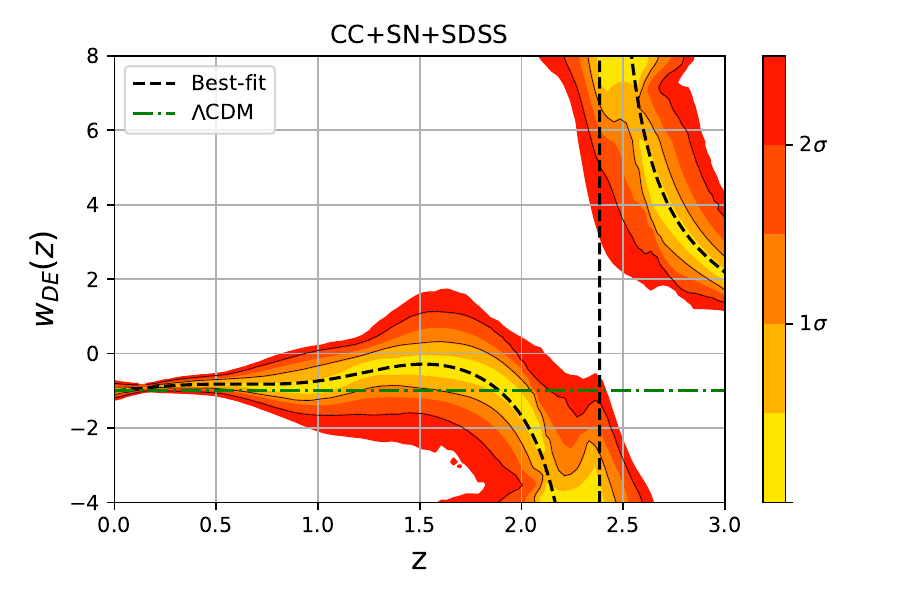}
      }
      \makebox[10cm][c]{
      \includegraphics[trim = 0mm  0mm 0mm 0mm, clip, width=8.9cm, height=5.3cm]{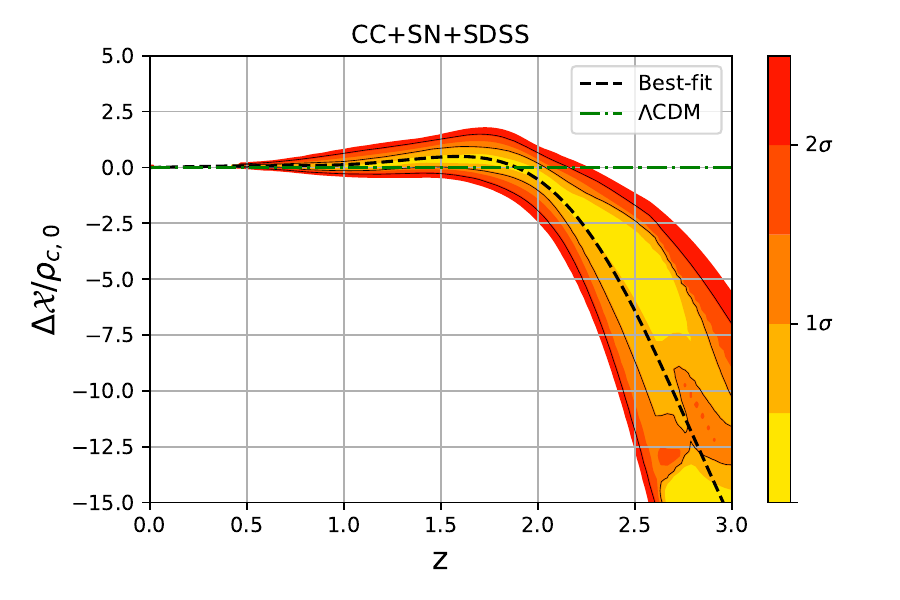}
      \includegraphics[trim = 0mm  0mm 0mm 0mm, clip, width=8.9cm, height=5.3cm]{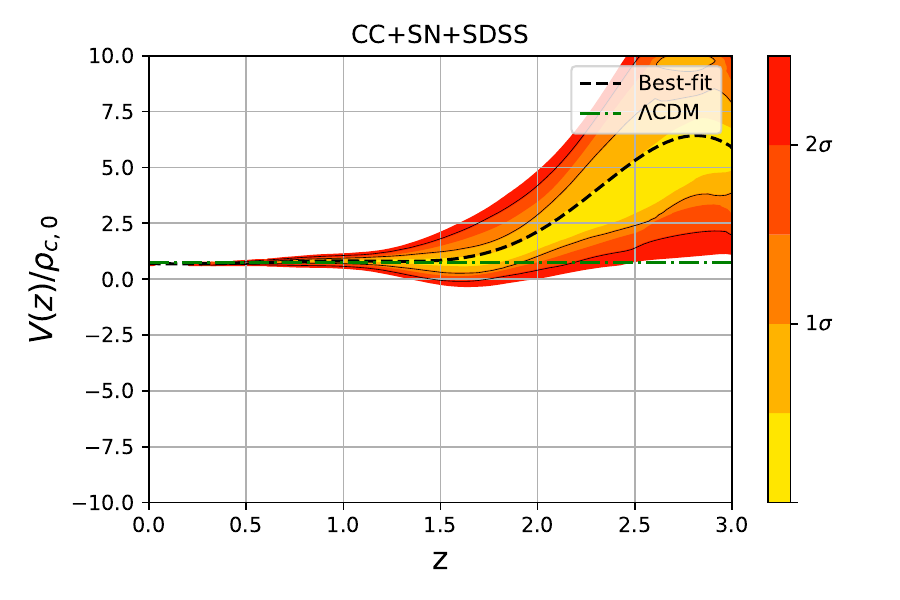}
      }
 \caption{Results of the reconstruction of $H(z)$ for the dataset combination of CC+SN+SDSS. From top-to-bottom and left-to-right we have: $H(z)$, $H(z)/(1+z)$, $q(z)$, $\Delta\mathcal{X} / \rho_{c,0}$, $\rho_{DE}/\rho_{c,0}$, $p_{DE}/\rho_{c,0}$, $w_{DE}$, and $V(z) / \rho_{c,0}$. An important thing to note and clarify is that the last node is located at $z=3.0$, which means that it cannot be constrained by data. As such, high-redshift results around this region should be taken as merely statistical noise. }\label{fig:GP_cc_sn_sdss}
 \end{figure*}

 \begin{figure*}[t!]
     \centering
       \makebox[10cm][c]{
      \includegraphics[trim = 0mm  0mm 0mm 0mm, clip, width=8.9cm, height=5.3cm]{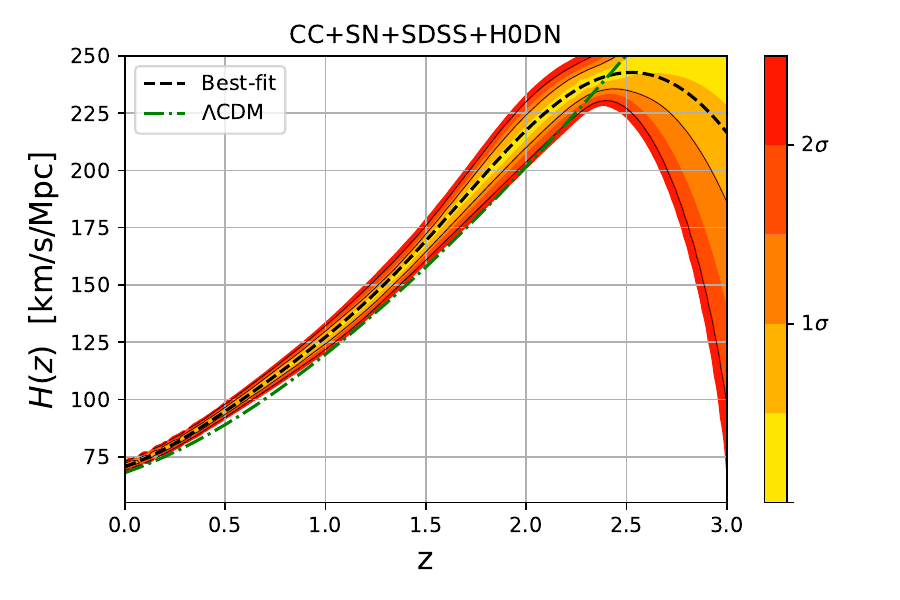}
      \includegraphics[trim = 0mm  0mm 0mm 0mm, clip, width=8.9cm, height=5.3cm]{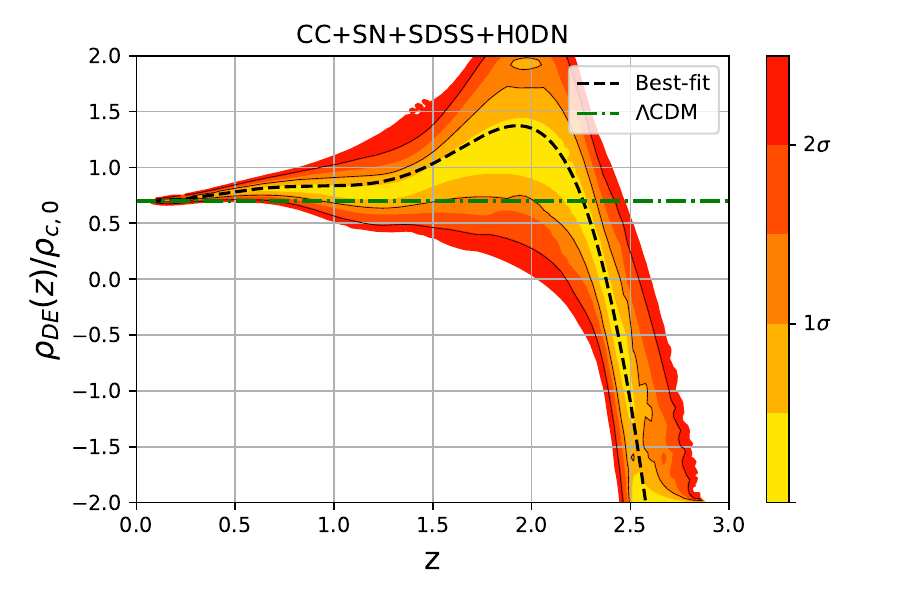}
      }
     \makebox[10cm][c]{
      \includegraphics[trim = 0mm  0mm 0mm 0mm, clip, width=8.9cm, height=5.3cm]{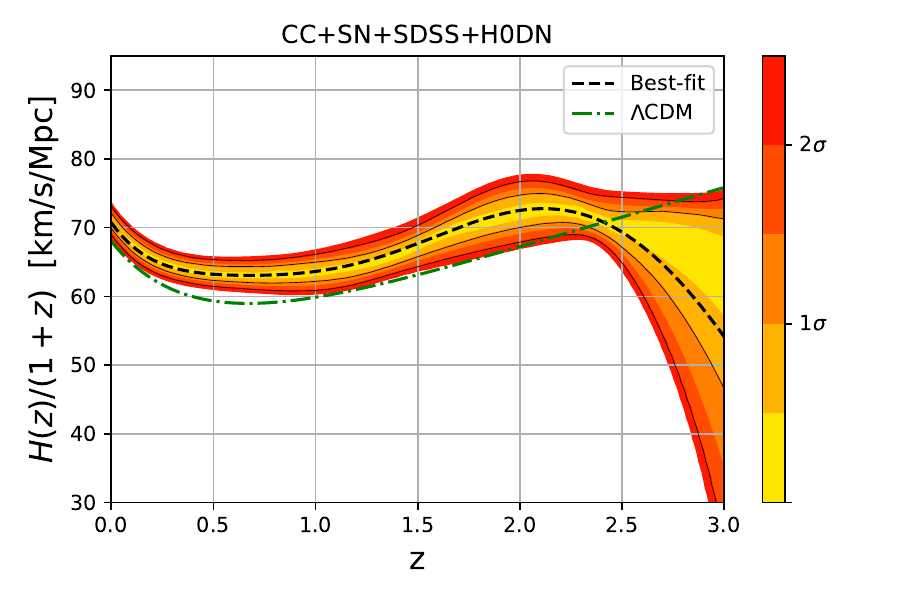}
      \includegraphics[trim = 0mm  0mm 0mm 0mm, clip, width=8.9cm, height=5.3cm]{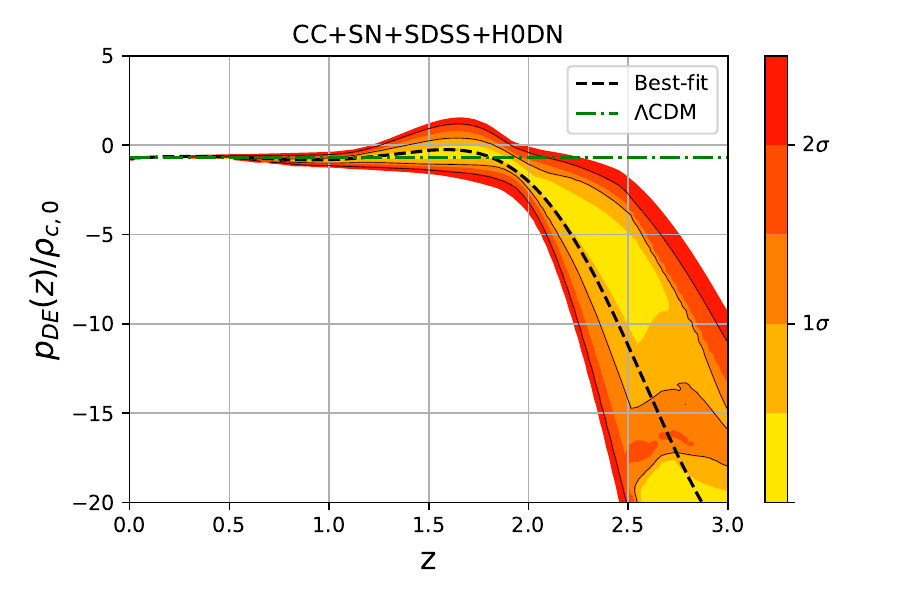}
      }
     \makebox[10cm][c]{
      \includegraphics[trim = 0mm  0mm 0mm 0mm, clip, width=8.9cm, height=5.3cm]{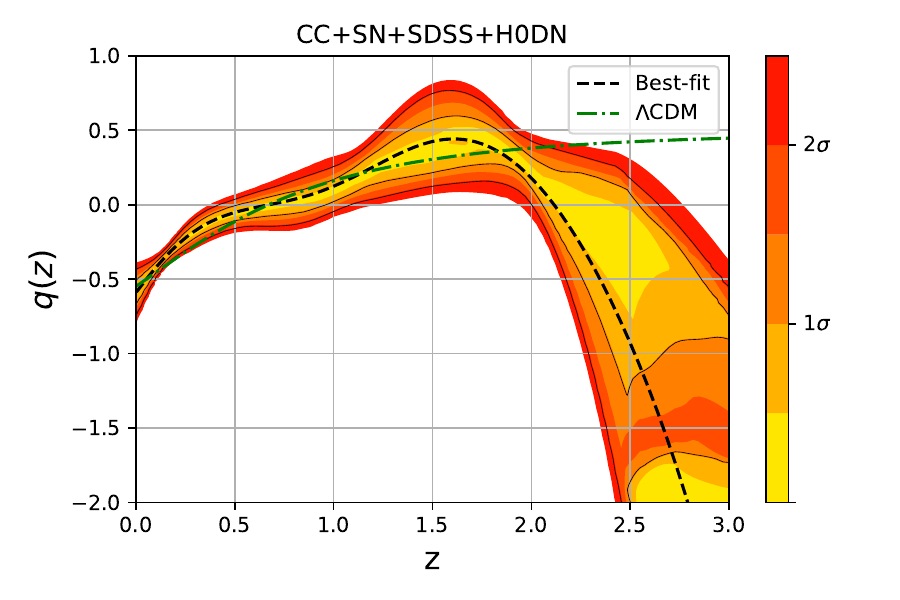}
      \includegraphics[trim = 0mm  0mm 0mm 0mm, clip, width=8.9cm, height=5.3cm]{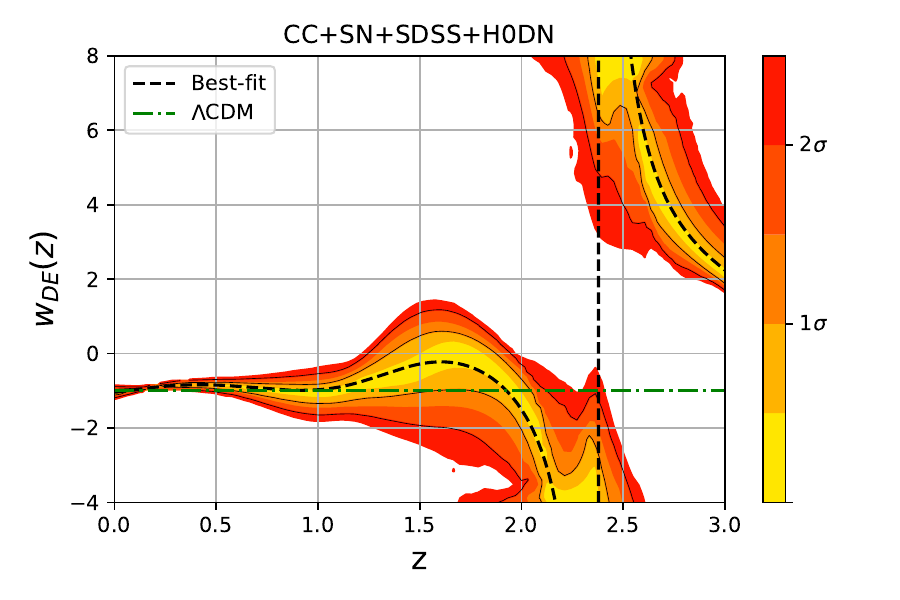}
      }
      \makebox[10cm][c]{
      \includegraphics[trim = 0mm  0mm 0mm 0mm, clip, width=8.9cm, height=5.3cm]{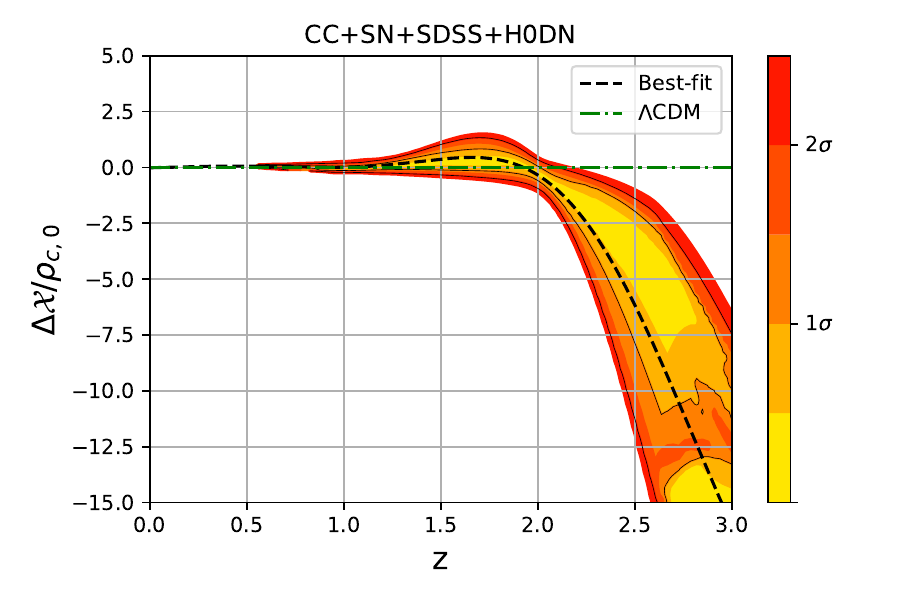}
      \includegraphics[trim = 0mm  0mm 0mm 0mm, clip, width=8.9cm, height=5.3cm]{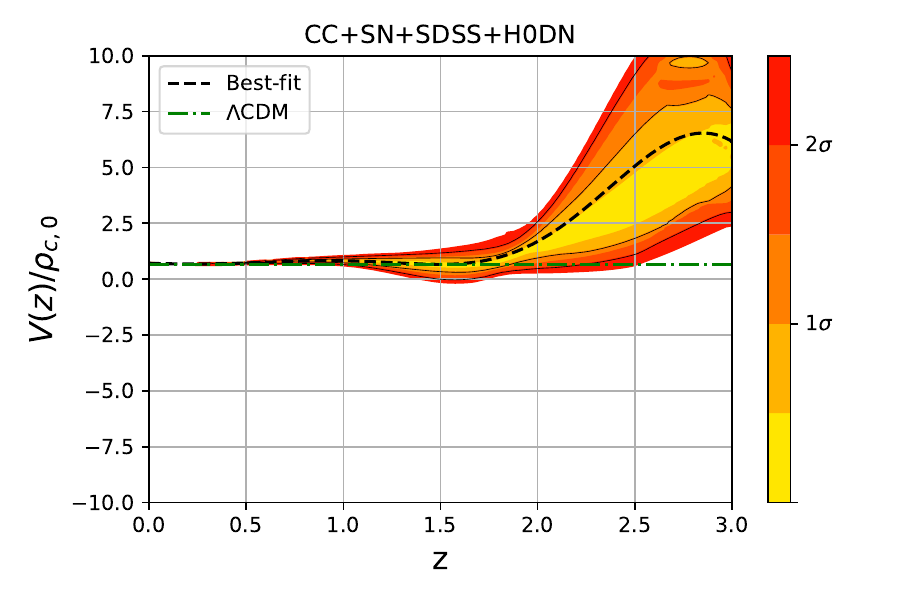}
      }
 \caption{Results of the reconstruction of $H(z)$ for the dataset combination of CC+SN+SDSS+H0DN. From top-to-bottom and left-to-right we have: $H(z)$, $H(z)/(1+z)$, $q(z)$, $\Delta\mathcal{X} / \rho_{c,0}$, $\rho_{DE}/\rho_{c,0}$, $p_{DE}/\rho_{c,0}$, $w_{DE}$, and $V(z) / \rho_{c,0}$. An important thing to note and clarify is that the last node is located at $z=3.0$, which means that it cannot be constrained by data. As such, high-redshift results around this region should be taken as merely statistical noise. }\label{fig:GP_cc_sn_sdss+h0dn}
 \end{figure*}

 \begin{figure*}[t!]
     \centering
       \makebox[10cm][c]{
      \includegraphics[trim = 0mm  0mm 0mm 0mm, clip, width=8.9cm, height=5.3cm]{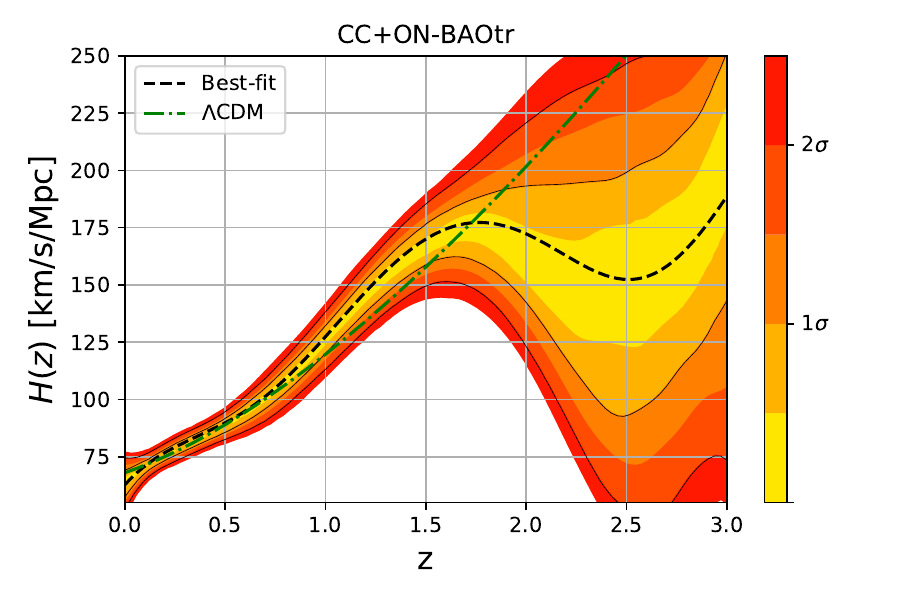}
      \includegraphics[trim = 0mm  0mm 0mm 0mm, clip, width=8.9cm, height=5.3cm]{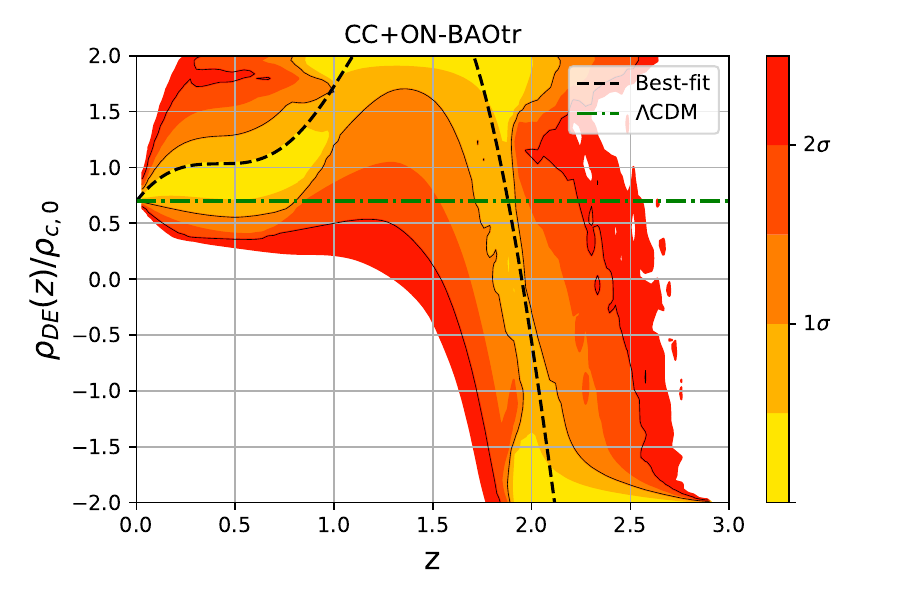}
      }
     \makebox[10cm][c]{
      \includegraphics[trim = 0mm  0mm 0mm 0mm, clip, width=8.9cm, height=5.3cm]{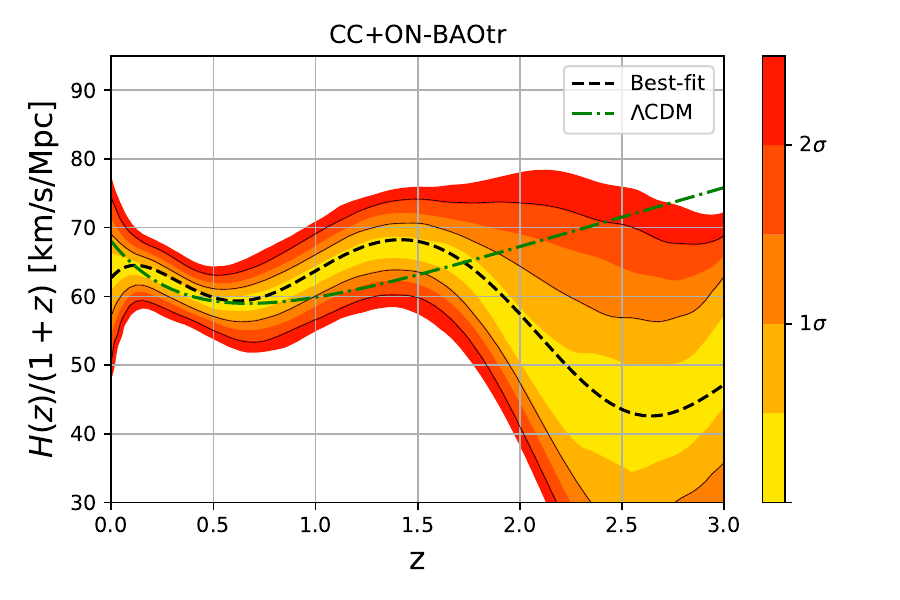}
      \includegraphics[trim = 0mm  0mm 0mm 0mm, clip, width=8.9cm, height=5.3cm]{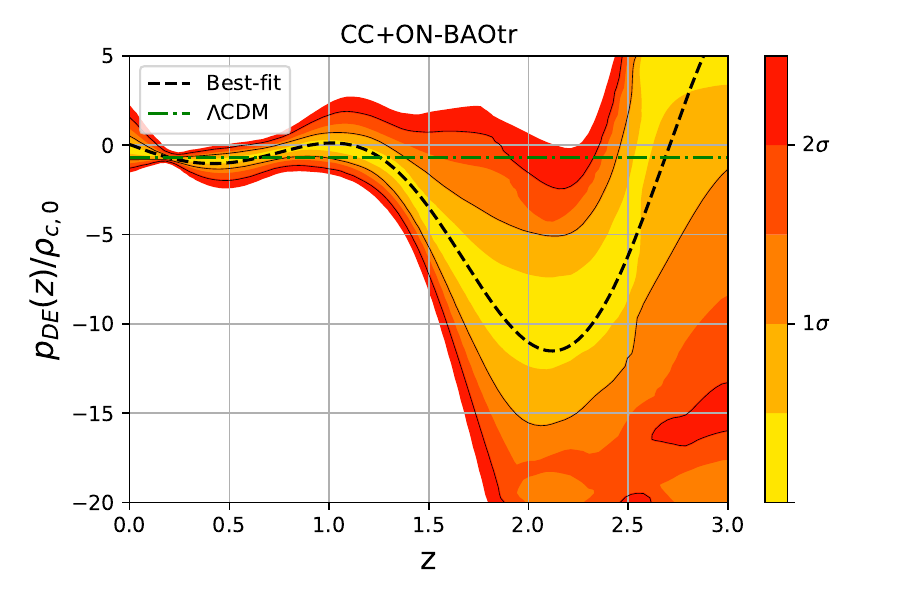}
      }
     \makebox[10cm][c]{
      \includegraphics[trim = 0mm  0mm 0mm 0mm, clip, width=8.9cm, height=5.3cm]{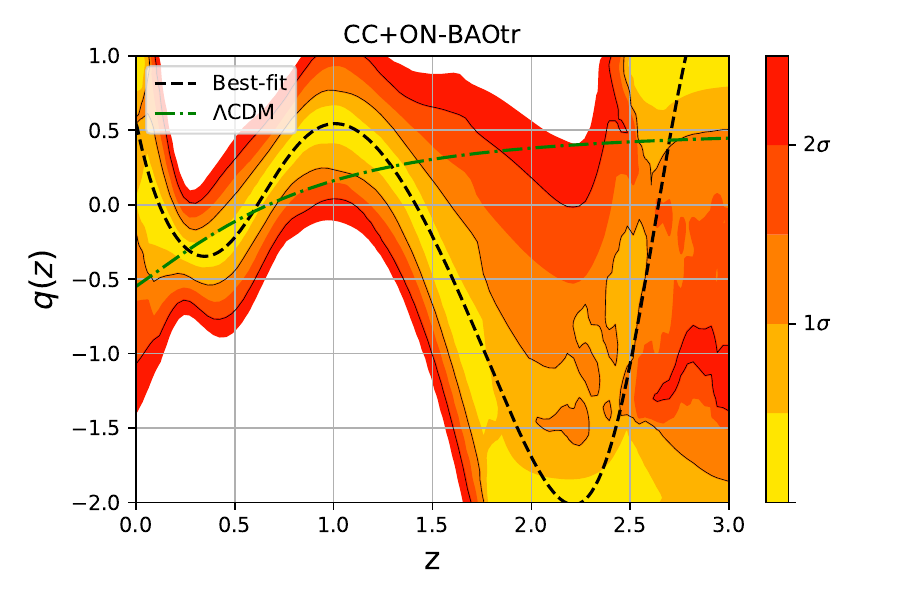}
      \includegraphics[trim = 0mm  0mm 0mm 0mm, clip, width=8.9cm, height=5.3cm]{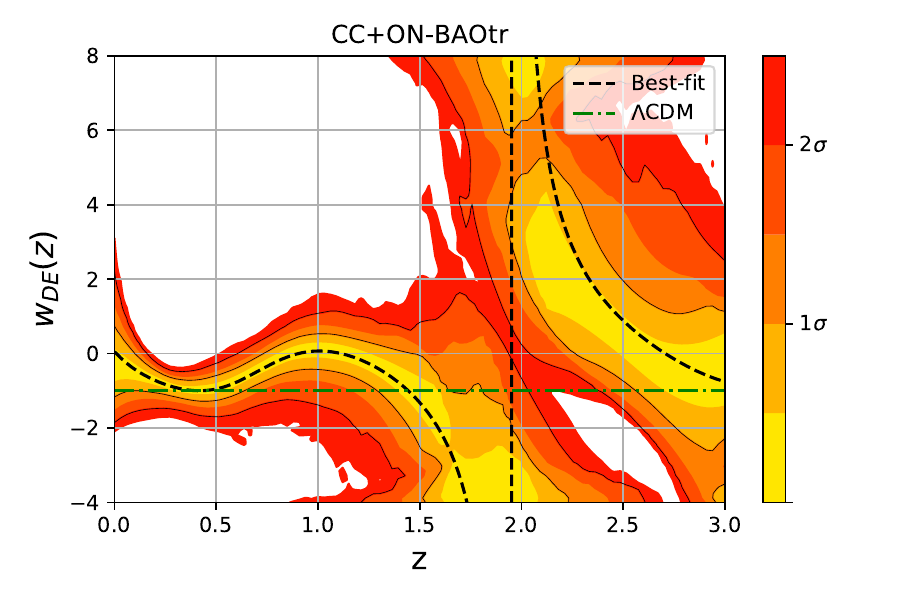}
      }
      \makebox[10cm][c]{
      \includegraphics[trim = 0mm  0mm 0mm 0mm, clip, width=8.9cm, height=5.3cm]{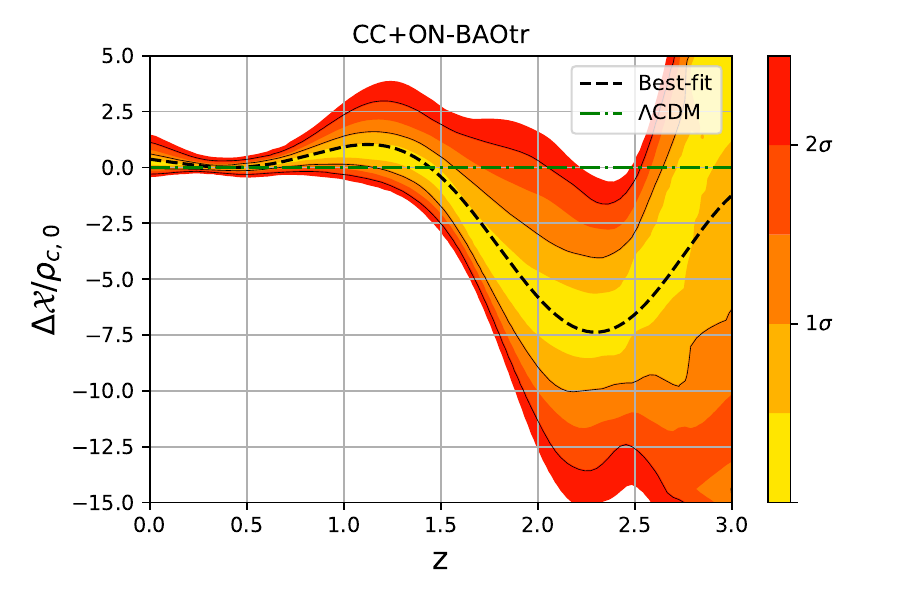}
      \includegraphics[trim = 0mm  0mm 0mm 0mm, clip, width=8.9cm, height=5.3cm]{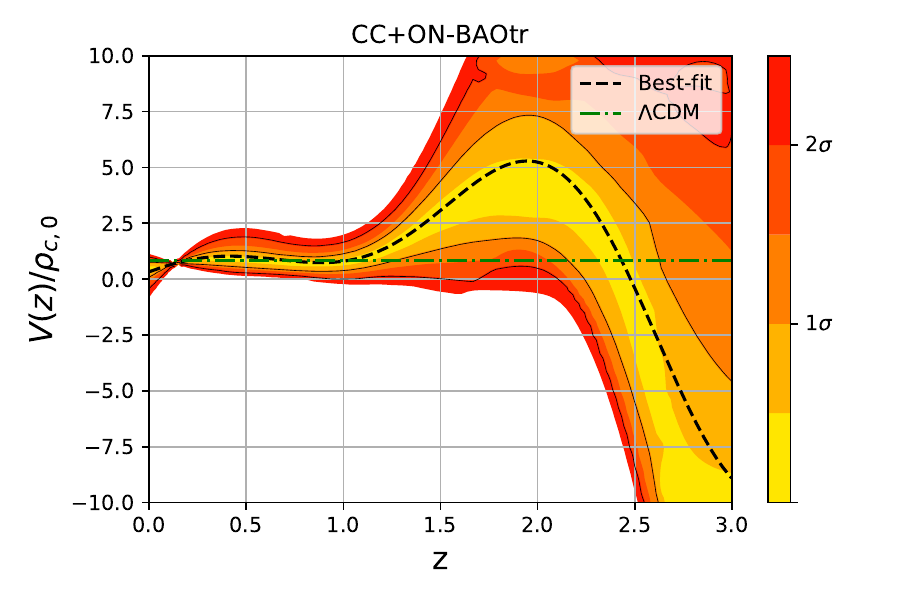}
      }
 \caption{Results of the reconstruction of $H(z)$ for the dataset combination of CC+ON-BAOtr. From top-to-bottom and left-to-right we have: $H(z)$, $H(z)/(1+z)$, $q(z)$, $\Delta\mathcal{X} / \rho_{c,0}$, $\rho_{DE}/\rho_{c,0}$, $p_{DE}/\rho_{c,0}$, $w_{DE}$, and $V(z) / \rho_{c,0}$. An important thing to note and clarify is that the last node is located at $z=3.0$, which means that it cannot be constrained by data. As such, high-redshift results around this region should be taken as merely statistical noise. }\label{fig:GP_cc_on2dbao}
 \end{figure*}

 \begin{figure*}[t!]
     \centering
       \makebox[10cm][c]{
      \includegraphics[trim = 0mm  0mm 0mm 0mm, clip, width=8.9cm, height=5.3cm]{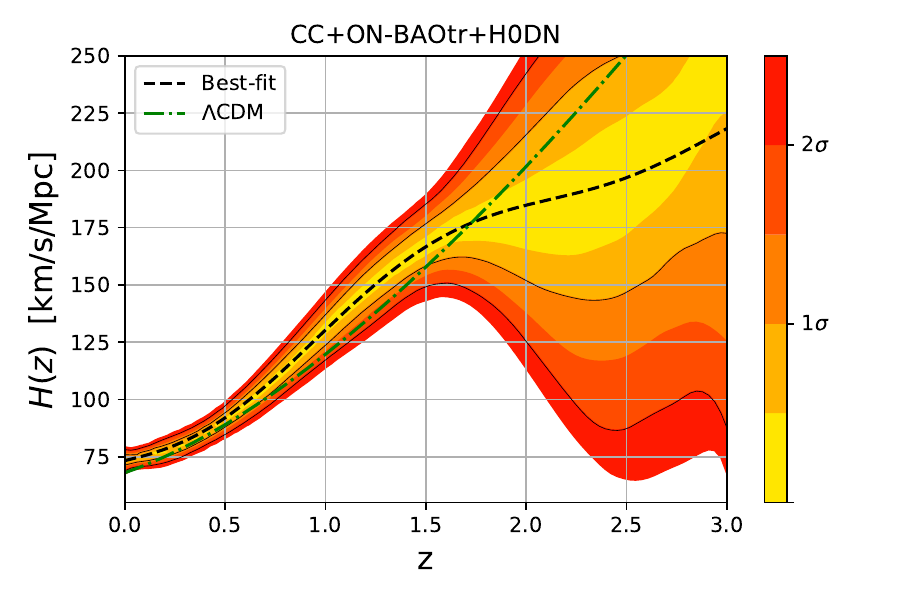}
      \includegraphics[trim = 0mm  0mm 0mm 0mm, clip, width=8.9cm, height=5.3cm]{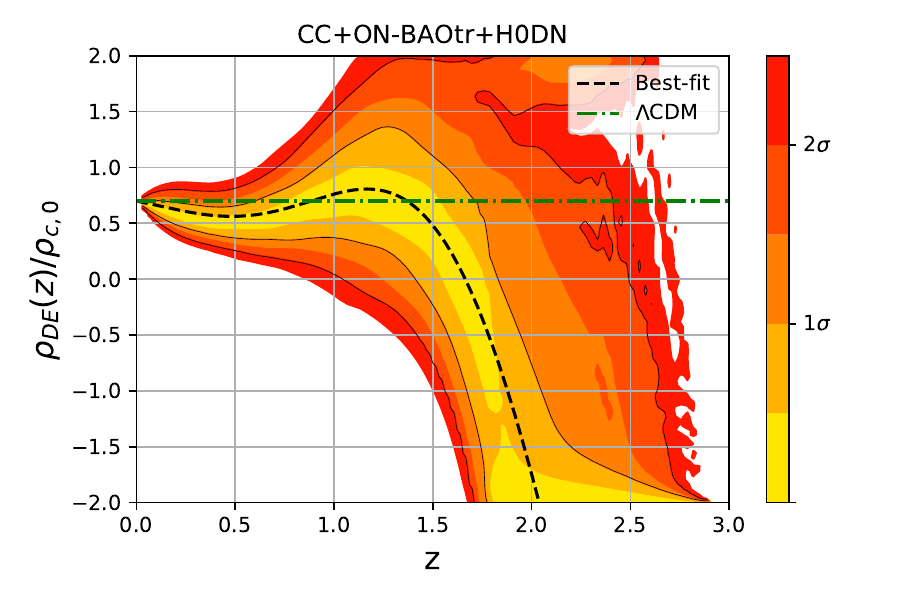}
      }
     \makebox[10cm][c]{
      \includegraphics[trim = 0mm  0mm 0mm 0mm, clip, width=8.9cm, height=5.3cm]{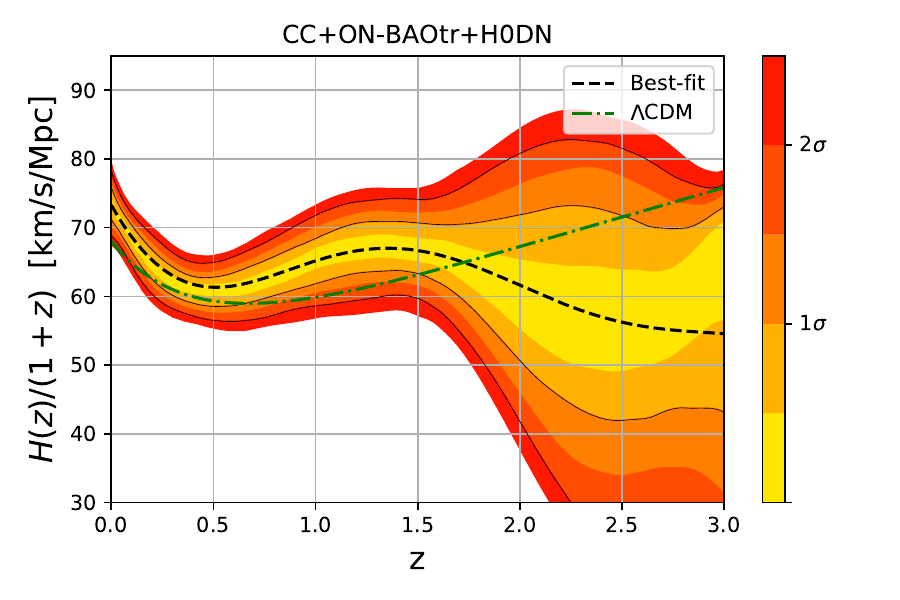}
      \includegraphics[trim = 0mm  0mm 0mm 0mm, clip, width=8.9cm, height=5.3cm]{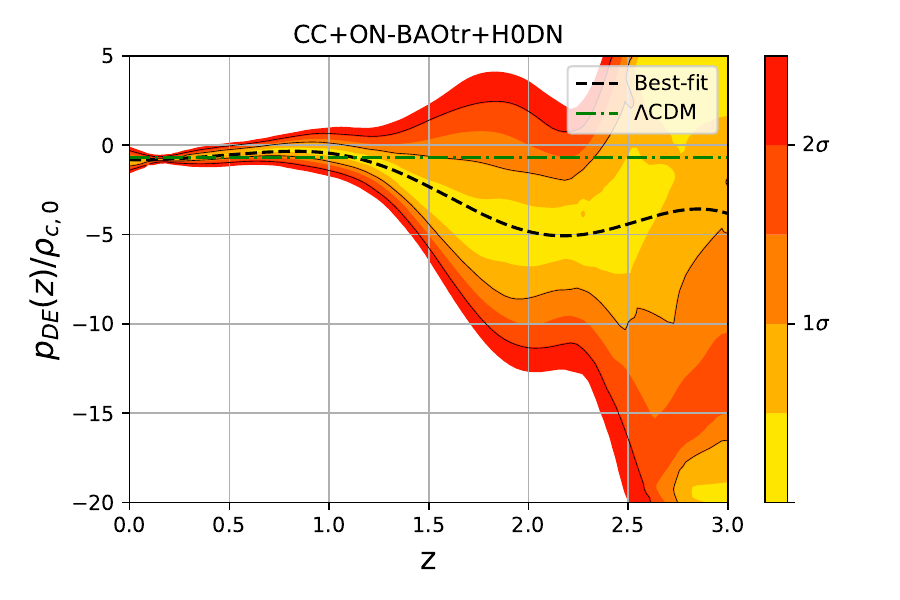}
      }
     \makebox[10cm][c]{
      \includegraphics[trim = 0mm  0mm 0mm 0mm, clip, width=8.9cm, height=5.3cm]{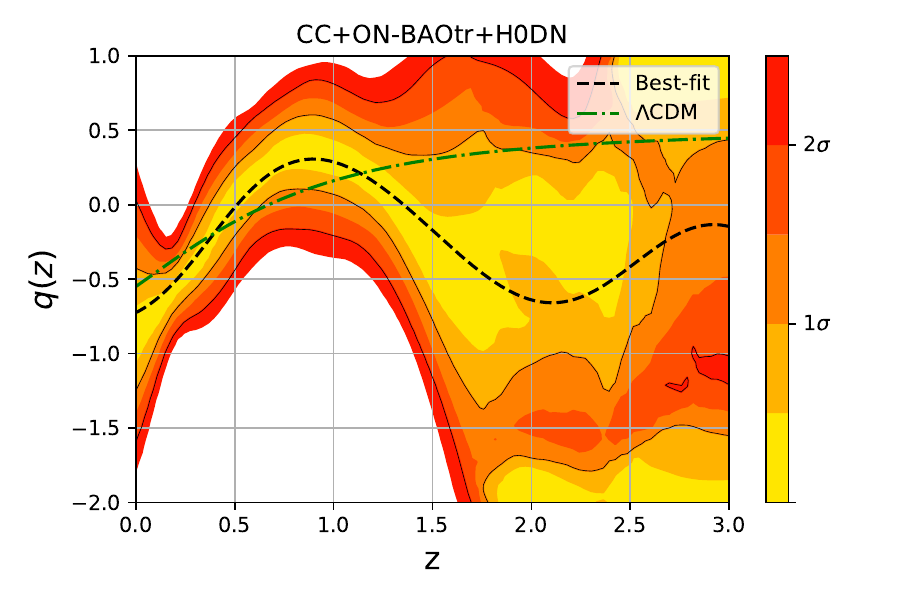}
      \includegraphics[trim = 0mm  0mm 0mm 0mm, clip, width=8.9cm, height=5.3cm]{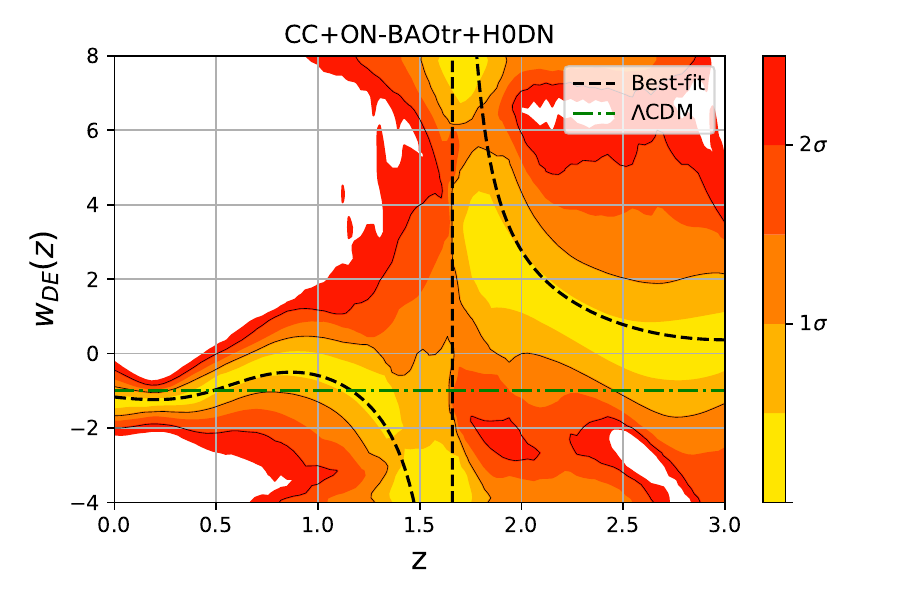}
      }
      \makebox[10cm][c]{
      \includegraphics[trim = 0mm  0mm 0mm 0mm, clip, width=8.9cm, height=5.3cm]{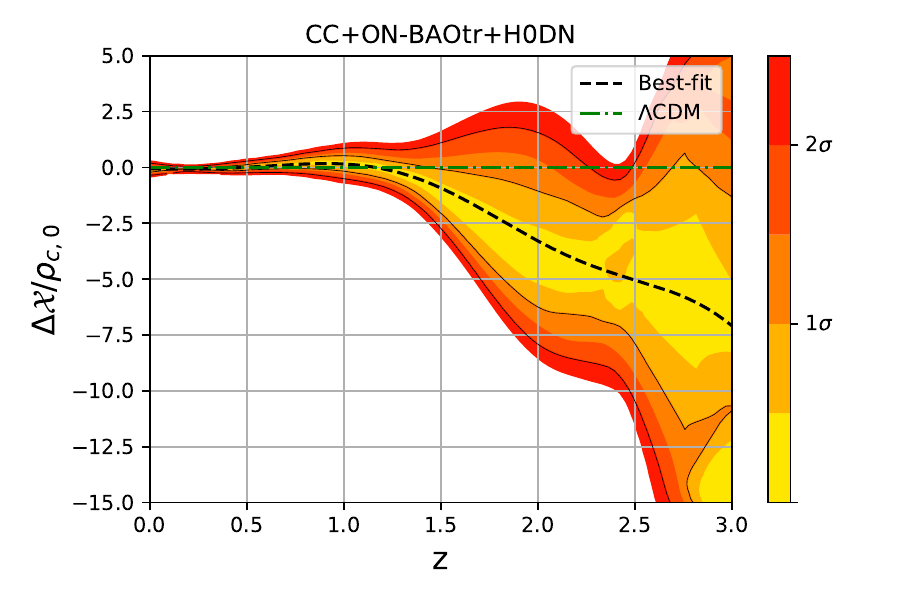}
      \includegraphics[trim = 0mm  0mm 0mm 0mm, clip, width=8.9cm, height=5.3cm]{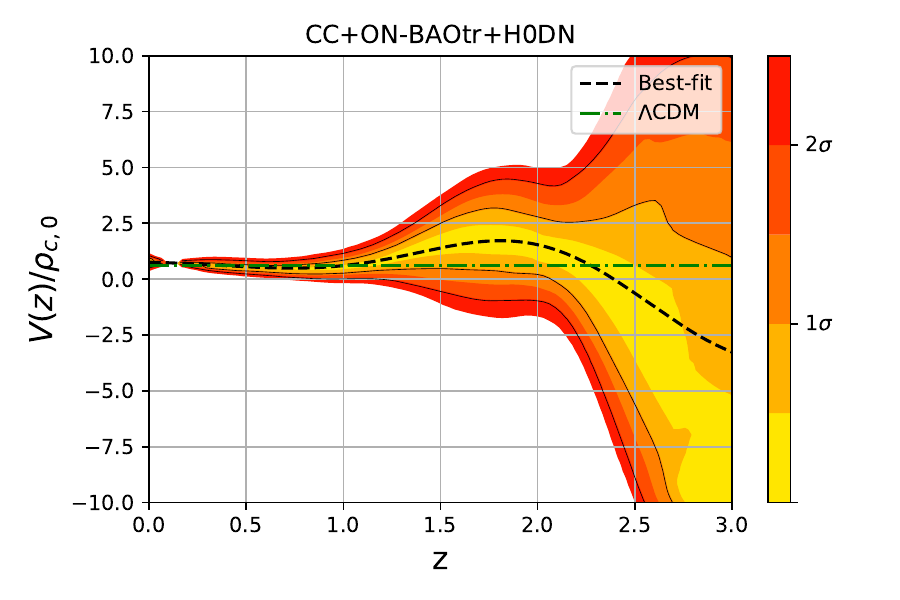}
     }
 \caption{Results of the reconstruction of $H(z)$ for the dataset combination of CC+ON-BAOtr+H0DN. From top-to-bottom and left-to-right we have: $H(z)$, $H(z)/(1+z)$, $q(z)$, $\Delta\mathcal{X} / \rho_{c,0}$, $\rho_{DE}/\rho_{c,0}$, $p_{DE}/\rho_{c,0}$, $w_{DE}$, and $V(z) / \rho_{c,0}$. An important thing to note and clarify is that the last node is located at $z=3.0$, which means that it cannot be constrained by data. As such, high-redshift results around this region should be taken as merely statistical noise. }\label{fig:GP_cc_on2dbao_h0dn}
 \end{figure*}

 \begin{figure*}[t!]
     \centering
       \makebox[10cm][c]{
      \includegraphics[trim = 0mm  0mm 0mm 0mm, clip, width=8.9cm, height=5.3cm]{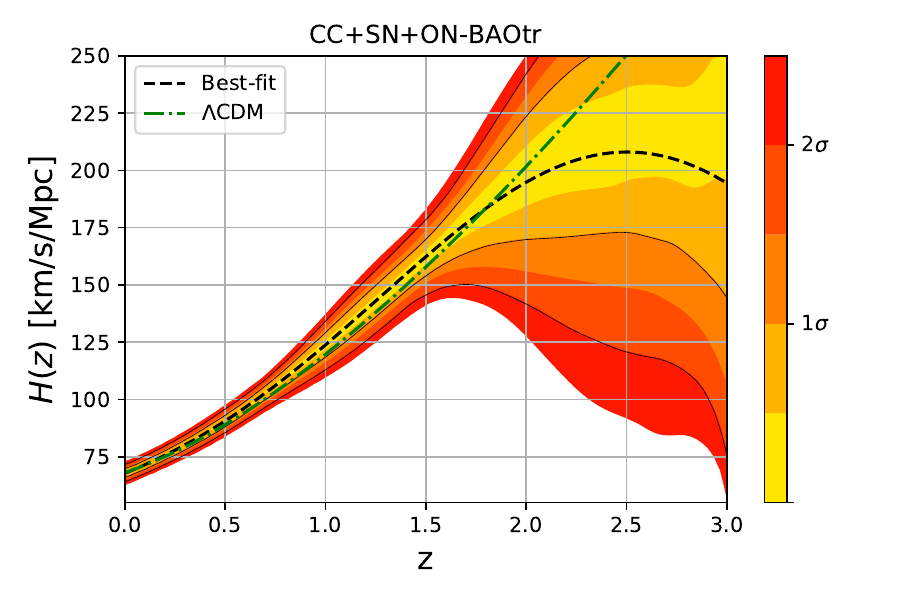}
      \includegraphics[trim = 0mm  0mm 0mm 0mm, clip, width=8.9cm, height=5.3cm]{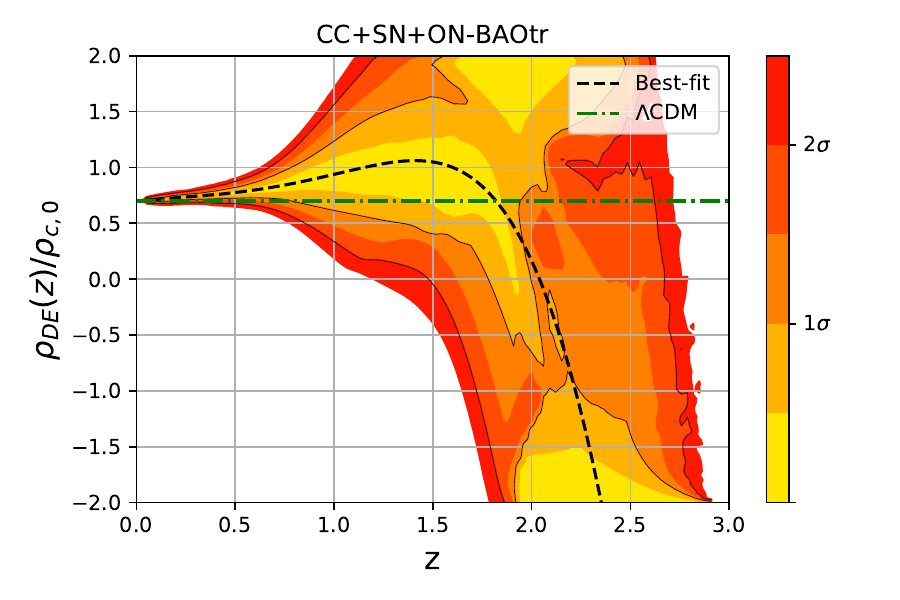}
      }
     \makebox[10cm][c]{
      \includegraphics[trim = 0mm  0mm 0mm 0mm, clip, width=8.9cm, height=5.3cm]{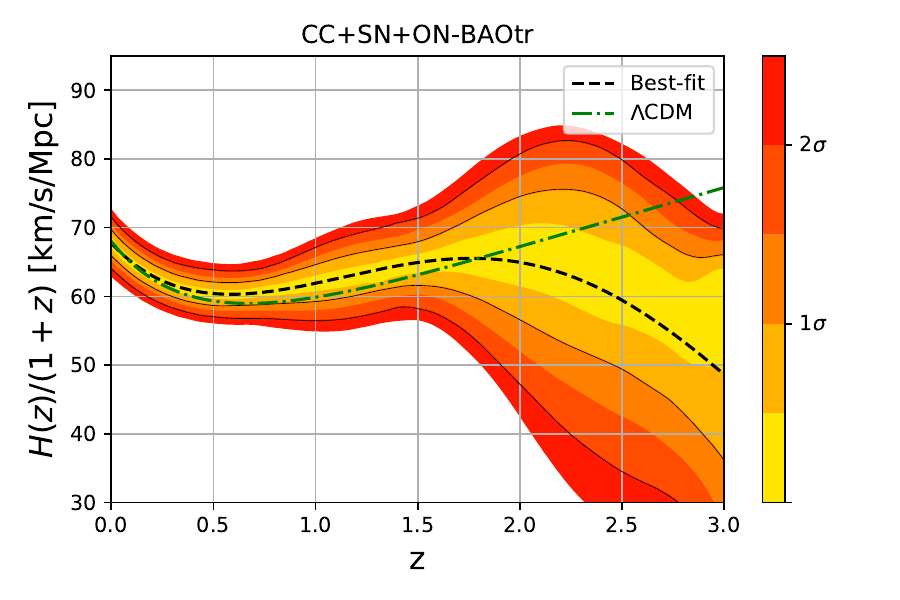}
      \includegraphics[trim = 0mm  0mm 0mm 0mm, clip, width=8.9cm, height=5.3cm]{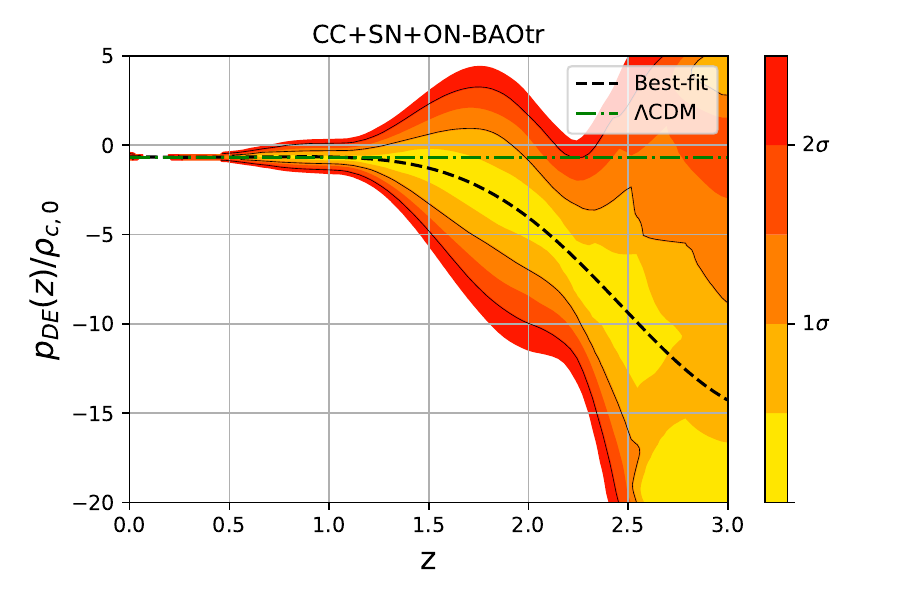}
      }
     \makebox[10cm][c]{
      \includegraphics[trim = 0mm  0mm 0mm 0mm, clip, width=8.9cm, height=5.3cm]{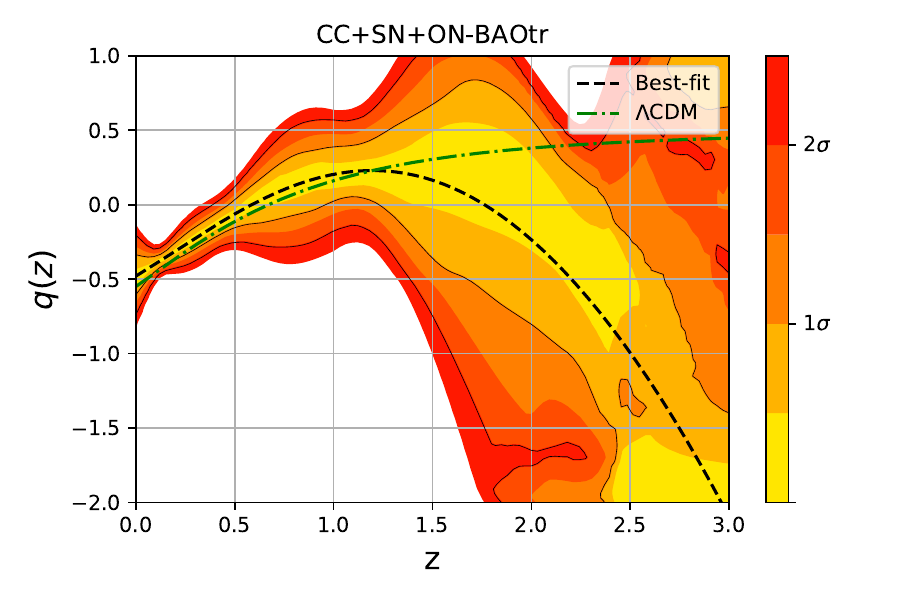}
      \includegraphics[trim = 0mm  0mm 0mm 0mm, clip, width=8.9cm, height=5.3cm]{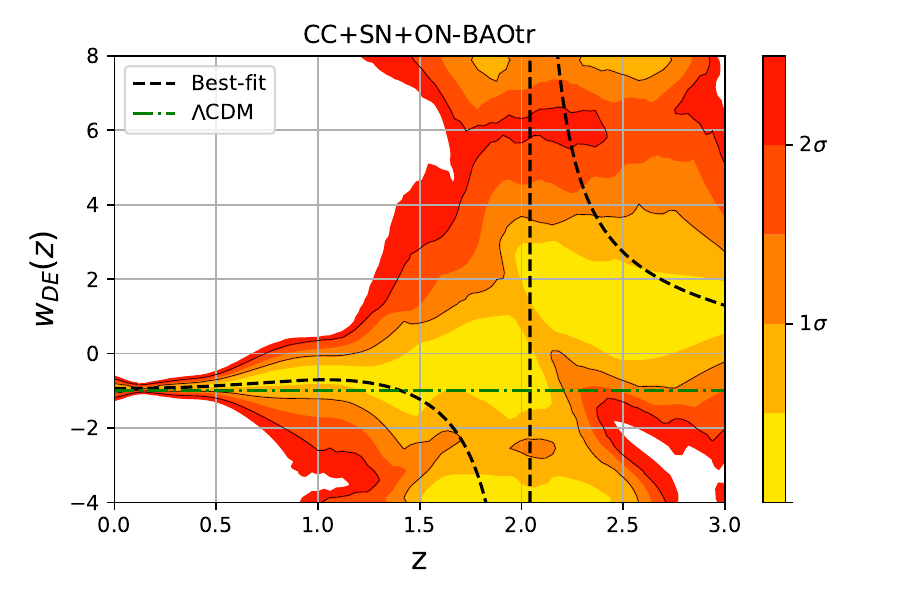}
      }
      \makebox[10cm][c]{
      \includegraphics[trim = 0mm  0mm 0mm 0mm, clip, width=8.9cm, height=5.3cm]{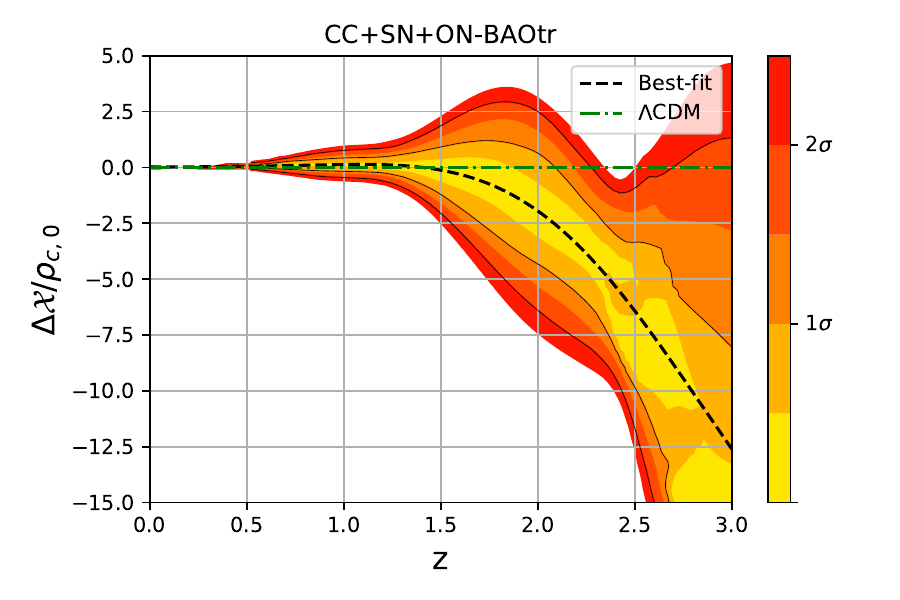}
      \includegraphics[trim = 0mm  0mm 0mm 0mm, clip, width=8.9cm, height=5.3cm]{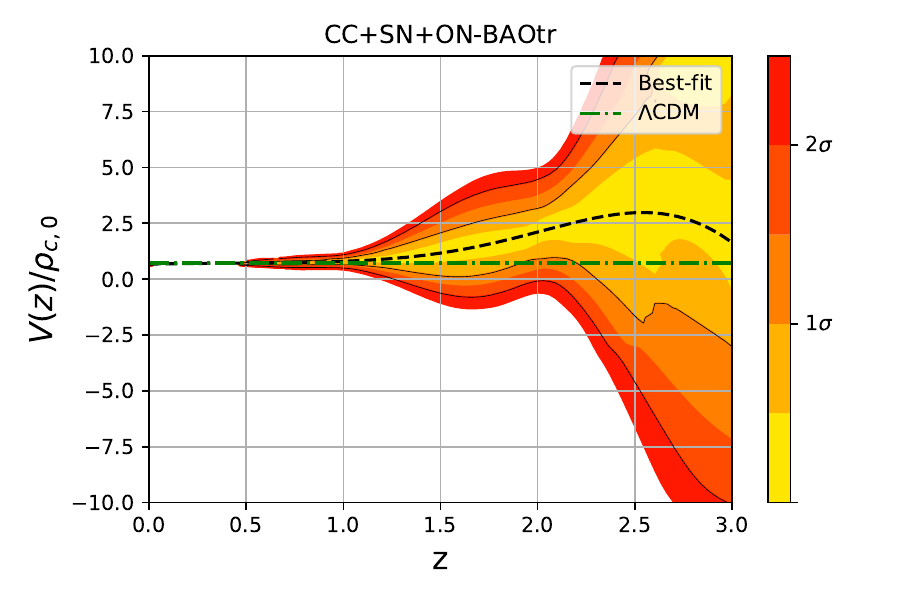}
      }
 \caption{Results of the reconstruction of $H(z)$ for the dataset combination of CC+SN+ON-BAOtr. From top-to-bottom and left-to-right we have: $H(z)$, $H(z)/(1+z)$, $q(z)$, $\Delta\mathcal{X} / \rho_{c,0}$, $\rho_{DE}/\rho_{c,0}$, $p_{DE}/\rho_{c,0}$, $w_{DE}$, and $V(z) / \rho_{c,0}$. An important thing to note and clarify is that the last node is located at $z=3.0$, which means that it cannot be constrained by data. As such, high-redshift results around this region should be taken as merely statistical noise. }\label{fig:GP_cc_sn_on2dbao}
 \end{figure*}

 \begin{figure*}[t!]
     \centering
       \makebox[10cm][c]{
      \includegraphics[trim = 0mm  0mm 0mm 0mm, clip, width=8.9cm, height=5.3cm]{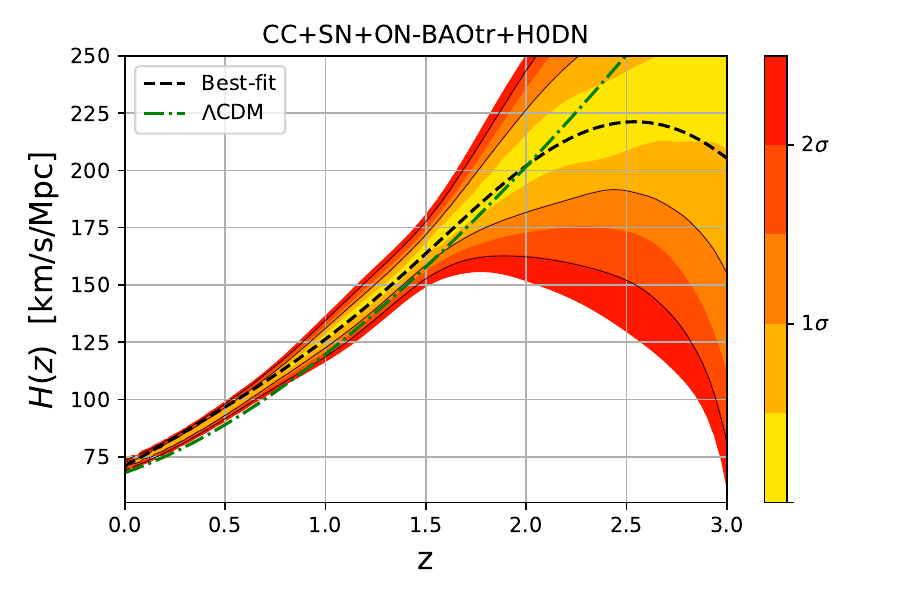}
      \includegraphics[trim = 0mm  0mm 0mm 0mm, clip, width=8.9cm, height=5.3cm]{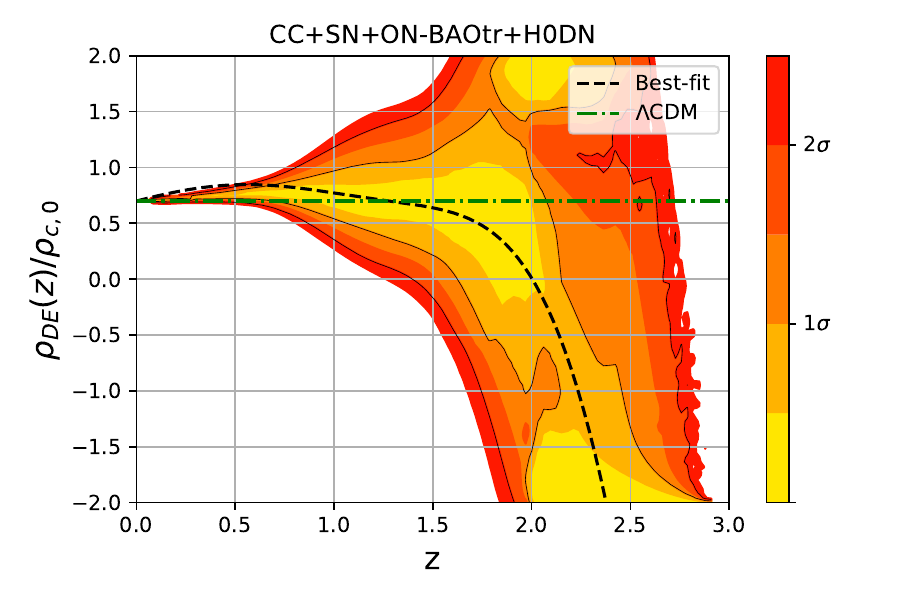}
      }
     \makebox[10cm][c]{
      \includegraphics[trim = 0mm  0mm 0mm 0mm, clip, width=8.9cm, height=5.3cm]{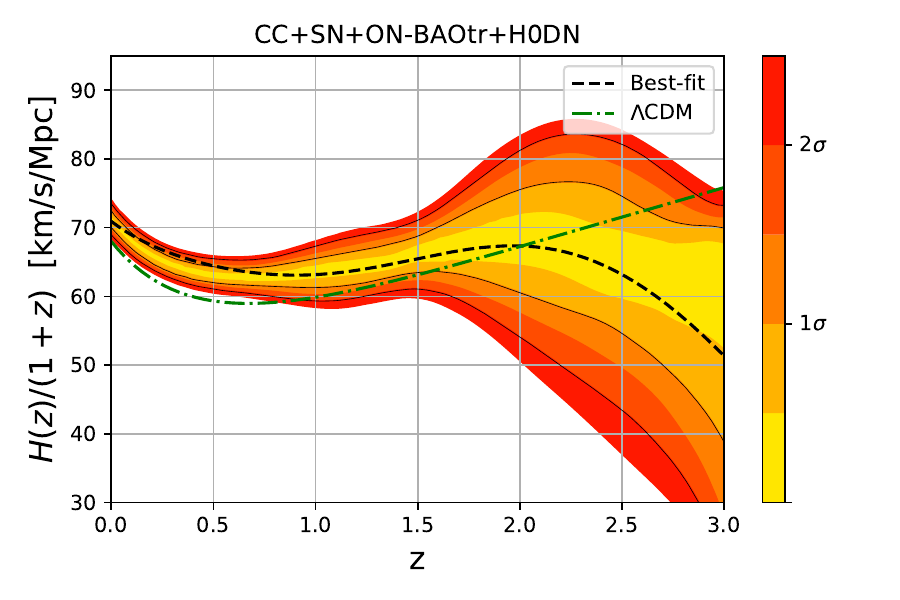}
      \includegraphics[trim = 0mm  0mm 0mm 0mm, clip, width=8.9cm, height=5.3cm]{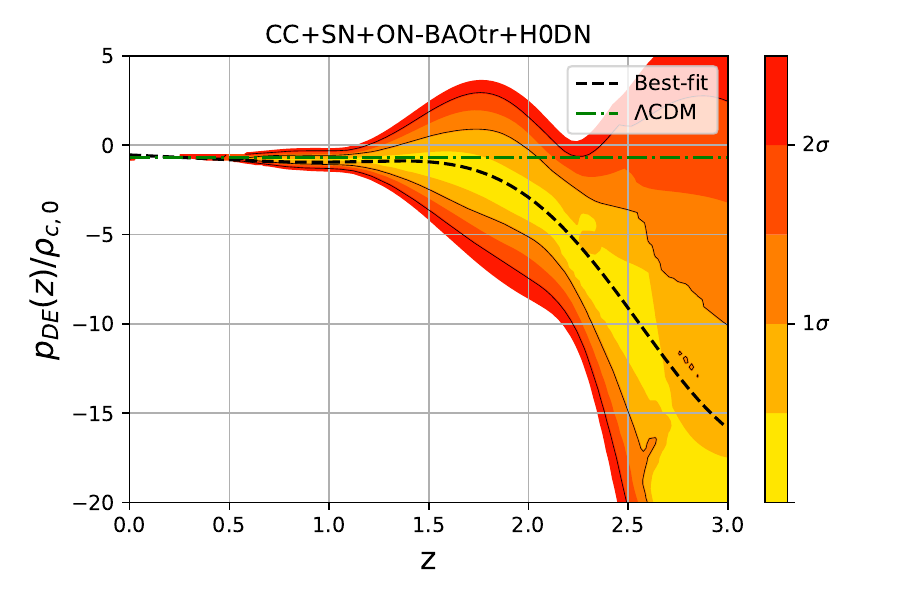}
      }
     \makebox[10cm][c]{
      \includegraphics[trim = 0mm  0mm 0mm 0mm, clip, width=8.9cm, height=5.3cm]{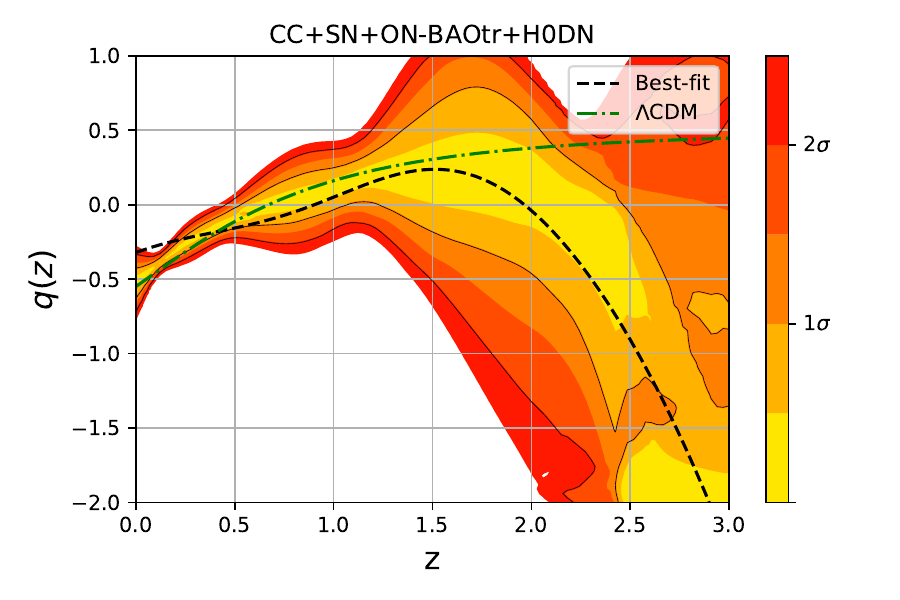}
      \includegraphics[trim = 0mm  0mm 0mm 0mm, clip, width=8.9cm, height=5.3cm]{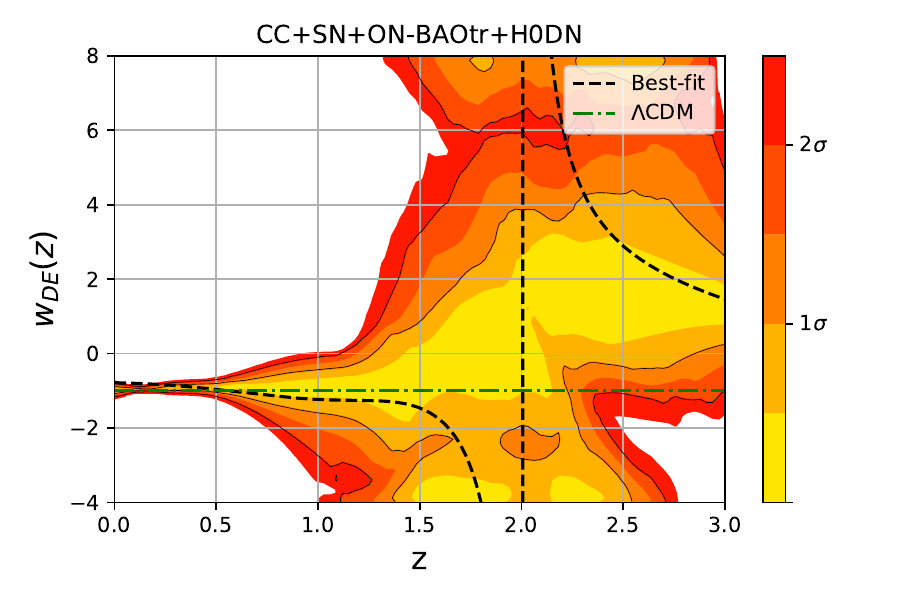}
      }
      \makebox[10cm][c]{
      \includegraphics[trim = 0mm  0mm 0mm 0mm, clip, width=8.9cm, height=5.3cm]{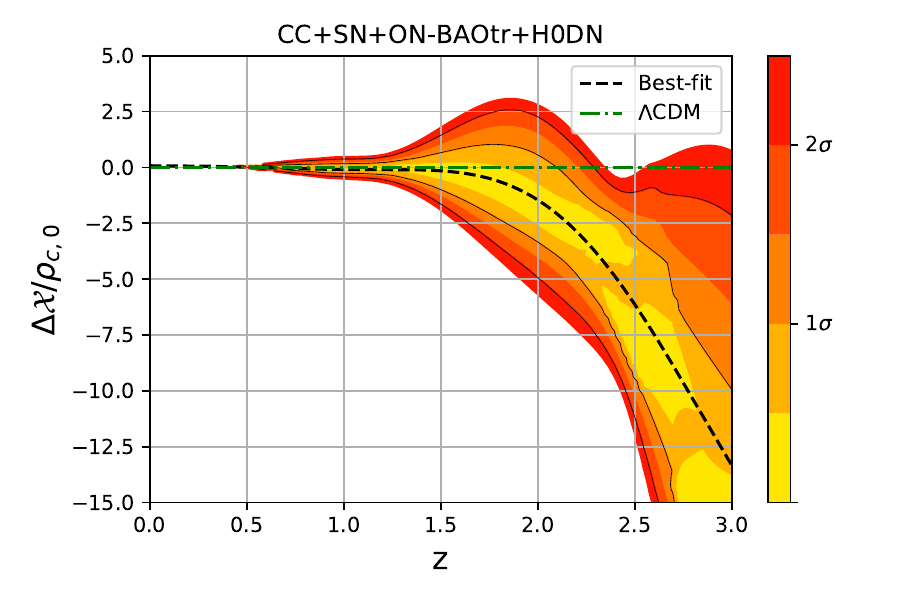}
      \includegraphics[trim = 0mm  0mm 0mm 0mm, clip, width=8.9cm, height=5.3cm]{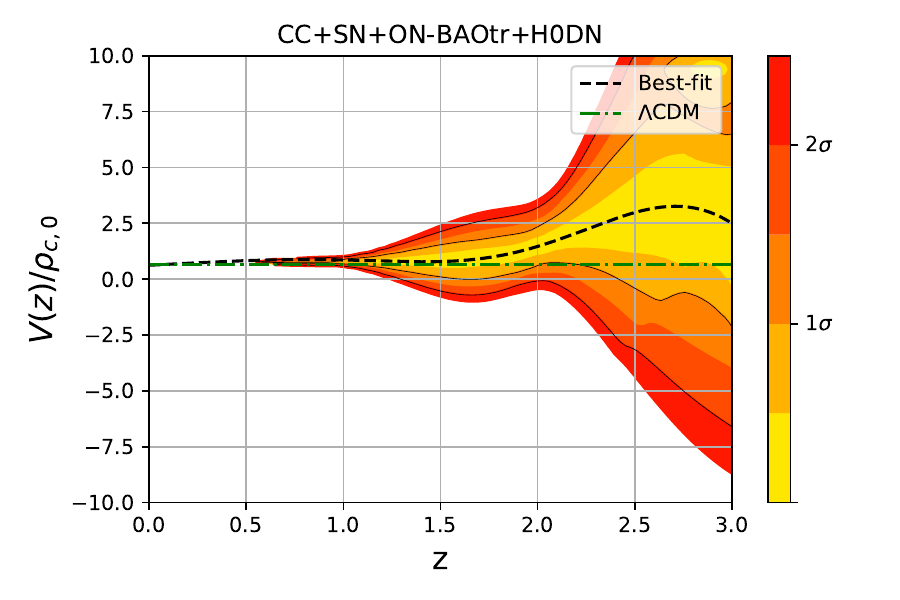}
     }
 \caption{Results of the reconstruction of $H(z)$ for the dataset combination of CC+SN+ON-BAOtr+H0DN. From top-to-bottom and left-to-right we have: $H(z)$, $H(z)/(1+z)$, $q(z)$, $\Delta\mathcal{X} / \rho_{c,0}$, $\rho_{DE}/\rho_{c,0}$, $p_{DE}/\rho_{c,0}$, $w_{DE}$, and $V(z) / \rho_{c,0}$. An important thing to note and clarify is that the last node is located at $z=3.0$, which means that it cannot be constrained by data. As such, high-redshift results around this region should be taken as merely statistical noise. }\label{fig:GP_cc_sn_on2dbao_h0dn}
 \end{figure*}

 \begin{figure*}[t!]
     \centering
       \makebox[10cm][c]{
      \includegraphics[trim = 0mm  0mm 0mm 0mm, clip, width=8.9cm, height=5.3cm]{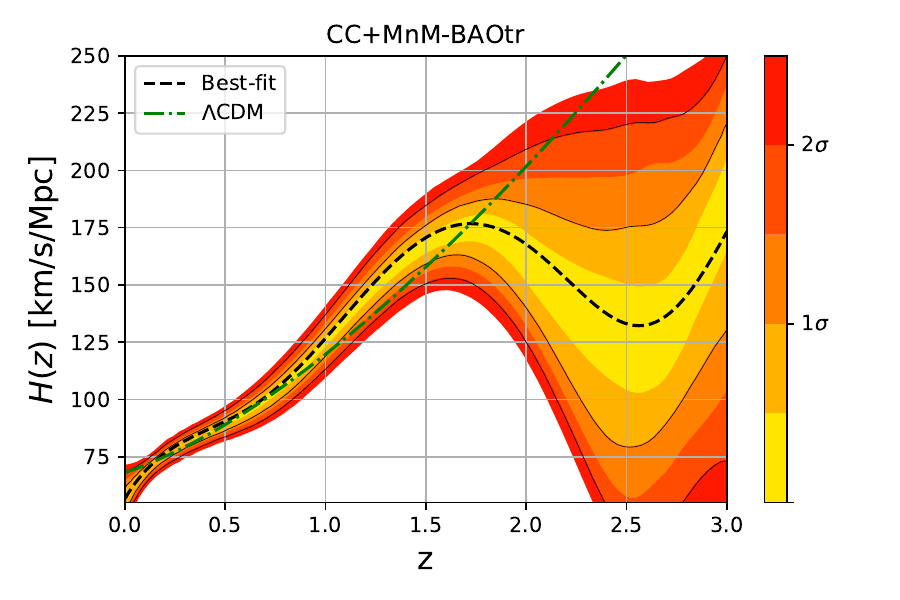}
      \includegraphics[trim = 0mm  0mm 0mm 0mm, clip, width=8.9cm, height=5.3cm]{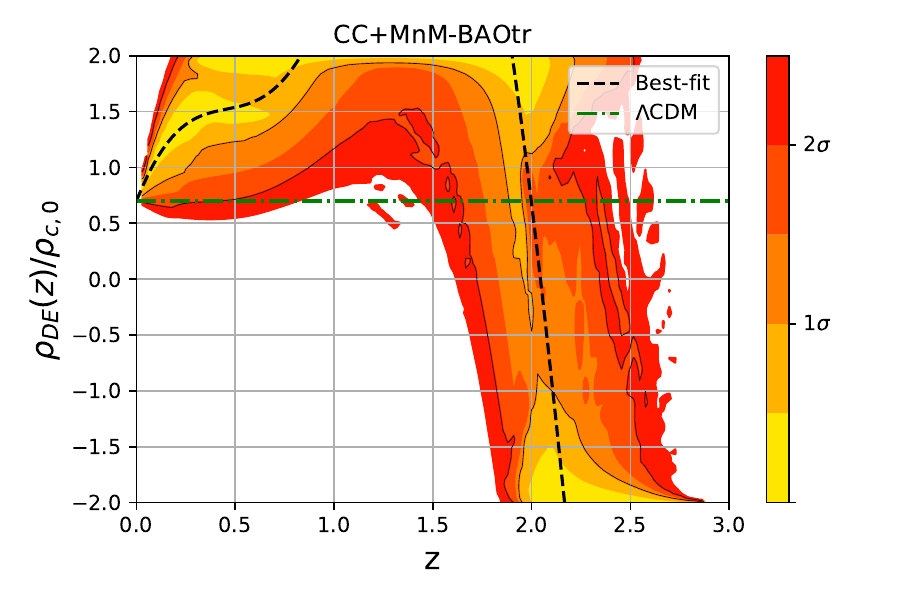}
      }
     \makebox[10cm][c]{
      \includegraphics[trim = 0mm  0mm 0mm 0mm, clip, width=8.9cm, height=5.3cm]{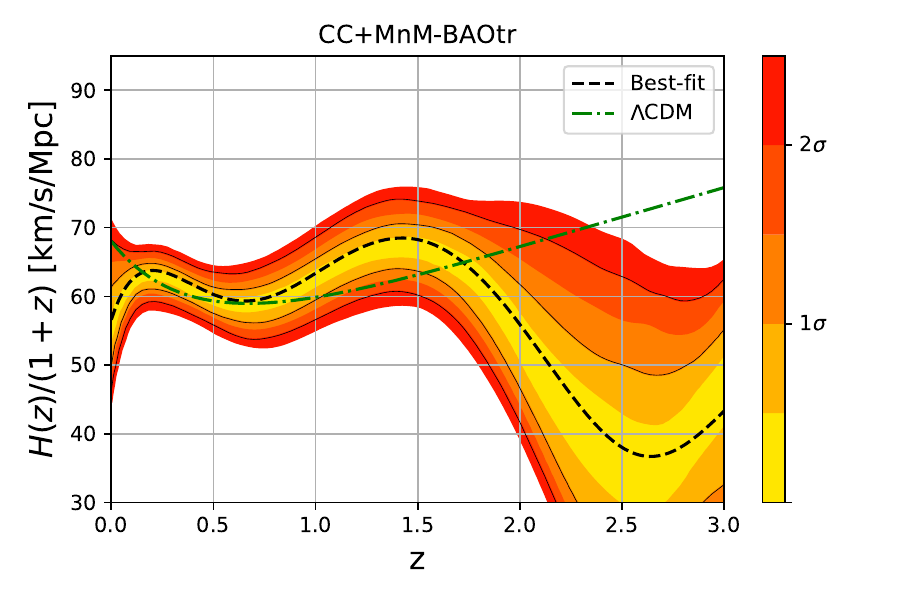}
      \includegraphics[trim = 0mm  0mm 0mm 0mm, clip, width=8.9cm, height=5.3cm]{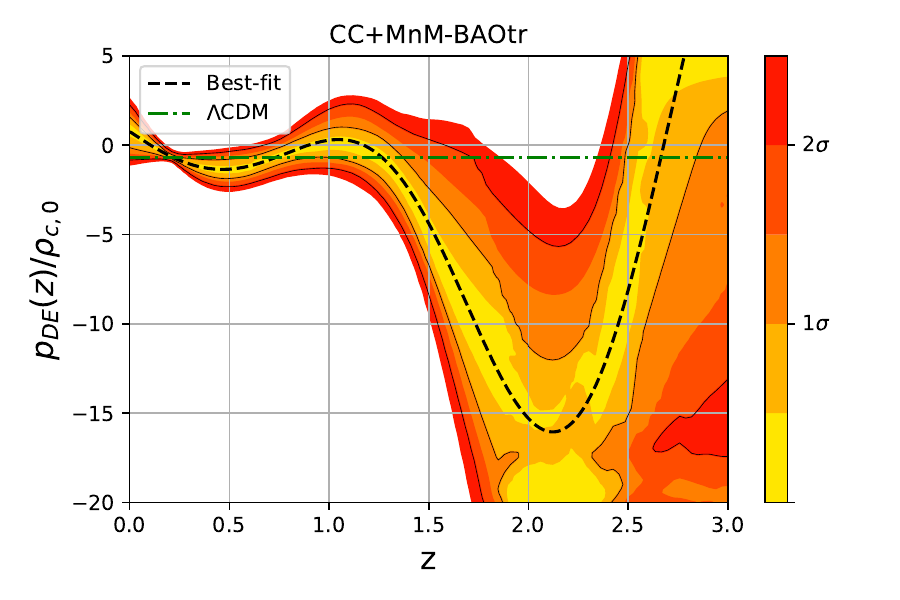}
      }
     \makebox[10cm][c]{
      \includegraphics[trim = 0mm  0mm 0mm 0mm, clip, width=8.9cm, height=5.3cm]{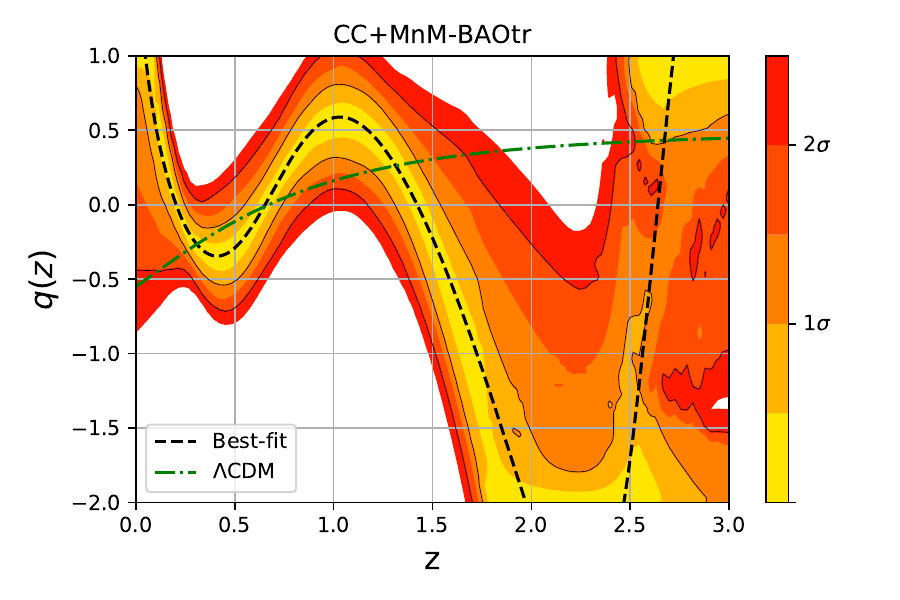}
      \includegraphics[trim = 0mm  0mm 0mm 0mm, clip, width=8.9cm, height=5.3cm]{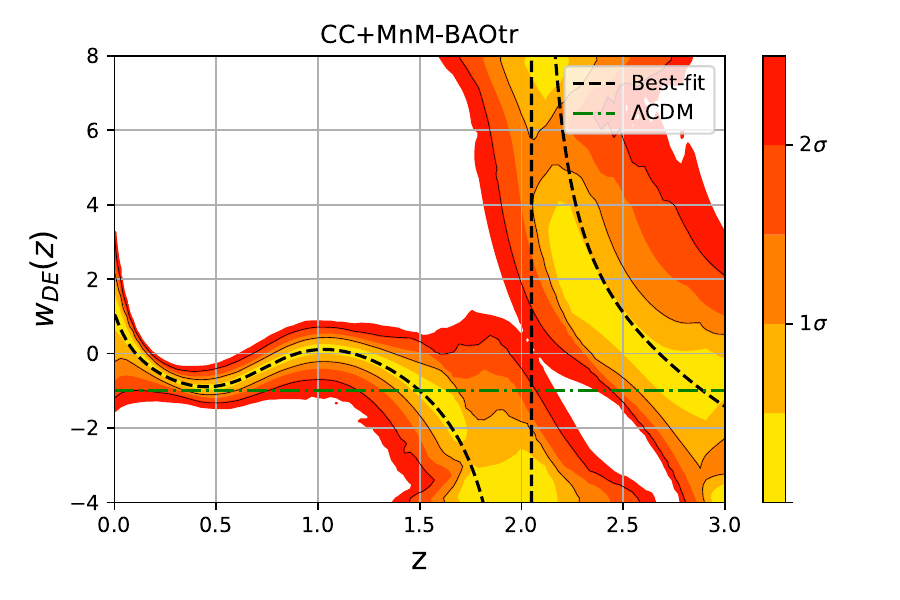}
      }
      \makebox[10cm][c]{
      \includegraphics[trim = 0mm  0mm 0mm 0mm, clip, width=8.9cm, height=5.3cm]{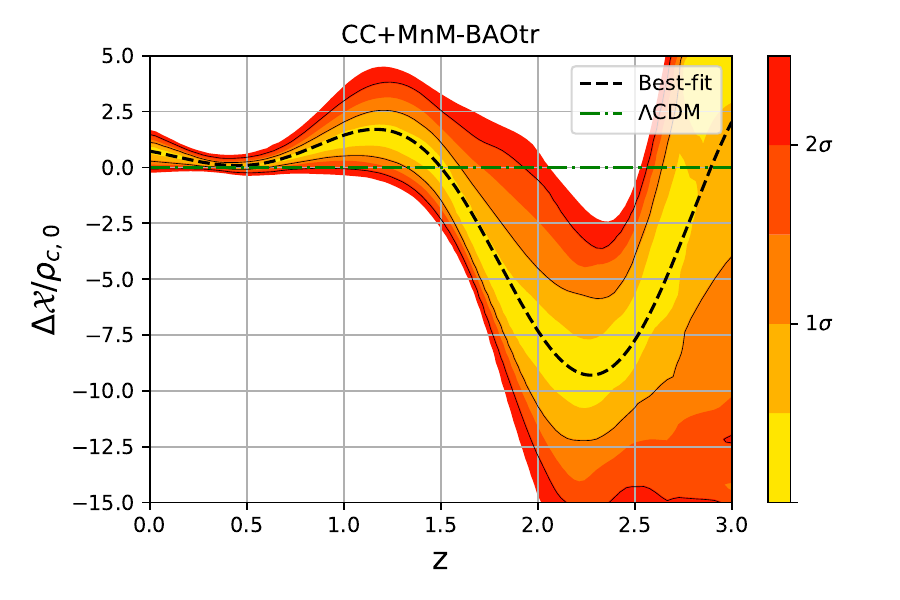}
      \includegraphics[trim = 0mm  0mm 0mm 0mm, clip, width=8.9cm, height=5.3cm]{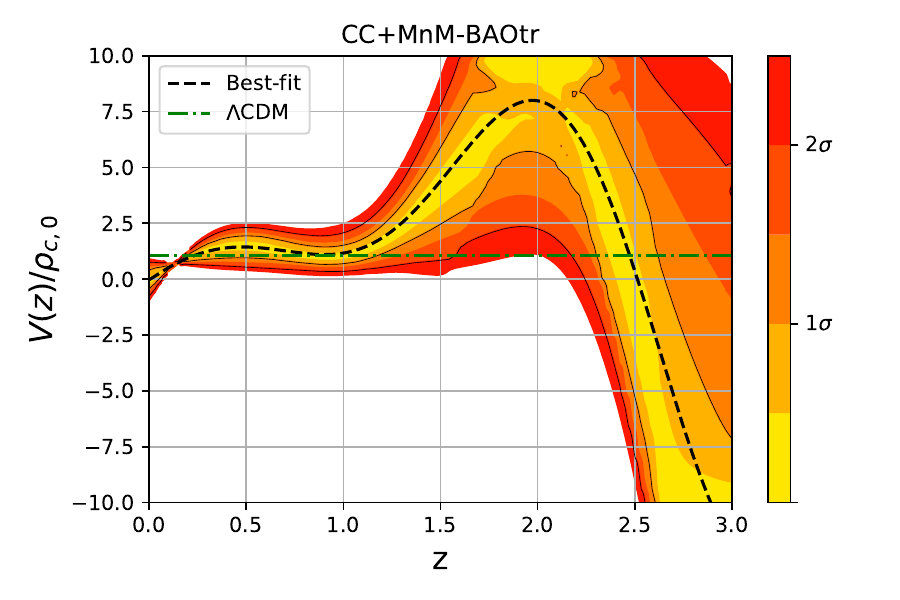}
      }
 \caption{Results of the reconstruction of $H(z)$ for the dataset combination of CC+MnM-BAOtr. From top-to-bottom and left-to-right we have: $H(z)$, $H(z)/(1+z)$, $q(z)$, $\Delta\mathcal{X} / \rho_{c,0}$, $\rho_{DE}/\rho_{c,0}$, $p_{DE}/\rho_{c,0}$, $w_{DE}$, and $V(z) / \rho_{c,0}$. An important thing to note and clarify is that the last node is located at $z=3.0$, which means that it cannot be constrained by data. As such, high-redshift results around this region should be taken as merely statistical noise. }\label{fig:GP_cc_mnm2bao}
 \end{figure*}

 \begin{figure*}[t!]
     \centering
       \makebox[10cm][c]{
      \includegraphics[trim = 0mm  0mm 0mm 0mm, clip, width=8.9cm, height=5.3cm]{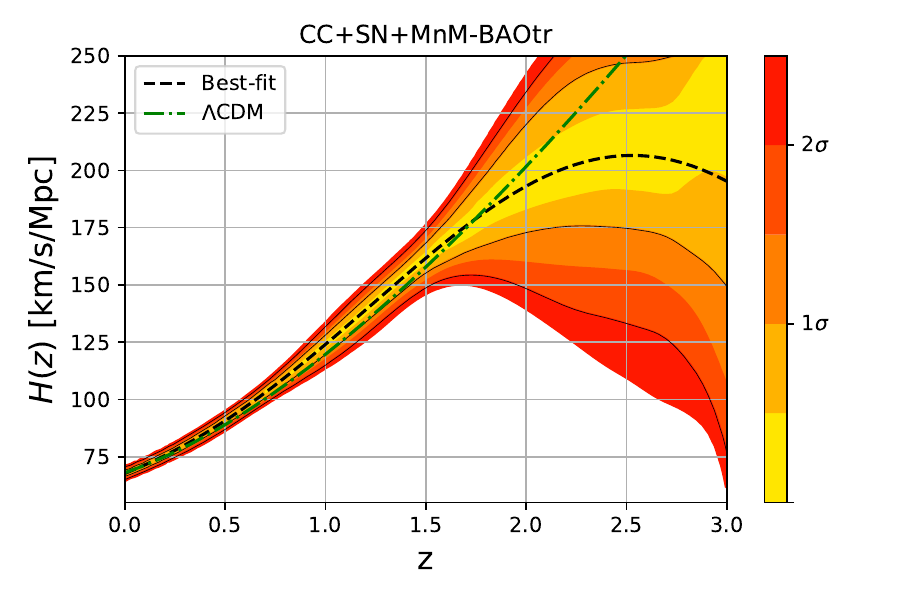}
      \includegraphics[trim = 0mm  0mm 0mm 0mm, clip, width=8.9cm, height=5.3cm]{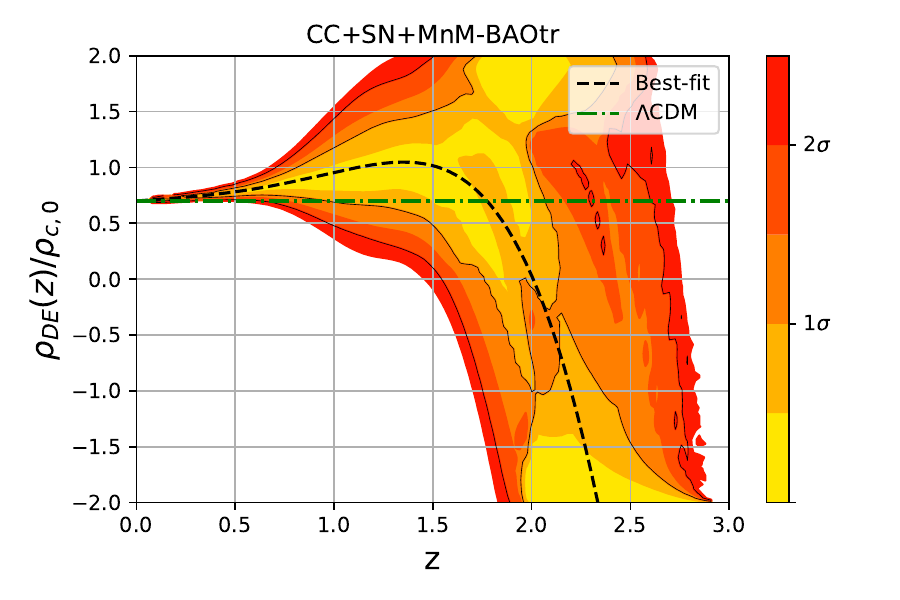}
      }
     \makebox[10cm][c]{
      \includegraphics[trim = 0mm  0mm 0mm 0mm, clip, width=8.9cm, height=5.3cm]{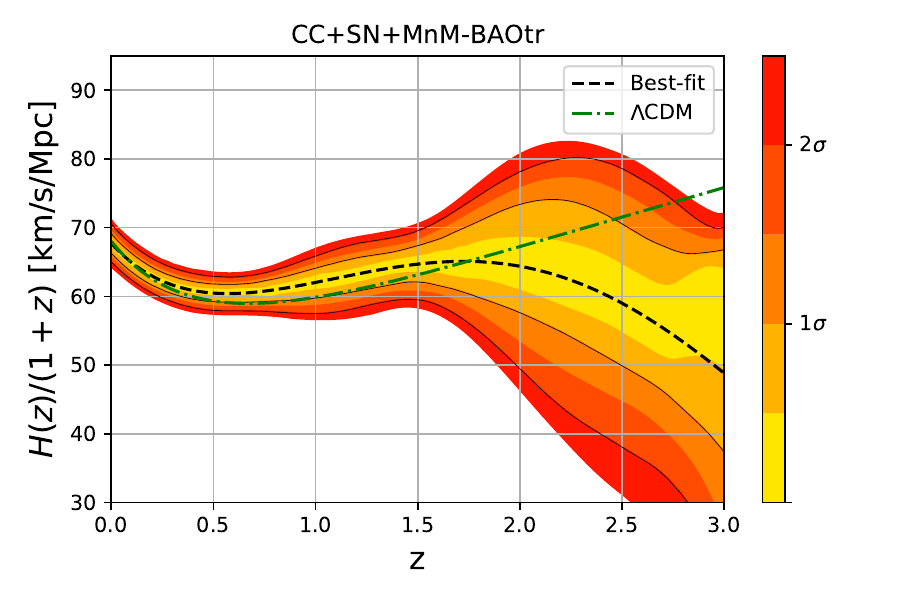}
      \includegraphics[trim = 0mm  0mm 0mm 0mm, clip, width=8.9cm, height=5.3cm]{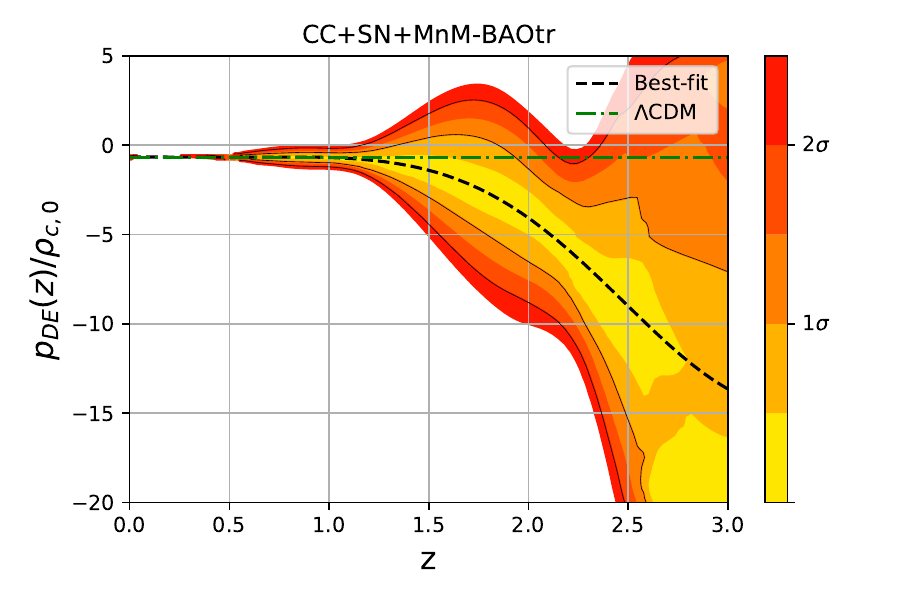}
      }
     \makebox[10cm][c]{
      \includegraphics[trim = 0mm  0mm 0mm 0mm, clip, width=8.9cm, height=5.3cm]{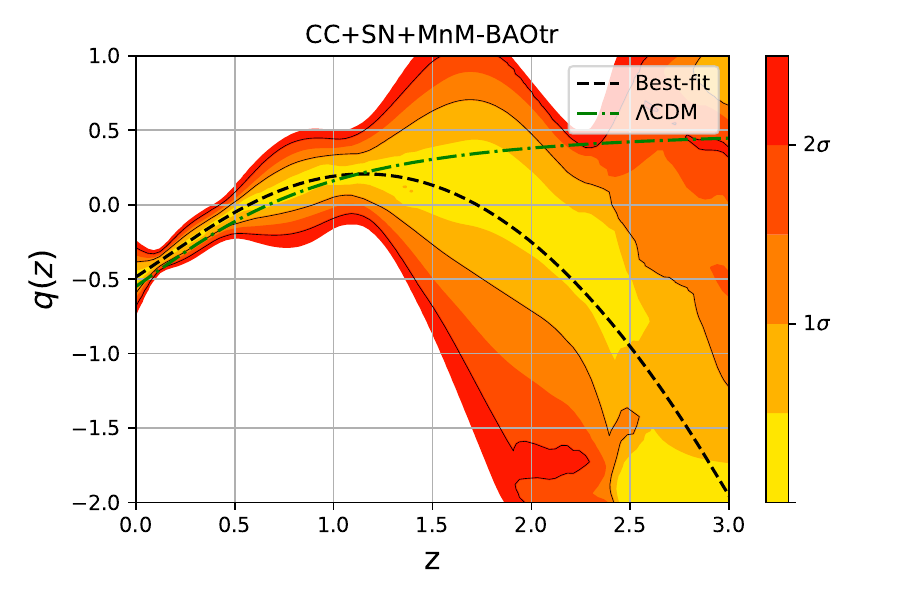}
      \includegraphics[trim = 0mm  0mm 0mm 0mm, clip, width=8.9cm, height=5.3cm]{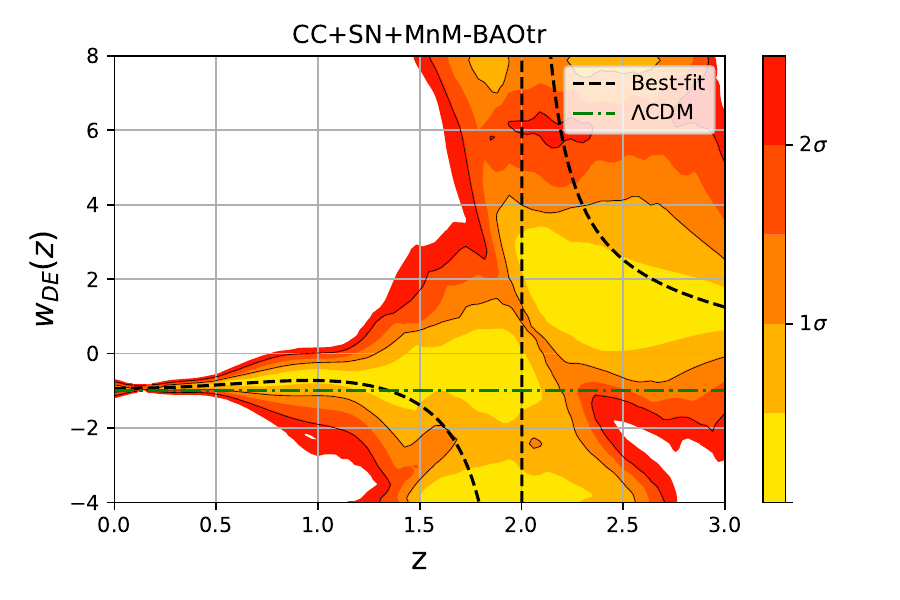}
      }
      \makebox[10cm][c]{
      \includegraphics[trim = 0mm  0mm 0mm 0mm, clip, width=8.9cm, height=5.3cm]{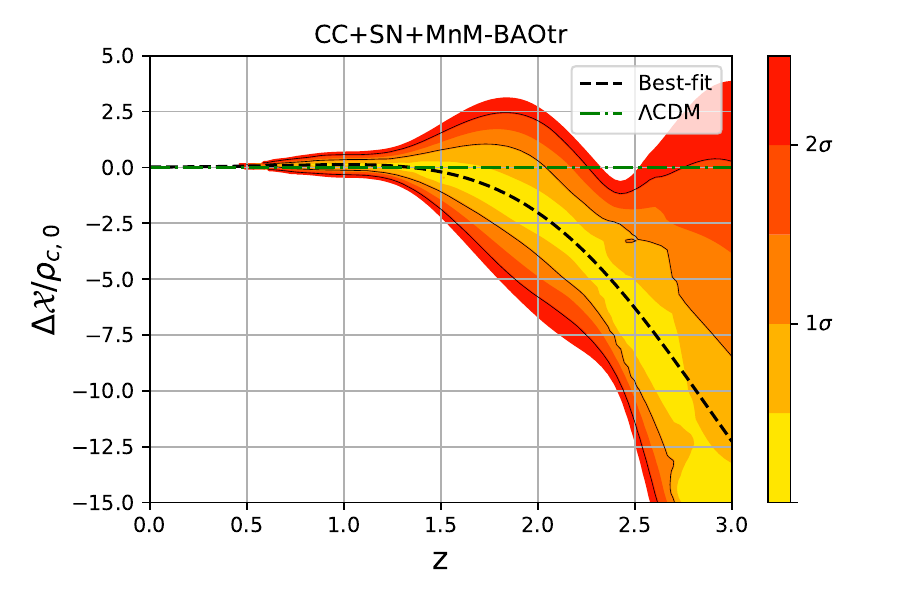}
      \includegraphics[trim = 0mm  0mm 0mm 0mm, clip, width=8.9cm, height=5.3cm]{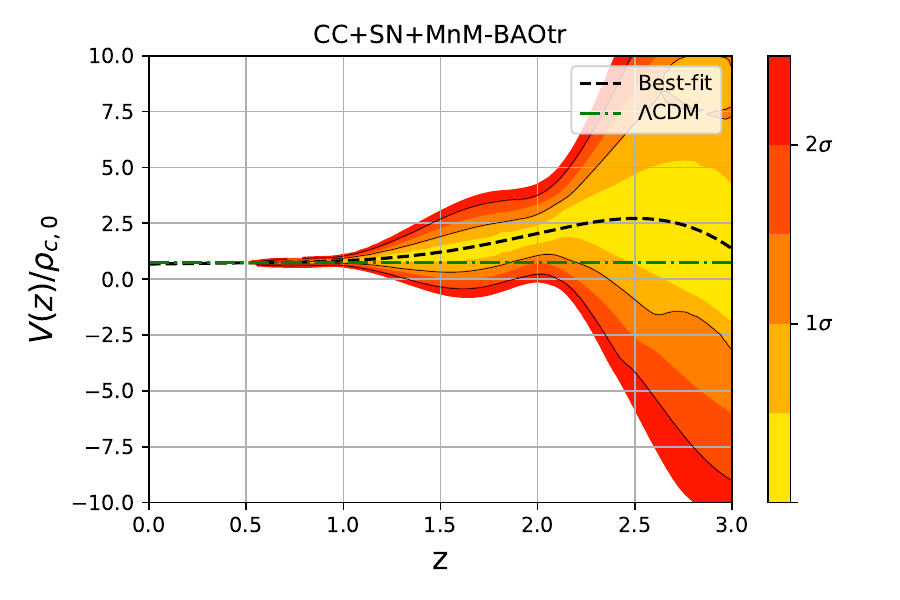}
      }
 \caption{Results of the reconstruction of $H(z)$ for the dataset combination of CC+SN+MnM-BAOtr. From top-to-bottom and left-to-right we have: $H(z)$, $H(z)/(1+z)$, $q(z)$, $\Delta\mathcal{X} / \rho_{c,0}$, $\rho_{DE}/\rho_{c,0}$, $p_{DE}/\rho_{c,0}$, $w_{DE}$, and $V(z) / \rho_{c,0}$. An important thing to note and clarify is that the last node is located at $z=3.0$, which means that it cannot be constrained by data. As such, high-redshift results around this region should be taken as merely statistical noise. }\label{fig:GP_cc_sn_mnm2dbao}
 \end{figure*}

 \begin{figure*}[t!]
     \centering
       \makebox[10cm][c]{
      \includegraphics[trim = 0mm  0mm 0mm 0mm, clip, width=8.9cm, height=5.3cm]{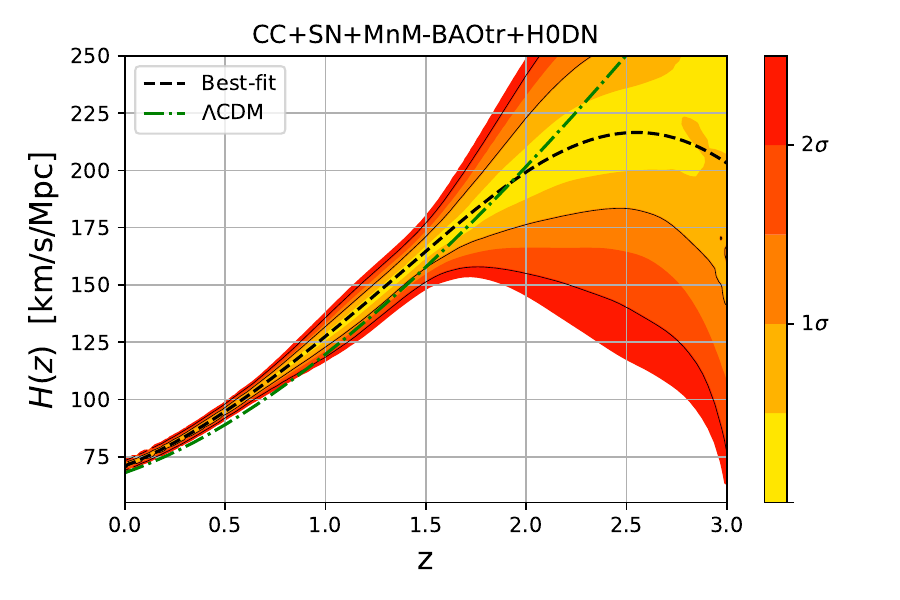}
      \includegraphics[trim = 0mm  0mm 0mm 0mm, clip, width=8.9cm, height=5.3cm]{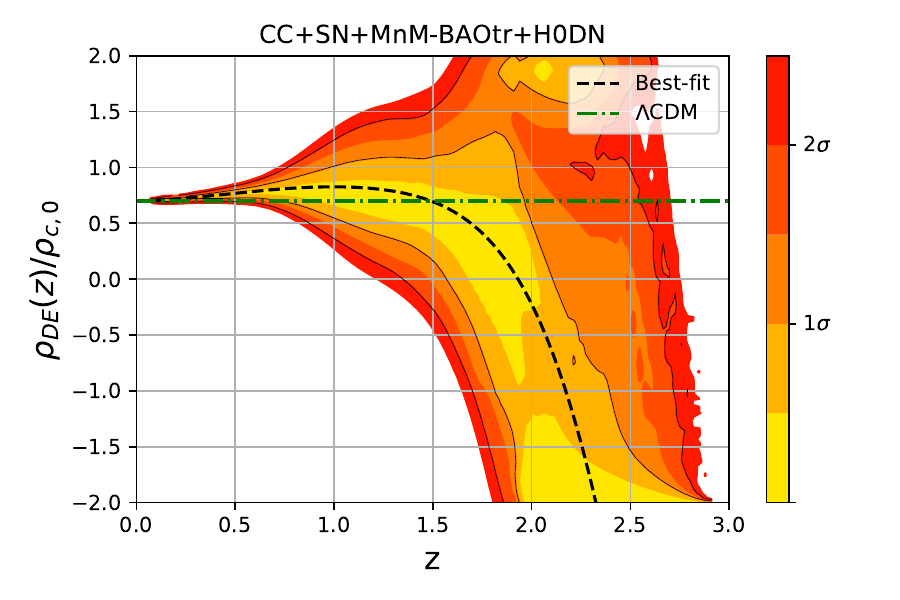}
      }
     \makebox[10cm][c]{
      \includegraphics[trim = 0mm  0mm 0mm 0mm, clip, width=8.9cm, height=5.3cm]{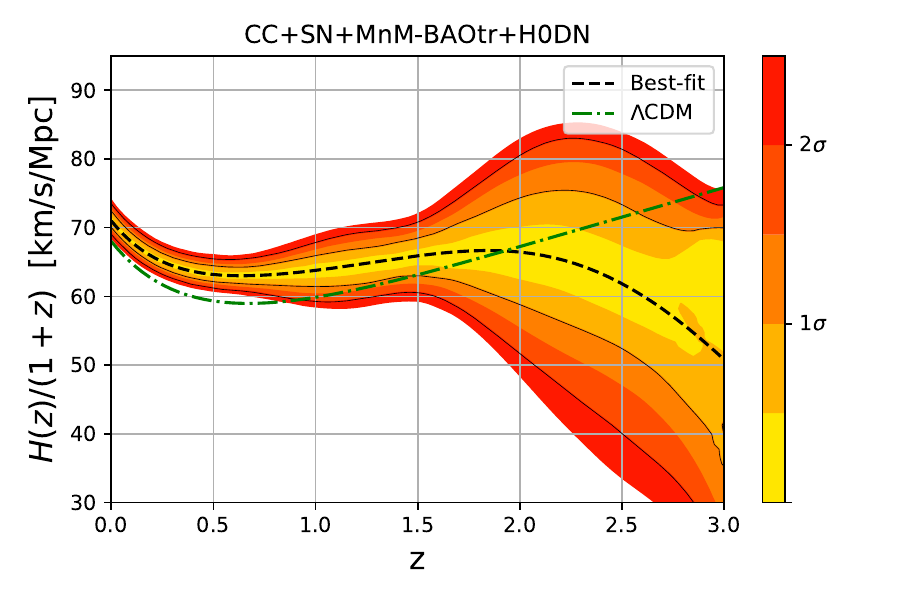}
      \includegraphics[trim = 0mm  0mm 0mm 0mm, clip, width=8.9cm, height=5.3cm]{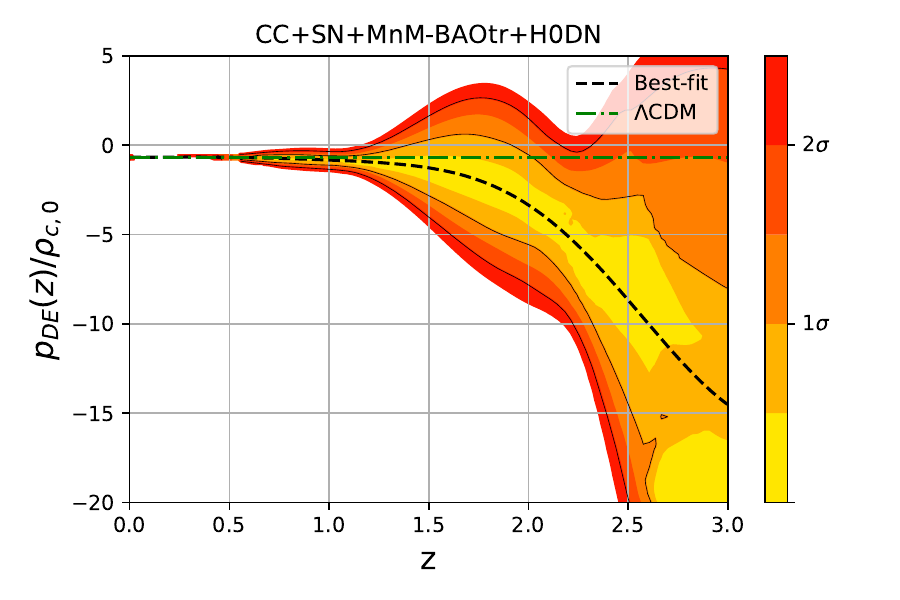}
      }
     \makebox[10cm][c]{
      \includegraphics[trim = 0mm  0mm 0mm 0mm, clip, width=8.9cm, height=5.3cm]{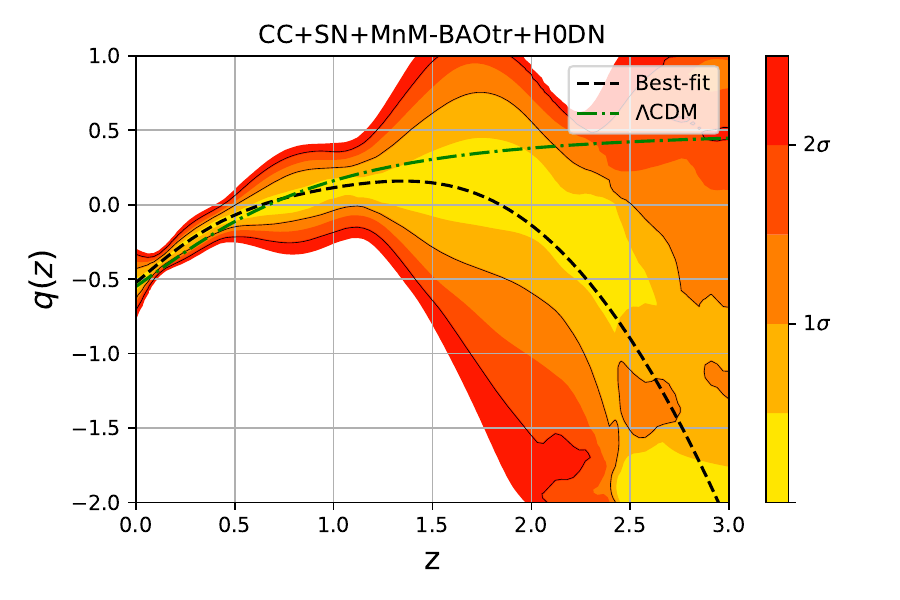}
      \includegraphics[trim = 0mm  0mm 0mm 0mm, clip, width=8.9cm, height=5.3cm]{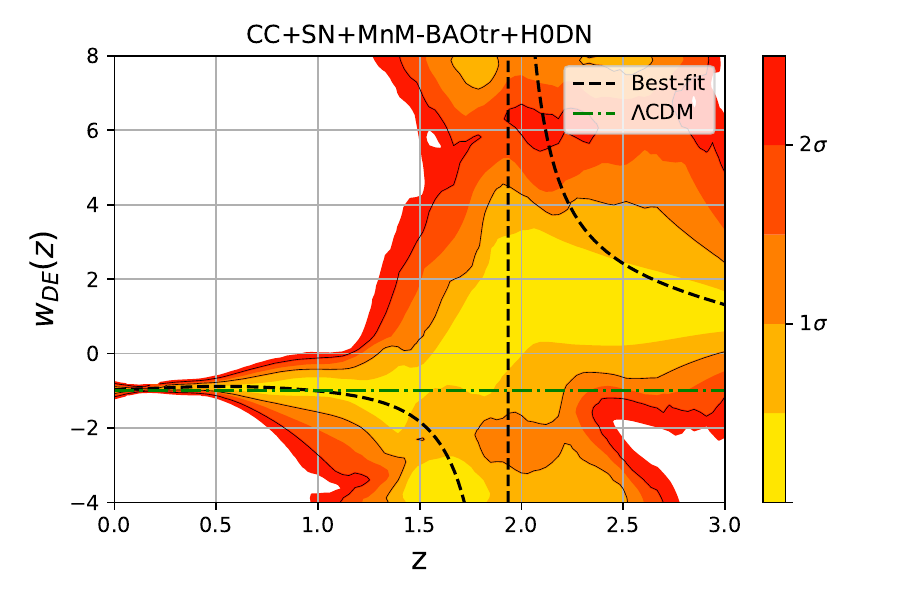}
      }
      \makebox[10cm][c]{
      \includegraphics[trim = 0mm  0mm 0mm 0mm, clip, width=8.9cm, height=5.3cm]{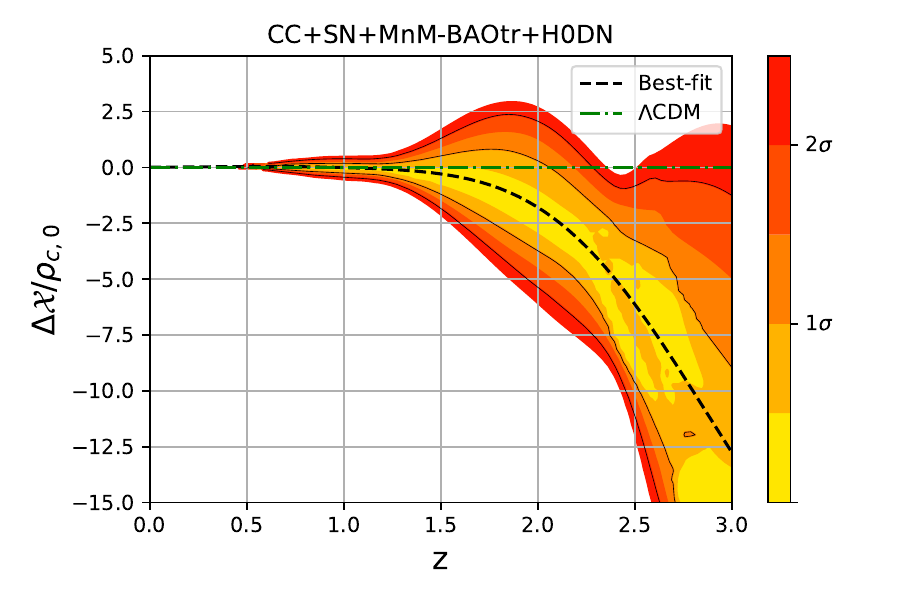}
      \includegraphics[trim = 0mm  0mm 0mm 0mm, clip, width=8.9cm, height=5.3cm]{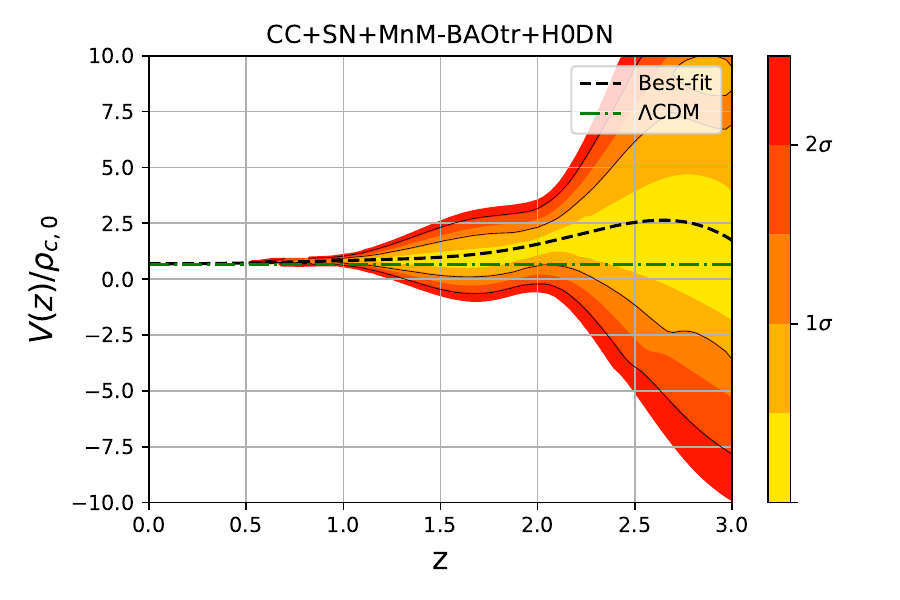}
      }
 \caption{Results of the reconstruction for the dataset combination of CC+SN+MnM-BAOtr+H0DN. From top-to-bottom and left-to-right we have: $H(z)$, $H(z)/(1+z)$, $q(z)$, $\Delta\mathcal{X} / \rho_{c,0}$, $\rho_{DE}/\rho_{c,0}$, $p_{DE}/\rho_{c,0}$, $w_{DE}$, and $V(z) / \rho_{c,0}$. An important thing to note is that the last node located at $z=3.0$ cannot be constrained by data. As such, results around this region should be taken as merely statistical noise. }\label{fig:GP_cc_sn_mnm2dbao_h0dn}
 \end{figure*}

\end{document}